# Design Principles for Fluid Molecular Ferroelectrics

Calum J. Gibb [1], Jordan Hobbs [2], William C. Ogle [1,2,3], Richard. J. Mandle *[1,2]

[1] School of Chemistry, University of Leeds, Leeds, UK, LS2 9JT.
[2] School of Physics and Astronomy, University of Leeds, Leeds, UK, LS2 9JT.
[3] School of Applied Mathematics, University of Leeds, Leeds, UK, LS29JT.

*Author for correspondence e-mail: r.mandle@leeds.ac.uk

**Abstract**

Fluid molecular ferroelectrics are a new class of organic materials where ferroelectricity is found in conjunction with 3D fluidity whilst still retaining spontaneous polarization values comparable to their traditional solid-state counterparts. One of the major challenges for soft condensed matter physics is predicting whether a fluid molecular material will form ferroelectric phase with nematic or smectic order. Through the synthesis of 45 systematically varied molecules, and by analogy to solid molecular ferroelectrics, is it shown that subtle hydrogen–fluorine (H/F) substitution(s) allows for tuneable *syn*-parallel pairing motifs resulting in either specific pairings leading too geometrically constrained lamellar order or diversified pairings stabilising nematic ordering. Large-scale, fully atomistic molecular dynamics simulations reveal that smectic ferroelectricity emerges from discrete lateral pairing modes, whereas nematic phases arise from a multiplicity of equivalent polar configurations. Together, these findings establish experimentally validated design principles for fluid molecular ferroelectrics and provide a predictive framework for engineering functional polar fluids.

**Introduction**

Organic molecular ferroelectric materials find use in emerging technologies such as energy harvesting [1], ultrasonic transducers for medical imaging [2], and actuators in robotics [3]. These solid-state materials have many advantages over their traditional inorganic counterparts [4,5] as they can be more lightweight, mechanically flexible and are synthetically more trivial to produce allowing for greater structural diversity [6]. Design driven synthesis of molecular ferroelectrics has advanced significantly and developed several key design mechanisms such as reduced molecular symmetry [7], introduction of homochirality [8], and hydrogen/fluorine (H/F) substitution [9] to tune the desired properties.

Discovered in 2017 [10,11], fluid molecular ferroelectrics are a new class of organic materials where ferroelectricity is found in conjunction with true 3D fluidity whilst still retaining spontaneous polarization values comparable to their traditional solid-state counterparts [12-18]. The addition of fluidity and lack of true crystal lattice gives fluid ferroelectrics distinct advantages over their solid-state counterparts (i.e. ease of monodomain fabrication, mechanically stable, structurally diverse), allowing for exploration of new physical phenomena and development of novel applications. As such, a wide variety of potential use cases for fluid ferroelectrics have already been suggested in in areas such as quantum optics [19,20], nanoscale non-linear electrooptical devices [21-24] and, fluidic capacitors [25].

Fluid molecular ferroelectrics are liquid crystalline, with the majority of materials possessing nematic ordering of their constituents; thus, this phase is termed the ferroelectric nematic, or $N_F$, phase (**Fig. 1a**). Given the low cost of elastic deformations in liquid crystals [26], further fluid ferroelectric sub-phases have also been discovered such as the ferroelectric smectic A (SmA$_F$) [26,27] (**Fig. 1b**), where the molecules form a lamellar structure with the molecules orthogonal to the layer normal, and the polar smectic C (SmC$_P$) [28-30] (**Fig. 1c**) where the molecules tilt with respect to layer normal.

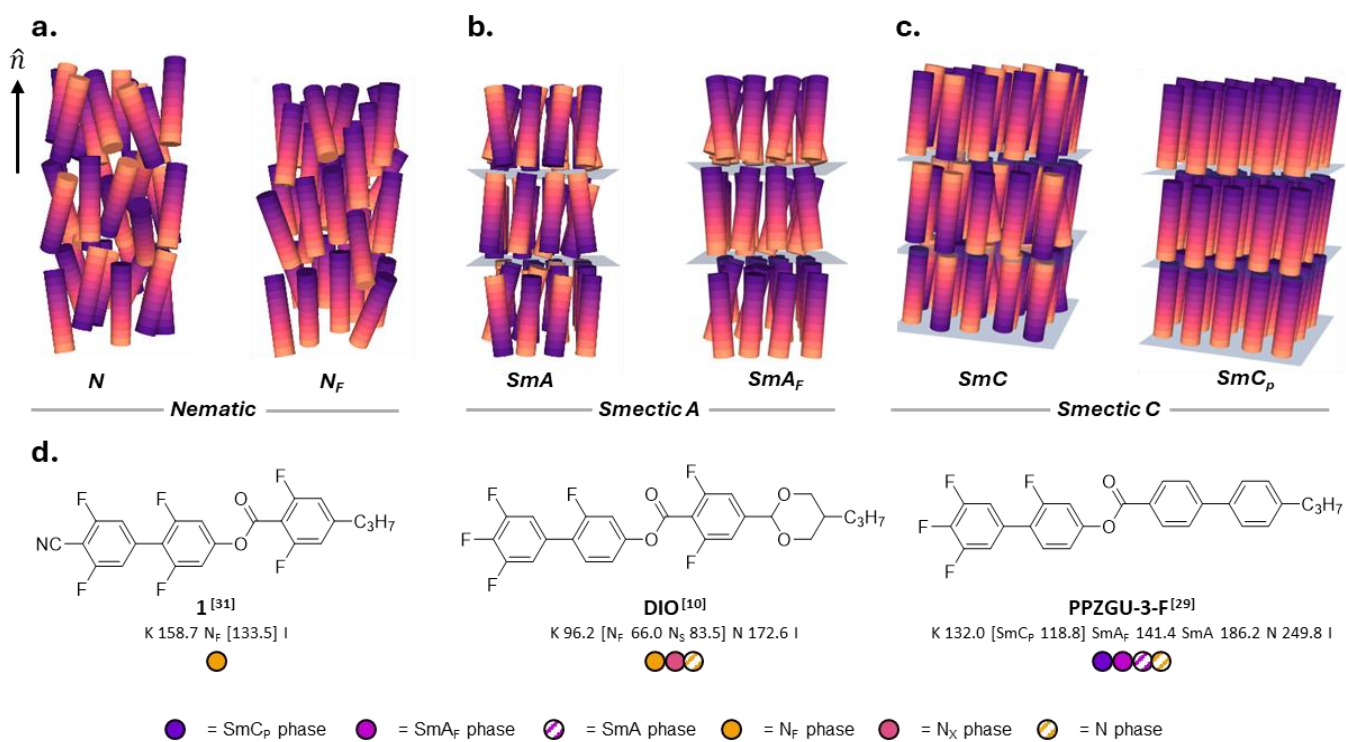

**Fig. 1**: Schematic representations of the para- and ferroelectric **(a)** nematic (N); **(b)** smectic C (SmC); and **(C)** smectic A (SmA) phases. Each rod is representative of a single molecule with the colour indicating the molecular direction. In fluid molecular ferroelectric phases, the polarisation vector is parallel to the liquid crystalline director ($\hat{n}$). Unlike in their conventional paraelectric analogues, in which the director is symmetry invariant ($\hat{n} = -\hat{n}$), the director in fluid molecular ferroelectric exhibits broken inversion symmetry due to the polarisation ($\hat{n} \neq -\hat{n}$) giving rise to bulk polar order. **(d)** examples of a typical small molecule fluid ferroelectric materials (i) **1** [31], (ii) **DIO** [10] and **PPZGU-3-F** [29] with their associated phase profiles and transition temperatures (°C) listed below. Stripped colours indicate paraelectric phases and solid colours indicate ferroelectric phases. All three molecules possess a rigid, fluorinated aromatic core appended by an EWG (either F or CN) and a short propyl aliphatic chain. Notably **F2100,** which exhibits ferroelectric smectic phases, has a lower degree of H/F substitution compared to **1** and **DIO**.

Fluid molecular ferroelectric materials are comprised of small organic molecules which currently exist in a constrained area of chemical space. Nominally this consists of a relatively rigid core unit made up of (fluorinated) aromatic rings flanked by a short aliphatic chain at one end and a polar electron withdrawing functionality (EWG) at the other (**Fig. 1d**). Generally, materials exhibiting such phases have large longitudinal molecular dipole moments (often achieved through H/F substitution) [26,31-33], although the relevance of the dipolar magnitude has been debated [34-37]. Madhusudana [38] suggested that favourable lateral electrostatic interactions along the length of the molecules can promote ferroelectricity with Gibb *et al.* expanding on this suggesting that uniformly distributed partial dipoles is a further requirement for phase formation [31,39-41]. However, moving beyond these postulates to predicting the onset of ferroelectric or indeed the type of phase itself from specific chemical modifications remains an unsolved problem. This makes assigning accurate design principles for the synthesis of fluid ferroelectrics difficult.

At present there are 438 reported fluid molecular ferroelectrics, 316 exhibit $N_F$ order with only around 74 displaying ferroelectricity with lamellar structures. Far less is currently known about the origins of smectic ferroelectricity, although as a general observation, ferroelectricity in these materials is normally achieved with fewer lateral fluorine atoms along the aromatic cores. A singular H/F substitution can be the tipping point between polar nematic and smectic ordering [10,26,41,42] and so the story is reminiscent of solid-state organic molecular ferroelectrics where such substitution can have a meaningful impact on the thermal stability of ferroelectricity [7]. Despite rapid materials discovery, a predictive framework for ferroelectric phase selection remains elusive. We envisaged that a selective study whereby the number and position of fluorine atoms across the length of a simple molecule known to exhibit polar smectic order varies could allow us to tune the delicate balance between nematic and lamellar ferroelectricity, gaining insight into the molecular design of fluid molecular ferroelectrics.

**Results and Discussion**

The initial investigation began with the synthesis of 15 analogues of a highly fluorinated fluid molecular ferroelectric whereby selectively varied the degree H/F substitution (**Fig. 2**) were performed. These materials are named with the acronym **FWXYZ** whereby F indicates the EWG of the molecule, and W, X, Y, and Z indicate the number of fluorine atoms *ortho* to the EWG or biphenyl linkages and are varied between 0 and 2. The chemical synthesis and structural characterisation of these materials is found in in the ESI to this article.

Of the 15 **FWXYZ** analogues, 9 materials exhibit ferroelectricity with the phase-type (i.e. nematic or smectic) varying with the degree of H/F substitution on the tail-group of the molecule (**Fig. 2**). Phase sequences of each analogue were determined by polarized optical microscopy (POM; **Fig. 3a**) and differential scanning calorimetry (DSC; **Fig. 3b**). X-ray scattering was used to confirm smectic order where implied by POM textures (**Fig. 3c** and **3d**) while the existence of ferroelectricity was confirmed by current reversal techniques (**Fig. 3e,f**), with a single peak being indicative of a ferroelectric phase. For the case of the SmC$_P$ phase, an additional small peak pre-voltage polarity reversal is observed which is indicative of the removal of molecular tilt [29,30]. Several materials did not exhibit polar phases in neat form due to their extreme proclivity for crystallisation. Therefore, extrapolated values for their transition temperatures were found by preparation of binary mixtures with **F2100**, allowing for determination of their phase sequences and associated transitions temperatures. (**Fig. S1-3**). Further data and details to all characterisation techniques are given in the ESI to this article.

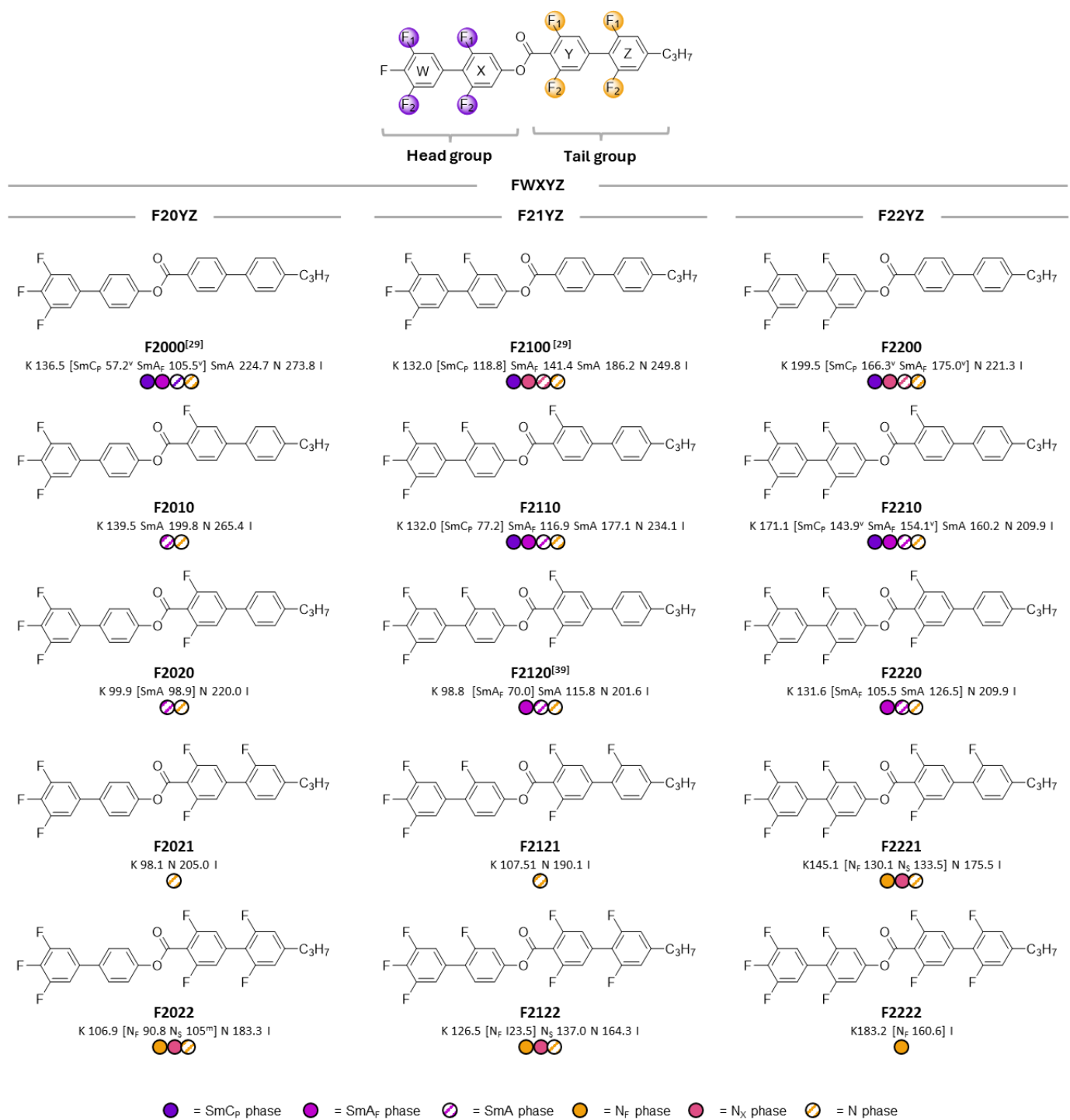

**Fig. 2:** The molecular structures, phase sequences and associated transition temperatures (°C) for the highly fluorinated **FWXYZ** materials. Beginning with the first ring on the head of the molecule, W, X, Y, and Z refer to the number of appended fluorine atoms on each ring and vary between 0 and 2. [ ] indicate a monotropic phase transition, $^m$ indicates a transition temperature determined by POM, and $^v$ is indicative of an extrapolated transition, determined by binary mixture studies between the material and **F2100** Stripped colours indicate apolar phases and solid colours indicate ferroelectric phases.

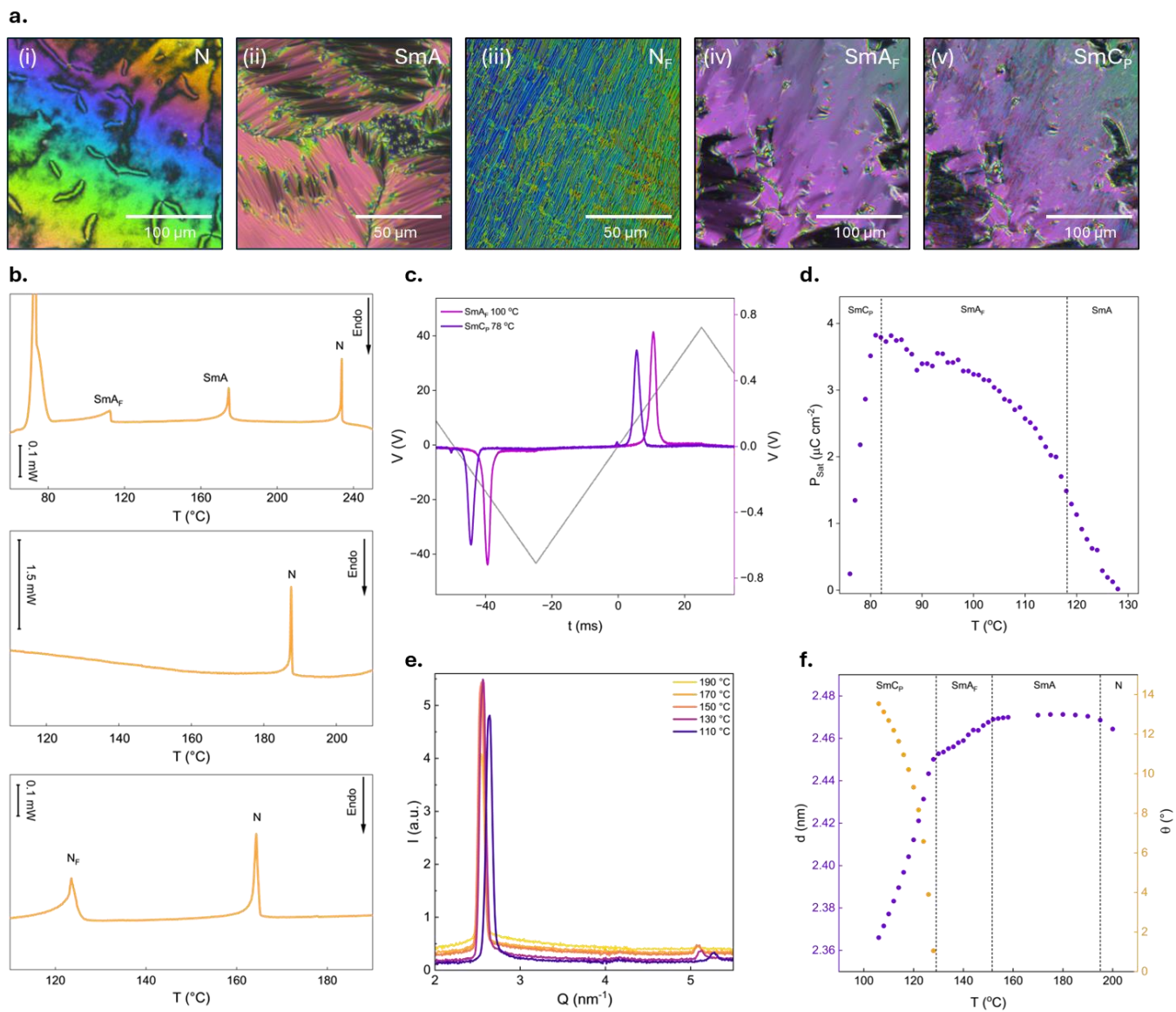

**Fig. 3:** **(a)** POM micrographs depicting characteristic textures of the **(i)** N, **(ii)** SmA, **(iii)** $N_F$, **(iv)** $SmA_F$, and **(v)** $SmC_P$ phases. All images are taken of thin samples sandwiched between untreated glass coverslips; **(b)** representative DSC thermogram depicting the phase sequence of (top) **F2110**, (middle) **F2121**, (bottom) **F2122**; **(c)** current response measurement of **F2110** at in the $SmA_F$ and $SmC_P$ phases measured at 10 Hz; **(d)** temperature dependence of the saturated polarization measured for F2110; **(e)** 1D X-ray scattering pattern measured for **F2100** across its 3 smectic phases. Moving of the peak to larger Q values indicates layer contraction due to molecular tilt; and **(f)** the temperature dependence of the layer spacing and molecular tilt obtained by SAXS measured for **F2100** in the SmA, $SmA_F$ and $SmC_P$ phases.

By increasing H/F substitution in front of the ester linkage (i.e. the head-group) results in an increase in the apolar-polar transition temperature with a corresponding increase of around 35 °C per additional F atom, irrespective of mesophase type observed (i.e. nematic or smectic). This reflects the resultant stronger lateral core-core interactions, afforded by the additional H/F substitution, between molecules and a more spatially uniform partial dipole organisation along the long molecular axis. This allows for a better packing of chessboard like electrostatic +ve and -ve charge interactions known to be favourable for the formation of fluid ferroelectric phases [31,43].

Whilst the degree of H/F substitution of the head-group seems to be vital to the observation of fluid ferroelectricity, fluorination behind the ester (i.e. the tail-group, **Fig. 2**) has quite a different pattern of effects. When the tail-group of the molecule has less than 2 appended F atoms (F < 2), smectic phases are observed, and the thermal stability of these phases decreases with increasing H/F substitution. When the tail-group possesses more than 2 appended F atoms (F > 2), $N_F$ phases are observed with the stability of the phases increasing with increasing H/F substitution. In the intermediate region (F = 2), only a very low temperature transition to a $SmA_F$ phase is observed in **F2220** and **F2120,** with no such phase exhibited by **F2020**. Effectively, this divides the **FWXYZ** analogues into two distinct groups providing us with two general observations. Firstly, H/F substitution of the head-group has a strong effect on the onset temperature for ferroelectric order while having little to no effect on the phase-type exhibited by a homologue. Secondly, while H/F substitution of the tail-group has only a modest effect on the onset temperature of ferroelectricity, it strongly influences molecular packing and dictates the type of phase observed (i.e. nematic or smectic).

In the case of paraelectric liquid crystal phases, the evolution from smectic to nematic phases with increasing lateral bulk is well understood: increasing lateral fluorination effectively forces molecules apart, decreasing π-π interactions between the aromatic units and hence destabilising the lamellar structure [44]. In the case of the **FWXYZ** analogues presented here, X-ray scattering measurements confirm increased lateral spacing upon increasing H/F substitution of the tail-group (**Fig. S5**), although they also demonstrate that materials that exhibit ferroelectric smectic phases general show stronger π-π interactions evidenced by a significant shoulder to the wide-angle scattering peak (**Fig. S5**). It has been suggested that the formation of ferroelectric smectic phases may be in part due to the microphase separation of fluorinated and un-fluorinated aromatic regions of the molecules (i.e. microphase separation of the front and back of molecules. Here that would be the WX and YZ regions.). Whilst microphase separation is known to occur when mixing fluorinated and un-fluorinated aliphatic moieties, the opposite is true for the mixing of their aromatic counterparts [45]. The most obvious example is the favourable interactions which occur between benzene and hexafluorobenzene, where binary mixtures of the two components can be isolated as crystalline solids near to room temperature when the pure material are liquids [46]. We therefore consider the possibility that such microphase separation drives the formation of ferroelectric smectic phases here to be improbable and therefore the molecular origins of lamellar ferroelectricity must be a result of different interactions between the molecules.

In-order to probe the interactions between molecules, we elected to perform large scale, long duration fully atomistic MD simulations (1000 molecules, > 1 µs) on select analogues (**F21YZ**, (**Fig. 2** [central column])) with full details of the setup configuration given in the ESI. For the 5 **F21YZ** analogues, each simulation reproduces the same phase type as observed experimentally where we see the evolution from ferroelectric smectic to ferroelectric nematic configurations via an intermediate region containing their apolar analogues (**Fig. 4 (i-iii)**). **F2100** forms a SmC$_P$ phase while **F2110** forms a SmA$_F$ phase where the calculated layer spacings and tilt angle (SmC$_P$ phase only) are in reasonable agreement with experimental values (**Table S4**). **F2120** forms an apolar SmA, **F2121** forms an apolar nematic and **F2122** forms a N$_F$ phase. In-order to understand interactions between the molecules in the different phase types, cylindrical distribution function (CDF) were computed for the **F21XY** analogues over the final 500 ns of the MD simulations.

Beginning with **F2100**, strong on-axis features at h=±24 are observed in the SmC$_P$ phase which correspond to head/tail arrangements of molecules in adjacent layers (**Fig. 4b**). The off-axis features at r ≈ 4.5 correspond to staggered pairs of molecules with face-face interactions between adjacent 3,4,5-trifluorobenzene units (features at h ≈ ±4, h ≈ ± 2). There is reasonable in-plane orientation in the simulated SmC$_P$ phase, with the "neighbours neighbour" giving corresponding signals, albeit more diffuse, at r ≈ 9, as well as diffuse areas for head-tail correlations at h=±20 and h=±28. Adding a single fluorine atom to afford **F2110** gives a comparable CDF, albeit more diffuse and with the features at h ≈ ±4, h ≈ ± 2 (staggered parallel pair) notably less prominent. The interaction between staggered 3,4,5-trifluorobenzene head-group units on adjacent molecules is significantly perturbed by the introduction of a distal fluorine atom, presumably via repulsion (*vide infra*). Experimentally, this change in structure is coincident with a 24.5 °C reduction in the onset temperature of ferroelectric smectic order and 9.1 °C reduction in apolar smectic order indicting the dual effect of reducing inhibiting layers both in the polar and apolar configurations.

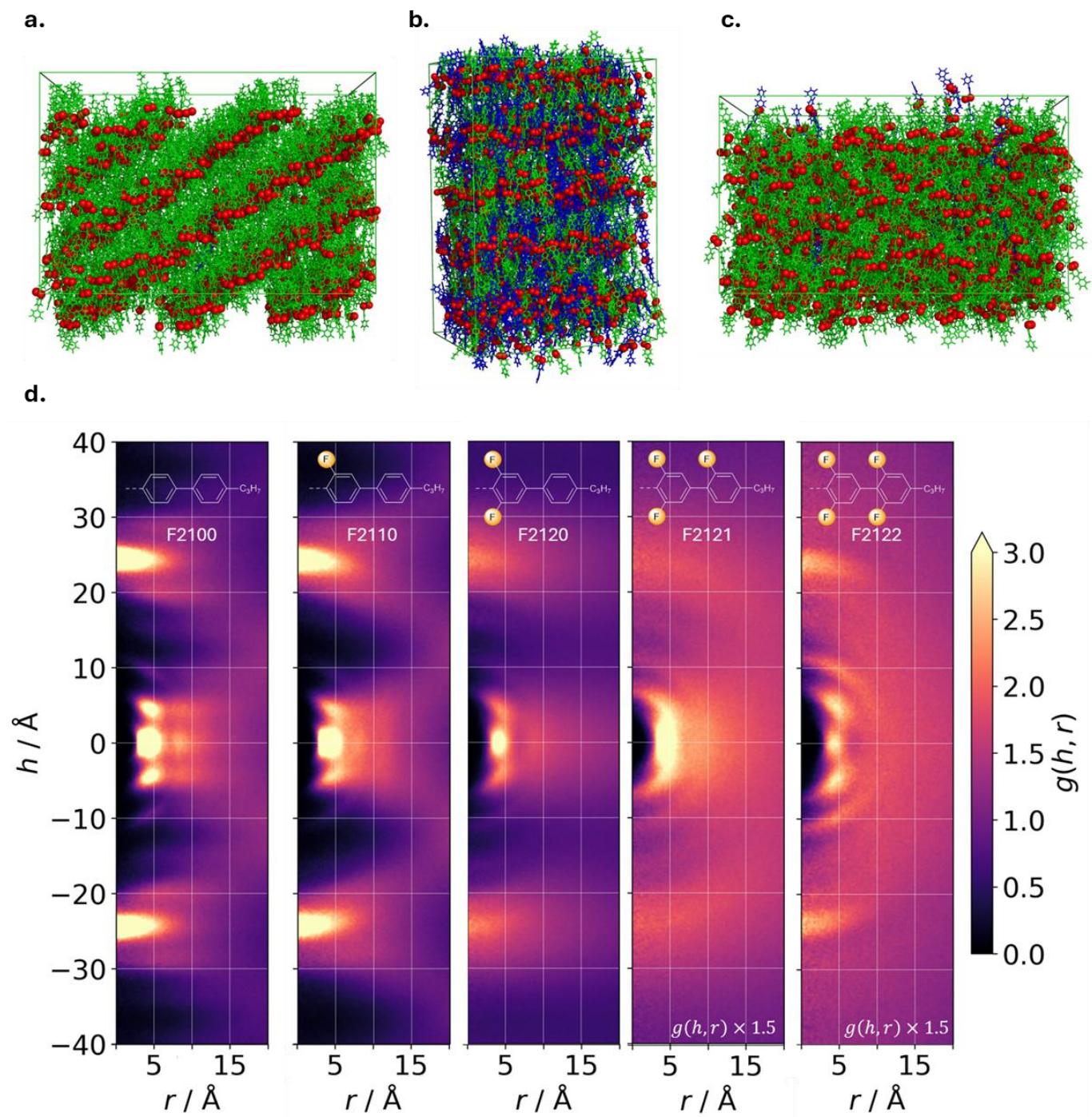

**Fig. 4:** **(a)** Instantaneous configurations of: **(i)** the SmC$_P$ phase of **F2100**; **(ii)** the apolar SmA phase of **F2120**; **(iii)** the polar N$_F$ phase of **F2122**. Molecules aligned parallel to the director are shown as green, molecules shown as antiparallel are shown as blue. The oxygen atoms are shown as spheres to aid visualisation of the layer structure in **F2100** and **F2120** (or its absence in **F2122**); and **(b)** Cylindrical distribution function (CDF) plots for compounds **F21XY** generated over the final 500 ns of each simulation trajectory.

For **F2120,** the simulation forms a SmA phase where the features around the molecular length (h ≈ 23), which result from the regular displacement due to the layer structure, are notably weaker and more diffuse than for the SmA$_F$ phase of **F2110** as they encode both head/tail, head/head and tail/tail interactions. For the N phase of **F2121,** the diffuse feature around the molecular length (h ≈ 23) reduces in prominence with the strongest feature being a broad lateral separation of molecules (r = 5, h -4 à 4). For both **F2120** and **F2121** the sharp features at h ≈ ±4, h ≈ ± 2 (staggered parallel pair) are reduced in prominence. Finally, for **F2122**, which forms an $N_F$ phase, shows a qualitatively different CDF dominated by arc-like intensity features off axis that open out in (h, r) away from the origin. These represent staggered side-by-side interactions between molecules, the diffusivity of which indicates a broad distribution of nearest-neighbour positions without fixed relative geometries. Such behaviour is consistent with other CDF analysis of MD trajectories of other materials exhibiting the $N_F$ phase [33,47] While the $N_F$ phase comprises rather unspecific pairing between molecules (and thus arc-like g(h,r), **Fig. 4b**), the ferroelectric smectic phases arise from specific geometric pairing motifs (thus spot like features in g(h,r), **Fig. 4b**). Examination of the Cambridge Crystallographic Database (CCDC) reveals that the 3,4,5-trifluorobenzene motif forms slipped parallel configurations [48]; consistent with the observed g(h,r) for the ferroelectric smectic phases observed for **F2100** to **F2120**.

In other words, the formation of ferroelectric smectic phases is driven, here at least, by specific preferred lateral pairing motifs. As the degree of fluorination begins to increase, these pairings are disrupted by electrostatic repulsion. As the degree of fluorination further increases beyond a certain number (in this case F > 2) the chemical similarity along the molecular length allows for a wider variety of polar pairing modes thus forming polar nematic phases. A logical extension of this is that the high dipoles of the molecules is not necessarily the key driver of ferroelectric smectic phase formation, rather, these could be generated in materials of even modest polarity by careful consideration and engineering of intermolecular interactions.

The balance between the formation of either ferroelectric nematic, ferroelectric smectic or their apolar analogues may then be explained by specific intermolecular pairing modes. Both ferroelectric nematic and smectic phases require a preference for syn-parallel pairing modes rather than anti-parallel. If the syn-parallel pairing has many possible pairing modes, then the phase formed is nematic-like which occurs for maximally fluorinated tail-groups (i.e. **FWX22**) as similar F-π interactions can form along the full length of the molecule. As the number of F atoms decreases and becomes locally segregated along the structure, the number of pairing modes reduces, increasing the tendency towards smectic order due to the reduction in competition. In the intermediate regime (i.e. F = 2) there is neither enough parallel pairing modes along the length to form an $N_F$ phase nor sufficient segregation to generate a dominant ferroelectric smectic mode and so apolar behaviour dominates. Experimental evidence strongly suggests competing modes occur in other fluid ferroelectric systems where a transition from an apolar SmA to the $N_F$ phase has been observed previously with a sharp change in packing [49]. This observation appears to be general to other fluid ferroelectric systems where H/F substitution is used to drive polar phase behaviour [26,31-33,41,42].

To further assess the generality of these observations, another set of analogous materials were synthesised; this time replacing the terminal F atom at the head of the material with a nitrile (CN) moiety. These 30 structures are denoted by **NCWXYZ.** Firstly, let us consider the analogues which possess a minimum of 3 F atoms on the head of the molecule of which 12 of the materials exhibit fluidic ferroelectricity (**Fig. 5**). In contrast to the **FWXYZ** materials discussed previously, all these **NCWXYZ** homologues display solely nematic ordering with no evolution in phase-type on increasing H/F substitution on the tail of the molecules.

Similarly to the **FWXYZ** materials, increasing the number of F atoms on the head of the molecule generally sees an increase the onset of polar order with there being only a small difference in the polar transition temperatures between the structural isomers (for example **NC2110** and **NC1210**). The onset of polar order displays a comparable trend to that seen for the FWXYZ materials, whereby increasing from 0 to 2 F atoms on the tail-group sees the gradual decrease in the polar transition before it begins to increases again for the **NCWX21** and **NCWX22** analogues. Atomistic MD simulations performed on the **NC22YZ** subset are conducted in the $N_F$ phase for these materials (**Fig. S6**) and the CDF computed (**Fig. S7**). The CDFs show arc-like intensity features off axis that open out in (h, r) away from the origin representing staggered side-by-side interactions between molecules, the diffusivity of which are consistent with broad distributions of neighbour-neighbour pairings with minimal fixed geometries (i.e nematic ordering). These features are consistent for the other **NC22YZ** analogues with increasing H/F substitution having minimal effect of the preferred pairing modes, thus, there is no tendency towards smectic phase formation. These results imply that the terminal nitrile is detrimental to the formation of lamellar fluid ferroelectrics likely due to its much larger electrostatic presence, dominating the molecular interactions, . Examination of the literature indicates that this may be a general observation with only 3 of the 74 known ferroelectric smectogens contain a CN moiety [50,51].

The final 15 **NCWXYZ** materials, in which only a single H atom is substituted for a F atom in the head-group of the molecule and up to 4 are substituted on the tail-group (**Fig. S8**), were surprisingly found to exhibit only paraelectric N and SmA phases. Many of these materials are structural isomers of the materials which exhibit fluid ferroelectricity shown in **Figure 5** (for example **NC2200** vs **NC0022**), and highlighting the importance of specific positional H/F substitution in the design of fluid molecular ferroelectrics [31]. Fluorination of the head-group of the molecule has a greater effect on the ferroelectricity, presumably reflecting the enhanced polarity of the head-group promoting the formation of *syn*-parallel pairing modes compared to *anti*-parallel. Further to this, this reinforces the previous observation that microphase segregation of fluorinated and non-fluorinated areas of the materials does not promote the formation lamellar fluidic ferroelectric phases when there is little fluorination in the head-group of the molecule compared to the tail-group.

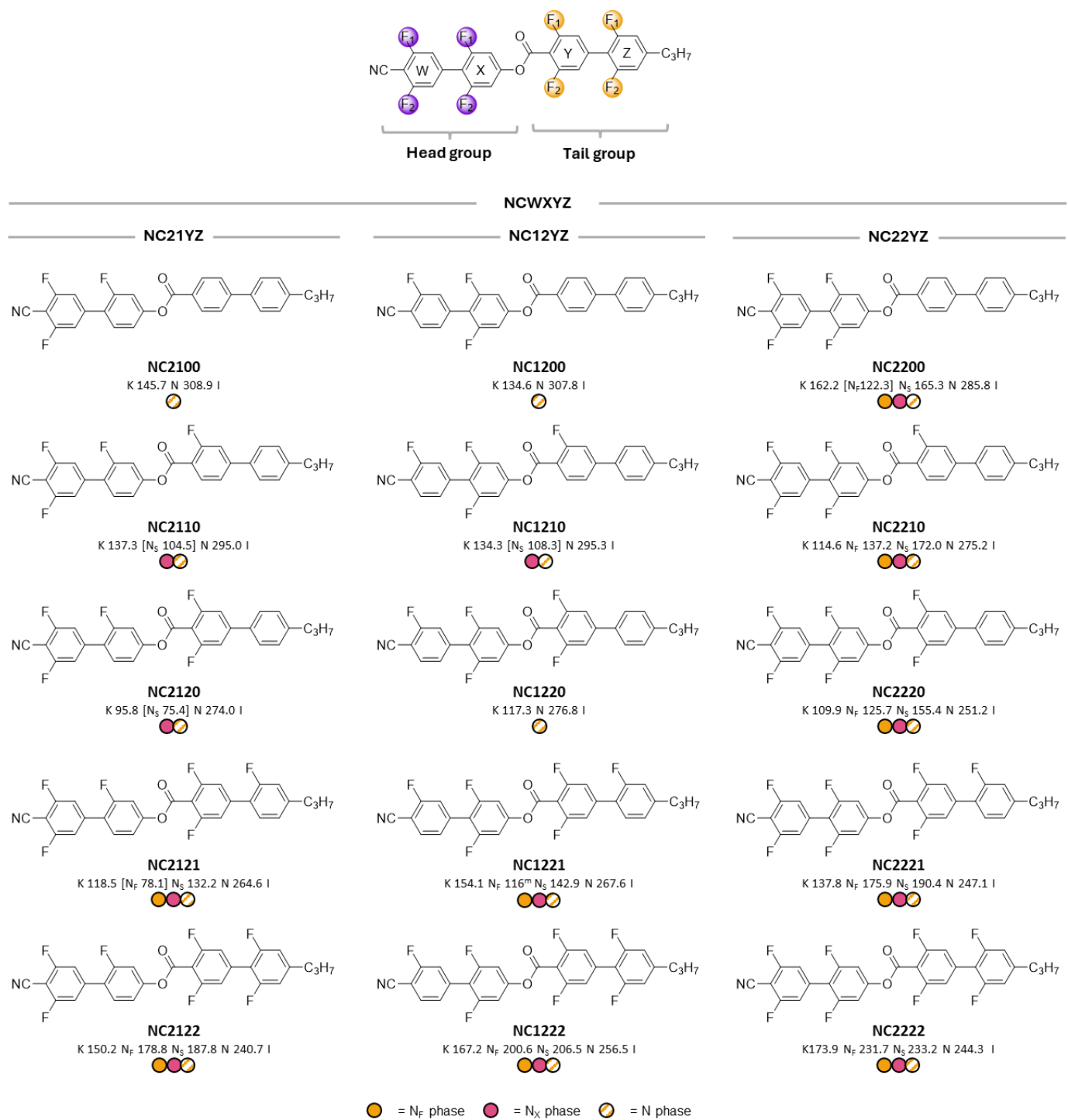

Fig. 5: The molecular structures, phase sequences and associated transition temperatures (°C) for the highly fluorinated **NCWXYZ** materials. Beginning with the left most ring on the head-group of the molecule, W, X, Y, and Z refer to the number of appended fluorine atoms on each ring and vary between 0 and 2. [ ] indicate a monotropic phase transition, $^m$ indicates a transition temperature determined by POM. Stripped colours indicate apolar phases and solid colours indicate polar phases.

## Conclusions

Analogous to solid state molecular ferroelectrics [6,7], H/F substitution has shown to be a powerful and versatile tool for controlling ferroelectric order in fluid molecular systems. In particular, specific fluorination patterns allow not only the thermal stability of the ferroelectric phase to be tuned, but also the nature of ferroelectric phase itself, determining whether nematic or smectic order is realised.

Across the materials studied, additional H/F substitution at the molecular head group generally enhances ferroelectric phase stability more effectively than the analogous substitution to the tail-group. However, head-group fluorination alone does not dictate the type of ferroelectric phase observed. Instead, the balance between nematic and smectic-like ordering is remarkably delicate, with a single H/F substitution capable of being the tipping point between nematic and smectic behaviour.

Molecular dynamics simulations reveal ferroelectric smectic order arises from specific syn-parallel pairing between neighbouring molecules, which promote lamellar organisation. These are easily perturbed by additional H/F substitution at the molecular tail, providing a microscopic explanation for the experimentally observed sensitivity of smectic ordering towards subtle changes to molecular structure. The nature of the EWD-group at the head of the molecule also influences these paring modes. Whereas a terminal fluorine can be efficacious towards the formation of ferroelectric smectic phases as well as nematics, nitriles in this study are solely nematogenic. This dependency maybe easily observed by computing the CDF from MD simulation and hence could be used to probe other EWG.

This study provides a set of design principles and a predictive framework which, together, can inform the design of future fluid molecular ferroelectrics with specific properties which, for the first time, can be tailored towards specific practical applications. Our results show that ferroelectric ordering in fluids is governed less by dipole magnitude than by the geometry of accessible pairing modes, reframing the molecular origins of spontaneous polar ordering in soft matter systems. The present results mark the transition from empirical discovery of fluid ferroelectric materials towards predictive, rational design.

## Data availability

The data associated with this paper are openly available from the University of Leeds Data Repository at https://doi.org/10.5518/1792.

## Author Contributions

C.J.G., W.C.O. and R.J.M. performed chemical synthesis, purification and structural characterisation. C.J.G performed mixture formulation and analysis. J.H and C.J.G. performed optical microscopy and calorimetry measurements. J.H. performed SAXS/WAXS and applied field studies. R.J.M. performed MD simulations and electronic structure calculations. All authors contributed to the conception of this work, to the interpretation of results, and to writing of the manuscript.

**Acknowledgements**

R.J.M. thanks UKRI for funding via a Future Leaders Fellowship, grant number MR/W006391/1, and the University of Leeds for funding via a University Academic Fellowship. R.J.M and J.H. thank the Royal Society for funding via Research Grant RGS\R2\242503. R.J.M. gratefully acknowledges support from Merck KGaA. Computational work was performed on ARC4 and AIRE**,** part of the high-performance computing facilities at the University of Leeds. The authors acknowledge EPSRC for funding the SAXS/WAXS system *via* a capital equipment grant EP/X0348011. W.C.O was supported by the EPSRC CDT in Soft Matter for Formulation and Industrial Innovation (SOFI$^2$), (EP/S023631/1).

**References**

1       Hales, A. & Jiang, X. A review of piezoelectric fans for low energy cooling of power electronics. *Appl. Energy* 215, 321-337 (2018).

2       Wang, W., Li, J., Liu, H. & Ge, S. Advancing Versatile Ferroelectric Materials Toward Biomedical Applications. *Adv. Sci.* 8, 2003074 (2021).

3       Zhang, S. et al. Piezo robotic hand for motion manipulation from micro to macro. *Nat. Commun.* 14, 500 (2023).

4       Scott, J. F. Applications of Modern Ferroelectrics. *Science* 315, 954-959 (2007).

5       Li, W. et al. Chemically diverse and multifunctional hybrid organic–inorganic perovskites. *Nat. Rev. Mater.* 2, 16099 (2017).

6       Horiuchi, S. & Tokura, Y. Organic Ferroelectrics. *Nat. Mater*. 7, 357 (2008).

7       Pan, Q. et al. The past 10 years of molecular ferroelectrics: structures, design, and properties. *Chem. Soc. Rev.* 53, 5781-5861 (2024).

8       Zhang, H.-Y., Tang, Y.-Y., Shi, P.-P. & Xiong, R.-G. Toward the Targeted Design of Molecular Ferroelectrics: Modifying Molecular Symmetries and Homochirality. *Acc. Chem. Res.* 52, 1928-1938 (2019).

9       Liu, H.-Y., Zhang, H.-Y., Chen, X.-G. & Xiong, R.-G. Molecular Design Principles for Ferroelectrics: Ferroelectrochemistry. *J. Am. Chem. Soc.* 142, 15205-15218 (2020).

10      Nishikawa, H. et al. A Fluid Liquid-Crystal Material with Highly Polar Order. *Adv. Mater.* 29, 1702354 (2017).

11      Mandle, R., Cowling, S. J. & Goodby, J. W. A Nematic to Nematic Transformation Exhibited by A Rod-Like Liquid Crystal. *Phys. Chem. Chem. Phys*. 19, 11429 (2017).

12      Sebastián, N. et al. Polarization patterning in ferroelectric nematic liquids via flexoelectric coupling. *Nat. Commun.* 14, 3029 (2023).

13      Li, J. et al. Development of ferroelectric nematic fluids with giant-ε dielectricity and nonlinear optical properties. *Sci. Adv.* 7, eabf5047 (2021).

14      Chen, X et al. First-principles experimental demonstration of ferroelectricity in a thermotropic nematic liquid crystal: Polar domains and striking electro-optics. *Proc. Natl. Acad. Sci.* 117, 14021 (2020).

15      Lavrentovich, O. D. Ferroelectric nematic liquid crystal, a century in waiting. *Proc. Natl. Acad. Sci.* 117, 14629 (2020).

16      Kumari, P., Basnet, B., Lavrentovich, M. O. & Lavrentovich, O. D. Chiral ground states of ferroelectric liquid crystals. *Science* 383, 1364 (2024).

17      Mertelj, A. et al. Splay Nematic Phase. *Phys. Rev. X* 8, 041025 (2018).

18      Sebastian, N. et al. Ferroelectric-Ferroelastic Phase Transition in a Nematic Liquid Crystal. *Phys. Rev. Lett.* 124, 037801 (2020).

19      Sultanov, V. et al. Tunable entangled photon-pair generation in a liquid crystal. *Nature* 631, 294-299 (2024).

20      Lovšin, M. et al. Ferroelectric Fluids for Nonlinear Photonics: Evaluation of Temperature Dependence of Second-Order Susceptibilities. *Adv. Opt. Mater.* 14, e03018 (2026).

21      Chen, X. et al. Polar In-Plane Surface Orientation of a Ferroelectric Nematic Liquid Crystal: Polar Monodomains and Twisted State Electro-Optics. *Proc. Natl. Acad. Sci.* 118, 1 (2021).

22      Kumar, M. P., Karcz, J., Kula, P., Karmakar, S. & Dhara, S. Giant Electroviscous Effects in a Ferroelectric Nematic Liquid Crystal. *Phys. Rev. Appl.* 19, 044082 (2023).

23      Sebastián, N., Mandle, R. J., Petelin, A., Eremin, A. & Mertelj, A. Electrooptics of mm-scale polar domains in the ferroelectric nematic phase. *Liq. Cryst.* 48, 2055-2071 (2021).

24      Himel, M. S. H. et al. Electrically Tunable Chiral Ferroelectric Nematic Liquid Crystal Reflectors. *Adv. Funct. Mater.*, 2413674 (2024).

25      Nishikawa, H., Sano, K. & Araoka, F. Anisotropic Fluid with Phototunable Dielectric Permittivity. *Nat. Commun*. 13, 1142 (2022).

26      Kikuchi, H. et al. Fluid Layered Ferroelectrics with Global C$_{\infty v}$ Symmetry. *Adv. Sci.* 9, 2202048 (2022).

27	Chen, X. et al. Observation of a uniaxial ferroelectric smectic A phase. *Proc. Natl. Acad. Sci.* 119, e2210062119 (2022).

28	Kikuchi, H. et al. Ferroelectric Smectic C Liquid Crystal Phase with Spontaneous Polarization in the Direction of the Director. *Adv. Sci.* 11, 2409827 (2024).

29	Hobbs, J. et al. Polar Order in a Fluid Like Ferroelectric with a Tilted Lamellar Structure – Observation of a Polar Smectic C ($SmC_P$) Phase. *Angew. Chem. In. Ed.* 64, e202416545 (2025).

30	Strachan, G. J., Górecka, E., Szydłowska, J., Makal, A. & Pociecha, D. Nematic and Smectic Phases with Proper Ferroelectric Order. *Adv. Sci.* 12, 2409754 (2025).

31	Gibb, C. J., Hobbs, J. & Mandle, R. J. Systematic Fluorination Is a Powerful Design Strategy toward Fluid Molecular Ferroelectrics. *J. Am. Chem. Soc.* 147, 4571-4577 (2025).

32	Strachan, G. J., Górecka, E., Hobbs, J. & Pociecha, D. Fluorination: Simple Change but Complex Impact on Ferroelectric Nematic and Smectic Liquid Crystal Phases. *J. Am. Chem. Soc.* 147, 6058-6066 (2025).

33	Gibb, C. J. et al. Spontaneous symmetry breaking in polar fluids. *Nat. Commun.* 15, 5845 (2024).

34	Li, J. et al. General phase-structure relationship in polar rod-shaped liquid crystals: Importance of shape anisotropy and dipolar strength. *Giant* 11, 100109 (2022).

35	Song, Y., Aya, S. & Huang, M. Updated view of new liquid-matter ferroelectrics with nematic and smectic orders. *Giant* 19, 100318 (2024).

36	Osipov, M. A. On the origin of the ferroelectric ordering in nematic liquid crystals and the electrostatic properties of ferroelectric nematic materials. *Liq. Cryst. Rev.* 12, 14-29 (2024).

37	Osipov, M. A. Dipole-dipole interactions and the origin of ferroelectric ordering in polar nematics. *Liq. Cryst.* 51, 2349–2355 (2024).

38	Madhusudana, N. V. Simple molecular model for ferroelectric nematic liquid crystals exhibited by small rodlike mesogens. *Phys. Rev. E* 104, 014704 (2021).

39	Hobbs, J., Gibb, C. J. & Mandle, R. J. Emergent Antiferroelectric Ordering and the Coupling of Liquid Crystalline and Polar Order. *Small Sci.* 4, 2400189 (2024).

40	Berrow, S. R., Hobbs, J. & Gibb, C. J. Reactive Fluid Ferroelectrics: A Gateway to the Next Generation of Ferroelectric Liquid Crystalline Polymer Networks. Small 21, 2501724 (2025).

41	Hobbs, J., Gibb, C. J. & Mandle, R. J. Ferri- and ferro-electric switching in spontaneously chiral polar liquid crystals. *Nat. Commun.* 16, 7510 (2025).

42	Strachan, G. J., Górecka, E. & Pociecha, D. Competition between mirror symmetry breaking and translation symmetry breaking in ferroelectric liquid crystals with increasing lateral substitution. *Mater. Horiz.* 13, 779-785 (2026).

43	Hobbs, J., Gibb, C. J. & Mandle, R. J. Polarity from the bottom up: a computational framework for predicting spontaneous polar order. *J. Mat. Chem. C* 13, 13367-13375 (2025).

44	Hird, M. Fluorinated liquid crystals – properties and applications. *Chem. Soc. Rev.* 36, 2070-2095 (2007).

45	Bruce, D. W. Employing fluorine for supramolecular control in self-assembled and self-organised molecular systems. *Chem. Sci.* 16, 22213-22230 (2025).

46	Patrick, C. R. & Prosser, G. S. A Molecular Complex of Benzene and Hexafluorobenzene. *Nature* 187, 1021-1021 (1960).

47	Chen, X. et al. First-principles experimental demonstration of ferroelectricity in a thermotropic nematic liquid crystal: Polar domains and striking electro-optics. *Proc. Natl. Acad. Sci.* 117, 14021-14031 (2020).

48	Kirchner, M. T., Blaser, D., Boese, R., Thakur, T. S. & Desiraju, G. R. 1,2,3-Trifluorobenzene. *Acta Cryst. E* 65, o2670 (2009).

49	Strachan, G. J. et al. Interplay of Polar Order and Positional Order in Liquid Crystals–Observation of Re-entrant Ferroelectric Nematic Phase. *Angew. Chem. In. Ed.* 64, e202516302 (2025).

50	Song, Y. et al. Emerging Ferroelectric Uniaxial Lamellar (Smectic $A_F$) Fluids for Bistable In-Plane Polarization Memory. *J. Phys. Chem Lett.* 13, 9983-9990 (2022).

51	Song, Y. et al. Ferroelectric Nematic Liquid Crystals Showing High Birefringence. *Adv. Sci.* 12, 2414317 (2025).

# Design Principles for Fluid Molecular Ferroelectrics

# Supplemental Information

Calum J. Gibb[1], Jordan Hobbs[2], William Ogle[1,3], Richard. J. Mandle*[1,2]

[1] School of Chemistry, University of Leeds, Leeds, UK, LS2 9JT.
[2] School of Physics and Astronomy, University of Leeds, Leeds, UK, LS2 9JT.
[3] School of Applied Mathematics, University of Leeds, Leeds, UK, LS29JT.

*Author for correspondence e-mail: r.mandle@leeds.ac.uk

## Contents

# 1 Supplementary Methods

## 1.1 Chemical Synthesis

Chemicals were purchased from commercial suppliers (Fluorochem, Merck, Apollo Scientific) and used as received. Solvents were purchased from Merck and used without further purification. Reactions were performed in standard laboratory glassware at ambient temperature and atmosphere and were monitored by TLC with an appropriate eluent and visualised with 254 nm or 365 nm light. Chromatographic purification was performed using a Combiflash NextGen 300+ System (Teledyne Isco) with a silica gel stationary phase and a hexane/ethyl acetate gradient as the mobile phase, with detection made in the 200-800 nm range. Chromatographed materials subjected to re-crystallisation from an appropriate solvent system.

## 1.2 Chemical Characterisation Methods

The structures of intermediates and final products were determined using $^1$H, $^{13}$C{$^1$H} and $^{19}$F NMR spectroscopy. NMR spectroscopy was performed using either a Bruker Avance III HDNMR spectrometer operating at 400 MHz, 100.5 MHz or 376.4 MHz ($^1$H, $^{13}$C{$^1$H} and $^{19}$F, respectively) or a Bruker AV4 NEO 11.75T spectrometer operating at 500 MHz or 125.5 MHz ($^1$H and $^{13}$C{$^1$H}, respectively). Unless otherwise stated, spectra were acquired as solutions in deuterated chloroform, coupling constants are quoted in Hz, and chemical shifts are quoted in ppm.

## 1.3 Mesophase Characterisation

Transition temperatures and measurement of associated latent heats were measured by differential scanning calorimetry (DSC) using a TA instruments Q2000 heat flux calorimeter with a liquid nitrogen cooling system for temperature control. Between 3-8 mg of sample was placed into T-zero aluminium DSC pans and then sealed. Samples were measured under a nitrogen atmosphere with 10 °C min$^{-1}$ heating and cooling rates. The transition temperatures and enthalpy values reported are averages obtained for duplicate runs. In general LC phase transition temperatures are measured on cooling from the onset of the transition while melt temperatures were measured on heating to avoid crystallization loops that can occur on cooling. Phase identification by polarised optical microscopy (POM) was performed using a Leica DM 2700 P polarised optical microscope equipped with a Linkam TMS 92 heating stage. Samples were studied sandwiched between two untreated glass coverslips.

### 1.4 X-ray Scattering

X-ray scattering measurements, both small angle (SAXS) and wide angle (WAXS) were recorded using an Anton Paar SAXSpoint 5.0. This was equipped with a primux 100 Cu X-ray source with a 2D EIGER2 R detector with a variable sample-detector distance. The X-rays had a wavelength of 0.154 nm. Samples were filled into either thin-walled quartz capillaries or held between Kapton tape. Temperature was controlled using an Anton Paar heated sampler with a range of 20∘C to 300∘C and the samples held in a chamber with an atmospheric pressure of >1 mbar. Background scattering patterns were recorded, scaled according to the sample transmission and then subtracted from the samples' obtained 2D SAXS pattern. 1D patterns were obtained by radially integrating the 2D SAXS patterns. Peak positions and FWHM were recorded and then converted into d spacing following Bragg's law. In tilted smectic phases, the tilt was obtained from:

$$\frac{d_C}{d_A} = \cos\theta$$

where $d_C$ is the layer spacing in the tilted smectic phase, $d_A$ is the extrapolated spacing from the non-tilted preceding phase, extrapolated to account for the weak temperature dependence of the preceding phases due to shifts in conformation and order, and θ the structural tilt angle.

### 1.5 Current Response Measurements

Current Response measurements are undertaken using the current reversal technique. Triangular waveform AC voltages are applied to the sample cells with an Agilent 33220A signal generator (Keysight Technologies), and the resulting current outflow is passed through a current-to-voltage amplifier and recorded on a RIGOL DHO4204 high-resolution oscilloscope (Telonic Instruments Ltd, UK). Heating and cooling of the samples during these measurements is achieved with an Instec HCS402 hot stage controlled to 10 mK stability by an Instec mK1000 temperature controller. The LC samples are held in 4μm thick cells with no alignment layer, supplied by Instec. The measurements consist of cooling the sample at a rate of 1 K min$^{-1}$ and applying a set voltage at a frequency of 20 Hz to the sample every 1 K. The voltage was set such that it would saturate the measured PS and was determined before final data collection. There are three contributions to the measured current trace: accumulation of charge in the cell ($I_c$), ion flow ($I_i$), and the current flow due to polarisation reversal ($I_p$). To obtain a $P_{Sat}$ value, we extract the latter, which manifests as one or multiple peaks in the current flow, and integrate as:

$$P_{Sat} = \int \frac{I_p}{2A} dt$$

where A is the active electrode area of the sample cell.

## 1.6 DFT Calculations

Electronic structure calculations were performed using Gaussian G16 revision C.02 [1] Click or tap here to enter text.and with a B3LYP-GD3BJ/cc-pVTZ [2-4] Click or tap here to enter text.basis set. Obtained structures were verified as a minimum from frequency calculations.

## 1.7 Molecular Dynamics

Parameters for the **F21YZ** and **NC22YZ** were modelled using the GAFF forcefield [5] with modifications for liquid crystalline molecules [6,7]. The initial geometries of compounds **F21YZ** and **NC22YZ** were obtained by optimisation (at the B3LYP-GD3BJ/cc-pVTZ level of DFT, as above). RESP charges were calculated at the HF/6-31G* levelClick or tap here to enter text.[8] using Gaussian G16.c02 [1] with the route section "iop(6/33=2) iop(6/42=6) pop(chelpg, regular)". Using Antechamber (part of AmberTools24 software package [9]) the Gaussian output file was converted to mol2 format; finally, this was parsed into Gromacs readable .itp format using the ACPYPE script [10]. Simulations were performed in Gromacs 2024.4 [11] using GPU acceleration (Nvidia L40s) *via* CUDA 13.6.2; bonded, non-bonded, PME and update tasks were offloaded to the GPU during the production run.

Initial coordinates were generated *pseudo* randomly using the Gromacs tool *gmx insert-molecules*; we construct a low-density (~ 0.1 g cm$^3$) initial configuration consisting of 1000 molecules with random positional and orientational order. To this we then perform energy minimisation *via* steepest descent, followed by equilibration in the NVE and NVT regimes (the later at T=600K). A brief (20 ns) compression simulation with an isotropic barostat ($P$ = 100 Bar) at T=600K affords a liquid-like density isotropic state, used as a starting configuration for subsequent simulations. In the case of polar simulations, a static electric field was applied across the x-axis of the simulation for 50 ns (at T=600K) to give a near saturated polar order parameter (<*P*1> > 0.95). This polar configuration was used as a starting point for all subsequent simulations which were performed at with an anisotropic barostat (P = 1 Bar, allowing the simulation box to deform). For simulations of the **FWXYZ** materials the simulation temperature was 373 K; for the **NCWXYZ** materials the temperature chosen was 25 K below the experimental ferroelectric onset temperature.

For each production MD simulation, we employ a total simulation time of 1 microsecond once stable values of <P1> and <P2> are achieved.

Simulations employed periodic boundary conditions in xyz. Bonds lengths were constrained to their equilibrium values with the LINCS algorithm [12]. Compressabilities in xyz dimensions were set to 4.5e-5, with the off-diagonal compressibilities were set to zero to ensure the simulation box remained rectangular. Long-range electrostatic interactions were calculated using the Particle Mesh Ewald method with a cut-off value of 1.2 nm. A van der Waals cut-off of 1.2 nm was used. A timestep of 2 fs was employed in all cases. Pressure coupling was via a Parinello-Rahman barostat, and temperature coupling via a velocity rescaling thermostat.

A simulation was considered liquid crystalline if the second-rank orientational order parameter (<*P*2>), calculated *via* the eigenvalues of the nematic Q-tensor using MdTraj v1.9.8[13] , was greater than 0.3; conversely, the simulation was judged to be isotropic if <*P2*> was below this value). The polar order parameter (<*P*1>) is found as the quotient of the total simulation dipole moment and the maximum possible dipole moment of the simulation[14]. Phase assignment was made by visual inspection of the trajectory. Cylindrical distribution functions were calculated using the cylindr program [15]. Analysis was conducted on each trajectory frame, with presented values are given as the average over all frames in the final 500 ns of the production MD simulation.

## 2    Supplementary Results

**Table S1.**    Tabulated phase sequences, including transition temperatures (°C) and associated enthalpies of transition (KJ mol$^{-1}$) for the **FWXYZ** materials. Data for **F2000**, **F2100**, and **F2120** have been published previously [16,17].

| Material | Phase sequence | |
|---|---|---|
| | Transition Temperatures (°C) | Enthalpies (KJ mol$^{-1}$) |
| **F2010** | K 139.5 SmA 199.8 N 256.4 I | K 29.5 SmA 0.33 N 0.26 I |
| **F2020** | K 99.9 [SmA 98.9] N 220.0 I | K 24.7 [SmA 0.06] N 0.58 I |
| **F2021** | K 98.1 N 205.0 I | K 27.3 N 0.27 I |
| **F2022** | K 106.9 [N$_F$ 90.8 N$_S$ 105$^m$] N 182.3 I | K 25.7 [N$_F$ 25.7 N$_S$ 105$^m$] N 0.43 I |
| **F2110** | K 91 [SmC$_P$ 77.2] SmA$_F$ 116.9 SmA 177.1 N 234.1 I | K 19.5 [SmC$_P$ 0.002] SmA$_F$ 0.69 SmA 0.30 N 0.39 I |
| **F2121** | K 107.5 N 190.1 I | K 30.7 N 0.32 I |
| **F2122** | K 126.5 [N$_F$ 123.5] N$_S$ 137$^m$ N 164.3 I | K 26.2 [N$_F$ 26.2] N$_S$ (N/A) N 0.36 I |
| **F2200** | K 199.5 [SmC$_P$ 166.3$^V$ SmA$_F$ 175.0$^V$] N 221.3 I | K 28.5 [SmC$_P$ (N/A) SmA$_F$ (N/A)] N 0.22 I |
| **F2210** | K 171.1 [SmC$_P$ 143.9$^V$ SmA$_F$ 153.9$^V$ SmA 160.2] N 209.9 I | K 26.0 [SmC$_P$ (N/A) SmA$_F$ (N/A) SmA 0.28] N 0.22 I |
| **F2220** | K 131.6 [SmA$_F$ 105.5 SmA 126.5] N 182.1 I | K 33.6 [SmA$_F$ 0.09 SmA 0.30] N 0.31 I |
| **F2221** | K 145.1 [N$_F$ 130.1 N$_S$ 133.5] N 175.5 I | K 33.9 [N$_F$ 0.25 N$_S$ 0.002] N 0.30 I |
| **F2222** | K 183.2 N$_F$ 160.6 I | K 43.6 N$_F$ 3.65 I |

**Table S2.** Tabulated phase sequences, including transition temperatures (°C) and associated enthalpies of transition (KJ mol$^{-1}$) for the **NCWXYZ** materials

| Material | Phase sequence | |
|---|---|---|
| | Transition Temperatures (°C) | Enthalpies (KJ mol$^{-1}$) |
| **NC2100** | K 145.7 N 308.9 I | K 39.1 N 1.40 I |
| **NC2110** | K 137.3 [N$_S$ 104.5] N 295.0 I | K 38.0 [N$_S$ 0.02] N 1.14 I |
| **NC2120** | K 95.8 [N$_S$ 75.4] N 274.0 I | K 22.7 [N$_S$ 0.01] N 0.56 I |
| **NC2121** | K 118.5 [N$_F$ 78.1] N$_S$ 132.2 N 264.6 I | K 31.3 [N$_F$ 0.11] N$_S$ 0.01 N 0.61 I |
| **NC2122** | K 150.2 N$_F$ 178.8 N$_S$ 187.8 N 240.7 I | K 36.3 N$_F$ 0.23 N$_S$ 0.01 N 0.65 I |
| **NC1200** | K 134.6 N 307.8 I | K 31.2 N 1.90 I |
| **NC1210** | K 134.3 [N$_S$ 108.3] N 295.3 I | K 41.2 [N$_S$ 0.02] N 1.74 I |
| **NC1220** | K 117.3 N 276.8 I | K 35.1 N 1.62 I |
| **NC1221** | K 154.1 [N$_F$ 116$^m$ N$_S$ 142.9] N 267.6 I | K 40.5 [N$_F$ N/A N$_S$ 0.010] N 0.97 I |
| **NC1222** | K 167.2 N$_F$ 200.6 N$_S$ 206.5 N 256.5 I | K 167.2 N$_F$ 200.6 N$_S$ 206.5 N 256.5 I |
| **NC2200** | K 162.2 [N$_F$ 122.3] N$_S$ 165.3 N 285.8 I | K 43.7 [N$_F$ 0.30] N$_S$ 0.011 N 2.03 I |
| **NC2210** | K 114.6 N$_F$ 137.2 N$_S$ 172.0 N 275.2 I | K 29.0 N$_F$ 0.64 N$_S$ 0.024 N 0.67 I |
| **NC2220** | K 109.9 N$_F$ 125.7 N$_S$ 155.4 N 251.2 I | K 29.4 N$_F$ 0.36 N$_S$ 0.014 N 0.60 I |
| **NC2221** | K 137.8 N$_F$ 175.9 N$_S$ 190.4 N 247.1 I | K 36.9 N$_F$ 0.41 N$_S$ 0.0.22 N 1.37 I |
| **NC2222** | K 173.9 N$_F$ 231.7 N$_S$ 233.2 N 244.3 I | K 139.1 N$_F$ 0.78 N$_S$ 0.017 N 1.84 I |

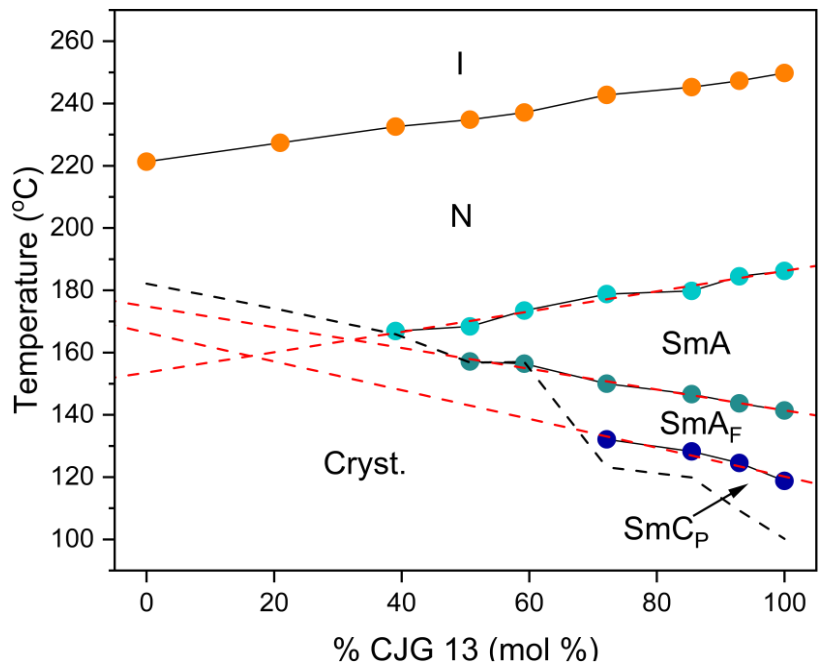

**Fig. S1.** Binary mixture diagram between **F2200** and **F2100**. The dashed line indicates the crystallisation temperature obtained on cooling from the relevant DSC thermograms.

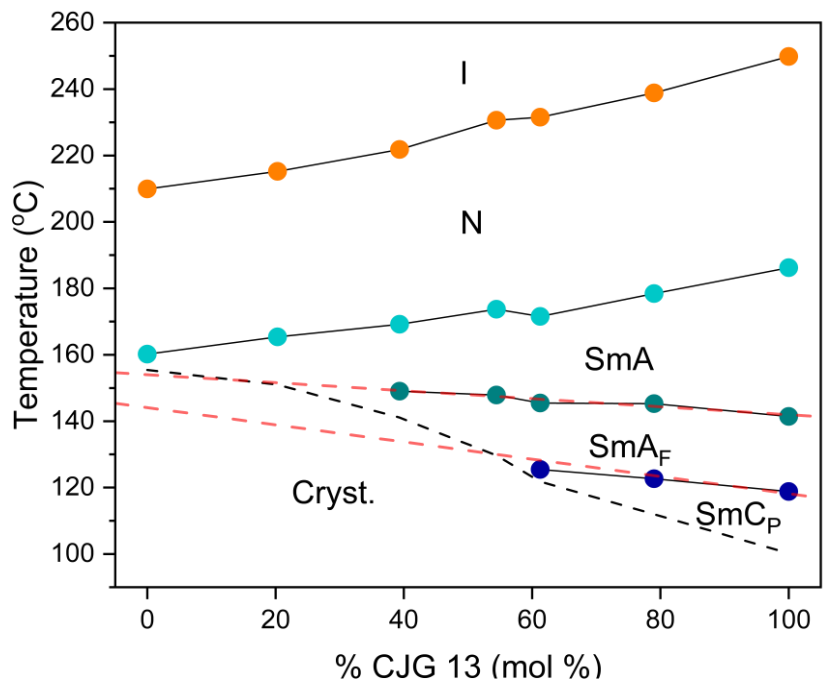

**Fig. S2.** Binary mixture diagram between **F2210** and **F2100**. The dashed line indicates the crystallisation temperature obtained on cooling from the relevant DSC thermograms.

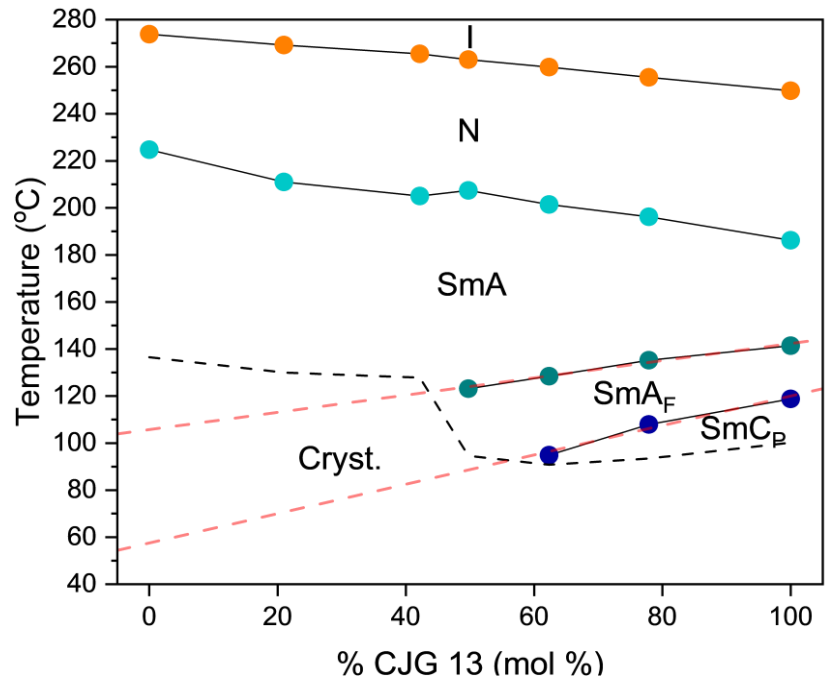

**Fig. S3.**     Binary mixture diagram between **F2000** and **F2100**. The dashed line indicates the crystallisation temperature obtained on cooling from the relevant DSC thermograms.

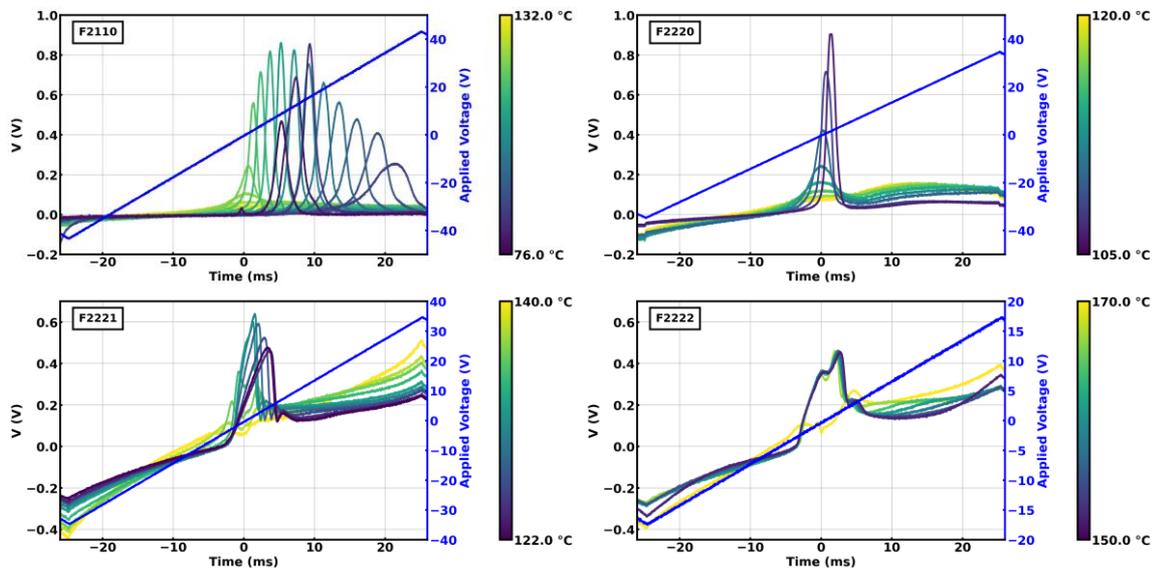

**Fig. S4.**     Current response measurements for **F2110**, **F2220**, **F2221** and **F2222.** All were taken using a 10 Hz triangular waveform with the maximum voltage as marked on the graphs.

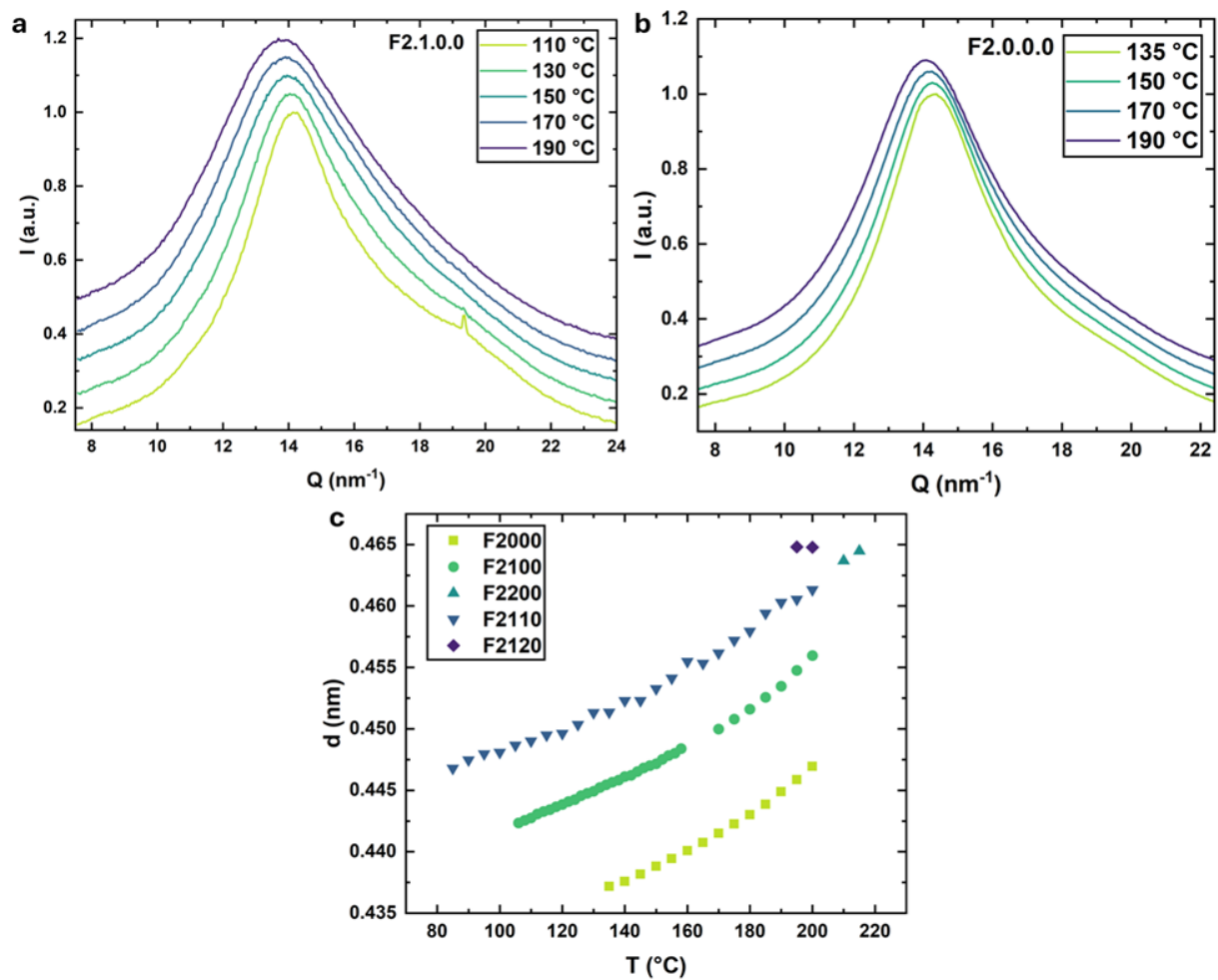

**Fig. S5.** WAXS patterns for **a. F2100** and **b. F2000** showing how the peak evolves with temperature. At around 19 nm$^{-1}$ a second peak can be observed that increases in intensity at lower temps indicating an increase in π-π stacking interactions. This is more prominent for **F2100** than for **F2000** due to its increase proclivity for ferroelectric phase formation. **c.** Temperature dependence of the lateral spacing for the molecules shown. The value seems to be a function of the number of fluorine substitutions irrespective of fluorination positioning.

**Table S3.** Tabulated data from molecular dynamics simulations of compounds **F21YZ** at 373 K. Given values are the average over the final 500 ns of the production MD simulation, with uncertainties given as one standard deviation. [#] Non polar phase as evidenced by <P1> and visualisation.

| Material | Phase | $\rho$ / g cm$^3$ | <P1> | <P2> | $\tau$ | d / nm (exp.) | $\theta$ / ° (exp.) |
|---|---|---|---|---|---|---|---|
| **F2100** | SmC$_P$ | 1.245 ± 0.003 | 0.93 ± 0.01 | 0.82 ± 0.04 | 0.39 ± 0.05 | 2.1±0.1 | 30.5±9.4 |
| **F2110** | SmC$_P$ | 1.273 ± 0.003 | 0.91 ± 0.01 | 0.81 ± 0.06 | 0.45 ± 0.05 | 2.1±0.2 | 30.5±10.7 |
| **F2120** | SmA[#] | 1.269 ± 0.005 | 0.04 ± 0.02 | 0.72 ± 0.10 | 0.25 ± 0.03 | 2.5±0.4 | n/a |
| **F2121** | N[#] | 1.297 ± 0.004 | 0.05 ± 0.04 | 0.62 ± 0.04 | 0.06 ± 0.04 | n/a | n/a |
| **F2222** | N$_F$ | 1.337 ± 0.004 | 0.89 ± 0.03 | 0.76 ± 0.03 | 0.08 ± 0.04 | n/a | n/a |

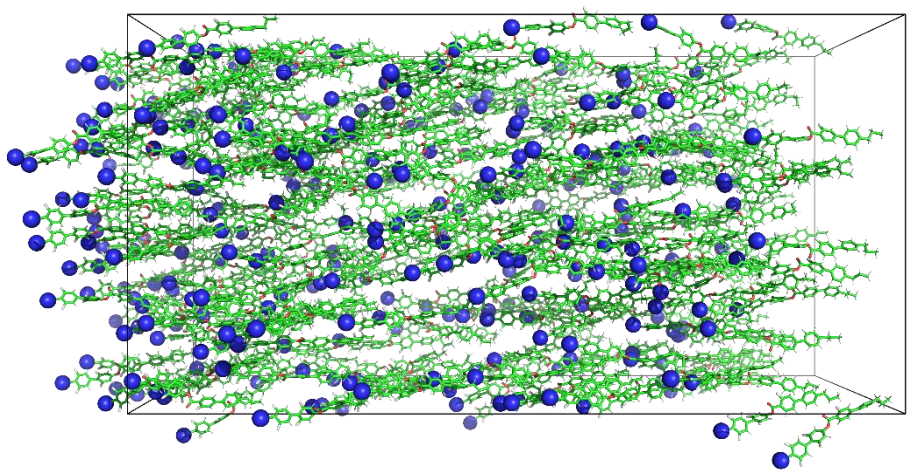

**Fig. S6.** Instantaneous configurations of **NC2200** in the polar N$_F$ phase at 383 K. Molecules aligned parallel to the director are shown as green. The nitrogen atoms are shown as spheres to aid visualisation.

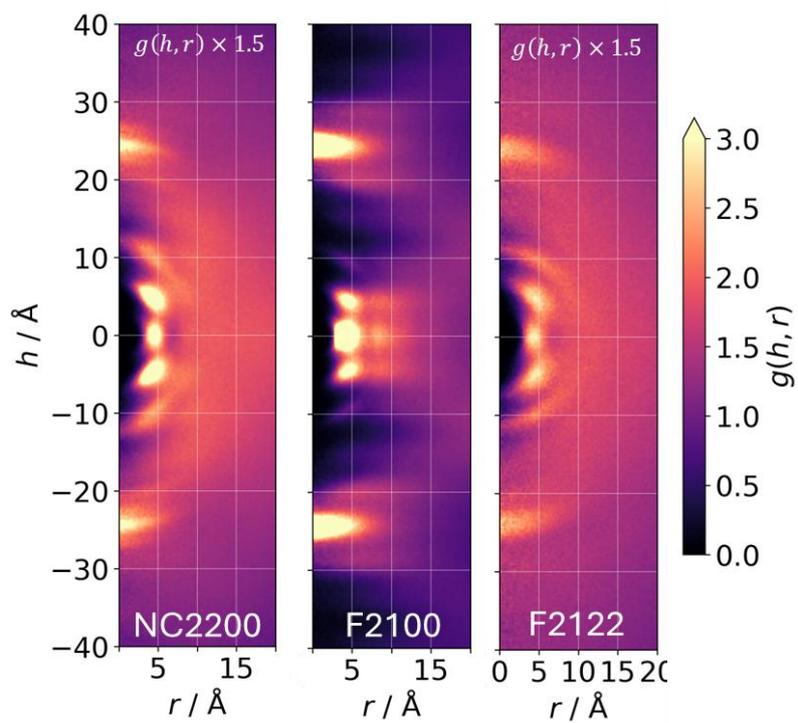

**Fig. S7.** Cylindrical distribution function (CDF) plots for **NC2200**, **F2200**, and **F2210** generated over the final 500 ns of each simulation trajectory.

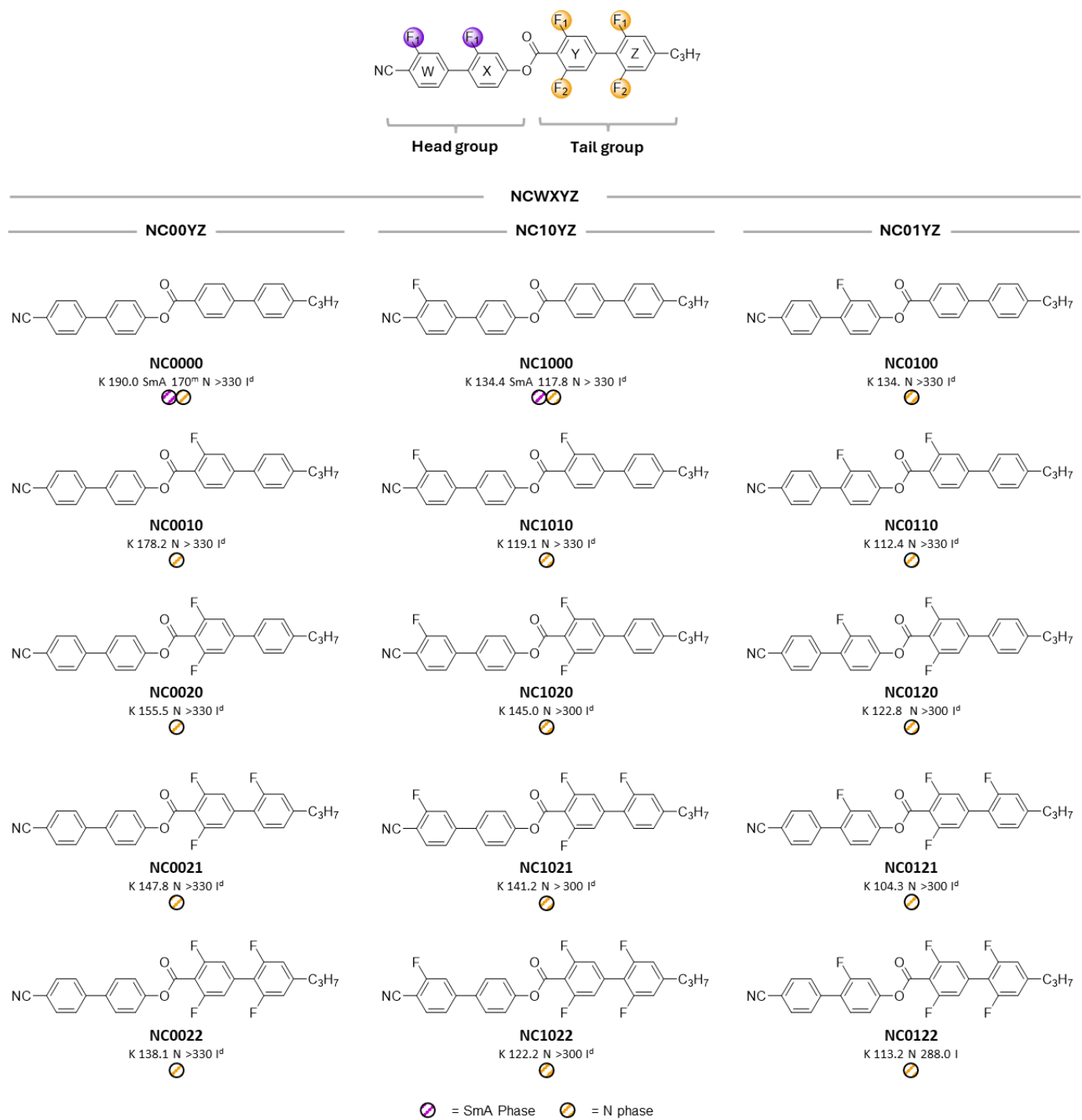

**Fig. S8.** The molecular structures, phase sequences and associated transition temperatures (°C) for the fluorinated **NCWXYZ** materials. Beginning with the first ring on the head of the molecule, W, X, Y, and Z refer to the number of appended fluorine atoms on each ring and vary between 0 and 2. [ ] indicate a monotropic phase transition, [d] indicates sample decomposition on reaching the isotropic phase on the first heating cycle. Stripped colours indicate paraelectric phases.

## 3    Organic synthesis

The total synthesis of materials **FWXYZ** and **NCWXYZ** is outlined in **Scheme S1**. The synthesis of *3,5-difluoro-4'-propyl-[1,1'-biphenyl]-4-carboxylic acid* is reported elsewhere [16]. The synthesis of the various phenolic starting materials has also been described previously within the literature [16,18,19].

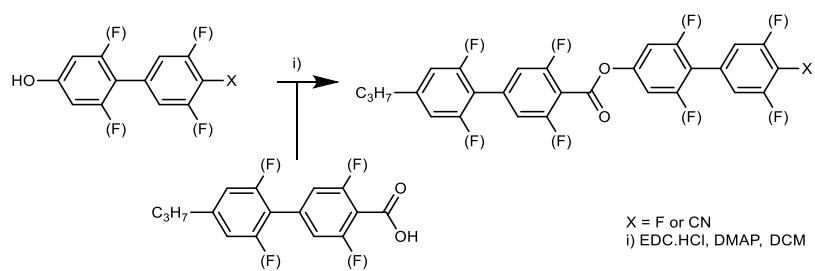

**Scheme S1.**    General esterification procedure for the synthesis of the **FWXYZ** and **NCWXYZ** materials.

## 3.1 Preparation of *4'-propyl-3-fluorobiphenyl-4-carboxylic acid*

The preparation of *4'-propyl-3-fluorobiphenyl-4-carboxylic acid* is described in **Scheme S2**. A reaction flask was charged with 4-*propylphenyl boronic acid* (5.41 g, 33 mmol) and dissolved in 200 mL of THF and 50 mL of 2M aqueous $K_2CO_3$. The resultant solution was sparged with $N_{2)}$ for 30 mins. *4-bromo-2-fluorobenzoic acid* (6.57 g, 30 mmol) was then added to the sparged solution which was then heated to 70 °C (external temperature) under reflux and under an atmosphere of dry nitrogen gas. Once at temperature, PdXPhos (G3) (0.05 mol %) was then added in a single portion and the reaction left for 18 h with constant stirring. The reaction was then cooled, acidifed using 2M HCl and extracted with EtOAc, the aqueous and organic layers separated, the organics being dried over $MgSO_4$. Removal of volatiles *in vacuo* afforded an off white solid which was re-crystalised from MeCN to afford the title compound as a fine, white powder.

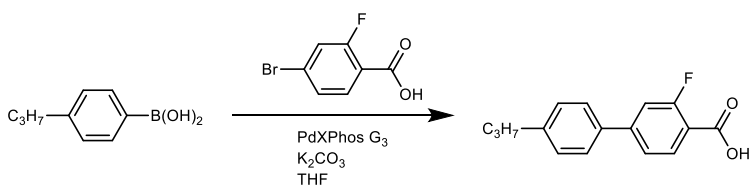

**Scheme S2.**     Synthesis of *4'-propyl-3-fluorobiphenyl-4-carboxylic acid*

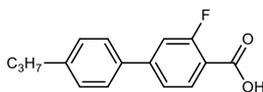

*4'-propyl-3-fluorobiphenyl-4-carboxylic acid*

| | |
|---|---|
| Yield: | (fine white powder) 7.1 g, 91% |
| $R_F$ (EtOAc): | 0.25 |
| $^1$H NMR (501 MHz, DMSO): | 13.21 (s, 1H, Ar-COO**H**), 7.92 (t, *J* = 8.2 Hz, 1H, Ar-**H**), 7.69 (ddd, *J* = 8.2, 2.2, 1.8 Hz, 2H, Ar-**H**), 7.65 – 7.58 (m, 2H, Ar-**H**)\*, 7.31 (ddd, *J* = 8.4, 2.0, 1.7 Hz, 2H, Ar-**H**), .61 (t, *J* = 7.4 Hz, 2H, Ar-C**H₂**-CH₂), 1.62 (h, *J* = 7.2 Hz, 2H, CH₂-C**H₂**-CH₃), 0.91 (t, *J* = 7.3 Hz, 3H, CH₂-C**H₃**) (\*Overlapping signals). |
| $^{13}$C{$^1$H} NMR (101 MHz, DMSO): | 165.33 (d, *J* = 3.2 Hz), 162.16 (d, *J* = 256.8 Hz), 146.88 (d, *J* = 8.7 Hz), 143.60, 135.46, 132.98, 129.56, 122.59 (d, *J* = 3.2 Hz), 117.91 (d, *J* = 10.3 Hz), 114.92 (d, *J* = 23.3 Hz), 37.33, 24.40, 14.09. |

¹⁹F NMR (376 MHz, DMSO): -109.81 (t, $J_{F-H}$ = 8.2 Hz, 1F, Ar-**F**).

## 3.2 Preparation of *2',3,5-trifluoro-4'-propyl-[1,1'-biphenyl]-4-carboxylic acid*

The preparation of *2',3,5-trifluoro-4'-propyl-[1,1'-biphenyl]-4-carboxylic acid* is described in **Scheme S3**.

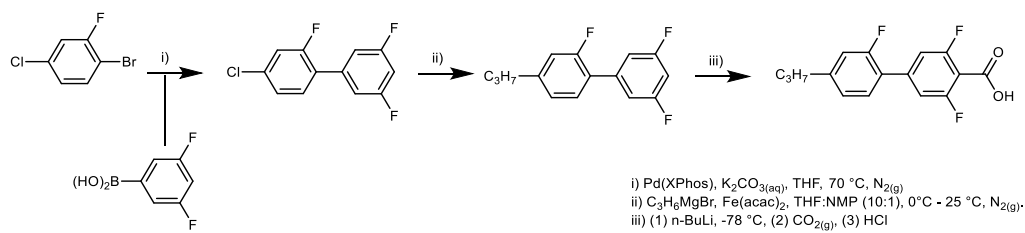

i) Pd(XPhos), $K_2CO_{3(aq)}$, THF, 70 °C, $N_{2(g)}$.
ii) $C_3H_6MgBr$, Fe(acac)$_2$, THF:NMP (10:1), 0°C - 25 °C, $N_{2(g)}$.
iii) (1) n-BuLi, -78 °C, (2) $CO_{2(g)}$, (3) HCl

**Scheme S3.** Preparation of *2',3,5-trifluoro-4'-propyl-[1,1'-biphenyl]-4-carboxylic acid*

### 3.2.1. Synthesis of *4-chloro-2,3',5'-trifluoro-1,1'-biphenyl*

A reaction flask was charged with *4-chloro, 2-fluro bromo benzene* (21.0 g, 100 mmol) and dissolved in 250 mL of THF and 60 mL of 2M aqueous $K_2CO_3$. The resultant solution was sparged with $N_{2(g)}$ for 30 mins. *2, 6-difluro benzene boronic acid* (17.4 g, 110 mmol) was then added to the sparged solution which was then heated to 70 °C (external temperature) under an atmosphere of dry nitrogen. Once at temperature, PdXPhos (G3) (0.05 mol %) was then added in a single portion and the reaction left for 18 h with constant stirring. The reaction was then cooled, extracted with EtOAc, the aqueous and organic layers separated, the organics being dried over MgSO₄. The suspension was filtered, concentrated *in vacuo* and purified by flash chromatography over silica gel using an isocratic run of hexane using a Combiflash NextGen300+ system. The resulting solid was then re-crystallised from hexane:toluene (10:1) to yield 4-chloro-2,3',5'-trifluoro-1,1'-biphenyl as a white, crystalline solid.

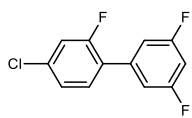

*4-chloro-2,3',5'-trifluoro-1,1'-biphenyl*

Yield: (white solid) 16.5 g, 68 %

$R_F$ (Hexanes): 0.33

¹H NMR (400 MHz): 7.39 (t, *J* = 8.3 Hz, 1H, Ar-**H**), 7.30 – 7.22 (m, 2H, Ar-**H**) †, 7.15 – 7.04 (m, 2H, Ar-**H**), 6.88 (tt, *J* = 8.9, 2.3 Hz, 1H, Ar-**H**). († Overlapping CDCl₃).

<sup>19</sup>F NMR (376 MHz):     -109.44 (t, $J_{F-H}$ = 7.9 Hz, 2F, Ar-**F**), -114.59 (t, $J_{F-H}$ = 9.4 Hz, 1F, Ar-**F**).

### 3.2.2. Synthesis of *2,3',5'-trifluoro-4-propyl-1,1'-biphenyl*

In a flame dried flask, *4-chloro-2,3',5'-trifluoro-1,1'-biphenyl* (11.0 g, 45 mmol) and Fe(acac)$_3$ (g, mmol) were dissolved in 250 mL dry THF:NMP (10:1). The solution was cooled over an ice bath before a solution of freshly prepared Grignard reagent (50 mmol) in dry Et$_2$O was added dropwise to the stirred solution. The progress of the reaction was followed by <sup>1</sup>H NMR with no further reaction occurring after 1.5 h. The solution was quenched with 2M HCl (crica. 50 mL) before being extracted with EtOAc, dried over MgSO$_4$ and a yellow oil was retrieved under reduced pressure. The oil was found to be a mixture of the solid starting material and the desired product. To purify the mixture, the remaining starting material was crystallized out of solution using hexanes and removed by filtration**.** The filtrate was concentrated reduced pressure to give *2,3',5'-trifluoro-4-propyl-1,1'-biphenyl* as a colourless oil.

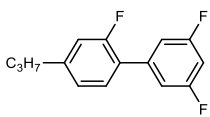

*2,3',5'-trifluoro-4-propyl-1,1'-biphenyl*

| | |
|---|---|
| Yield: | (colourless oil) 11.4 g, 91 % |
| R$_F$ (Hexanes): | 0.33 |
| <sup>1</sup>H NMR (400 MHz): | 7.31 (t, *J* = 8.1 Hz, 1H, Ar-**H**), 7.12 – 6.96 (m, 4H, Ar-**H**)\*, 6.80 (tt, *J* = 9.0, 2.3 Hz, 1H, Ar-**H**), 2.63 (t, *J* = 7.6 Hz, 2H, Ar-C**H₂**-CH$_2$), 1.68 (h, *J* = 7.6 Hz, 2H,m CH$_2$-CH$_2$-CH$_3$), 0.98 (t, *J* = 7.4 Hz, 3H, CH$_2$-C**H₃**). *Overlapping Signals. |
| <sup>19</sup>F NMR (376 MHz): | -110.23 (t, $J_{F-H}$ = 8.2 Hz, 2F, Ar-**F**), -118.30 (t, $J_{F-H}$ = 9.9 Hz, 1F, Ar-**F**). |

### 3.2.3 Preparation of *2',3,5-trifluoro-4'-propyl-[1,1'-biphenyl]-4-carboxylic acid*

A solution of *n-butyl lithium* (1.6 M in hexanes, 45 mmol, 28 mL) was added dropwise to a stirred, cooled (-78 °C) solution of *2,3',5'-trifluoro-4-propyl-1,1'-biphenyl* (10.0 g, 40 mmol) in anhydrous THF (150 mL) under an atmosphere of dry nitrogen. The aryl lithium solution was allowed to stir for 1.5 h at this same temperature before adding a large excess of gaseous carbon dioxide which was pre-dried by bubbling through conc. H$_2$SO$_4$. The solution was then allowed too slowly warm to ambient temperature (~ 2 h). The basic solution was acidified with 2M HCl (~100 ml) and extracted with EtOAc. The organics were then dried over MgSO$_4$ and a white solid retrieved under reduced pressure. These were then purified by recrystallization from MeCN to give *2',3,5-trifluoro-4'-propyl-[1,1'-biphenyl]-4-carboxylic acid* as a fine white solid.

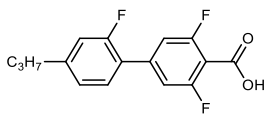

*2',3,5-trifluoro-4'-propyl-[1,1'-biphenyl]-4-carboxylic acid*

| | |
|---|---|
| Yield: | (fine white solid) 10.1 g, 86 % |
| $R_F$ (DCM): | 0.00 |
| $^1$H NMR (400 MHz, DMSO): | 7.54 (t, *J* = 8.2 Hz, 1H, Ar-**H**), 7.40 (ddd, *J* = 9.4, 4.6, 2.8 Hz, 2H, Ar-**H**), 7.23 – 7.13 (m, 2H, Ar-**H**)*, 2.61 (t, *J* = 7.5 Hz, 2H, Ar-C**H₂**-CH₂), 1.61 (h, *J* = 7.3 Hz, 2H, CH₂-C**H₂**-CH₃), 0.90 (t, *J* = 7.3 Hz, 3H, CH₂-C**H₃**). *Overlapping signals. |
| $^{19}$F NMR (376 MHz, DMSO): | -111.81 (d, $J_{F-H}$ = 10.0 Hz, 2F, Ar-**F**), -118.21 (t, $J_{F-H}$ = 10.0 Hz, 1F, Ar-**F**). |

## 3.3  Preparation of *2',3,5,6'-tetrafluoro-4'-propyl-[1,1'-biphenyl]-4-carboxylic acid*

The preparation of *2',3,5,6'-tetrafluoro-4'-propyl-[1,1'-biphenyl]-4-carboxylic acid* is described in **Scheme S4**.

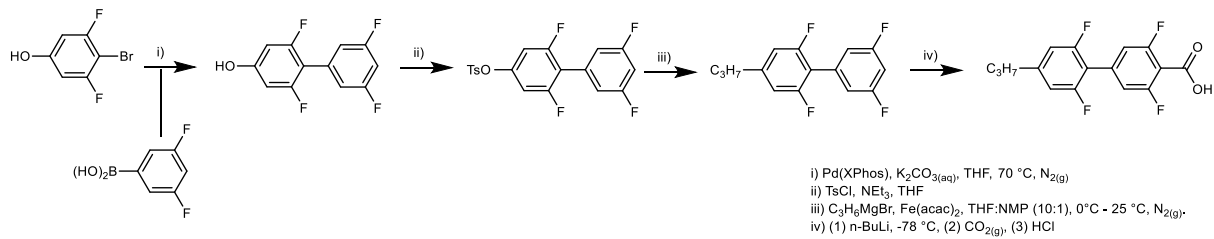

i) Pd(XPhos), K₂CO₃(aq), THF, 70 °C, N₂(g)
ii) TsCl, NEt₃, THF
iii) C₃H₆MgBr, Fe(acac)₂, THF:NMP (10:1), 0°C - 25 °C, N₂(g).
iv) (1) n-BuLi, -78 °C, (2) CO₂(g), (3) HCl

**Scheme S4.**    Preparation of *2',3,5,6'-tetrafluoro-4'-propyl-[1,1'-biphenyl]-4-carboxylic acid*.

### 3.3.1. Synthesis of *2,3',5',6-tetrafluoro-[1,1'-biphenyl]-4-ol*

A reaction flask was charged with 4-bromo-3,5-difluorophenol (10 g, 48 mmol) and dissolved in 250 mL of THF and 60 mL of 2M aqueous K₂CO₃. The resultant solution was sparged with N₂ for 30 mins. 2, 6-difluro benzene boronic acid (8.32 g, 53 mmol) was then added to the sparged solution which was then heated to 70 °C (external temperature) under an atmosphere of dry nitrogen.. Once at temperature, PdXPhos (G3) (0.05 mol %) was then added in a single portion and the reaction left for 18 h with constant stirring. The reaction was then cooled, extracted with EtOAc, the aqueous and organic layers separated, the organics being dried over MgSO₄. The suspension was filtered, concentrated *in* and purified by flash chromatography over silica gel using an isocratic run of hexane using a Combiflash NextGen300+ system to yield *3',5',6-tetrafluoro-[1,1'-biphenyl]-4-ol* as a white, crystalline solid.

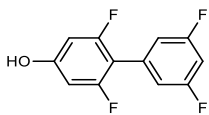

*2,3',5',6-tetrafluoro-[1,1'-biphenyl]-4-ol*

| | |
|---|---|
| Yield: | (white crystalline solid) 10.8 g, 93 % |
| R$_F$ (DCM): | 0.20 |
| ¹H NMR (400 MHz, DMSO): | 10.64 (s, 1H, Ar-O**H**), 7.26 (tt, *J* = 9.4, 2.4 Hz, 1H, Ar-**H**), 7.18 – 7.07 (m, 2H, Ar-**H**), 6.59 (ddd, *J* = 10.3, 6.6, 3.1 Hz, 2H, Ar-**H**). |
| ¹⁹F NMR (376 MHz, DMSO): | -110.18 (t, *J*$_{F-H}$ = 8.6 Hz, 2F, Ar-**F**), -114.71 (d, *J*$_{F-H}$ = 10.5 Hz, 2F, Ar-**F**). |

### 3.3.2. Synthesis of *2,3',5',6-tetrafluoro-[1,1'-biphenyl]-4-yl 4-methylbenzenesulfonate*

*2,3',5',6-tetrafluoro-[1,1'-biphenyl]-4-ol* (9.68 g, 40 mmol) was dissolved in 100 mL of dry THF and dry NEt₃ (12 mL, 80 mmol) under a dry nitrogen atmosphere. The flask was then cooled over an ice-water bath before TsCl (11.46 g, 60 mmol), dissolved in 50 mL of dry THF, was added dropwise to the solution over 30 minutes before the ice bath was removed and the reaction allowed to progress until complete consumption of the phenolic component as judged by TLC (circa 14 h). The reaction was then extracted with 100 mL of EtOAc and washed 3x 50 mL of 2M NaOH. The organics were combined, dried over MgSO₄ and concentrated under reduced pressure before being purified by flash chromatography over silica gel using a gradient of hexanes:DCM using a Combiflash NextGen300+ system to yield *2,3',5',6-tetrafluoro-[1,1'-biphenyl]-4-yl 4-methylbenzenesulfonate* as a white, crystalline solid.

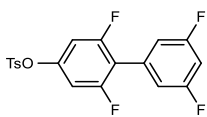

*2,3',5',6-tetrafluoro-[1,1'-biphenyl]-4-yl 4-methylbenzenesulfonate*

| | |
|---|---|
| Yield: | (White crystalline solid) 15.1 g, 95 % |
| R<sub>F</sub> (DCM): | 0.84 |
| ¹H NMR (400 MHz): | 7.92 (ddd, *J* = 8.4, 2.4, 1.8 Hz, 2H, Ar-**H**), 7.41 (d, *J* = 8.5 Hz, 2H, Ar-**H**), 2.49 (s, 3H, Ar-C**H₃**). |
| ⁹F NMR (376 MHz): | -109.57 (t, *J*<sub>F-H</sub> = 7.9 Hz, 2F, Ar-**F**), -110.96 (d, *J*<sub>F-H</sub> = 8.3 Hz, 2F, Ar-**F**). |

### 3.3.3. Synthesis of *2,3',5',6-tetrafluoro-4-propyl-1,1'-biphenyl*

In a flame dried flask, *2,3',5',6-tetrafluoro-[1,1'-biphenyl]-4-yl 4-methylbenzenesulfonate* (14.65 g, 37 mmol) and Fe(acac)₃ (1.31 g, 3.7 mmol) were dissolved in 250 mL dry THF:NMP (10:1). The solution was cooled over an ice bath before a solution of freshly prepared Grignard reagent (55 mmol) in dry Et₂O was added dropwise to the stirred solution. The progress of the reaction was followed by ¹H NMR with no further reaction occurring after 1.5 h. The solution was quenched with HCl (crica. 50 mL) before being extracted with EtOAc, dried over MgSO₄ and a yellow oil was retrieved under reduced pressure. The oil was purified by flash chromatography over silica gel using a gradient of hexanes:DCM using a Combiflash NextGen300+ system, yielding *2,3',5',6-tetrafluoro-4-propyl-1,1'-biphenyl* as a colourless oil.

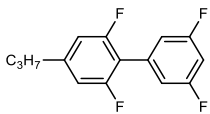

*2,3',5',6-tetrafluoro-4-propyl-1,1'-biphenyl*

| | |
|---|---|
| Yield: | (colourless oil) 7.74 g, 78 % |
| R<sub>F</sub> (EtOAc): | 0.94 |
| ¹H NMR (400 MHz): | 7.01 (m, 2H, Ar-**H**), 6.82 (ddd, *J* = 8.7, 3.1, 2.9 Hz, 2H, Ar-**H**), 2.61 (t, *J* = 7.6 Hz, 2H, Ar-C**H₂**-CH₂), 1.68 (h, *J* = 7.4 Hz, 2H, CH₂-C**H₂**-CH₃), 0.98 (t, *J* = 7.3 Hz, 2H, CH₂-C**H₃**). |
| ¹⁹F NMR (376 MHz): | -110.37 (t, *J*<sub>F-H</sub> = 8.1 Hz, 2F, Ar-**F**), -115.31 (d, *J*<sub>F-H</sub> = 9.2 Hz, 2F, Ar-**F**). |

### 3.3.4. Synthesis of *2',3,5,6'-tetrafluoro-4'-propyl-[1,1'-biphenyl]-4-carboxylic acid*

A solution of *n-butyl lithium* (1.6 M in hexanes, 25 mmol, 15 mL) was added dropwise to a stirred, cooled (-78 °C) solution of *2,3',5',6-tetrafluoro-4-propyl-1,1'-biphenyl* (4.4 g, 16.6 mmol) in anhydrous THF (150 mL) under an atmosphere of dry nitrogen. The aryl lithium solution was allowed to stir for 1.5 h at this same temperature before adding a large excess of gaseous carbon dioxide which was pre-dried by bubbling through conc. H₂SO₄. The solution was then allowed to slowly warm to ambient temperature (~ 2 h). The basic solution was acidified with 2M HCl (~100 ml) and extracted with EtOAc. The organics were then dried over MgSO₄ and a white solid retrieved under reduced pressure. These were then purified by recrystallization from MeCN to give 2',3,5,6'-tetrafluoro-4'-propyl-[1,1'-biphenyl]-4-carboxylic acidacid as a fine white solid.

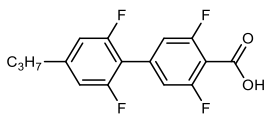

*2',3,5,6'-tetrafluoro-4'-propyl-[1,1'-biphenyl]-4-carboxylic acid*

| | |
|---|---|
| Yield: | (white needles) 4.35g, 84 % |
| R$_F$ (DCM): | 0.00 |
| ¹H NMR (400 MHz, DMSO): | 8.94 (d, *J* = 8.9 Hz, 2H, Ar-**H**), 8.70 (ddd, *J* = 9.1, 3.1, 1.9 Hz, 2H, Ar-**H**), 4.19 (t, *J* = 7.3 Hz, 2H, Ar-C**H₂**-CH₂), 3.20 (h, *J* = 7.4 Hz, 2H, CH₂-C**H₂**-CH₃), 2.48 (t, *J* = 7.3 Hz, 3H, CH₂-C**H₃**). |
| ¹⁹F NMR (376 MHz, DMSO): | -112.26 (d, *J*$_{F-H}$ = 9.5 Hz, 2F, Ar-**F**), -115.40 (d, *J*$_{F-H}$ = 9.6 Hz, 2F, Ar-**F**). |

## 3.4  General esterification procedure used in the preparation of materials FWXYZ and NCWXYZ

The general synthetic procedure is described in **Scheme S1**. A small vial was charged with the appropriate phenol (0.5 mmol, 1.0 eq), benzoic acid (0.6 mmol, 1.1 eq.), EDC.HCl (0.75 mmol, 1.5 eqv.) and DMAP (~ 2 mol%). Dichloromethane was added (conc. ~ 0.1 M) and the suspension stirred until complete consumption of the phenol as judged by TLC. Once complete, the reaction solution was concentrated and purified by flash chromatography over silica gel with a gradient of hexane/DCM using a Combiflash NextGen300+ system. To remove ionic impurities, a pre-column containing a small amount of neutral alumina was used before samples entered the silica gel column. The chromatographed material was dissolved into the minimum quantity of DCM, filtered through a 0.2 micron PTFE filter, concentrated to dryness and finally recrystalised from either MeCN or EtOH as indicated to afford the title materials as white solids.

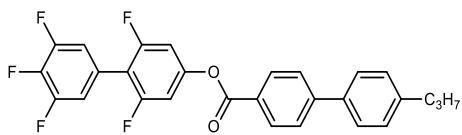

**F2200**

*2,3',4',5',6-pentafluoro-[1,1'-biphenyl]-4-yl 4'-(prop-1-yn-1-yl)-[1,1'-biphenyl]-4-carboxylate*

| | |
|---|---|
| Yield: | (crystalline white solid) 183 mg, 76 % |
| Re-crystallisation solvent: | MeCN |
| $^1$H NMR (400 MHz): | 8.25 (ddd, *J* = 8.3, 2.1, 1.7 Hz, 2H, Ar-**H**), 7.77 (ddd, *J* = 8.2, 1.7, 1.7 Hz, 2H, Ar-**H**), 7.61 (ddd, *J* = 8.1, 2.2, 1.5 Hz, 2H, Ar-**H**), 7.34 (d, *J* = 8.0 Hz, 2H, Ar-**H**), 7.16 (t, *J* = 7.2 Hz, 2H, Ar-**H**), 7.07 – 6.98 (m, 2H, Ar-**H**), 2.69 (t, *J* = 7.8 Hz, 2H, Ar-C**H$_2$**-CH$_2$), 1.73 (h, *J* = 7.5 Hz, 2H, CH$_2$-C**H$_2$**-CH$_3$), 1.02 (t, *J* = 7.3 Hz, 3H, CH2-C**H$_3$**). |
| $^{13}$C{$^1$H} NMR (101 MHz): | 164.24, 159.77 (dd, *J* = 250.1, 8.6 Hz), 152.33 (ddd, *J* = 250.1, 9.2, 4.5 Hz), 151.54 (t, *J* = 14.2 Hz), 147.01, 143.44, 141.04 (t, *J* = 15.9 Hz), 136.91, 130.86, 129.22, 127.18, 126.78, 124.49 – 124.20 (m), 114.83 (dd, *J* = 21.4, 6.4 Hz), 107.04 – 106.21 (m), 37.74, 24.52, 13.86. |
| $^{19}$F NMR (376 MHz): | -112.54 (d, *J$_{F-H}$* = 8.9 Hz, 2F, Ar-**F**), -134.36 (dd, *J$_{F-F}$* = 20.5 Hz, *J$_{F-H}$* = 8.5 Hz, 2F, Ar-**F**), -160.04 (tt, *J$_{F-F}$* = 20.4 Hz, *J$_{F-H}$* = 6.4 Hz, 1F, Ar-**F**). |

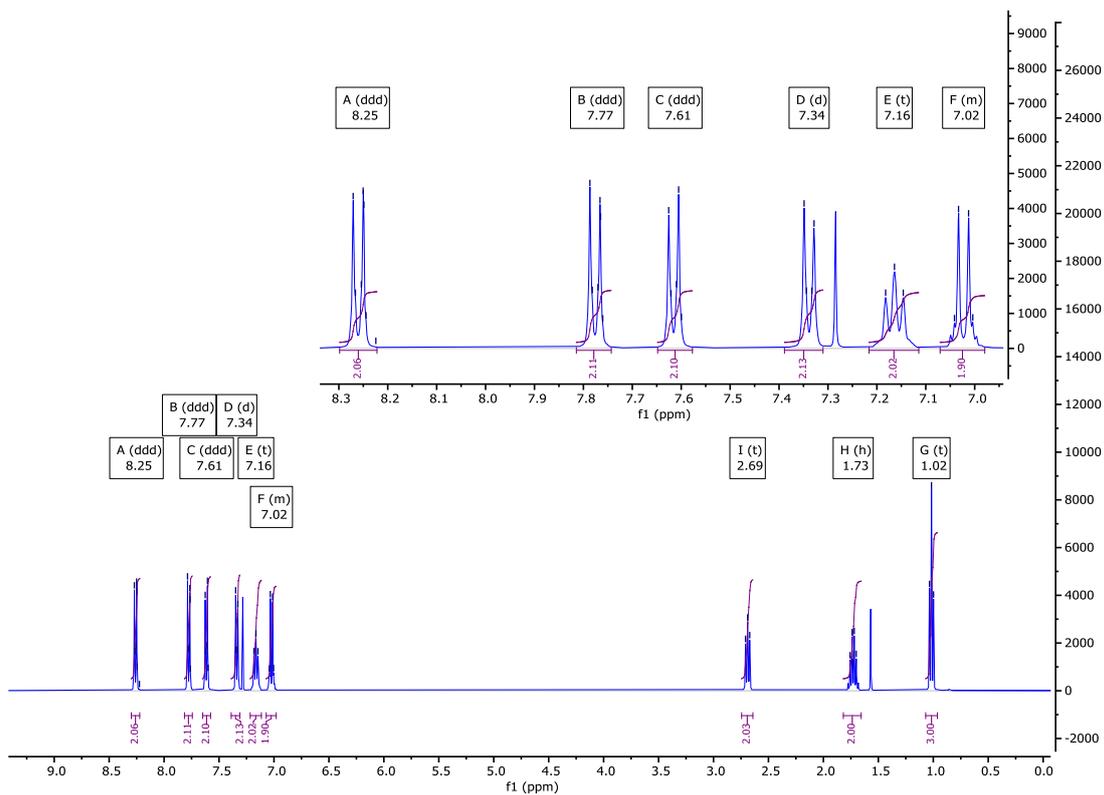

**Fig. S9** $^1$H NMR spectra of **F2200** in CDCl$_3$.

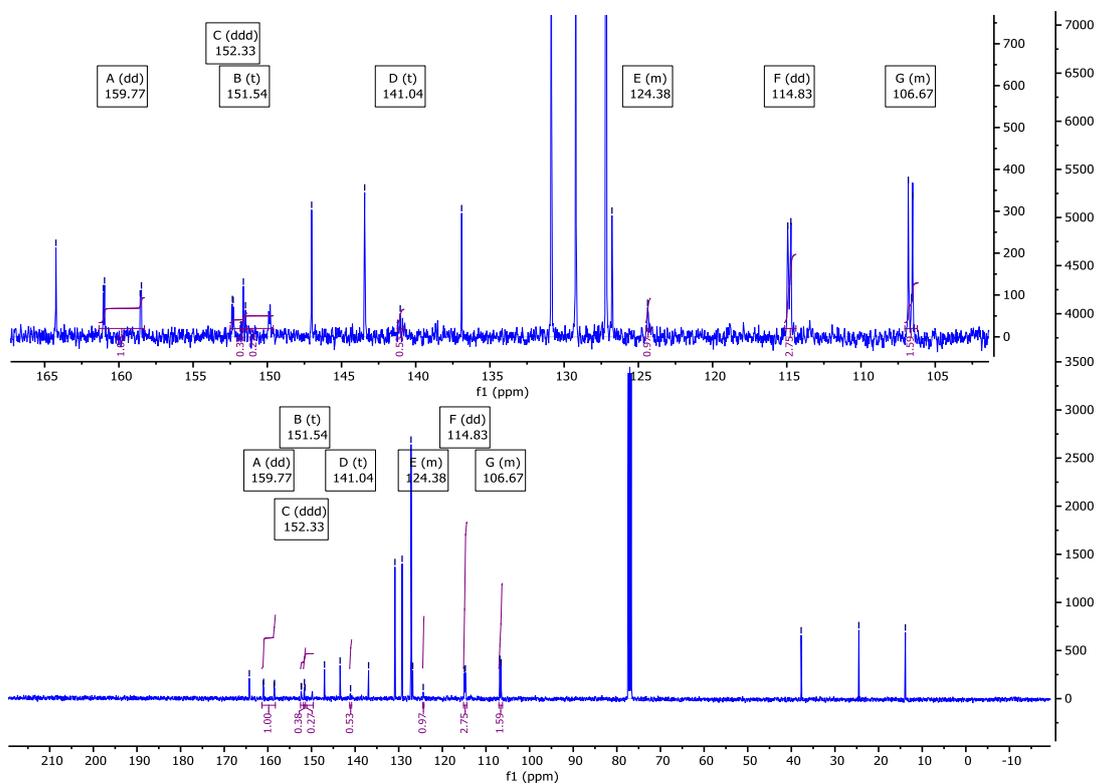

**Fig. S10** $^{13}$C{$^1$H} NMR spectra of **F2200** in CDCl$_3$.

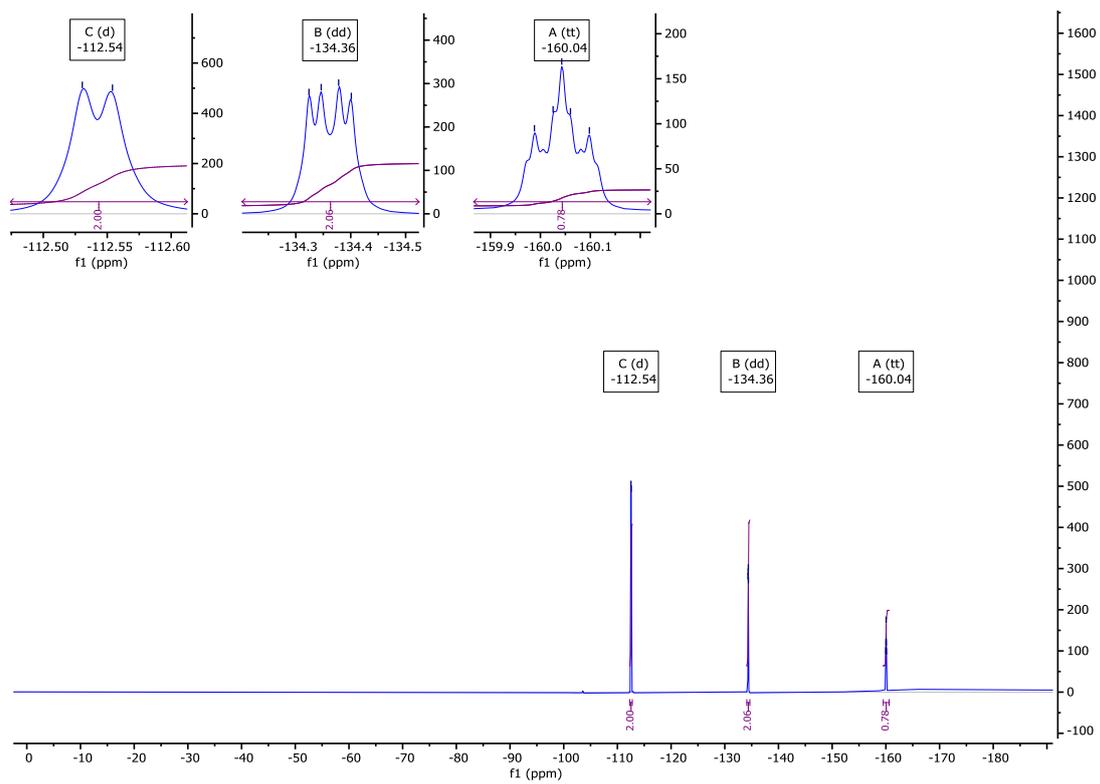

**Fig. S11** $^{19}$F NMR spectra of **F2200** in CDCl$_3$.

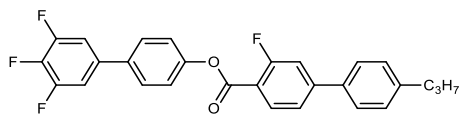

**F2010**

*3',4',5'-trifluoro-[1,1'-biphenyl]-4-yl 3-fluoro-4'-(prop-1-yn-1-yl)-[1,1'-biphenyl]-4-carboxylate*

| | |
|---|---|
| Yield: | (crystalline white solid) 188 mg, 81 % |
| Re-crystallisation solvent: | MeCN |
| $^1$H NMR (400 MHz): | 8.16 (t, *J* = 7.9 Hz, 1H, Ar-**H**), 7.62 – 7.54 (m, 4H, Ar-**H**)*, 7.52 (dd, *J* = 8.2, 1.7 Hz, 1H, Ar-**H**), 7.44 (dd, *J* = 12.1, 1.7 Hz, 1H, Ar-**H**), 7.38 – 7.28 (m, 4H, Ar-**H**), 7.20 (dd, *J* = 8.4, 6.3 Hz, 2H, Ar-**H**), 2.66 (t, *J* = 7.6 Hz, 2H, Ar-C**H$_2$**-CH$_2$), 1.70 (h, *J* = 7.4 Hz, 2H, CH$_2$-C**H$_2$**-CH$_3$), 0.99 (t, *J* = 7.3 Hz, 3H, CH$_2$-C**H$_3$**). *Overlapping Signals.. |
| $^{13}$C{$^1$H} NMR (101 MHz): | 162.75 (d, *J* = 261.5 Hz), 162.61 (d, *J* = 3.9 Hz), 151.81 (ddd, *J* = 249.8, 9.2, 3.7 Hz), 150.93, 148.81 (d, *J* = 8.8 Hz), 143.99, 136.49 (d, *J* = 4.3 Hz), 135.93 (d, *J* = 33.2 Hz), 132.98, 129.30, 128.04, 127.06, 122.45, 115.82 (t, *J* = 5.6 Hz), 115.26 (d, *J* = 23.3 Hz), 111.33 – 110.95 (m), 37.74, 24.48, 13.84. |
| $^{19}$F NMR (376 MHz): | -107.57 (dd, 1F, *J$_{F-H}$* = 12.2 Hz, *J$_{F-H}$* = 7.5 Hz, Ar-**F**), -133.92 (dd, 2F, *J$_{F-F}$* = 20.6 Hz, *J$_{F-H}$* = 8.9 Hz, Ar-**F**), -162.43 (tt, 1F, *J$_{F-F}$* = 20.3 Hz, *J$_{F-H}$* = 6.3 Hz, Ar-**F**). |

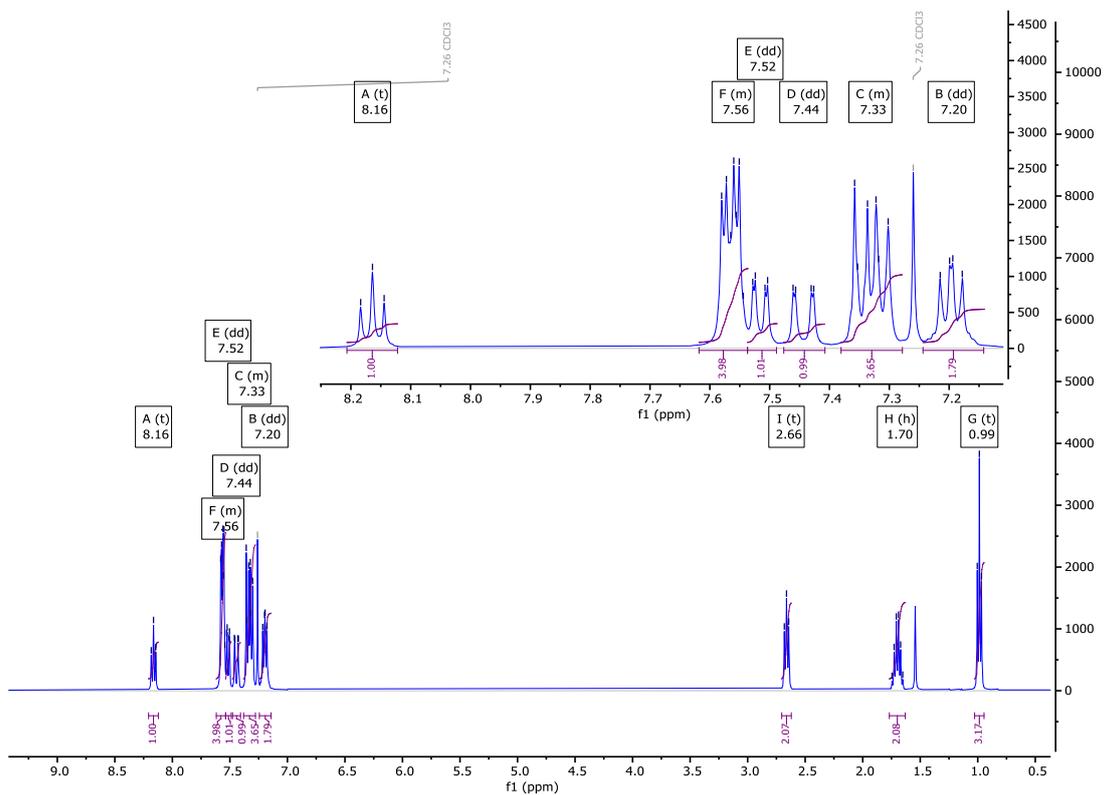

**Fig. S12**  $^1$H NMR spectra of **F2010** in CDCl$_3$.

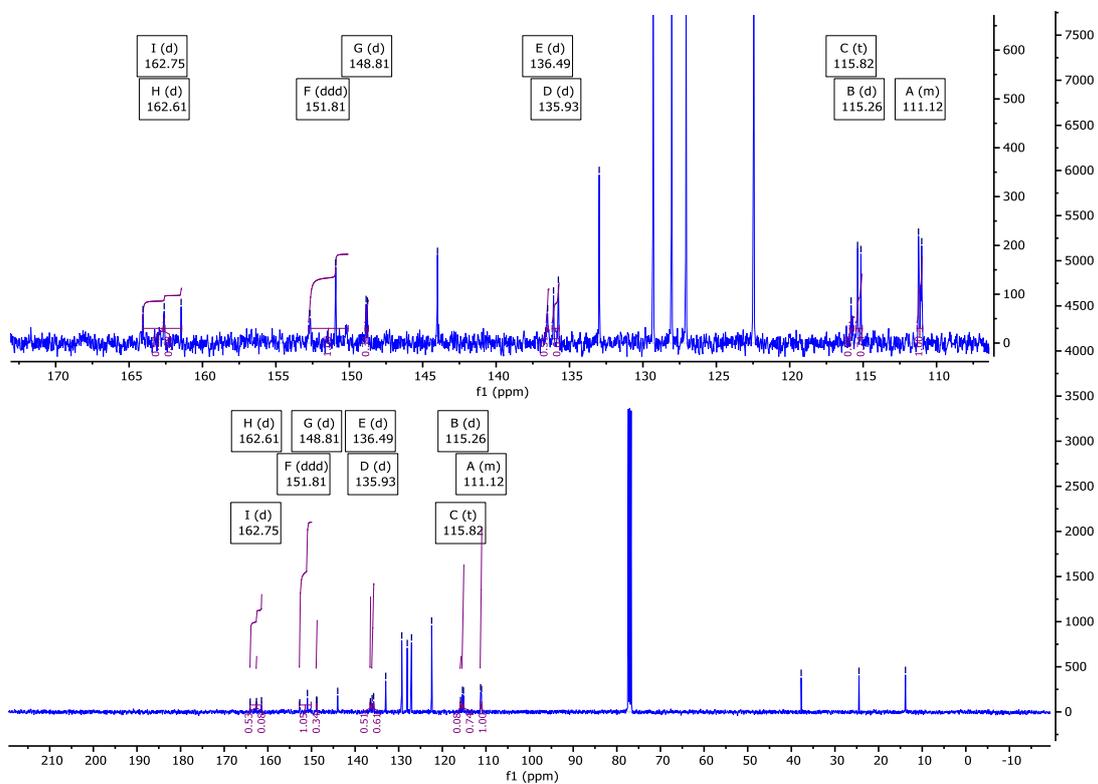

**Fig. S13**  $^{13}$C{$^1$H} NMR spectra of **F2010** in CDCl$_3$.

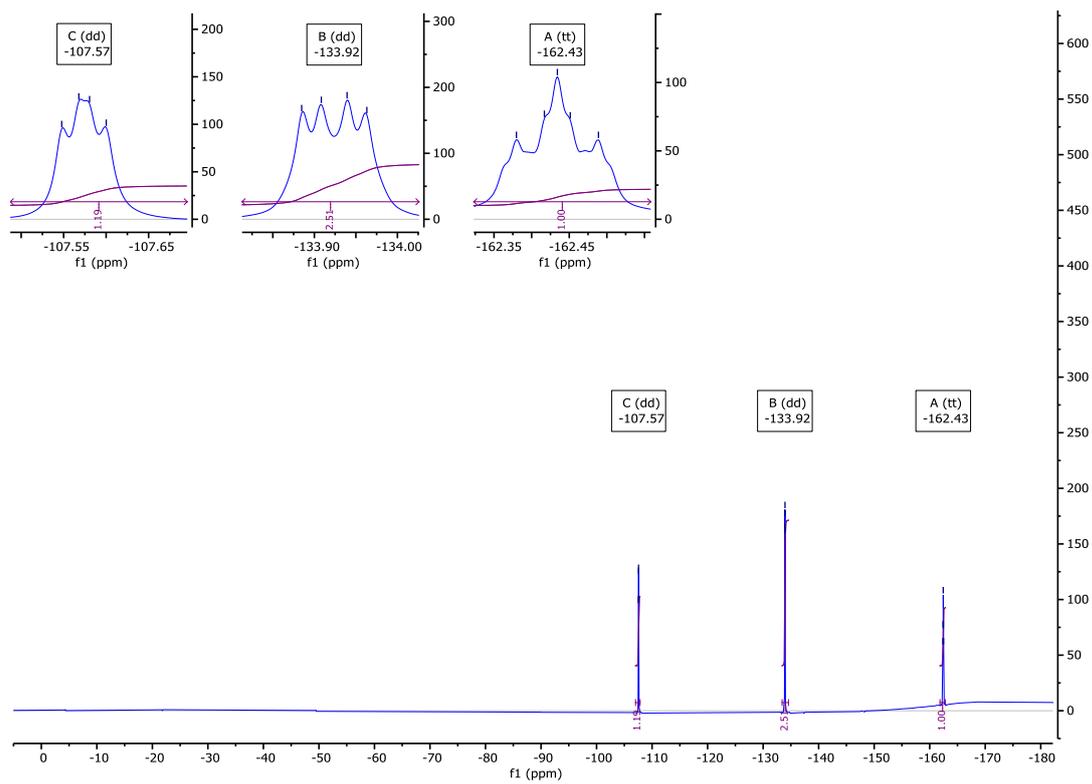

**Fig. S14**   $^{19}$F NMR spectra of **F2010** in CDCl$_3$.

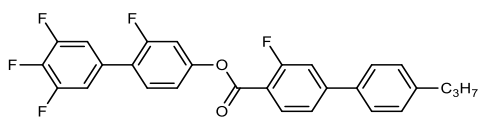

**F2110**

*2,3',4',5'-tetrafluoro-[1,1'-biphenyl]-4-yl 3-fluoro-4'-propyl-[1,1'-biphenyl]-4-carboxylate*

| | |
|---|---|
| Yield: | (white solid) 193 mg, 80 % |
| Re-crystallisation solvent: | MeCN |
| $^1$H NMR (400 MHz: | 8.15 (t, *J* = 7.9 Hz, 1H, Ar-**H**), 7.57 (d, *J* = 7.8 Hz, 2H, Ar-**H**), 7.54 – 7.40 (m, 3H)*, 7.31 (d, *J* = 7.8 Hz, 2H, Ar-**H**), 7.24 – 7.13 (m, 4H, Ar-**H**)*, 2.66 (t, *J* = 7.7 Hz, 2H, Ar-C**H$_2$**-CH$_2$), 1.69 (h, *J* = 7.5 Hz, 2H, CH$_2$-C**H$_2$**-CH$_3$), 0.99 (t, *J* = 7.4 Hz, 3H, CH$_2$-C**H$_3$**). *Overlapping Signals. |
| $^{13}$C{$^1$H} NMR (101 MHz): | 162.79 (d, *J* = 261.9 Hz), 162.18 (d, *J* = 4.3 Hz), 159.40 (d, *J* = 251.3 Hz), 151.43 (d, *J* = 11.1 Hz), 151.22 (ddd, *J* = 249.8, 9.9, 4.2 Hz), 149.08 (d, *J* = 9.2 Hz), 135.67, 132.99, 131.19 – 130.81 (m), 130.59 (d, *J* = 3.9 Hz), 129.32, 127.06, 124.13 – 123.86 (m), 122.52 (d, *J* = 3.3 Hz), 118.29 (d, *J* = 3.7 Hz), 115.30 (d, *J* = 22.7 Hz), 113.48 – 112.97 (m), 110.77 (d, *J* = 25.7 Hz), 37.74, 24.48, 13.84. |
| $^{19}$F NMR (376 MHz): | -107.41 (t, 1F, *J$_{F-H}$* = 8.5 Hz, Ar-**F**), -114.57 (t, 1F *J$_{F-H}$* = 9.8 Hz, Ar-**F**), -134.21 (dd, 2F *J$_{F-F}$* = 20.4 Hz, *J$_{F-H}$* = 8.9 Hz, Ar-**F**), -161.25 (tt, 1F, *J$_{F-F}$* = 20.5 Hz, *J$_{F-H}$* = 6.5 Hz, Ar-**F**). |

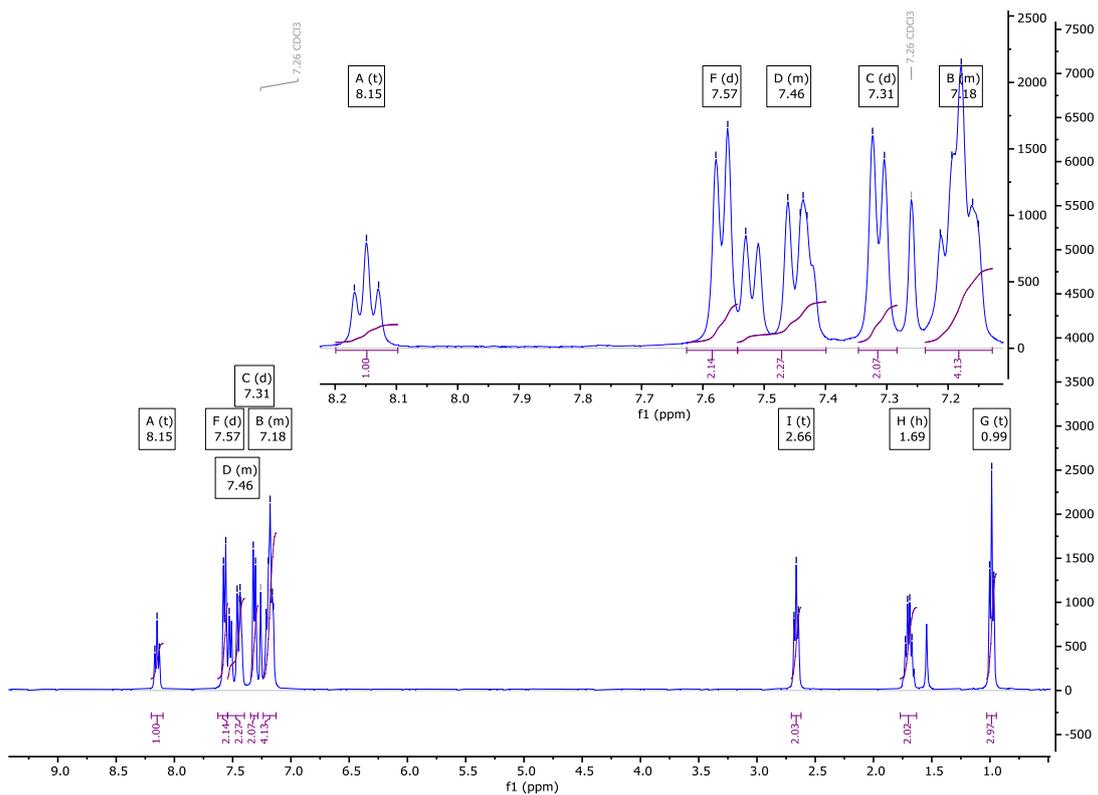

**Fig. S15** $^1$H NMR spectra of **F2110** in CDCl$_3$.

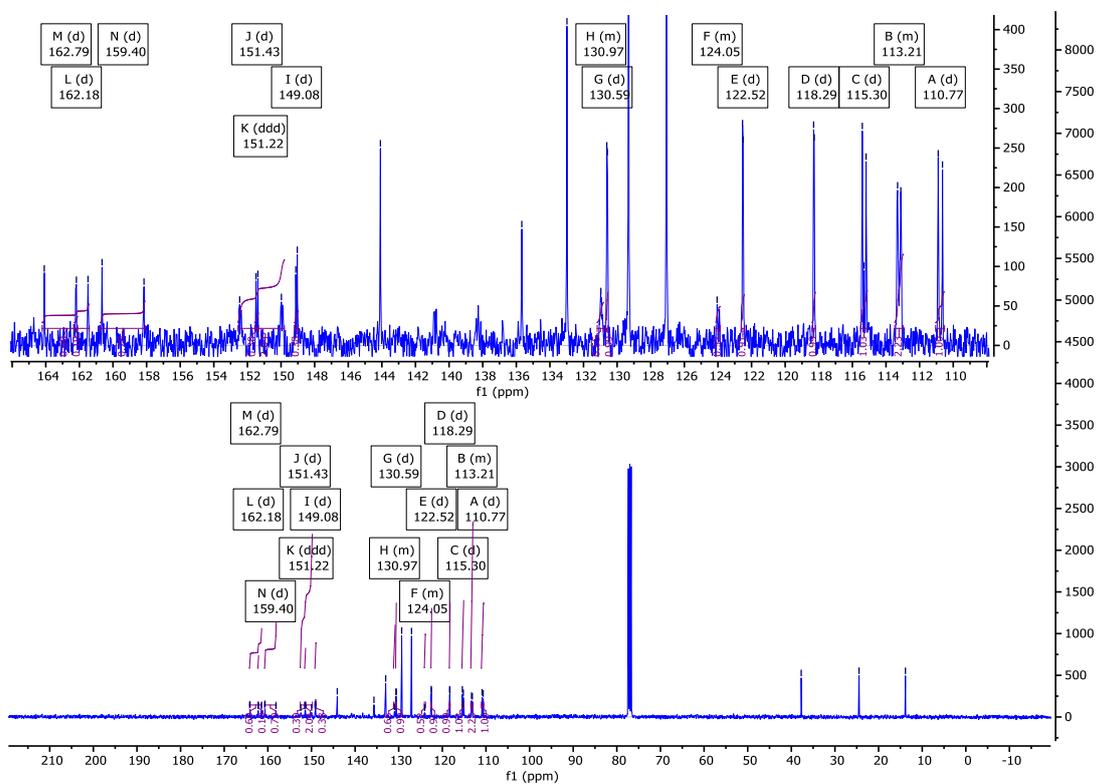

**Fig. S16** $^{13}$C{$^1$H} NMR spectra of **F2110** in CDCl$_3$.

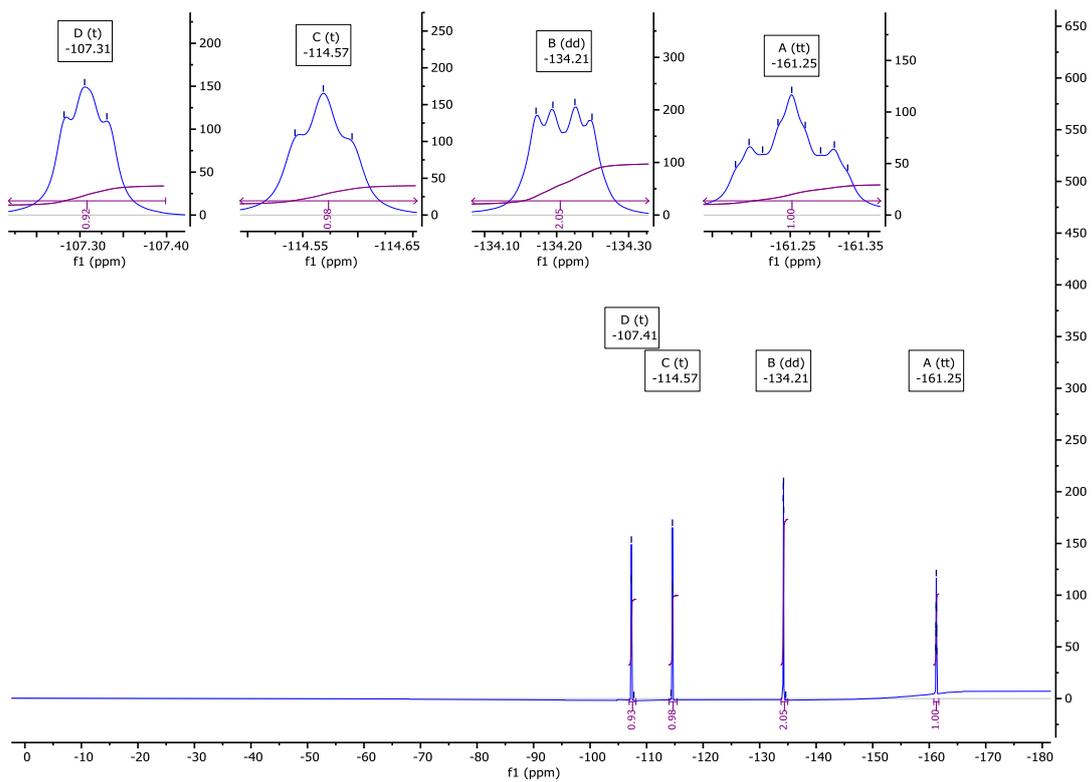

**Fig. S17**  ¹⁹F NMR spectra of **F2110** in CDCl$_3$.

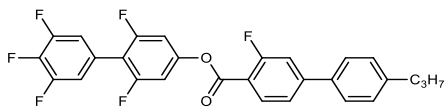

**F2210**

*2,3',4',5',6-pentafluoro-[1,1'-biphenyl]-4-yl 3-fluoro-4'-propyl-[1,1'-biphenyl]-4-carboxylate*

| | |
|---|---|
| Yield: | (white needles) 183 mg, 73 % |
| Re-crystallization solvent: | MeCN |
| $^1$H NMR (400 MHz): | δ 8.13 (t, *J* = 8.2 Hz, 1H, Ar-**H**), 7.57 (ddd, *J* = 8.3, 1.9, 1.8 Hz, 2H, Ar-**H**), 7.52 (dd, *J* = 8.2, 1.7 Hz, 1H, Ar-**H**), 7.45 (dd, *J* = 12.2, 1.7 Hz, 1H, Ar-**H**), 7.32 (ddd, *J* = 8.2, 1.8, 1.8 Hz, 2H, Ar-**H**), 7.14 (t, *J* = 7.4 Hz, 2H, Ar-**H**), 7.07 – 6.97 (m, 2H, Ar-**H**), 2.67 (t, *J* = 6.7 Hz, 2H, Ar-C**H$_2$**-CH$_2$), 1.70 (h, *J* = 7.3 Hz, 2H, CH$_2$-C**H$_2$**-CH$_3$), 0.99 (t, *J* = 7.3 Hz, 3H, CH$_2$-C**H$_3$**). |
| $^{13}$C{$^1$H} NMR (101 MHz): | 162.81 (d, *J* = 261.9 Hz), 161.78 (d, *J* = 4.0 Hz), 159.74 (dd, *J* = 250.3, 8.6 Hz), 151.22 (t, *J* = 14.3 Hz), 151.07 (ddd, *J* = 250.0, 9.8, 3.9 Hz), 149.34 (d, *J* = 8.9 Hz), 144.18, 139.79 (dt, *J* = 253.6, 15.7 Hz), 135.56, 132.99, 129.34, 127.06, 124.32 (dd, *J* = 5.1, 3.4 Hz), 122.57 (d, *J* = 3.2 Hz), 115.32 (d, *J* = 23.0 Hz), 115.11 – 114.52 (m), 113.58 (t, *J* = 17.9 Hz), 106.99 – 106.28 (m), 37.74, 24.47, 13.83. |
| $^{19}$F NMR (376 MHz): | -107.02 (dd, *J$_{F-H}$* = 12.3 Hz, *J$_{F-H}$* = 7.7 Hz, 1F, Ar-**F**), -112.47 (d, *J$_{F-H}$* = 8.9 Hz, 2F, Ar-**F**), -134.34 (dd, *J$_{F-F}$* = 20.7 Hz, *J$_{F-H}$* = 8.5 Hz, 2F, Ar-**F**), -160.01 (tt, *J$_{F-F}$* = 20.7 Hz, *J$_{F-H}$* = 6.6 Hz, 1F, Ar-**F**). |

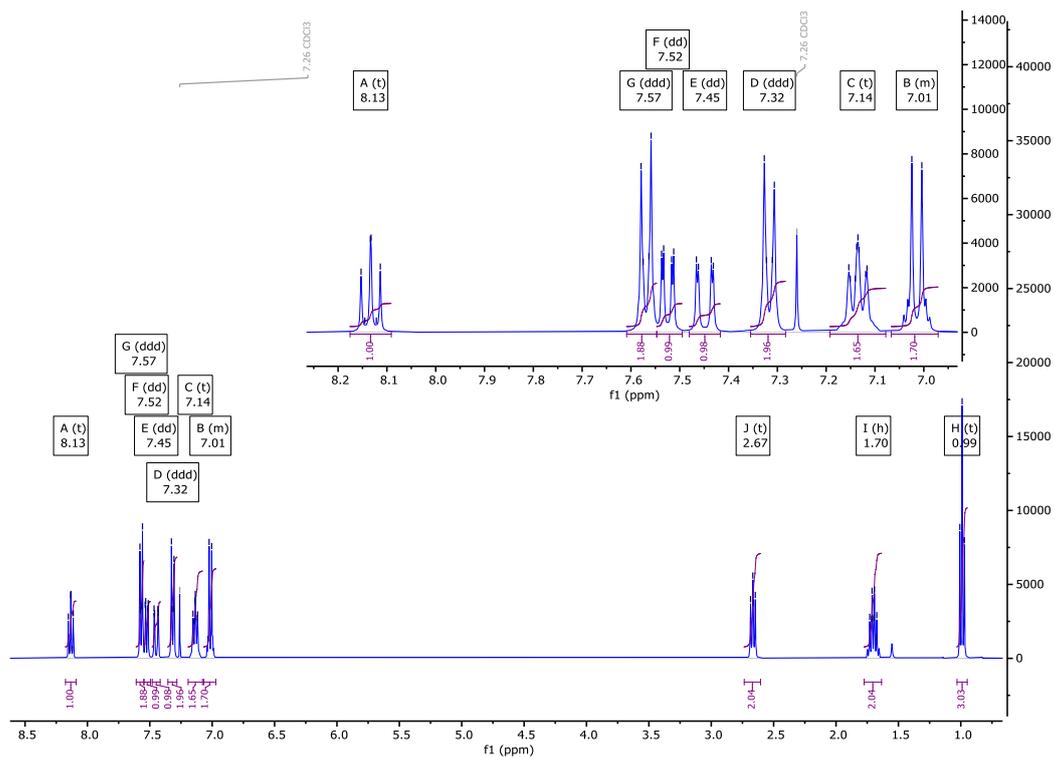

**Fig. S18** $^1$H NMR spectra of **F2210** in CDCl$_3$.

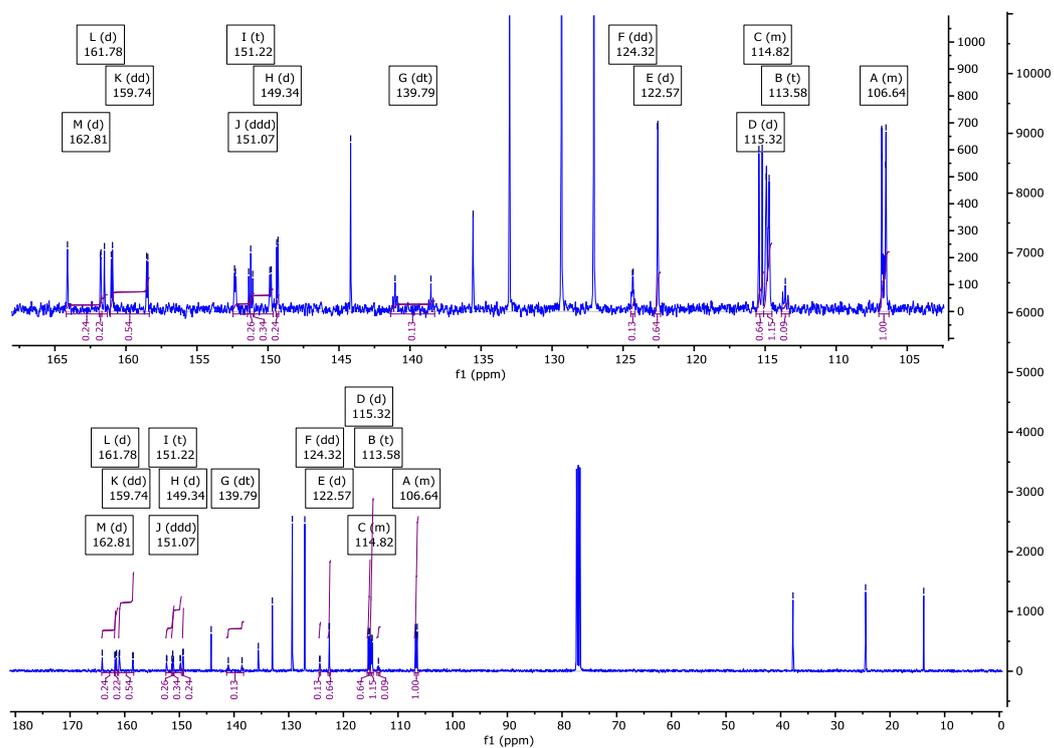

**Fig. S19** $^{13}$C{$^1$H} NMR spectra of **F2210** in CDCl$_3$.

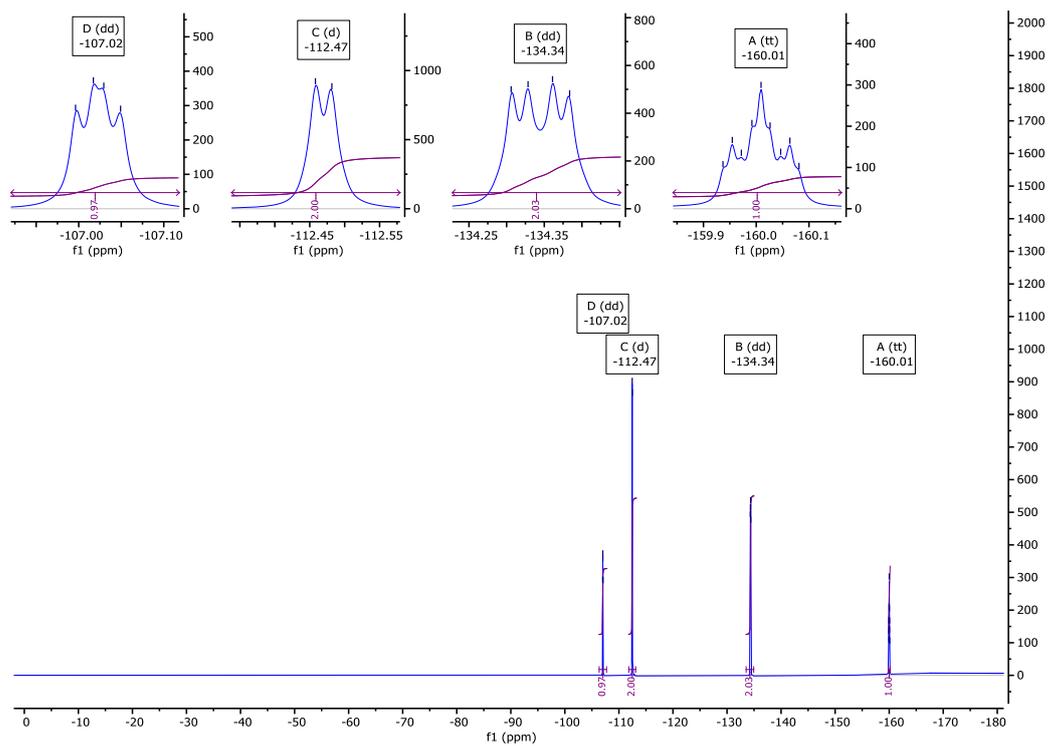

**Fig. S20**  $^{19}$F NMR spectra of **F2210** in CDCl$_3$.

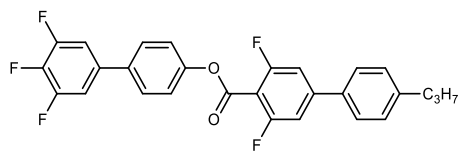

**F2020**

*3',4',5'-trifluoro-[1,1'-biphenyl]-4-yl 3,5-difluoro-4'-propyl-[1,1'-biphenyl]-4-carboxylate*

| | |
|---|---|
| Yield: | (white crystalline solid) 186 mg, 77 % |
| Re-crystallisation solvent: | MeCN |
| $^1$H NMR (400 MHz): | 7.56 (ddd, *J* = 8.6, 3.0, 1.8 Hz, 2H, Ar-**H**), 7.53 (ddd, *J* = 8.1, 2.2, 1.3 Hz, 2H, Ar-**H**), 7.37 (ddd, *J* = 8.7, 2.6, 2.1 Hz, 2H, Ar-**H**), 7.31 (ddd, *J* = 8.0, 2.2, 1.9 Hz, 2H, Ar-**H**), 7.29 – 7.24 (m, 2H, Ar-**H**)†, 7.23 – 7.16 (m, 2H, Ar-**H**), 2.66 (t, *J* = 7.5 Hz, 2H, Ar-C**H$_2$**-CH$_2$), 1.69 (h, *J* = 7.4 Hz, 2H, CH$_2$-C**H$_2$**-CH$_3$), 0.98 (t, *J* = 7.3 Hz, 3H, CH$_2$-C**H$_3$**). † Overlapping CDCl$_3$. |
| $^{13}$C{$^1$H} NMR (101 MHz): | 161.46 (dd, *J* = 258.6, 5.7 Hz), 158.74 (dd, *J* = 250.7, 8.5 Hz), 151.05 (ddd, *J* = 250.2, 10.6, 4.9 Hz), 148.26 (dt, *J* = 261.0, 15.0 Hz), 144.68, 134.76, 129.43, 126.89, 124.39 – 124.03 (m), 114.84 (dd, *J* = 16.8, 5.6 Hz), 110.52 (dd, *J* = 22.9, 2.4 Hz), 107.22 – 106.08 (m), 37.73, 24.43, 13.81. |
| $^{19}$F NMR (376 MHz): | -108.55 (d, *J$_{F-H}$* = 10.4 Hz, 2F, Ar-**F**), -133.87 (dd, *J$_{F-F}$* = 20.4 Hz, *J$_{F-H}$* = 8.9 Hz, 2F, Ar-**F**), -162.33 (tt, *J$_{F-F}$* = 20.6 Hz, *J$_{F-H}$* = 6.5 Hz, 1F, Ar-**F**). |

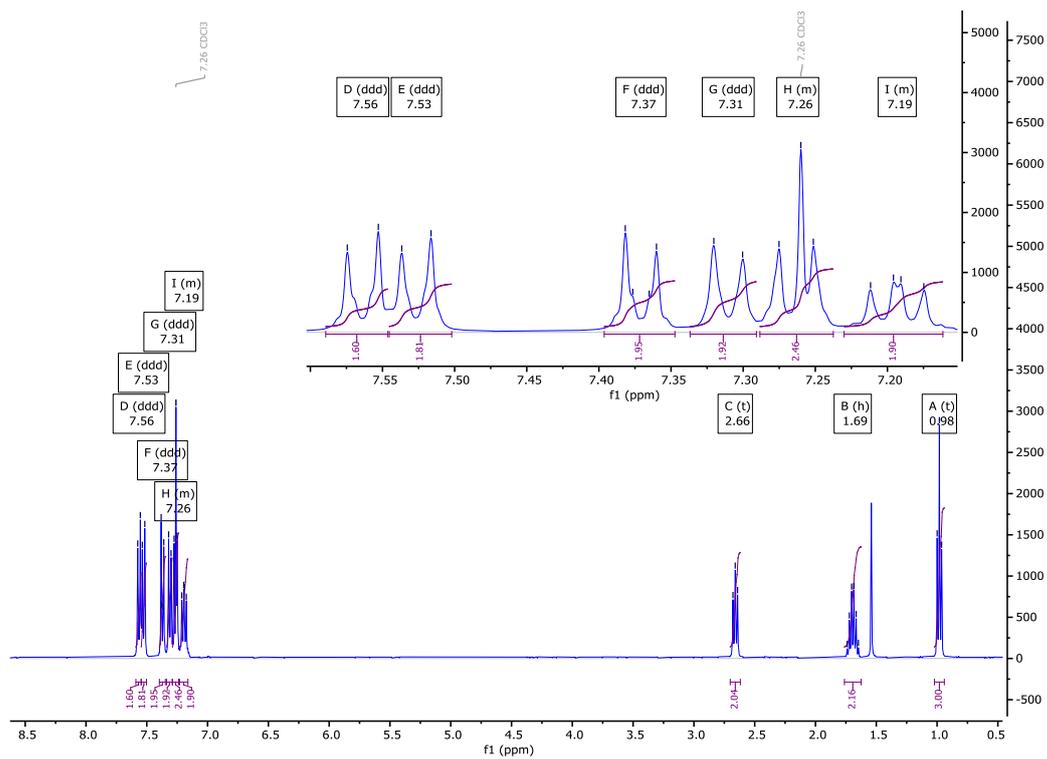

**Fig. S21** $^1$H NMR spectra of **F2020** in CDCl$_3$.

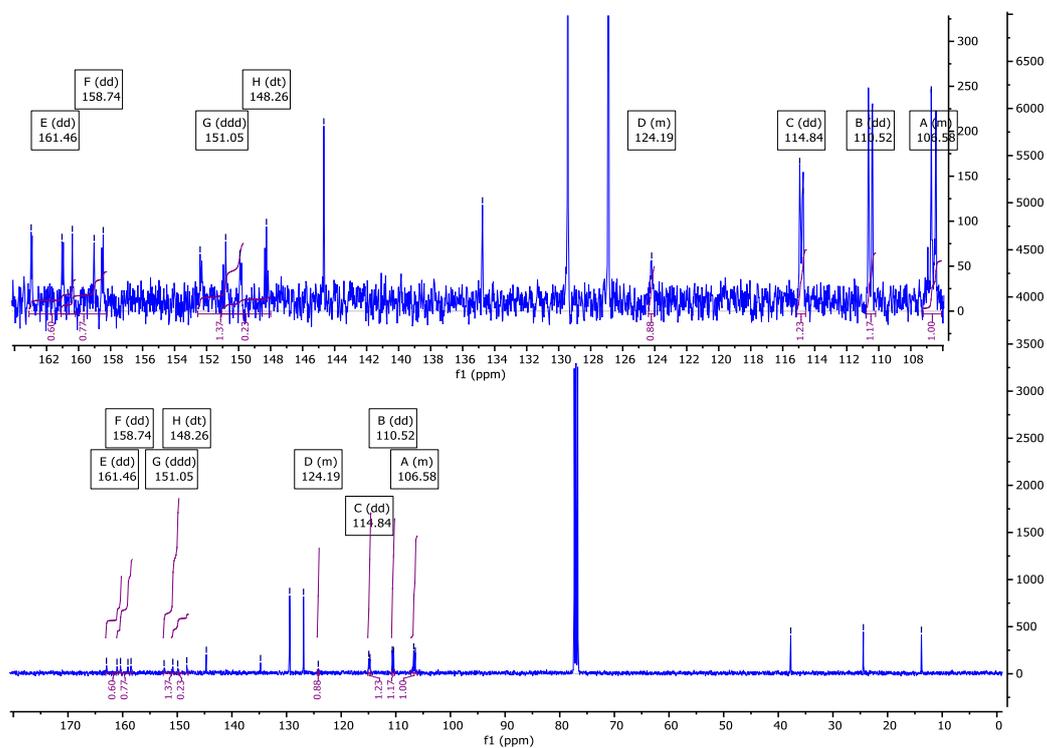

**Fig. S22** $^{13}$C{$^1$H} NMR spectra of **F2020** in CDCl$_3$.

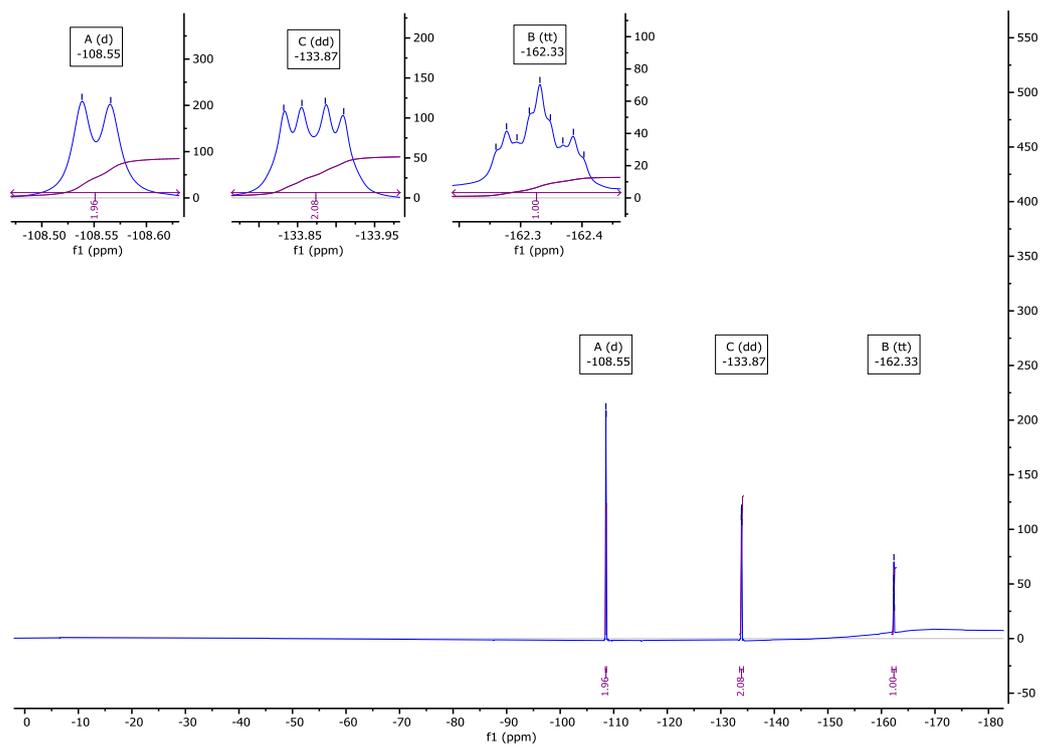

**Fig. S23** $^{19}$F NMR spectra of **F2020** in CDCl$_3$.

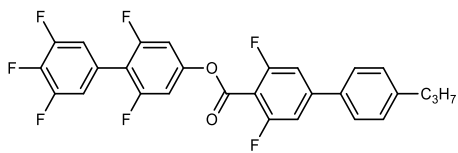

**F2220**

*2,3',4',5',6-pentafluoro-[1,1'-biphenyl]-4-yl 3,5-difluoro-4'-propyl-[1,1'-biphenyl]-4-carboxylate*

| | |
|---|---|
| Yield: | (white needles) 173 mg, 69 % |
| Re-crystallisation solvent: | MeCN |
| $^1$H NMR (400 MHz): | 7.53 (d, *J* = 7.8 Hz, 2H, Ar-**H**), 7.36 – 7.23 (m, 4H, Ar-**H**)*†, 7.13 (t, *J* = 7.3 Hz, 2H, Ar-**H**), 7.04 (d, *J* = 8.2 Hz, 2H, Ar-**H**), 2.66 (t, *J* = 7.6 Hz, 2H, Ar-C**H₂**-CH₂), 1.69 (h, *J* = 7.4 Hz, 2H, CH₂-C**H₂**-CH₃), 0.98 (t, *J* = 7.3 Hz, 3H, CH₂-C**H₃**). *Overlapping Signals, overlapping CDCl₃. |
| $^{13}$C{$^1$H} NMR (101 MHz): | 161.46 (dd, *J* = 258.6, 5.7 Hz), 159.03, 158.74 (dd, *J* = 250.7, 8.5 Hz), 151.05 (ddd, *J* = 250.2, 10.6, 4.9 Hz), 148.26 (dt, *J* = 261.0, 15.0 Hz), 144.68, 134.76, 129.43, 126.89, 124.39 – 124.03 (m), 114.84 (dd, *J* = 16.8, 5.6 Hz), 110.52 (dd, *J* = 22.9, 2.4 Hz), 107.22 – 106.08 (m), 37.73, 24.43, 13.81. |
| $^{19}$F NMR (376 MHz): | -107.97 (d, *J*$_{F-H}$ = 10.8 Hz, 2F, Ar-**F**), -112.20 (d, *J*$_{F-H}$ = 8.8 Hz, 2F, Ar-**F**), -134.30 (dd, *J*$_{F-F}$ = 20.6 Hz, *J*$_{F-H}$ = 8.7 Hz, 2F, Ar-**F**), -159.92 (tt, *J*$_{F-F}$ = 20.5 Hz, *J*$_{F-H}$ = 6.5 Hz, 1F, Ar-**F**). |

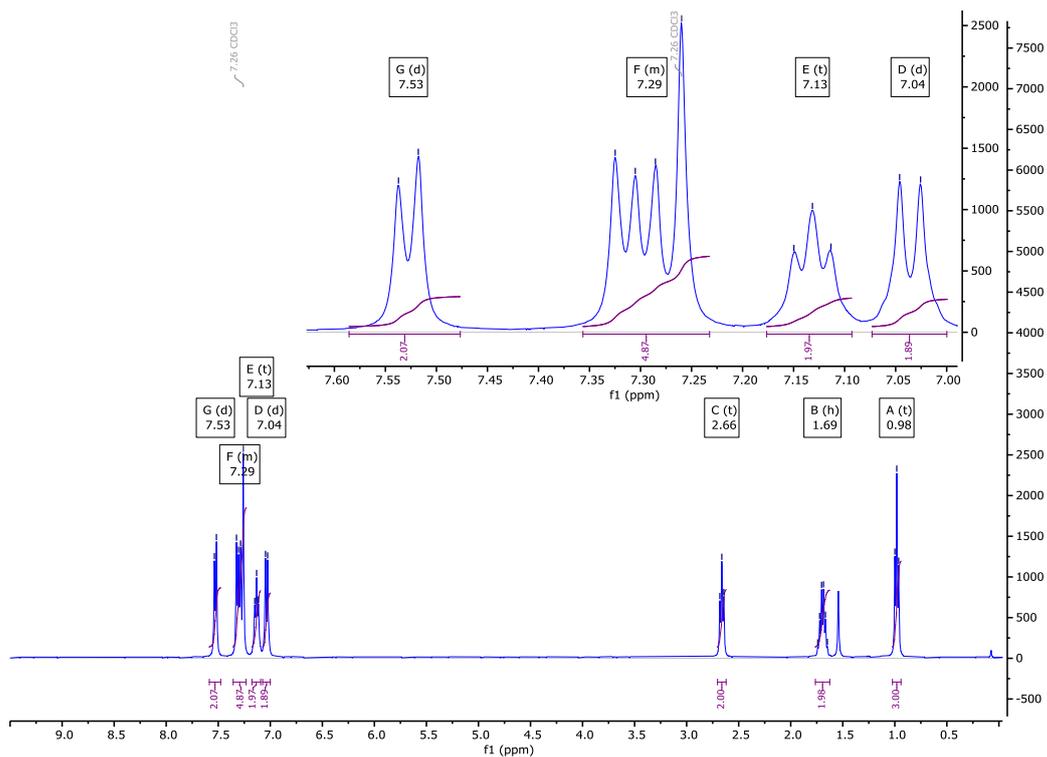

**Fig. S24** $^1$H NMR spectra of **F2220** in CDCl$_3$.

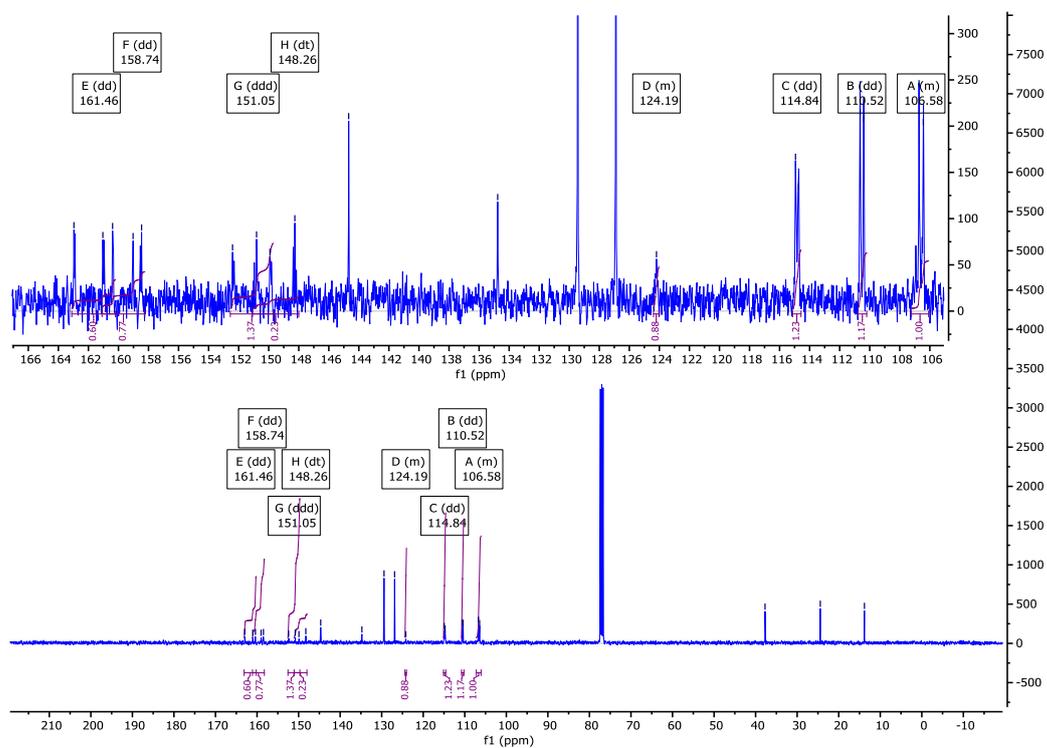

**Fig. S25** $^{13}$C{$^1$H} NMR spectra of **F2220** in CDCl$_3$.

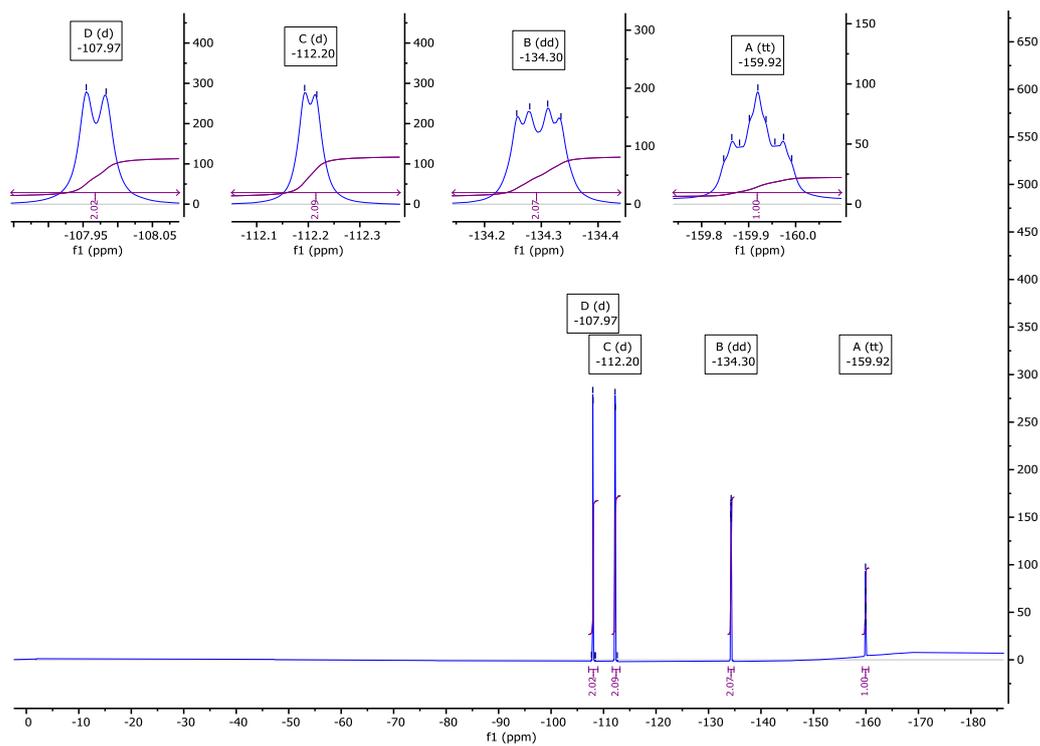

**Fig. S26** $^{19}$F NMR spectra of **F2220** in CDCl$_3$.

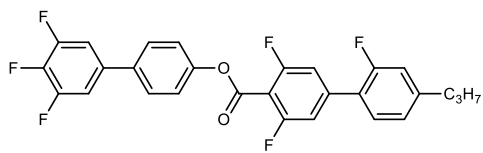

**F2021**

*3',4',5'-trifluoro-[1,1'-biphenyl]-4-yl 2',3,5-trifluoro-4'-propyl-[1,1'-biphenyl]-4-carboxylate*

| | |
|---|---|
| Yield: | (white crystalline solid) 155 mg, 31 % |
| Re-crystallisation solvent: | MeCN |
| $^1$H NMR (400 MHz: | 7.56 (ddd, *J* = 8.5, 3.0, 1.8 Hz, 2H, Ar-**H**), 7.41 – 7.33 (m, 3H, Ar-**H**)*, 7.30 – 7.23 (m, 2H, Ar-**H**)†, 7.23 – 7.15 (m, 2H, Ar-**H**)*, 7.09 (dd, *J* = 8.0, 1.6 Hz, 1H, Ar-**H**), 7.03 (dd, *J* = 12.0, 1.6 Hz, 1H, Ar-**H**), 2.65 (t, *J* = 7.6 Hz, 2H, Ar-C**H₂**-CH₂), 1.69 (h, *J* = 7.3 Hz, 2H, CH₂-C**H₂**-CH₃), 0.99 (t, *J* = 7.3 Hz, 3H, CH₂-C**H₃**). † Overlapping CDCl₃, *Overlapping Signals. |
| $^{13}$C{$^1$H} NMR (101 MHz): | 162.42 – 159.57 (m)*, 159.65 (d, *J* = 250.2 Hz), 151.68 (ddd, *J* = 249.9, 9.9, 4.5 Hz), 150.61, 146.93 (d, *J* = 7.8 Hz), 142.28 (t, *J* = 10.8 Hz), 139.39 (dt, *J* = 252.2, 15.1 Hz), 136.40, 129.77 (d, *J* = 3.0 Hz), 128.10, 125.01 (d, *J* = 3.0 Hz), 122.85 (dt, *J* = 12.8, 2.1 Hz) 122.33, 116.55, 116.33, 112.63 (dt, *J* = 22.5, 3.3 Hz), 111.36 – 110.97 (m), 108.46 (t, *J* = 17.1 Hz), 37.54, 24.14, 13.72. *Overlapping Signals. |
| $^{19}$F NMR (376 MHz): | -109.08 (d, 2F, *J*$_{F-H}$ = 10.4 Hz, Ar-**F**), -117.58 (t,1F *J*$_{F-H}$ = 9.1 Hz , Ar-**F**), -133.86 (dd, 2F *J*$_{F-F}$ = 20.6 Hz, *J*$_{F-H}$ = 8.7 Hz, Ar-**F**), -162.32 (tt, 1F, *J*$_{F-F}$ = 14.1 Hz, *J*$_{F-H}$ 6.6 Hz, Ar-**F**). |

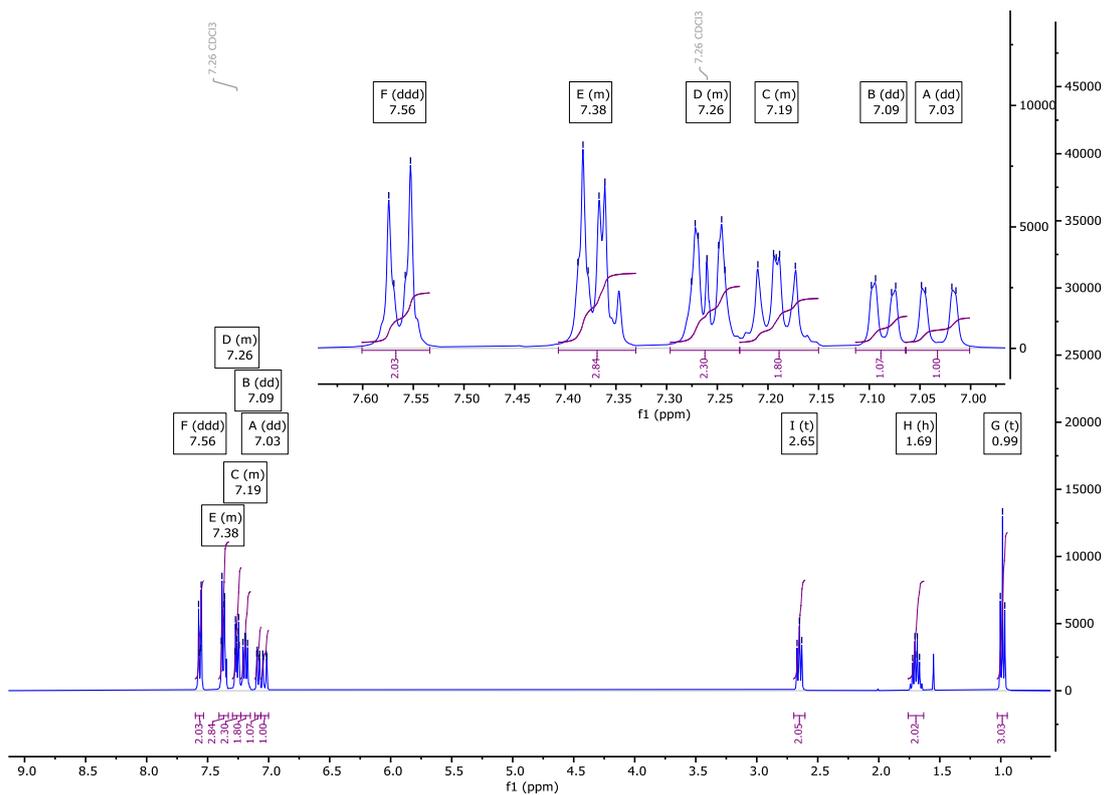

**Fig. S27** $^1$H NMR spectra of **F2021** in CDCl$_3$.

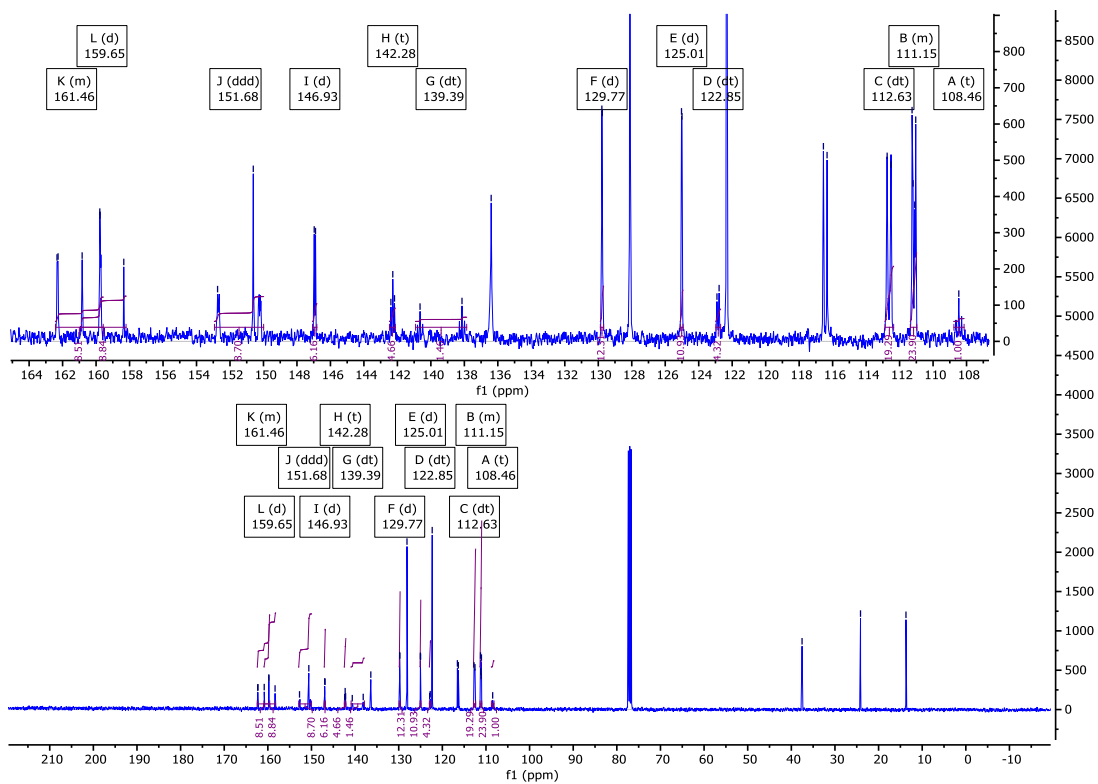

**Fig. S28** $^{13}$C{$^1$H} NMR spectra of **F2021** in CDCl$_3$.

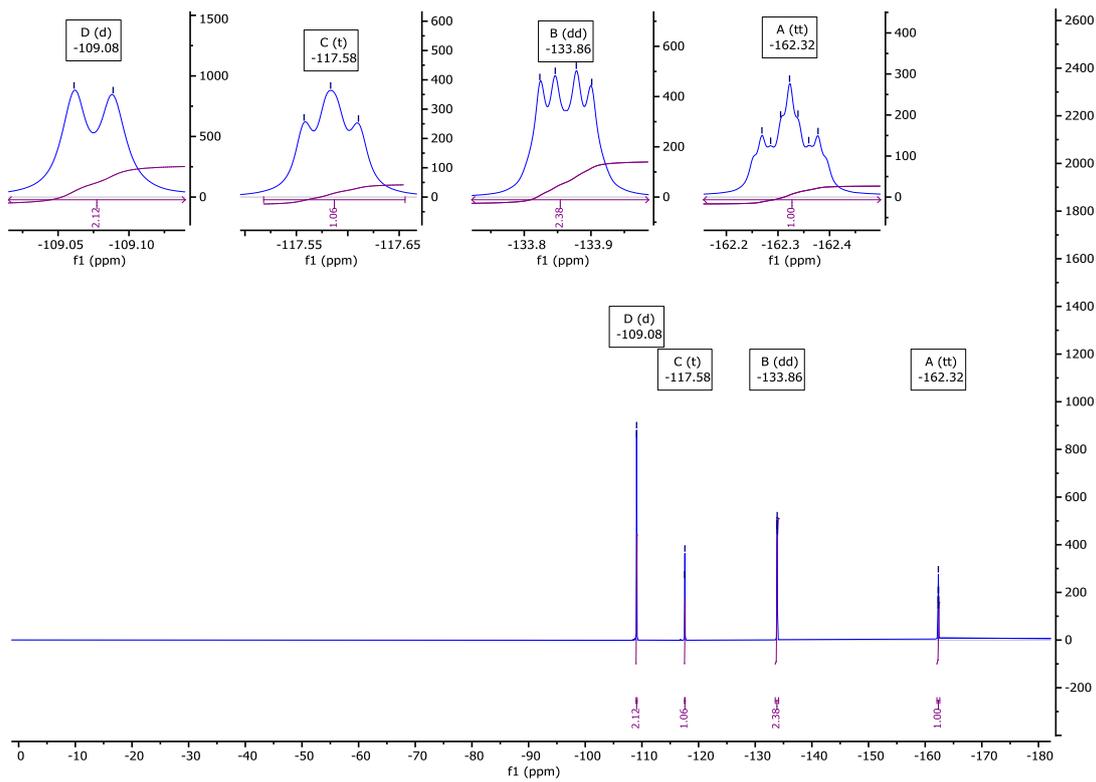

**Fig. S29**     $^{19}$F NMR spectra of **F2021** in CDCl$_3$.

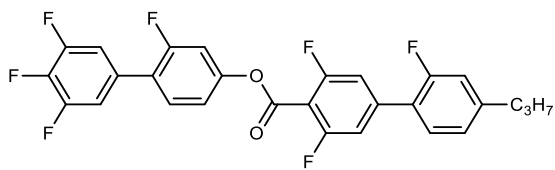

**F2121**

*2,3',4',5'-tetrafluoro-[1,1'-biphenyl]-4-yl 2',3,5-trifluoro-4'-propyl-[1,1'-biphenyl]-4-carboxylate*

| | |
|---|---|
| Yield: | (white crystalline solid) 161 mg, 62 % |
| Re-crystallisation solvent: | MeCN |
| $^{1}$H NMR (400 MHz): | 7.45 (t, *J* = 8.6 Hz, 1H, Ar-**H**), 7.37 (t, *J* = 8.0 Hz, 1H, Ar-**H**), 7.29 – 7.23 (m, 2H, Ar-**H**)†, 7.23 – 7.14 (m, 4H, Ar-**H**)*, 7.09 (dd, *J* = 7.9, 1.7 Hz, 1H, Ar-**H**), 7.03 (dd, *J* = 11.9, 1.6 Hz, 1H, Ar-**H**), 2.65 (t, *J* = 7.4 Hz, 2H, Ar-C**H₂**-CH₂), 1.69 (h, *J* = 7.4 Hz, 2H, CH₂-C**H₂**-CH₃), 0.98 (t, *J* = 7.3 Hz, 3H, CH₂-C**H₃**). † Overlapping CDCl₃ peak; *Overlapping Signals. |
| $^{13}$C{$^{1}$H} NMR (101 MHz): | 162.40 (dd, *J* = 258.3, 6.8 Hz), 160.27 (dd, *J* = 250.8, 19.6 Hz), 159.33, 152.47 (dd, *J* = 9.9, 4.1 Hz), 151.03 (d, *J* = 11.0 Hz), 149.92 (t, *J* = 4.0 Hz), 147.02 (d, *J* = 7.7 Hz), 142.59 (t, *J* = 11.4 Hz), , 130.67 (d, *J* = 3.7 Hz), 129.76 (d, *J* = 3.0 Hz), 125.04 (d, *J* = 3.2 Hz), 124.31 (d, *J* = 13.9 Hz), 122.71 (d, *J* = 11.9 Hz), 118.16 (d, *J* = 3.7 Hz), 116.57, 116.35, 113.42 – 113.01 (m), 112.69 (tt, *J* = 23.5, 3.6 Hz), 110.79, 110.54, 108.01 (t, *J* = 16.8 Hz), 37.55, 24.13, 13.72. |
| $^{19}$F NMR (376 MHz): | -108.80 (d, 2F, *J*$_{F-H}$ = 10.3 Hz, Ar-**F**), -114.32 (t, 1F, *J*$_{F-H}$ = 9.7 Hz, Ar-**F**), -117.56 (dd, 1F, *J*$_{F-H}$ = 11.8 Hz, *J*$_{F-H}$ = 8.1 Hz, Ar-**F**), -134.15 (dd, 2F, *J*$_{F-F}$ = 20.7 Hz, *J*$_{F-H}$ = 8.7 Hz, Ar-**F**), -161.14 (tt, 1F, *J*$_{F-F}$ = 20.6 Hz, *J*$_{F-H}$ = 6.5 Hz, Ar-**F**). |

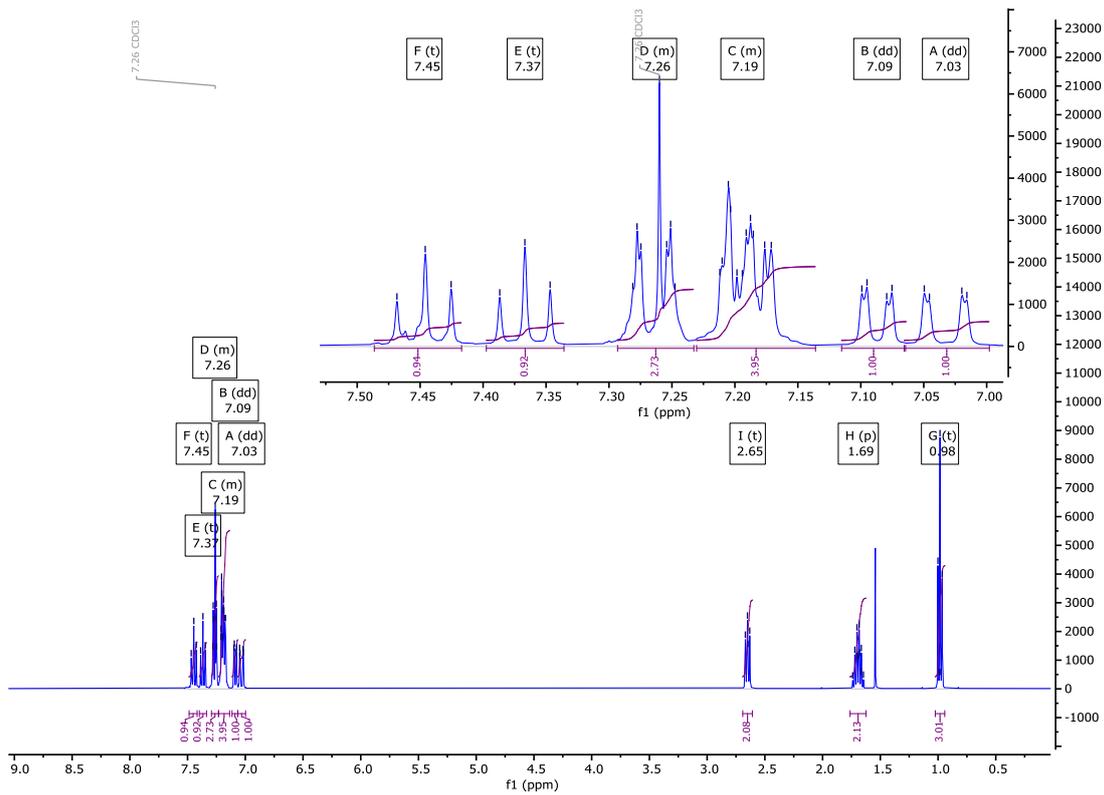

**Fig. S30**  $^1$H NMR spectra of **F2121** in CDCl$_3$.

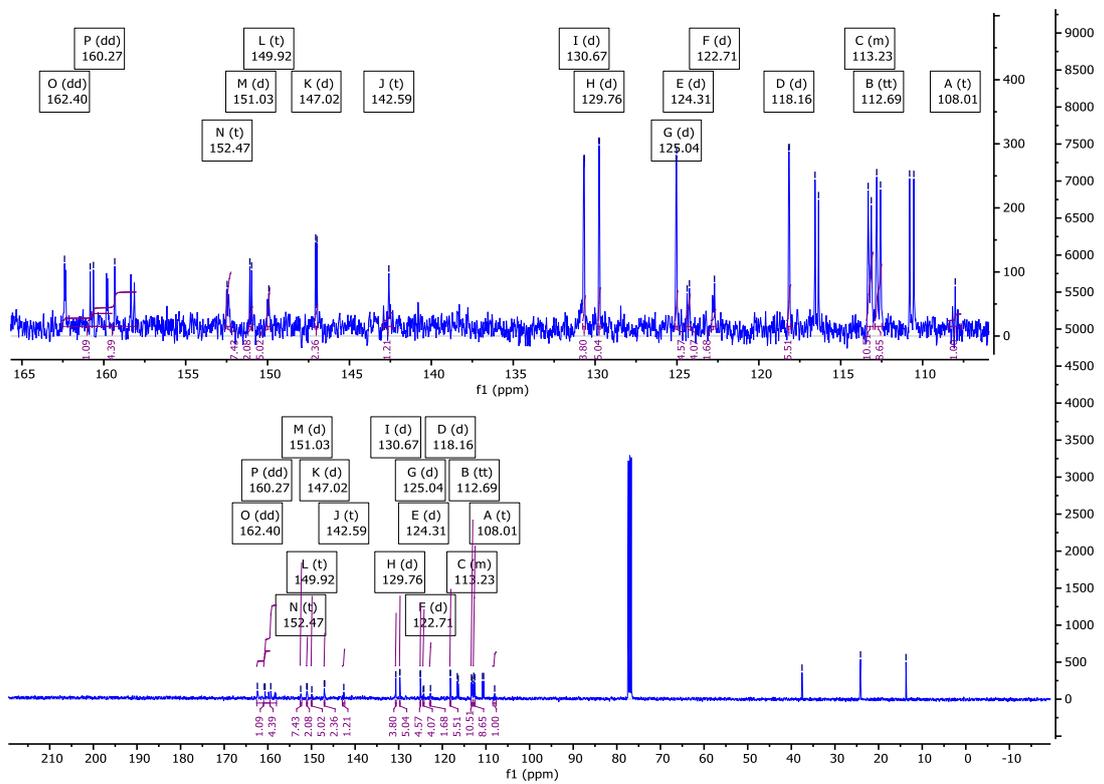

**Fig. S31**  $^{13}$C{$^1$H} NMR spectra of **F2121** in CDCl$_3$.

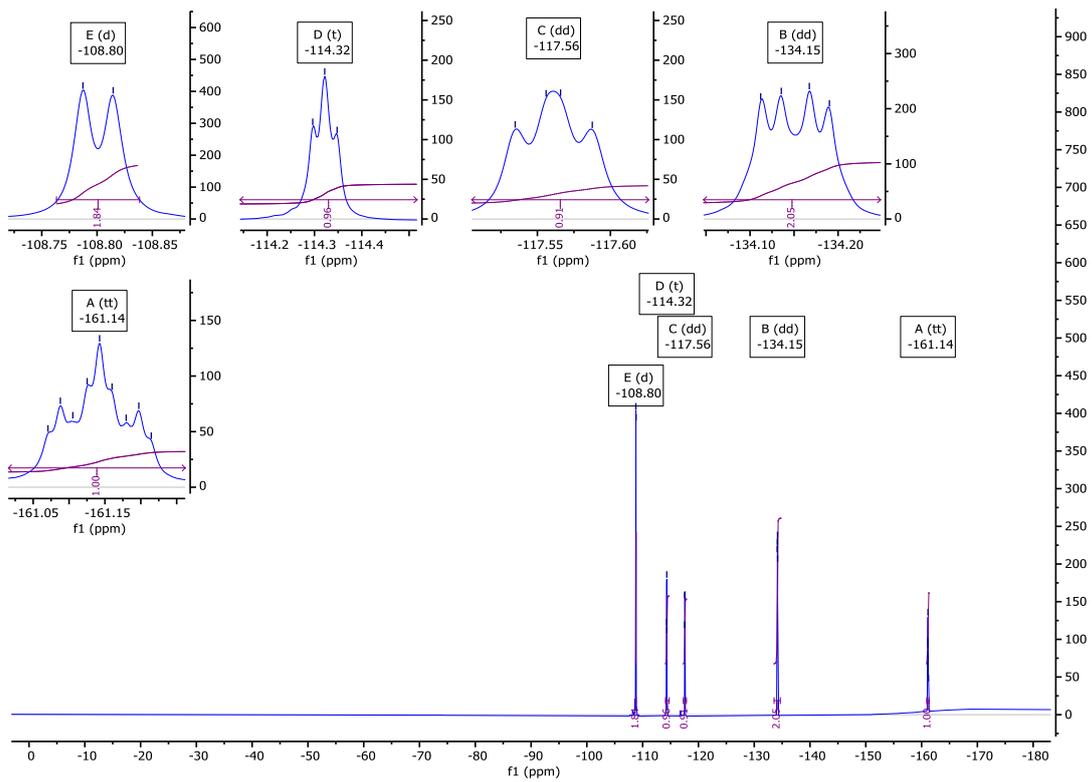

**Fig. S32**  $^{19}$F NMR spectra of **F2121** in CDCl$_3$.

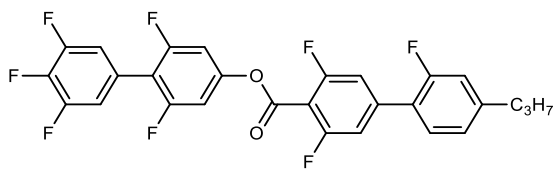

**F2221**

*2,3',4',5',6-pentafluoro-[1,1'-biphenyl]-4-yl 2',3,5-trifluoro-4'-propyl-[1,1'-biphenyl]-4-carboxylate*

| | |
|---|---|
| Yield: | (white needles) 145 mg, 54 % |
| Re-crystallisation solvent: | MeCN |
| $^1$H NMR (400 MHz): | 7.37 (t, *J* = 8.0 Hz, 1H, Ar-**H**), 7.31 – 7.24 (m, 2H, Ar-**H**)†, 7.18 – 6.99 (m, 6H, Ar-**H**)*, 2.65 (t, *J* = 7.7 Hz, 2H, Ar-C**H$_2$**-CH$_2$), 1.69 (h, *J* = 7.5 Hz, 2H, CH$_2$-C**H$_2$**-CH$_3$), 0.98 (t, *J* = 7.3 Hz, 3H, CH$_2$-C**H$_3$**). † Overlapping CDCl$_3$ peak; *Overlapping Signals. |
| $^{13}$C{$^1$H} NMR (101 MHz): | 162.46 (dd, *J* = 258.8, 5.8 Hz), 160.31 (dd, *J* = 250.5, 8.8 Hz), 159.59 (d, *J* = 250.5 Hz), 158.92, 151.10 (ddd, *J* = 243.9, 9.7, 3.6 Hz), 150.77 (t, *J* = 14.3 Hz), 147.13 (d, *J* = 8.0 Hz), 142.92 (d, *J* = 9.9 Hz), 138.56 (dd, *J* = 254.5, 16.4 Hz), 129.74 (d, *J* = 3.1 Hz), 125.06 (d, *J* = 3.1 Hz), 124.18 (d, *J* = 8.9 Hz), 122.75 (d, *J* = 12.5 Hz), 116.59, 116.37, 115.01 – 114.64 (m$_{apparent}$), 112.79 (dt, *J* = 23.2, 3.4 Hz), 107.57 (t, *J* = 17.6 Hz), 106.86 – 106.31 (m$_{apparent}$), 37.55, 24.13, 13.71. |
| $^{19}$F NMR (376 MHz): | -108.52 (d, 2F, *J$_{F-H}$* = 10.5 Hz, Ar-**F**), -112.15 (d, 2F *J$_{F-H}$* = 8.8 Hz, Ar-**F**), -117.52 (dd, 1F *J$_{F-H}$* = 11.9 Hz, *J$_{F-H}$* 8.1 Hz, Ar-**F**), -134.29 (dd, 2F *J$_{F-F}$* = 20.6 Hz, *J$_{F-H}$* = 8.4 Hz, Ar-**F**), -159.91 (tt, 1F, *J$_{F-H}$* = 20.6, *J$_{F-H}$* = 6.6 Hz, Ar-**F**). |

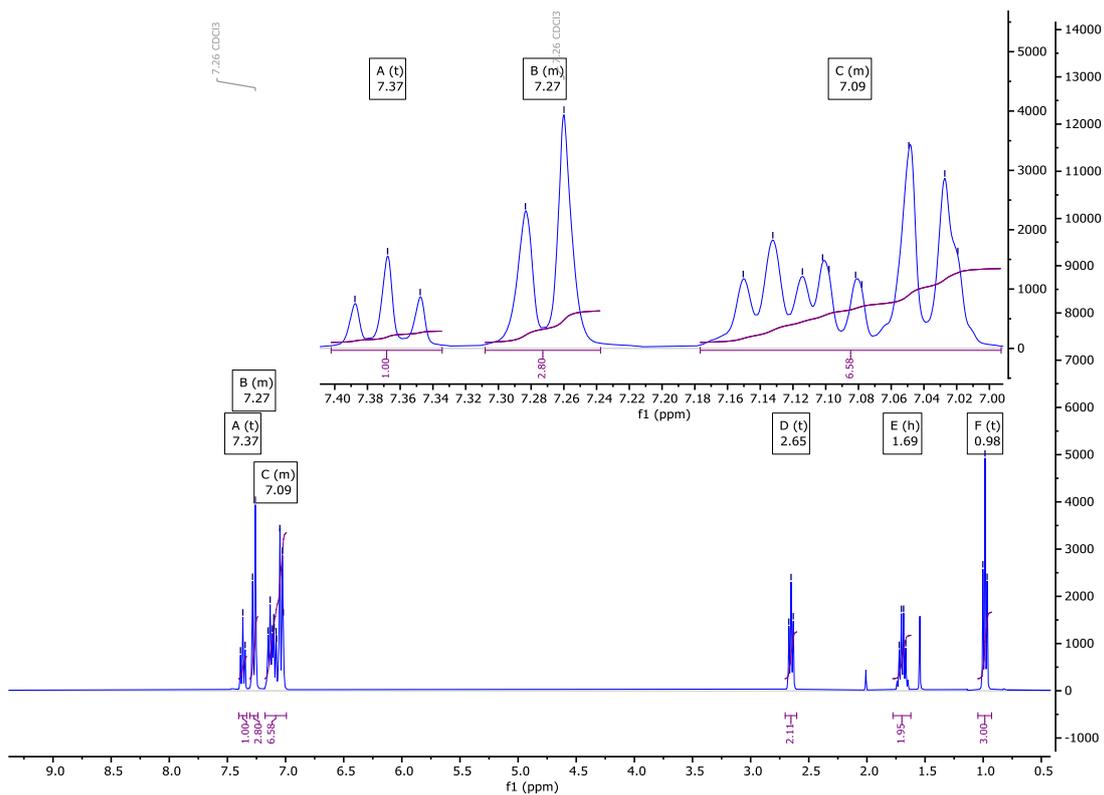

**Fig. S33** $^1$H NMR spectra of **F2221** in CDCl$_3$.

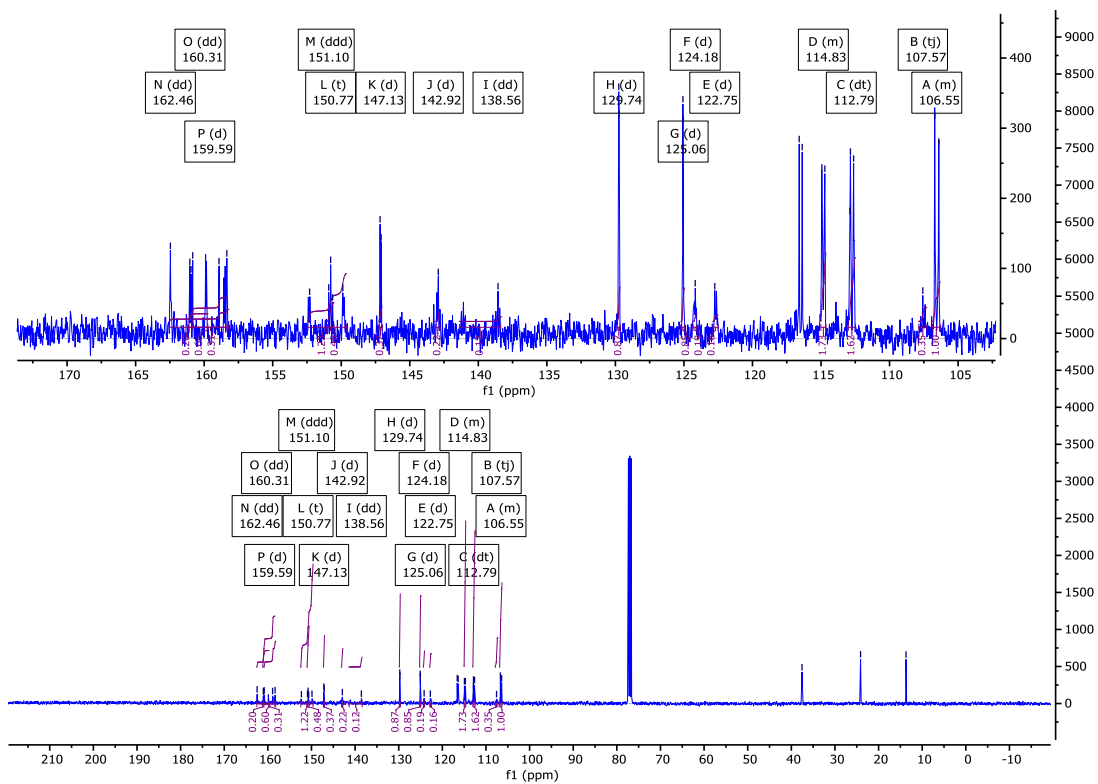

**Fig. S34** $^{13}$C{$^1$H} NMR spectra of **F2221** in CDCl$_3$.

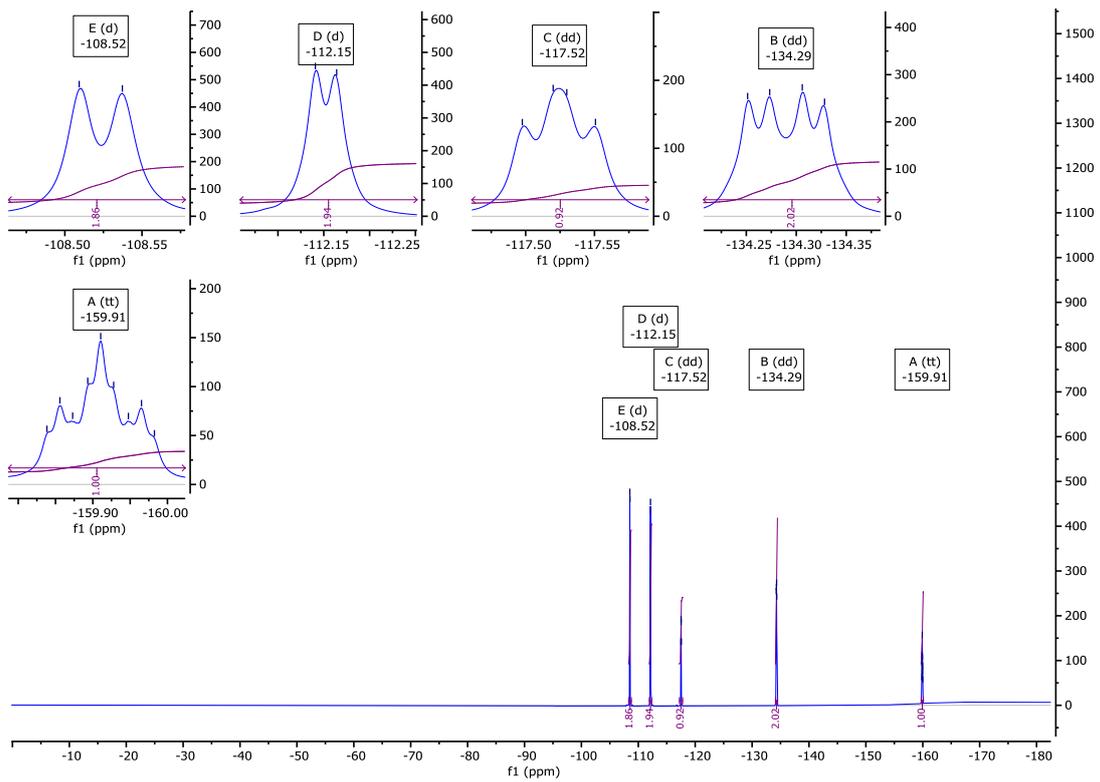

**Fig. S35** $^{19}$F NMR spectra of **F2221** in CDCl$_3$.

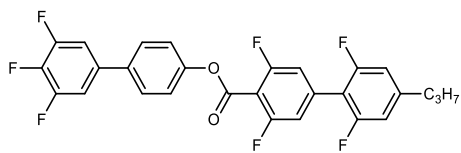

**F2022**

*3',4',5'-trifluoro-[1,1'-biphenyl]-4-yl 2',3,5,6'-tetrafluoro-4'-propyl-[1,1'-biphenyl]-4-carboxylate*

| | |
|---|---|
| Yield: | (white crystalline solid) 130 mg, 50 % |
| Re-crystallisation solvent: | MeCN |
| $^1$H NMR (400 MHz): | 7.57 (ddd, *J* = 8.7, 2.8, 2.1 Hz, 2H, Ar-**H**), 7.37 (ddd, *J* = 8.8, 2.7, 2.2 Hz, 2H, Ar-**H**), 7.23 – 7.15 (m, 4H, Ar-**H**)*, 6.87 (ddd, *J* = 9.0, 2.8, 2.2 Hz, 2H, Ar-**H**), 2.63 (t, *J* = 7.4 Hz, 2H, Ar-C**H₂**-CH₂), 1.69 (h, *J* = 7.3 Hz, 2H, CH₂-C**H₂**-CH₃), 0.99 (t, *J* = 7.3 Hz, 3H, CH₂-C**H₃**). *Overlapping signals. |
| $^{13}$C{$^1$H} NMR (101 MHz): | 160.54 (dd, *J* = 257.1, 7.0 Hz), 159.43, 158.23 (dd, *J* = 250.7, 6.4 Hz), 150.93 (ddd, *J* = 249.8, 10.1, 4.5 Hz), 147.15 (t, *J* = 9.6 Hz), 138.93 (dt, *J* = 252.6, 15.6 Hz), 136.63 – 136.22 (m), 135.85 (t, *J* = 11.2 Hz), 128.13, 122.30, 114.53, 114.43 (dd, *J* = 23.8, 2.6 Hz), 112.51 (t, *J* = 17.8 Hz), 111.97 (dd, *J* = 19.1, 6.0 Hz), 111.16 (dd, *J* = 15.7, 5.9 Hz), 109.27 (t, *J* = 17.2 Hz), 37.68, 23.86, 13.63. |
| $^{19}$F NMR (376 MHz): | -109.54 (d, *J*$_{F-H}$ = 10.0 Hz, 2F, Ar-**F**), -114.95 (d, *J*$_{F-H}$ = 9.5 Hz, 2F, Ar-**F**), -133.86 (dd, *J*$_{F-F}$ = 20.4 Hz, *J*$_{F-H}$ = 8.8 Hz, 2F, Ar-**F**), -162.32 (tt, *J*$_{F-F}$ = 20.4 Hz, *J*$_{F-H}$ = 6.5 Hz, 1F, Ar-**F**). |

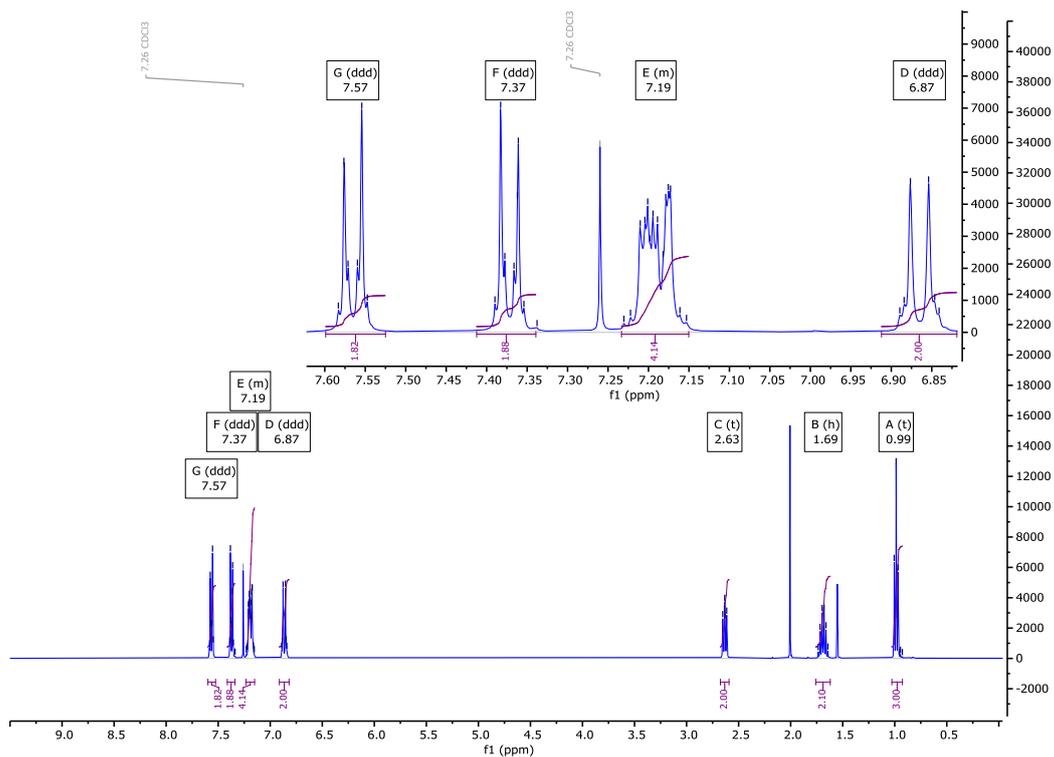

**Fig. S36**                 $^1$H NMR spectra of **F2022** in CDCl$_3$.

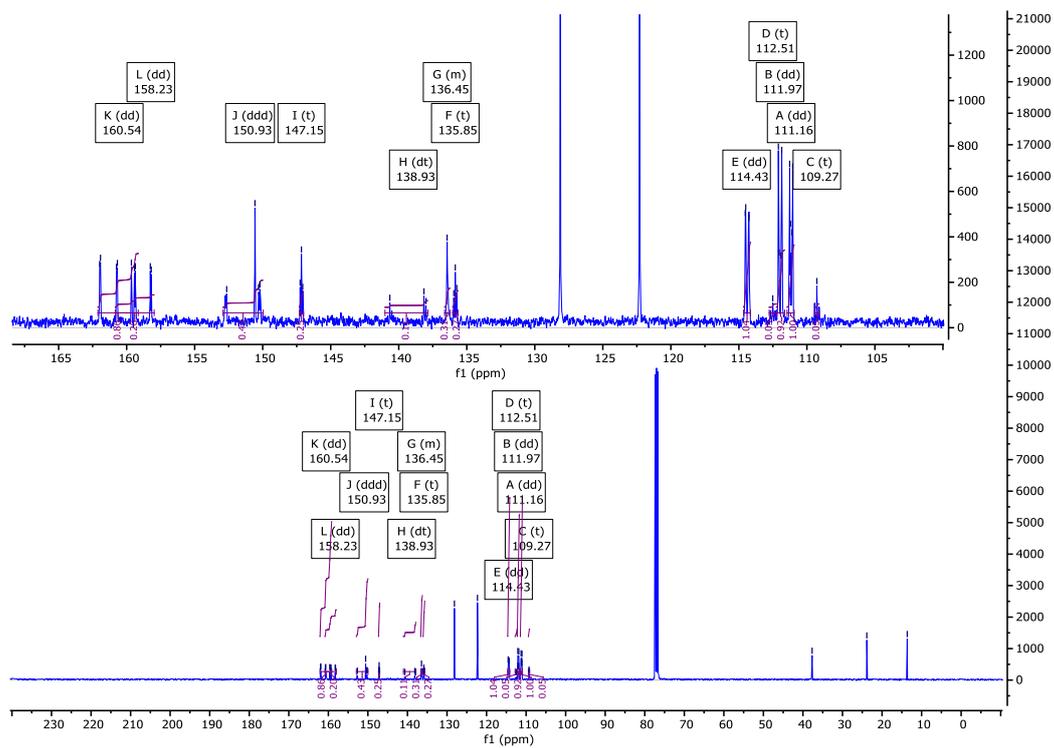

**Fig. S37**                 $^{13}$C{$^1$H} NMR spectra of **F2022** in CDCl$_3$.

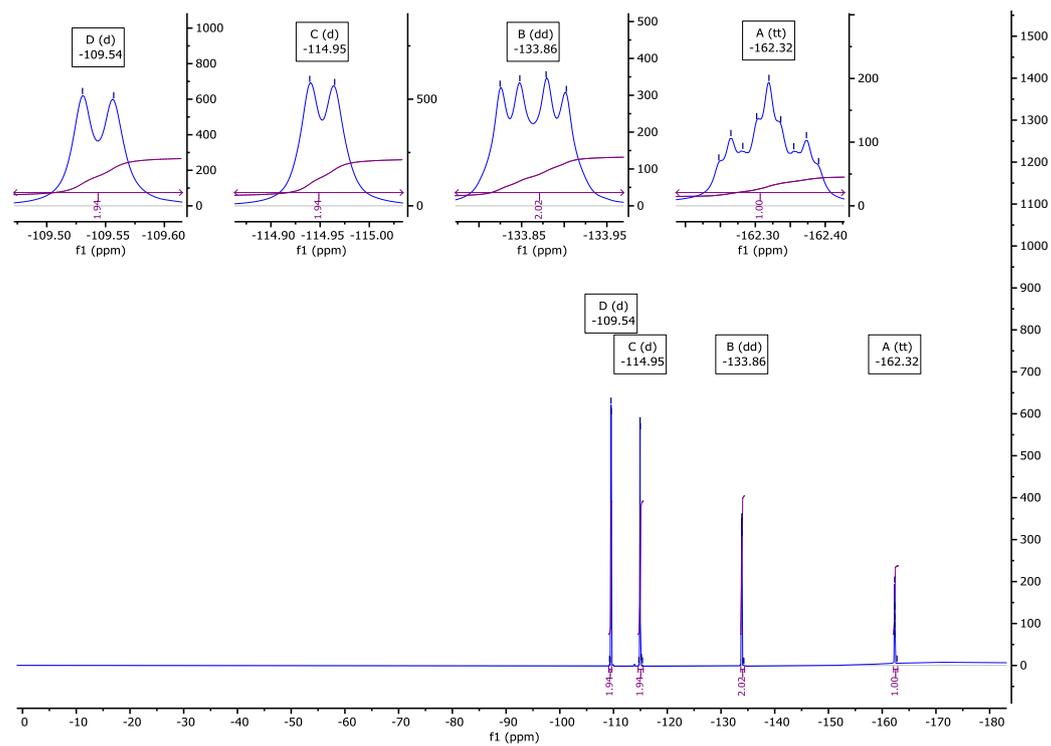

**Fig. S38**     $^{19}$F NMR spectra of **F2022** in CDCl$_3$.

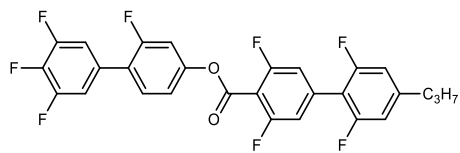

**F2122**

*2,3',4',5'-tetrafluoro-[1,1'-biphenyl]-4-yl 2',3,5,6'-tetrafluoro-4'-propyl-[1,1'-biphenyl]-4-carboxylate*

| | |
|---|---|
| Yield: | (white solid) 140 mg, 52 % |
| Re-crystallisation solvent: | MeCN |
| $^1$H NMR (400 MHz): | 7.44 (t, *J* = 8.3 Hz, 1H, Ar-**H**), 7.23 – 7.15 (m, 6H, Ar-**H**)\*, 6.87 (ddd, *J* = 9.1, 5.3, 2.1 Hz, 2H, Ar-**H**), 2.64 (t, *J* = 7.5 Hz, 2H, Ar-C**H$_2$**-CH$_2$), 1.69 (h, *J* = 7.6 Hz, 2H, CH$_2$-C**H$_2$**-CH$_3$), 0.99 (t, *J* = 7.3 Hz, 3H, CH$_2$-C**H$_3$**). \*Overlapping Signals. |
| $^{13}$C{$^1$H} NMR (101 MHz): | 160.74 (dd, *J* = 258.1, 6.5 Hz), 159.22, 158.74 (d, *J* = 250.3 Hz), 158.14 (dd, *J* = 250.9, 3.3 Hz), 151.23 (ddd, *J* = 250.0, 9.9, 4.1 Hz), 150.99 (d, *J* = 11.0 Hz), 147.25 (t, *J* = 9.5 Hz), 139.56 (dt, *J* = 252.8, 15.5 Hz), 136.18 (t, *J* = 11.5 Hz), 130.70 (d, *J* = 3.8 Hz), 124.61 – 124.17 (m), 118.14 (d, *J* = 3.8 Hz), 114.45 (dd, *J* = 23.6, 2.3 Hz), 113.42 – 113.06 (m), 112.43 (t, *J* = 10.8 Hz), 111.99 (dd, *J* = 19.1, 6.1 Hz), 110.65 (d, *J* = 26.1 Hz), 108.81 (t, *J* = 16.8 Hz), 37.68, 23.85, 13.63. |
| $^{19}$F NMR (376 MHz): | -109.27 (d, *J$_{F-H}$* = 10.0 Hz, 2F, Ar-**F**), -114.26 (t, *J$_{F-H}$* = 9.4 Hz, 1F, Ar-**F**), -114.94 (d, *J$_{F-H}$* = 9.6 Hz, 2F, Ar-**F**), -134.14 (dd, *J$_{F-F}$* = 20.6 Hz, *J$_{F-H}$* = 8.7 Hz, 2F, Ar-**F**), -161.13 (tt, *J$_{F-F}$* = 20.6 Hz, *J$_{F-H}$* = 6.5 Hz, 1F, Ar-**F**). |

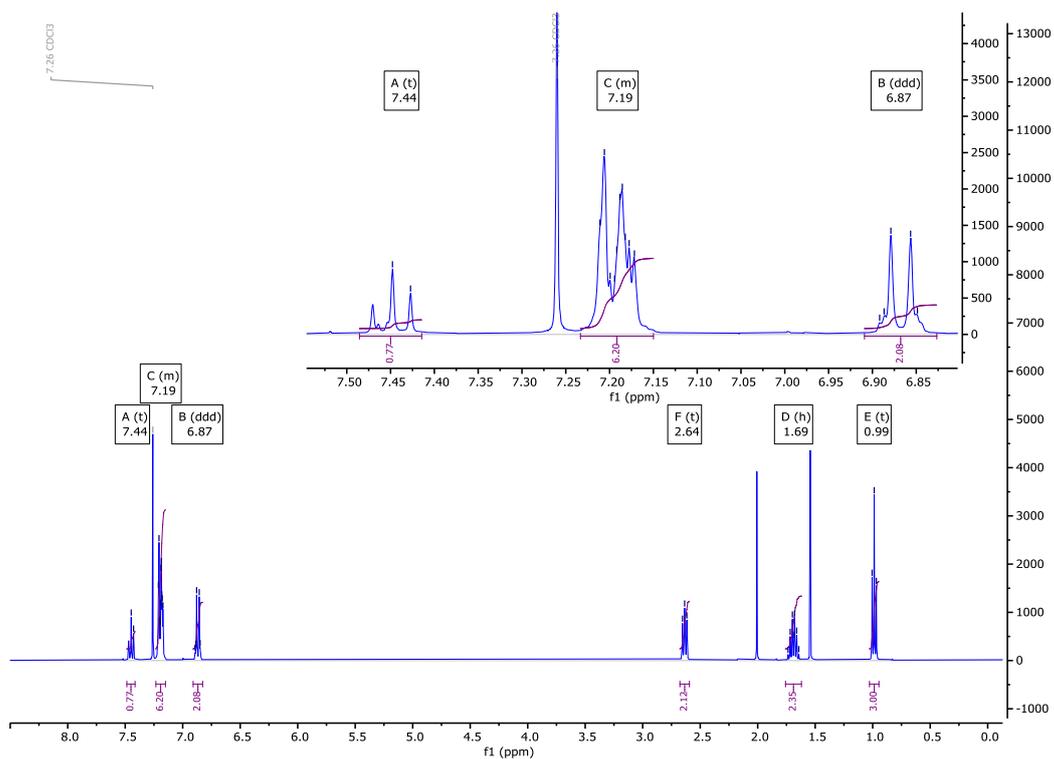

**Fig. S39**   $^1$H NMR spectra of **F2122** in CDCl$_3$.

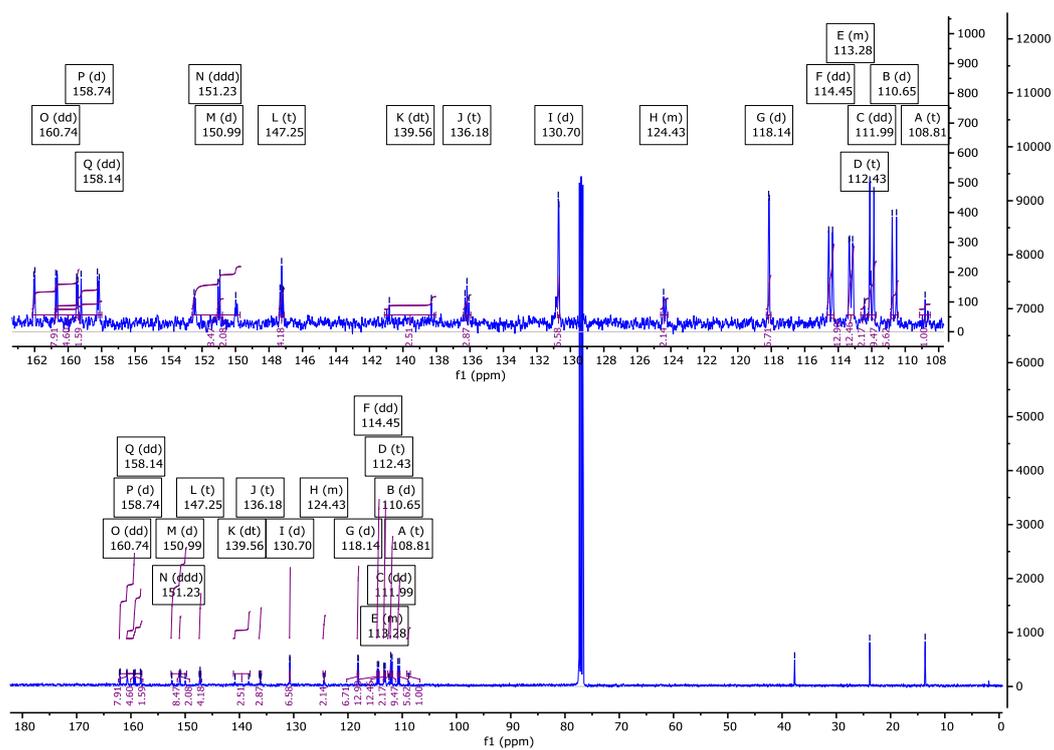

**Fig. S40**   $^{13}$C{$^1$H} NMR spectra of **F2122** in CDCl$_3$.

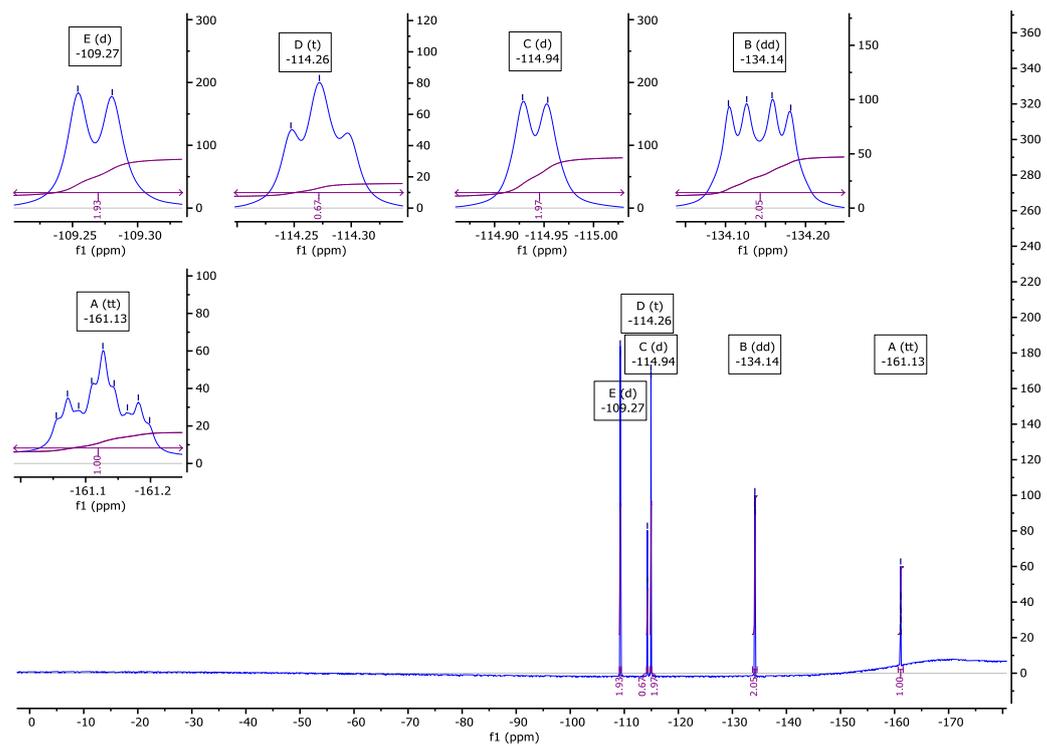

**Fig. S41**  $^{19}$F NMR spectra of **F2122** in CDCl$_3$.

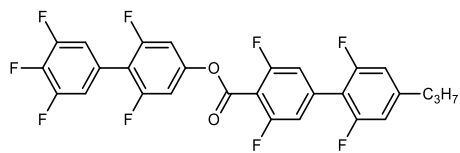

**F2222**

*2,3',4',5',6-pentafluoro-[1,1'-biphenyl]-4-yl 2',3,5,6'-tetrafluoro-4'-propyl-[1,1'-biphenyl]-4-carboxylate*

| | |
|---|---|
| Yield: | (white needles) 131 mg, 47 % |
| Re-crystallisation solvent: | MeCN |
| $^1$H NMR (400 MHz): | 7.20 (d, *J* = 9.1 Hz, 2H, Ar-**H**), 7.13 (t, *J* = 7.2 Hz, 2H, Ar-**H**), 7.03 (ddd, *J* = 8.0, 3.3, 3.2 Hz, 2H, Ar-**H**), 6.87 (ddd, *J* = 9.1, 4.8, 2.4 Hz, 2H, Ar-**H**), 2.64 (t, *J* = 7.4 Hz, 2H, Ar-C**H$_2$**-CH$_2$), 1.69 (h, *J* = 7.4 Hz, 2H, CH$_2$-C**H$_2$**-CH$_3$), 0.99 (t, *J* = 7.3 Hz, 3H, CH$_2$-C**H$_3$**). |
| $^{13}$C{$^1$H} NMR (101 MHz): | 161.21 (dd, *J* = 139.3, 6.1 Hz), 158.84 (dd, *J* = 250.7, 8.3 Hz), 158.80, 158.24 (dd, *J* = 250.6, 6.1 Hz), 151.13 (ddd, *J* = 249.3, 8.8, 3.9 Hz), 150.73 (t, *J* = 14.4 Hz), 147.34 (t, *J* = 9.5 Hz), 139.83 (dt, *J* = 253.9, 15.1 Hz), 136.51 (t, *J* = 11.2 Hz), 124.17 (m), 114.67 (m), 113.95 (d, *J* = 16.8 Hz), 112.37 (t, *J* = 12.7 Hz), 112.01 (dd, *J* = 24.3, 5.5 Hz), 108.38 (t, *J* = 16.5 Hz), 106.54 (dd, *J* = 28.3, 8.7 Hz), 37.68, 23.84, 13.63. |
| $^{19}$F NMR (376 MHz): | -109.00 (d, *J$_{F-H}$* = 10.1 Hz, 2F, Ar-**F**), -112.10 (d, *J$_{F-H}$* = 8.8 Hz, 2F, Ar-**F**), -114.94 (d, *J$_{F-H}$* = 9.8 Hz, 2F, Ar-**F**), -134.29 (dd, *J$_{F-F}$* = 20.7 Hz, *J$_{F-H}$* = 8.4 Hz, 2F, Ar-**F**), -159.90 (tt, *J$_{F-F}$* = 20.7 Hz, *J$_{F-H}$* = 6.5 Hz, 1F, Ar-**F**). |

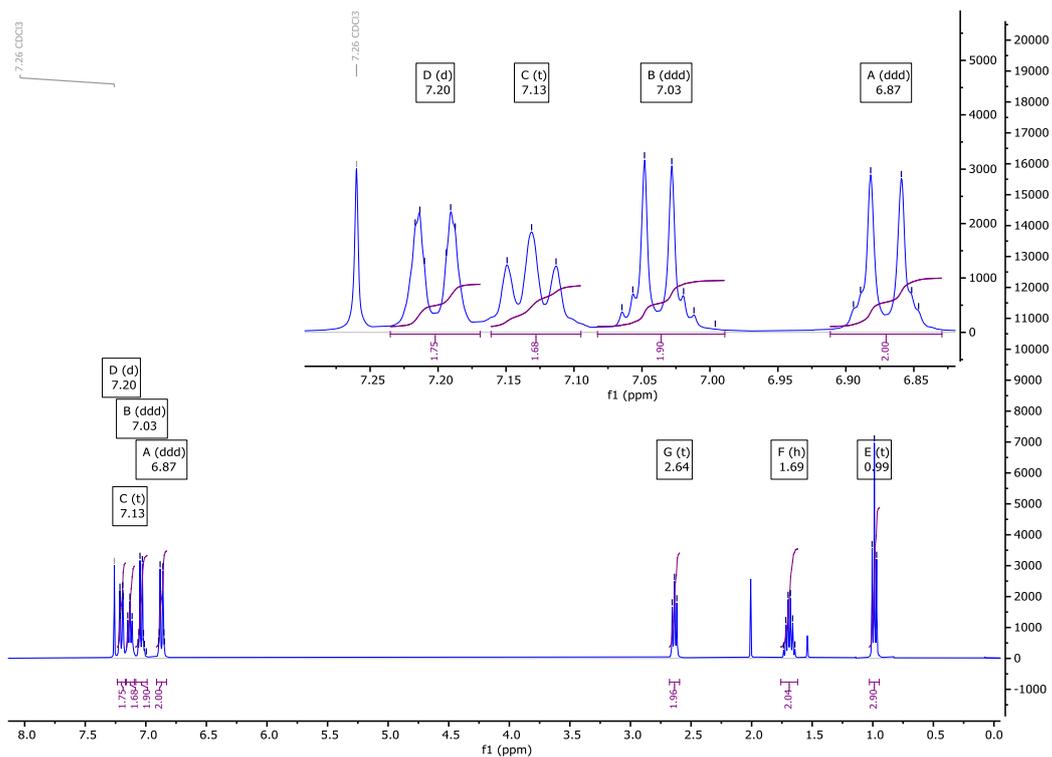

**Fig. S42**                  $^1$H NMR spectra of **F2222** in CDCl$_3$.

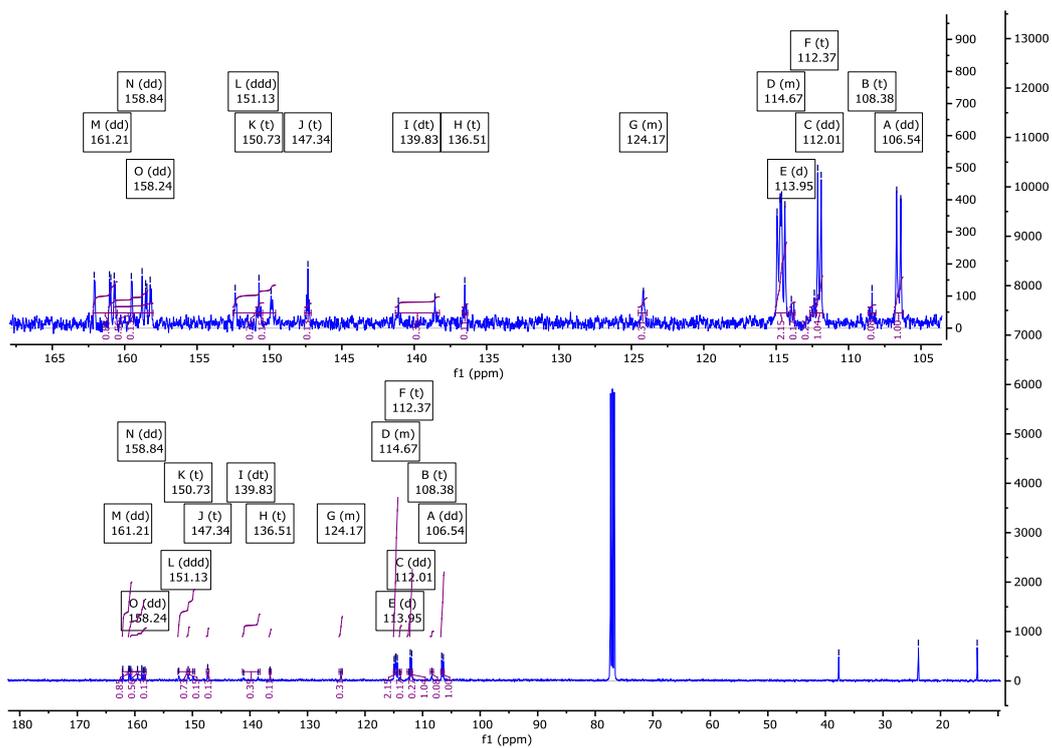

**Fig. S43**                  $^{13}$C{$^1$H} NMR spectra of **F2222** in CDCl$_3$.

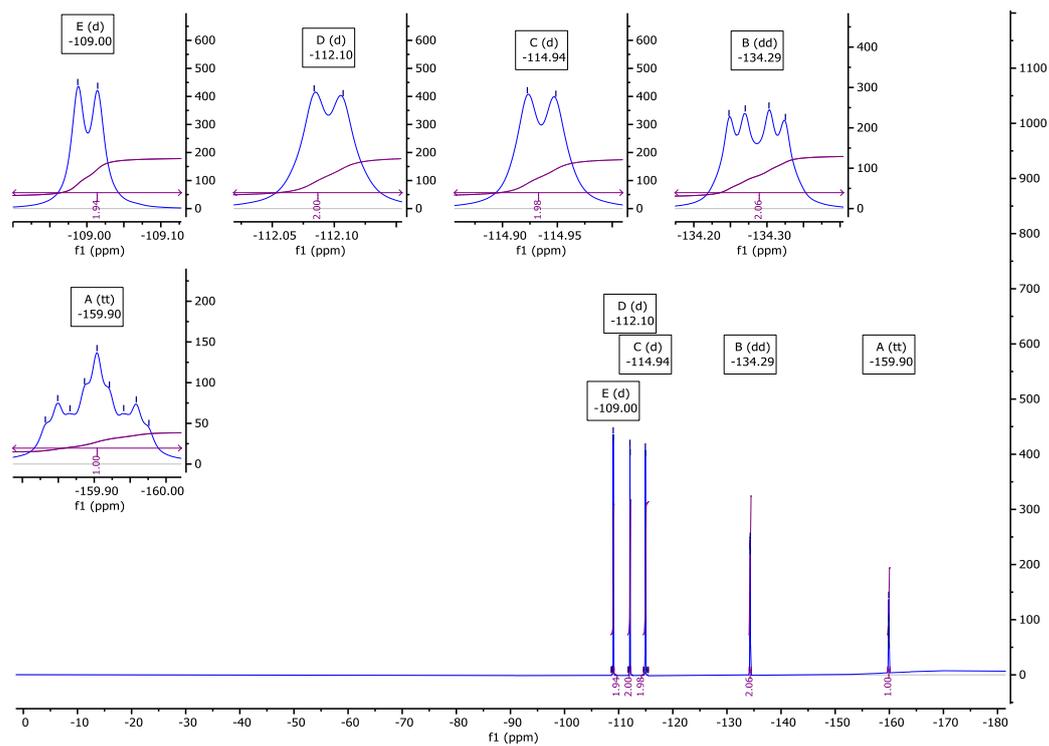

**Fig. S44** 19F NMR spectra of **F2222** in CDCl$_3$.

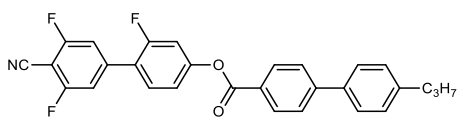

**NC2100**

*4'-cyano-2,3',5'-trifluoro-[1,1'-biphenyl]-4-yl 4'-propyl-[1,1'-biphenyl]-4-carboxylate*

| | |
|---|---|
| Yield: | (white crystalline solid) 193 mg, 82 % |
| Re-crystallisation solvent: | MeCN |
| $^1$H NMR (400 MHz): | 8.27 (d, *J* = 8.0 Hz, 2H, Ar-**H**), 7.78 (d, *J* = 8.1 Hz, 2H, Ar-**H**), 7.62 (d, *J* = 7.8 Hz, 2H, Ar-**H**), 7.52 (t, *J* = 8.6 Hz, 1H, Ar-**H**), 7.38 – 7.18 (m, 6H, Ar-**H**)*†, 2.69 (t, *J* = 7.6 Hz, 2H, Ar-C**H₂**-CH₂), 1.73 (h, *J* = 7.4 Hz, 2H, CH₂-C**H₂**-CH₃), 1.02 (t, *J* = 7.3 Hz, 3H, CH₂-C**H₃**). *Overlapping signals, †Overlapping CDCl₃. |
| $^{13}$C{$^1$H} NMR (126 MHz): | 164.47, 163.08 (dd, *J* = 260.9, 5.2 Hz), 159.56 (d, *J* = 252.8 Hz), 152.99 (d, *J* = 11.2 Hz), 146.92, 143.44, 143.16 (t, *J* = 10.2 Hz), 136.92, 130.85, 130.55 (d, *J* = 3.6 Hz), 129.22, 127.17 (d, *J* = 2.8 Hz), 126.96, 122.83 (dt, *J* = 12.3, 2.2 Hz), 118.76 (d, *J* = 3.6 Hz), 112.61 (dt, *J* = 20.7, 3.7 Hz), 111.14 (d, *J* = 25.7 Hz), 109.17, 91.42 (t, *J* = 19.3 Hz), 37.75, 24.53, 13.87. |
| $^{19}$F NMR (376 MHz): | -103.53 (d, *J*$_{F-H}$ = 9.4 Hz, 2F, Ar-**F**), -113.45 (t, *J*$_{F-H}$ = 9.8 Hz, 1F, Ar-**F**). |

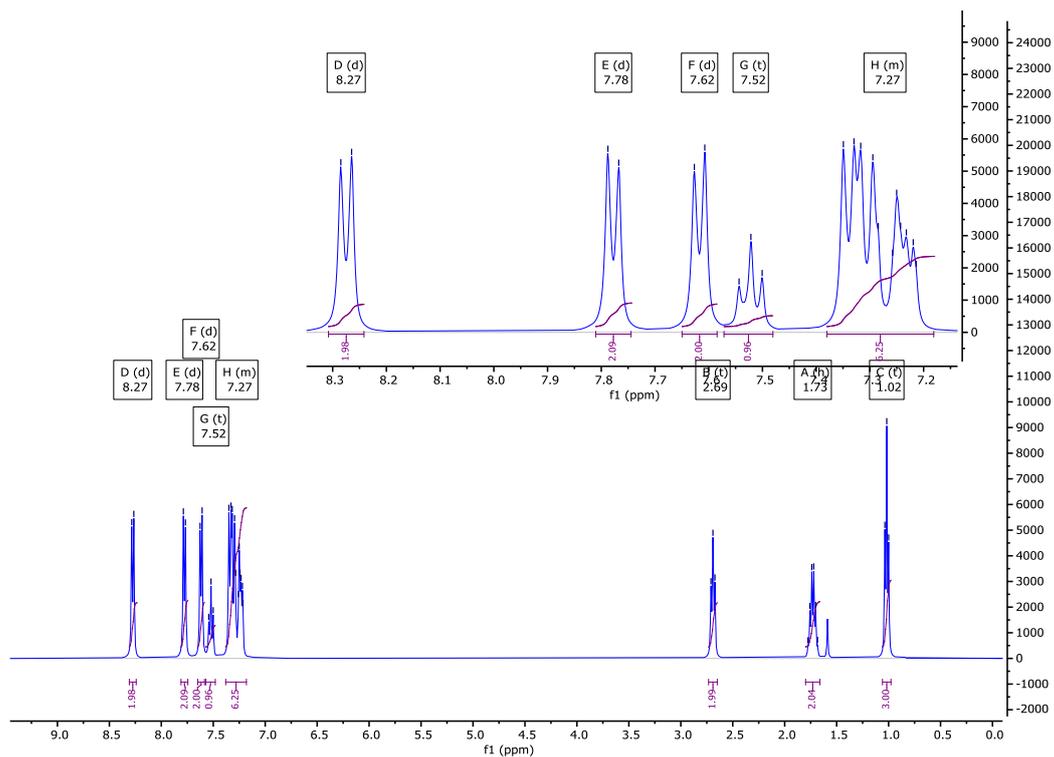

**Fig. S45**  $^1$H NMR spectra of **NC2100** in CDCl$_3$.

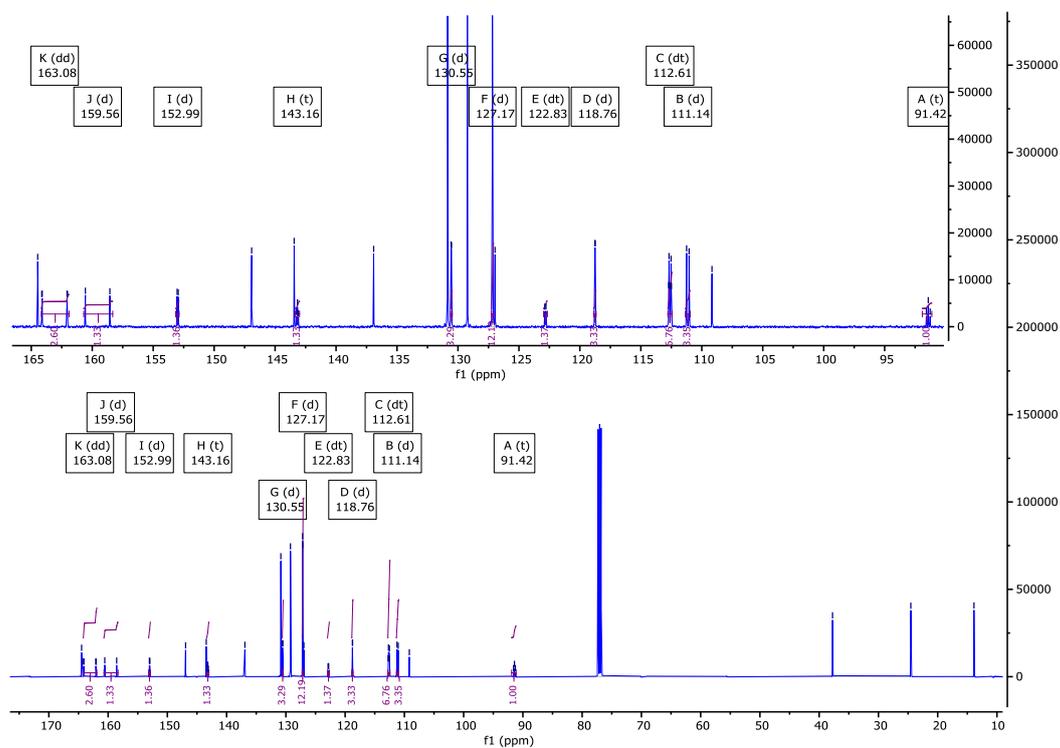

**Fig. S46**  $^{13}$C{$^1$H} NMR spectra of **NC2100** in CDCl$_3$.

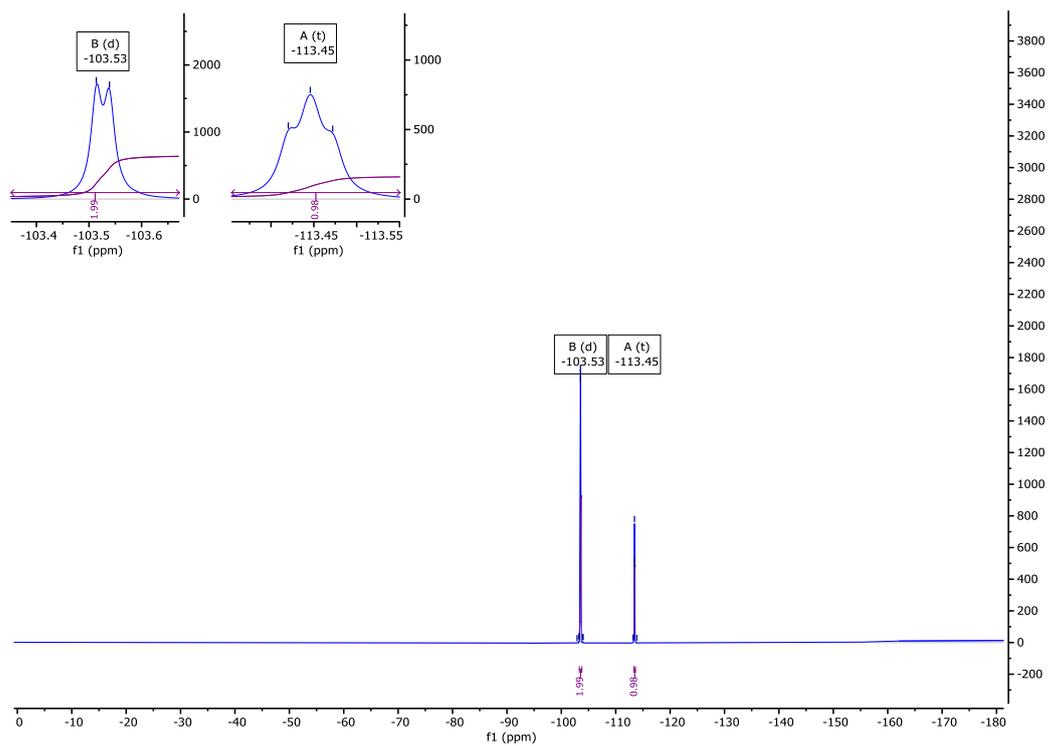

**Fig. S47** $^{19}$F NMR spectra of **NC2100** in CDCl$_3$.

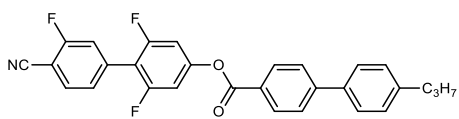

**NC1200**

*4'-cyano-2,3',6-trifluoro-[1,1'-biphenyl]-4-yl 4'-propyl-[1,1'-biphenyl]-4-carboxylate*

| | |
|---|---|
| Yield: | (white crystalline solid) 205 mg, 87 % |
| Re-crystallisation solvent: | MeCN |
| $^1$H NMR (501 MHz): | 8.24 (ddd, *J* = 8.3, 2.1, 1.7 Hz, 2H, Ar-**H**), 7.76 (dt, *J* = 8.3, 1.8 Hz, 2H, Ar-**H**), 7.73 (t, *J* = 6.9 Hz, 1H, Ar-**H**), 7.59 (ddd, *J* = 8.0, 2.0, 1.9 Hz, 2H, Ar-**H**), 7.44 – 7.36 (m, 2H, Ar-**H**), 7.31 (ddd, *J* = 8.1, 2.1, 2.1 Hz, 2H, Ar-**H**), 7.03 (ddd, *J* = 8.2, 3.2, 2.7 Hz, 2H, Ar-**H**), 2.66 (t, *J* = 6.8 Hz, 2H, Ar-C**H₂**-CH₂), 1.70 (h, *J* = 7.3 Hz, 2H, CH₂-C**H₂**-CH₃), 0.99 (t, *J* = 7.3 Hz, 3H, CH₂-C**H₃**). |
| $^{13}$C{$^1$H} NMR (126 MHz): | 164.18, 162.83 (d, *J* = 259.2 Hz), 159.72 (dd, *J* = 251.2, 8.7 Hz), 152.24 (t, *J* = 14.3 Hz), 147.09, 143.49, 136.86 136.05 (d, *J* = 8.7 Hz), 133.25, 130.89, 129.23, 127.19 (d, *J* = 1.5 Hz), 126.99 – 126.79 (m), 118.52 (dt, *J* = 20.8, 2.0 Hz), 113.40 (td, *J* = 18.0, 2.0 Hz), 107.14 – 106.49 (m), 101.25 (d, *J* = 15.5 Hz), 37.75, 24.53, 13.87. |
| $^{19}$F NMR (376 MHz): | -106.13 (t, *J*$_{F\text{-}H}$ = 8.3 Hz, 1F, Ar-**F**), -112.22 (d, *J*$_{F\text{-}H}$ = 9.0 Hz, 2F, Ar-**F**). |

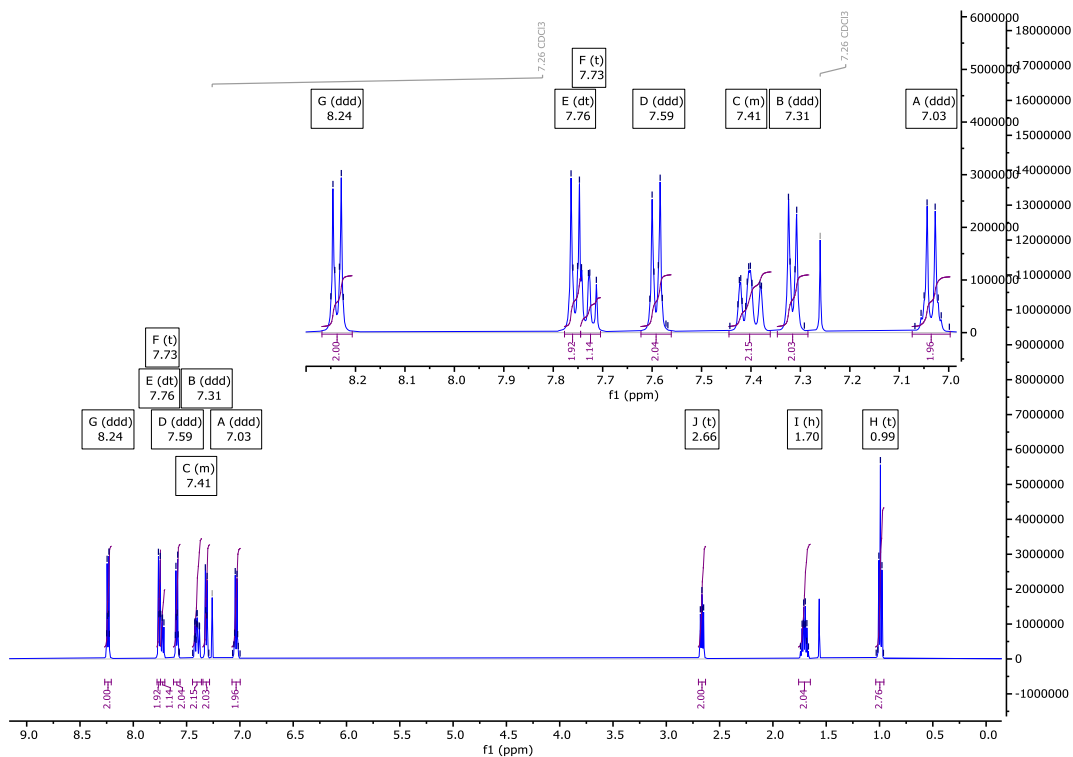

**Fig. S48** $^1$H NMR spectra of **NC1200** in CDCl$_3$.

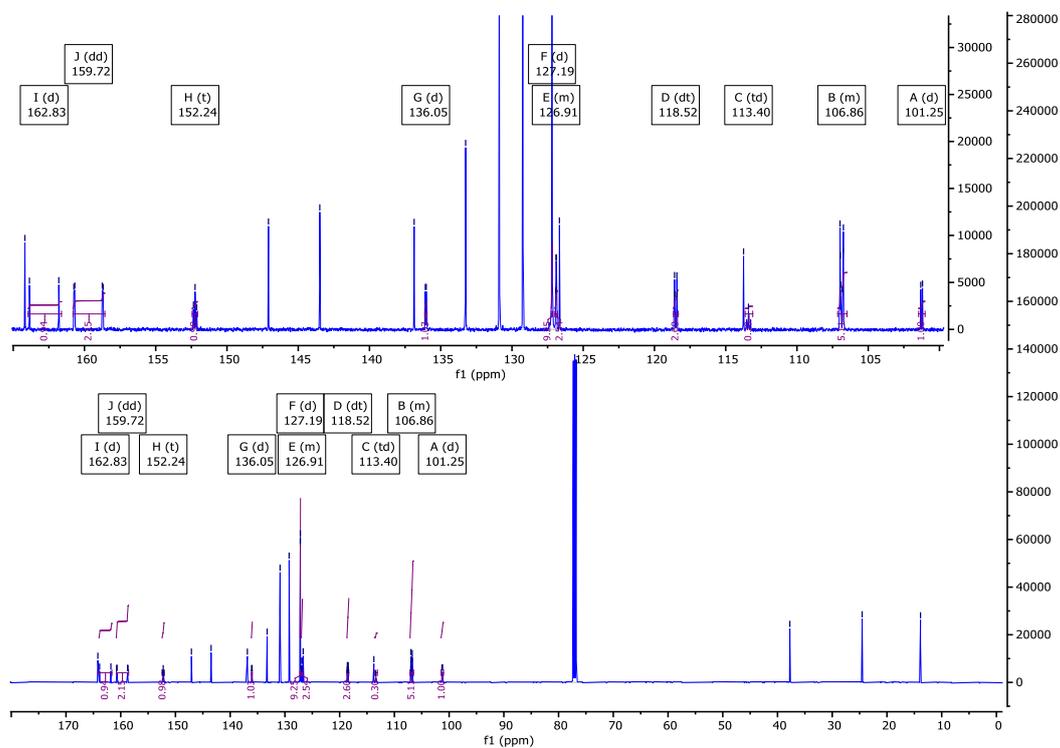

**Fig. S49** $^{13}$C{$^1$H} NMR spectra of **NC1200** in CDCl$_3$.

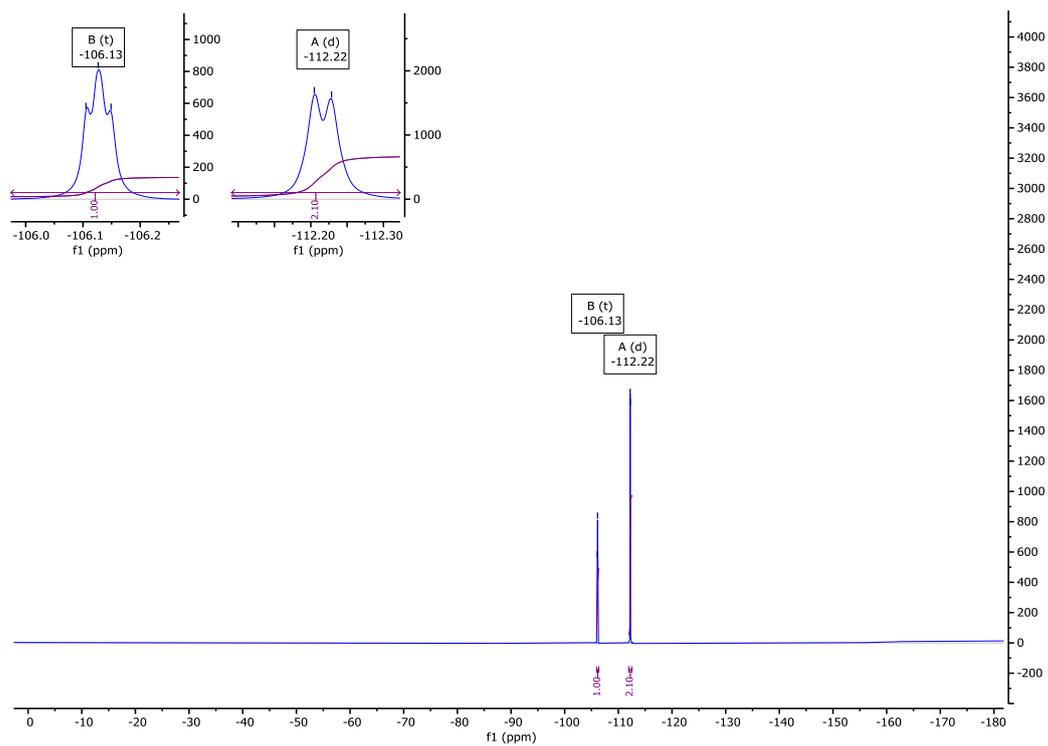

**Fig. S50** $^{19}$F NMR spectra of **NC1200** in CDCl$_3$.

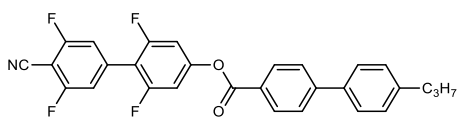

**NC2200**

*4'-cyano-2,3',5',6-tetrafluoro-[1,1'-biphenyl]-4-yl 4'-propyl-[1,1'-biphenyl]-4-carboxylate*

| | |
|---|---|
| Yield: | (white solid) 196 mg, 80 % |
| Re-crystallisation solvent: | MeCN |
| $^1$H NMR (400 MHz): | 8.24 (ddd, *J* = 8.4, 1.9, 1.8 Hz, 2H, Ar-**H**), 7.76 (ddd, *J* = 8.3, 1.6, 1.6 Hz, 2H, Ar-**H**), 7.59 (ddd, *J* = 8.2, 1.8, 1.5 Hz, 2H, Ar-**H**), 7.32 (ddd, *J* = 8.1, 1.9, 1.8 Hz, 2H, Ar-**H**), 7.23 (d, *J* = 8.3 Hz, 2H, Ar-**H**), 7.05 (ddd, *J* = 8.3, 1.4, 1.3 Hz, 2H, Ar-**H**), 2.66 (t, *J* = 7.4 Hz, 2H, Ar-C**H$_2$**-CH$_2$), 1.70 (h, *J* = 7.4 Hz, 2H, CH$_2$-C**H$_2$**-CH$_3$), 0.99 (t, *J* = 7.3 Hz, 3H, CH$_2$-C**H$_3$**). |
| $^{13}$C{$^1$H} NMR (101 MHz): | 164.18, 162.85 (dd, *J* = 261.3, 5.2 Hz), 159.62 (dd, *J* = 251.8, 8.3 Hz), 152.77 (t, *J* = 14.0 Hz), 147.15, 143.51, 136.87, 136.83 (t, *J* = 10.9 Hz), 130.90, 129.23, 127.21, 127.18, 126.56, 114.28 (dd, *J* = 21.0, 2.6 Hz), 112.58 (t, *J* = 17.9 Hz), 107.32 – 106.74 (m), 92.19 (t, *J* = 19.1 Hz), 37.74, 24.52, 13.86. |
| $^{19}$F NMR (376 MHz): | -103.69 (d, *J$_{F-H}$* = 8.8 Hz, 2F, Ar-**F**), -111.90 (d, *J$_{F-H}$* = 9.5 Hz, 2F, Ar-**F**). |

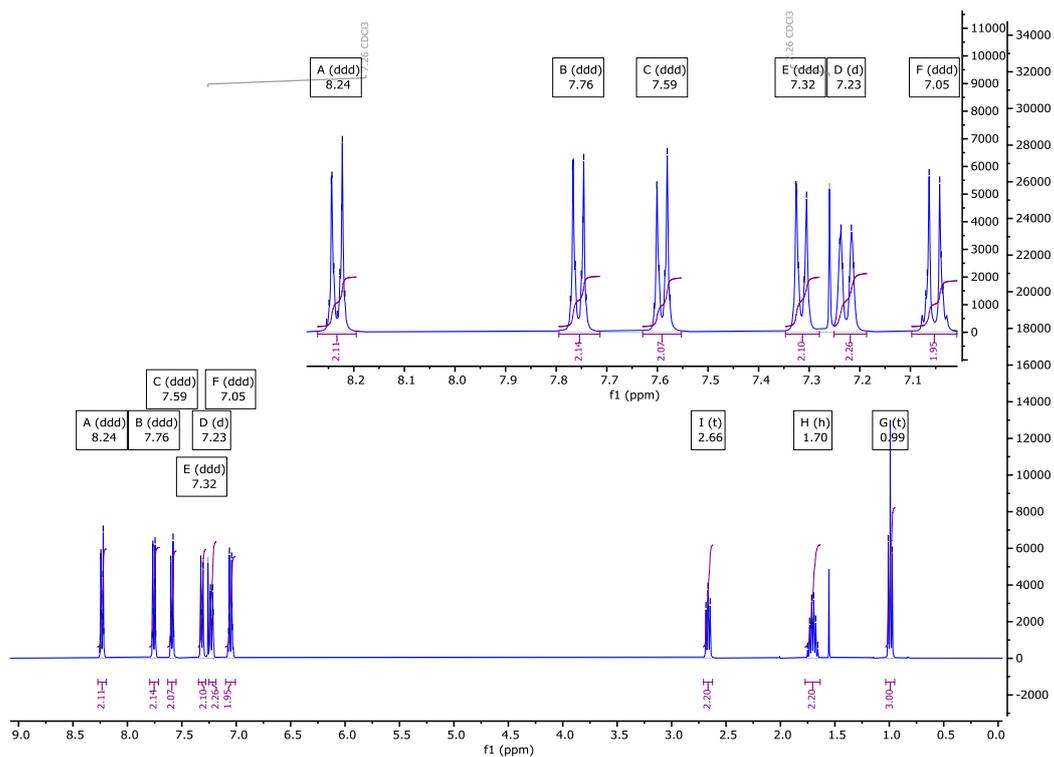

**Fig. S51** $^1$H NMR spectra of **NC2200** in CDCl$_3$.

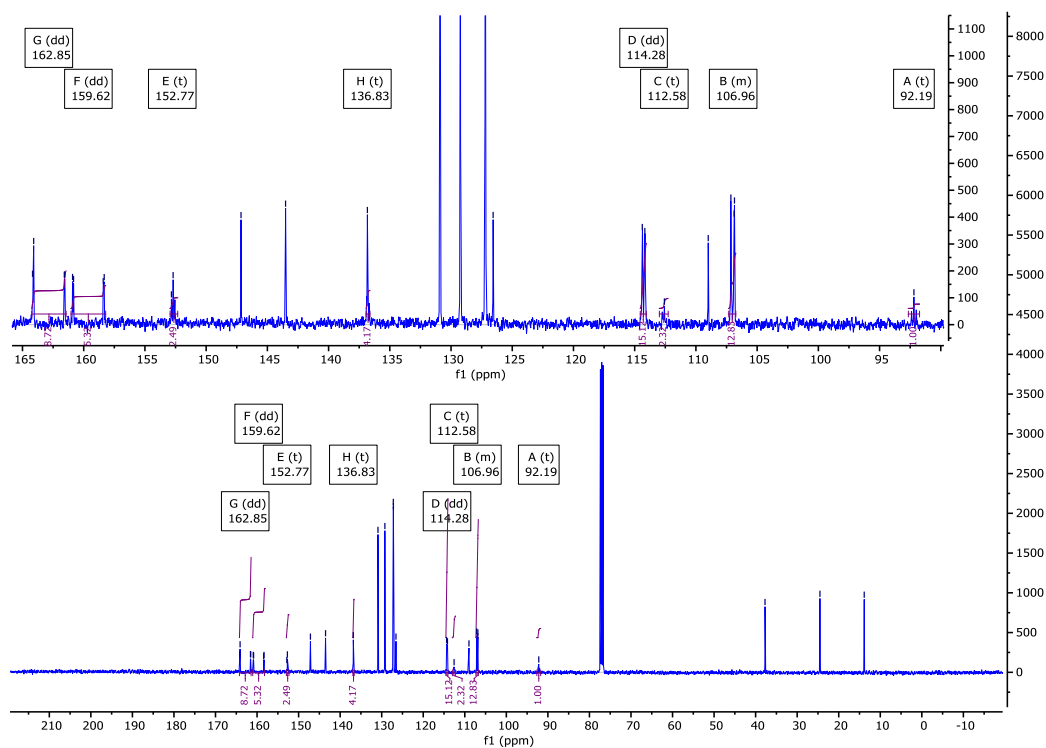

**Fig. S52** $^{13}$C{$^1$H} NMR spectra of **NC2200** in CDCl$_3$.

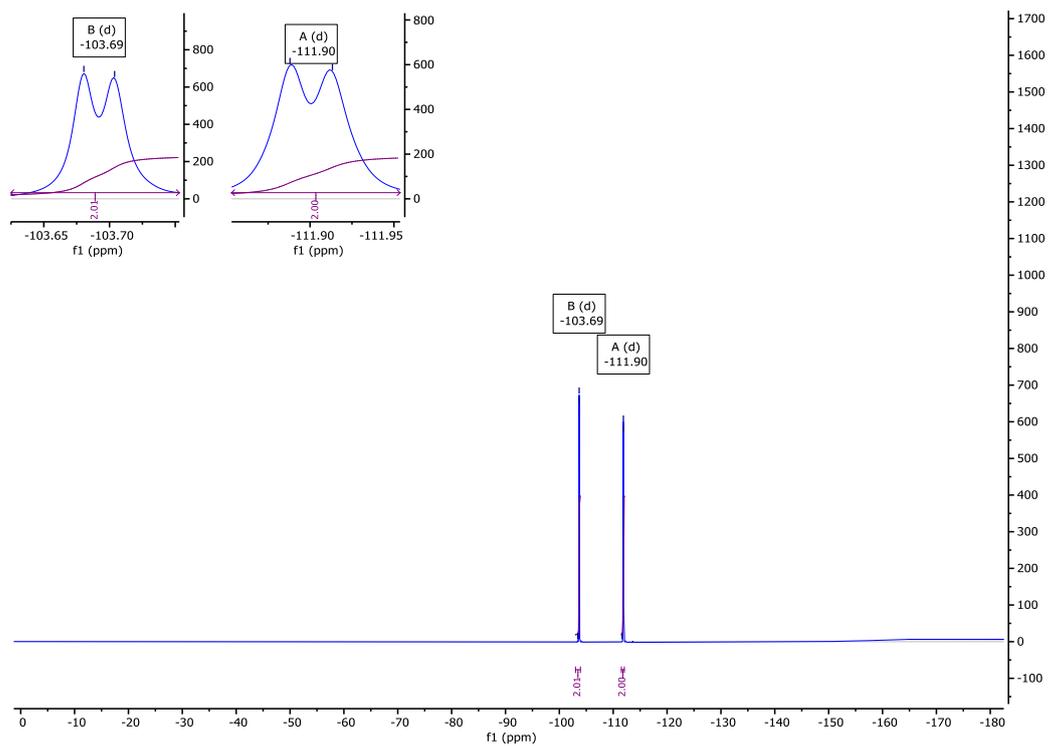

**Fig. S53** $^{19}$F NMR spectra of **NC2200** in CDCl$_3$.

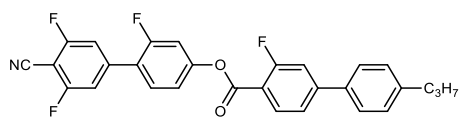

**NC2110**

*4'-cyano-2,3',5'-trifluoro-[1,1'-biphenyl]-4-yl 3-fluoro-4'-propyl-[1,1'-biphenyl]-4-carboxylate*

| | |
|---|---|
| Yield: | (white crystalline solid) 193 mg, 79 % |
| Re-crystallisation solvent: | MeCN |
| $^1$H NMR (400 MHz): | 8.08 (t, *J* = 7.9 Hz, 1H, Ar-**H**), 7.50 (d, *J* = 8.1 Hz, 2H, Ar-**H**), 7.48 – 7.34 (m, 2H, Ar-**H**)*, 7.30 – 7.10 (m, 6H, Ar-**H**)*†, 2.60 (t, *J* = 7.6 Hz, 2H, Ar-C**H$_2$**-CH$_2$), 1.63 (h, *J* = 7.4 Hz, 2H, CH$_2$-C**H$_2$**-CH$_3$), 0.92 (t, *J* = 7.3 Hz, 3H, CH$_2$-CH$_3$). * Overlapping Signals, †Overlapping CDCl$_3$. |
| $^{13}$C{$^1$H} NMR (101 MHz): | 164.36, 164.11, 164.23 (dd, *J* = 260.6, 5.4 Hz), 162.04 (d, *J* = 4.1 Hz), 161.70 (d, *J* = 261.6 Hz), 159.53 (d, *J* = 252.7 Hz), 152.59 (d, *J* = 11.4 Hz), 149.25 (d, *J* = 9.1 Hz), 144.16, 143.11 (t, *J* = 9.9 Hz), 135.59, 133.00, 130.54 (d, *J* = 3.7 Hz), 129.34, 127.06, 122.92 (d, *J* = 12.5 Hz), 122.56 (d, *J* = 3.2 Hz), 118.71 (d, *J* = 3.7 Hz), 115.32 (d, *J* = 23.2 Hz), 112.62 (dt, *J* = 21.1, 3.4 Hz), 111.12 (d, *J* = 25.7 Hz), 91.46 (t, *J* = 19.6 Hz), 37.74, 24.47, 13.84. |
| $^{19}$F NMR (376 MHz): | -103.50 (d, *J$_{F-H}$* = 9.4 Hz, 2F, Ar-**F**), -107.17 (dd, *J$_{F-H}$* = 12.3 Hz, *J$_{F-H}$* = 7.6 Hz, 1F, Ar-**F**), -113.40 (t, *J$_{F-H}$* = 9.8 Hz, 1F, Ar-**F**). |

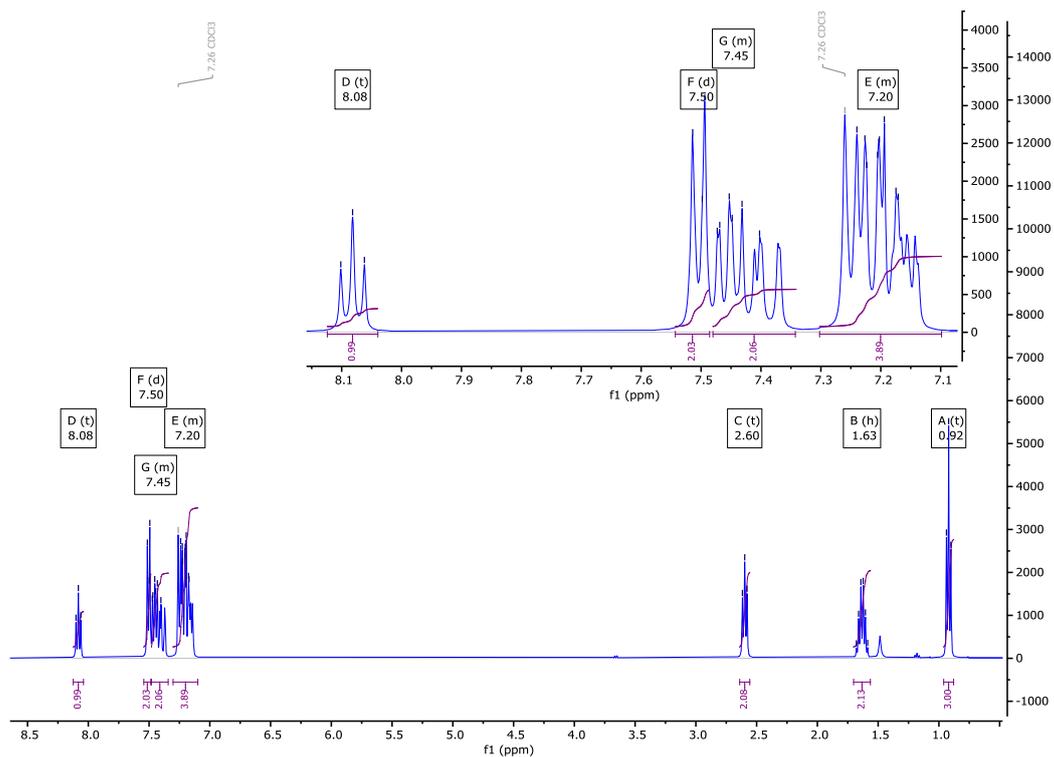

**Fig. S54** ¹H NMR spectra of **NC2110** in CDCl₃.

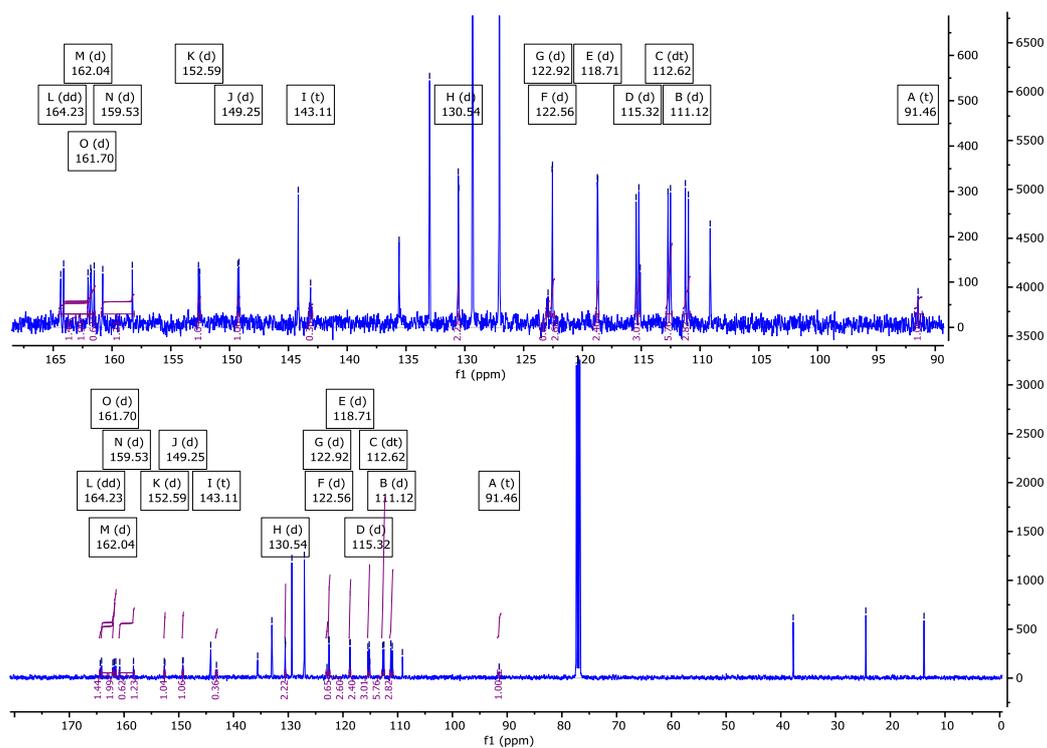

**Fig. S55** ¹³C{¹H} NMR spectra of **NC2110** in CDCl₃.

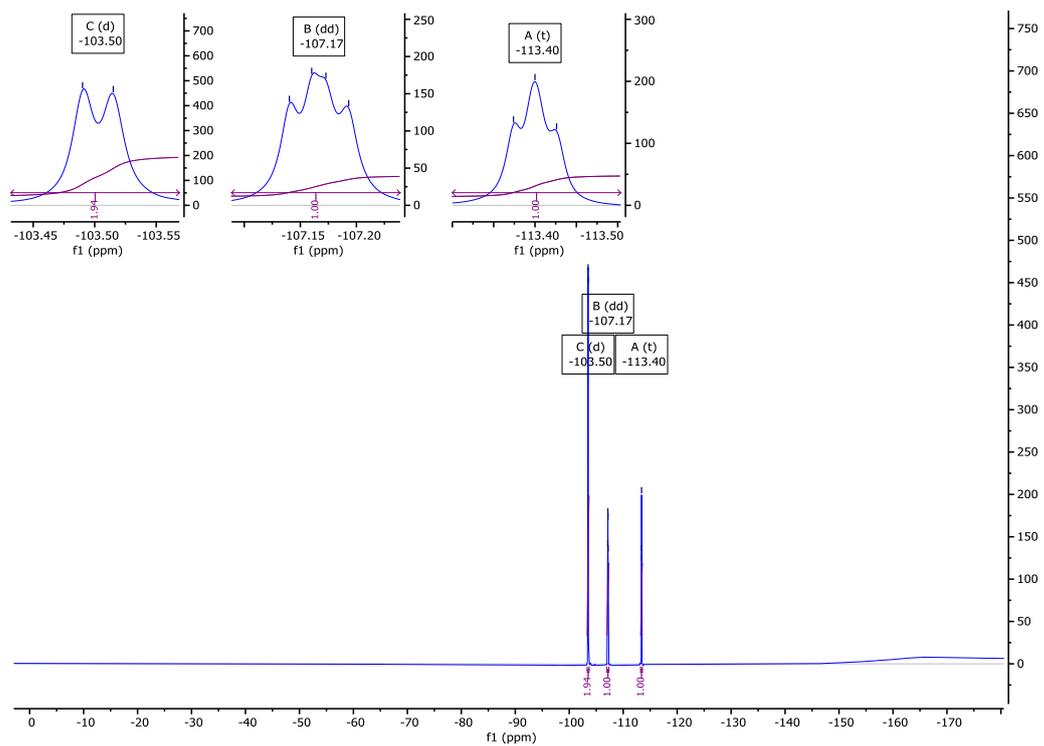

**Fig. S56** [19]F NMR spectra of **NC2110** in CDCl$_3$.

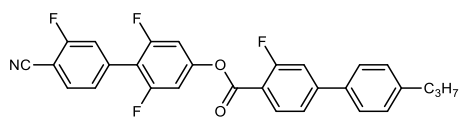

**NC1210**

*4'-cyano-2,3',6-trifluoro-[1,1'-biphenyl]-4-yl 3-fluoro-4'-propyl-[1,1'-biphenyl]-4-carboxylate*

| | |
|---|---|
| Yield: | (white solid) 183 mg, 75 % |
| Re-crystallisation solvent: | MeCN |
| $^1$H NMR (400 MHz): | 8.14 (t, *J* = 7.8 Hz, 1H, Ar-**H**), 7.73 (t, *J* = 7.3 Hz, 1Hm Ar-**H**), 7.62 – 7.35 (m, 6H, Ar-**H**)*, 7.32 (d, *J* = 7.8 Hz, 2H, Ar-**H**), 7.05 (d, *J* = 8.4 Hz, 2H, Ar-**H**), 2.66 (t, *J* = 7.6 Hz, 2H, Ar-C**H$_2$**-CH$_2$), 1.70 (h, *J* = 7.4 Hz, 2H, CH$_2$-C**H$_2$**-CH$_3$), 0.99 (t, *J* = 7.3 Hz, 3H, CH$_2$-C**H$_3$**). * Overlapping Signals. |
| $^{13}$C{$^1$H} NMR (101 MHz): | 162.83 (d, *J* = 260.8 Hz), 161.70 (d, *J* = 4.5 Hz), 159.69 (dd, *J* = 251.2, 8.6 Hz), 151.83 (t, *J* = 14.3 Hz), 149.43 (d, *J* = 9.0 Hz), 144.22, 136.04, 136.00 (d, *J* = 8.7 Hz), 133.25, 133.00, 129.35, 127.07, 126.98 (d, *J* = 18.5 Hz)*, 122.60 (d, *J* = 3.3 Hz), 118.51 (d, *J* = 20.7 Hz), 115.35 (d, *J* = 23.0 Hz), 114.85 (d, *J* = 9.3 Hz), 113.73, 107.19 – 106.36 (m), 101.28 (d, *J* = 15.5 Hz), 37.74, 24.47, 13.83. * Overlapping Signals. |
| $^{19}$F NMR (376 MHz): | -106.12 (t, *J$_{F-H}$* = 8.3 Hz, 1F, Ar-**F**), -106.96 (dd, *J$_{F-H}$* = 12.2 Hz, *J$_{F-H}$* = 7.6 Hz, 1F. Ar-**F**), -112.14 (d, *J$_{F-H}$* = 9.3 Hz, 2F, Ar-**F**). |

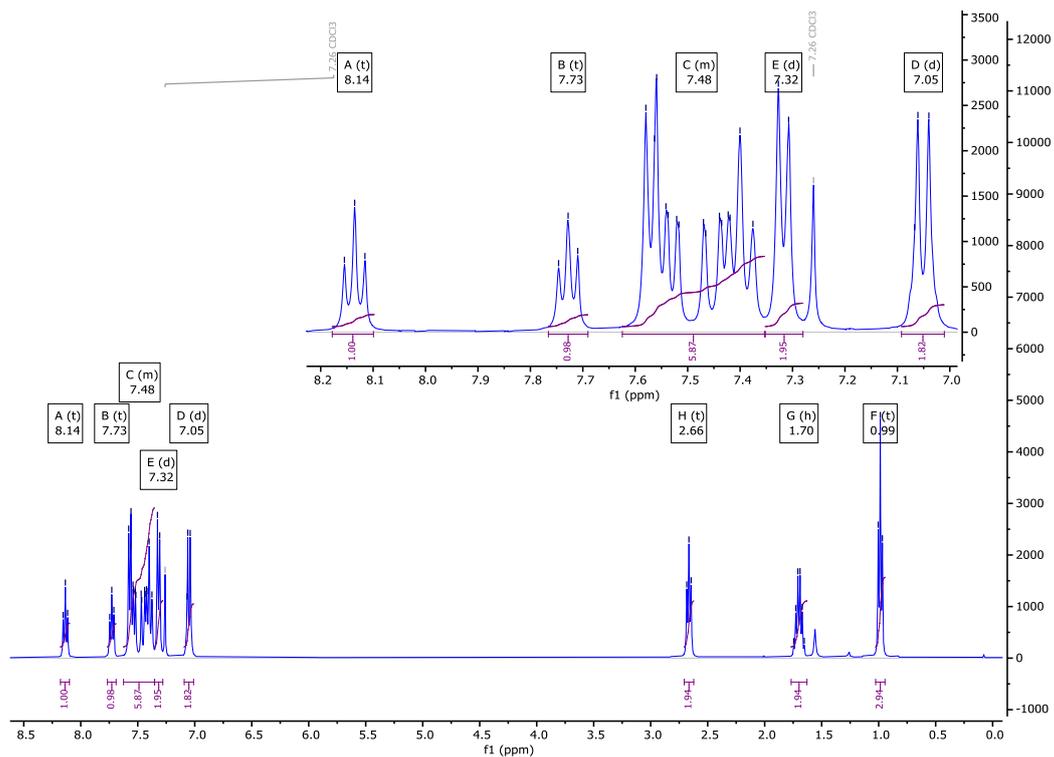

**Fig. S57**  $^1$H NMR spectra of **NC2110** in CDCl$_3$.

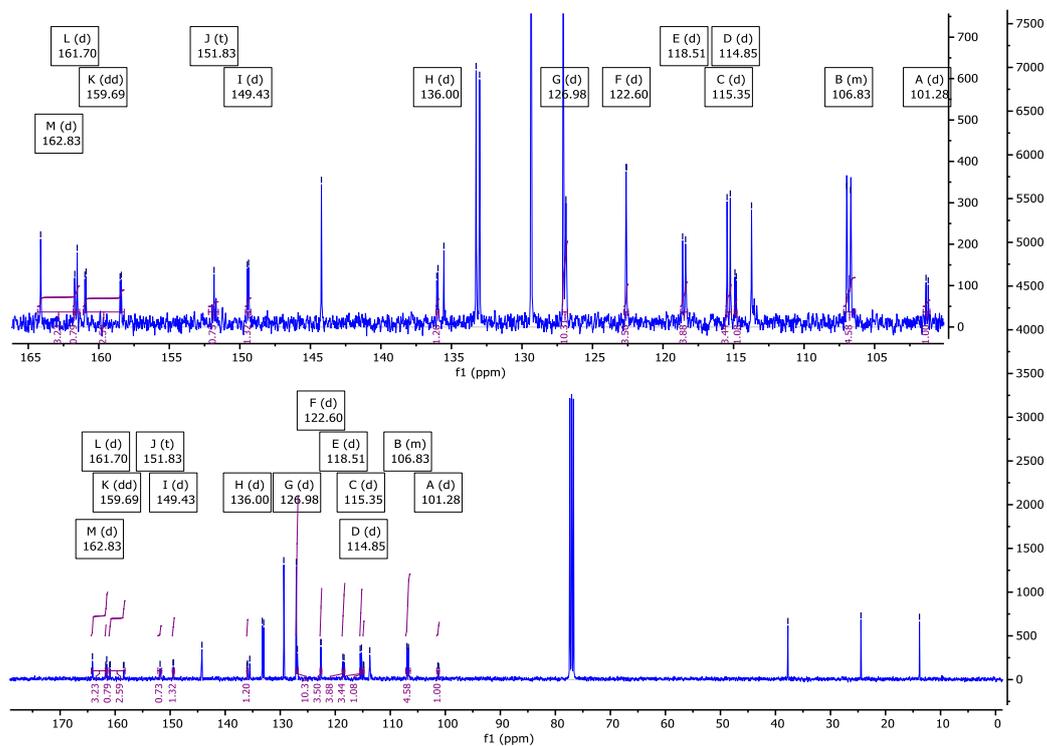

**Fig. S58**  $^{13}$C{$^1$H} NMR spectra of **NC2110** in CDCl$_3$.

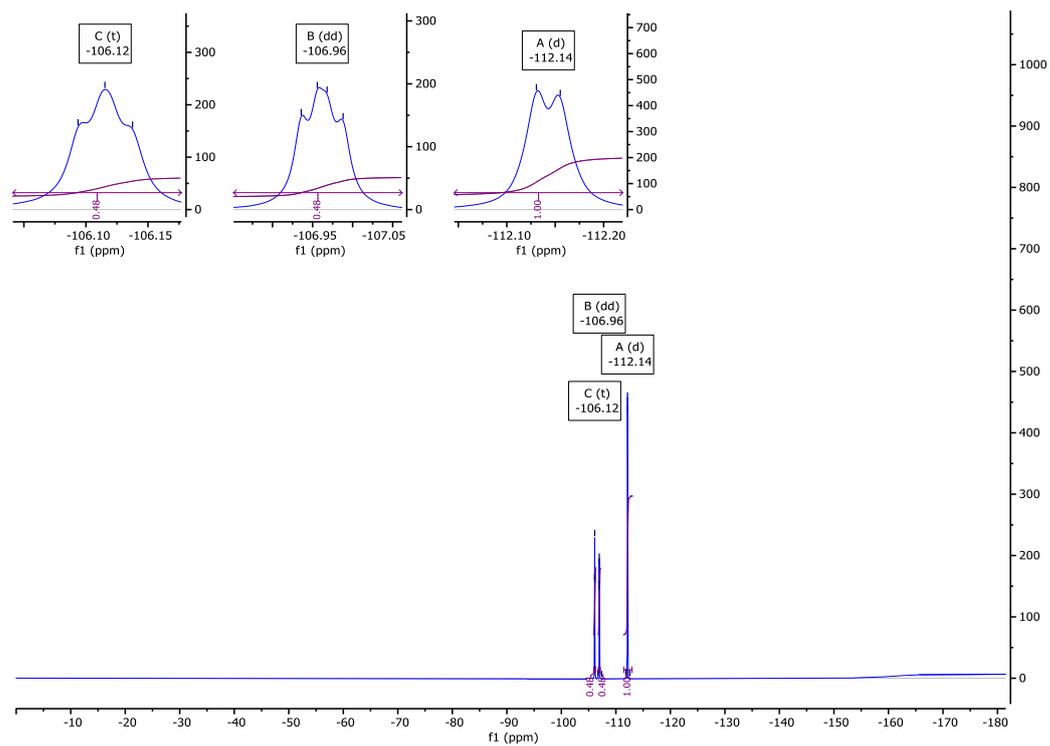

**Fig. S59**     $^{19}$F NMR spectra of **NC2110** in CDCl$_3$.

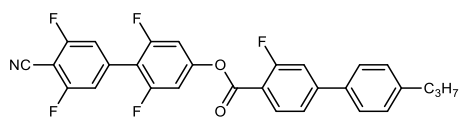

**NC2210**

*4'-cyano-2,3',5',6-tetrafluoro-[1,1'-biphenyl]-4-yl 3-fluoro-4'-propyl-[1,1'-biphenyl]-4-carboxylate*

| | |
|---|---|
| Yield: | (white solid) 190 mg, 75 % |
| Re-crystallisation solvent: | MeCN |
| $^1$H NMR (400 MHz): | 8.13 (t, *J* = 7.9 Hz, 1H, Ar-**H**), 7.57 (d, *J* = 8.1 Hz, 2H, Ar-**H**), 7.53 (dd, *J* = 8.3, 1.7 Hz, 1H, Ar-**H**), 7.45 (dd, *J* = 12.1, 1.7 Hz, 1H, Ar-**H**), 7.32 (d, *J* = 7.9 Hz, 2H, Ar-**H**), 7.23 (d, *J* = 8.8 Hz, 2H, Ar-**H**), 7.07 (ddd, *J* = 8.4, 5.6, 3.0 Hz, 2H, Ar-**H**), 2.66 (t, *J* = 7.6 Hz, 2H, Ar-C**H$_2$**-CH$_2$), 1.70 (h, *J* = 7.4 Hz, 2H, CH$_2$-C**H$_2$**-CH$_3$), 0.98 (t, *J* = 7.3 Hz, 3H, CH$_2$-C**H$_3$**). |
| $^{13}$C{$^1$H} NMR (101 MHz): | 162.84 (dd, *J* = 261.6, 5.0 Hz), 161.59 (d, *J* = 3.9 Hz), 159.60 (dd, *J* = 252.0, 8.2 Hz), 152.20 (t, *J* = 14.6 Hz), 149.51 (d, *J* = 8.9 Hz), 144.25, 136.72 (t, *J* = 10.6 Hz), 135.52, 133.01, 129.35, 127.07, 122.62 (d, *J* = 3.3 Hz), 115.36 (d, *J* = 22.9 Hz), 114.75 (d, *J* = 9.6 Hz), 114.28 (d, *J* = 21.1 Hz), 112.71 (t, *J* = 18.8 Hz), 108.98, 107.28 – 106.69 (m), 92.22 (t, *J* = 18.7 Hz), 37.73, 24.47, 13.83. |
| $^{19}$F NMR (376 MHz): | -103.67 (d, *J$_{F-H}$* = 9.2 Hz, 2F, Ar-**F**), -106.90 (dd, *J$_{F-H}$* = 12.2 Hz, *J$_{F-H}$* = 7.6 Hz, 1F), -111.82 (d, *J$_{F-H}$* = 9.5 Hz, 2F, Ar-**F**). |

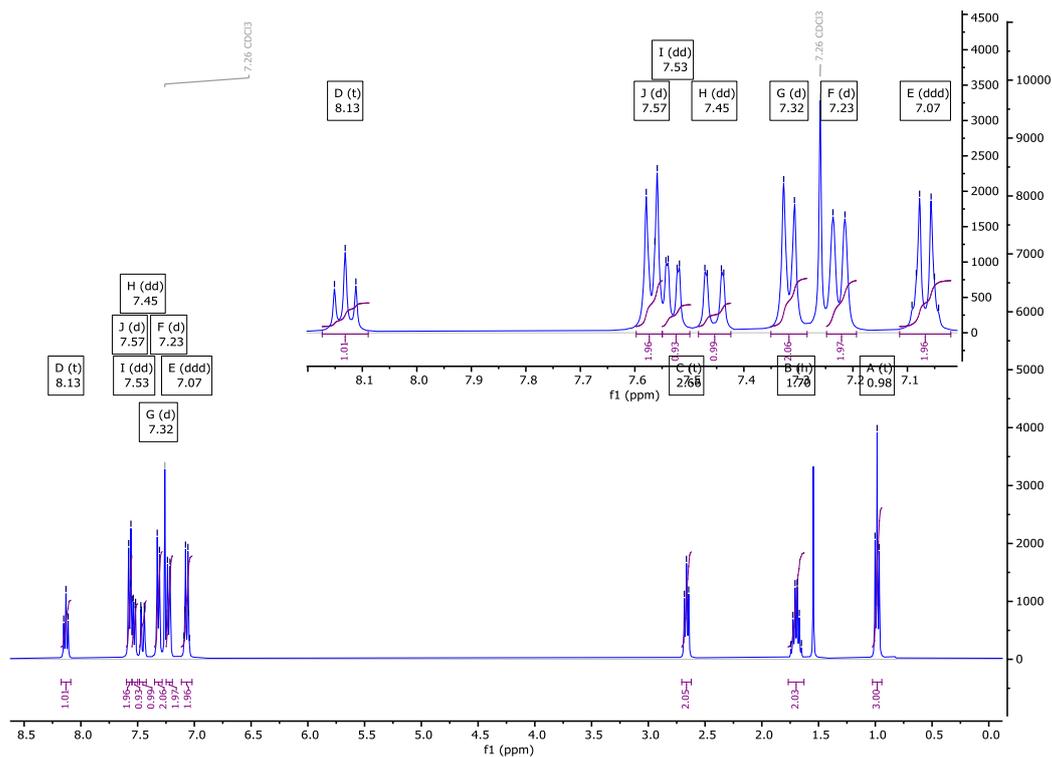

**Fig. S60**	$^1$H NMR spectra of **NC2210** in CDCl$_3$.

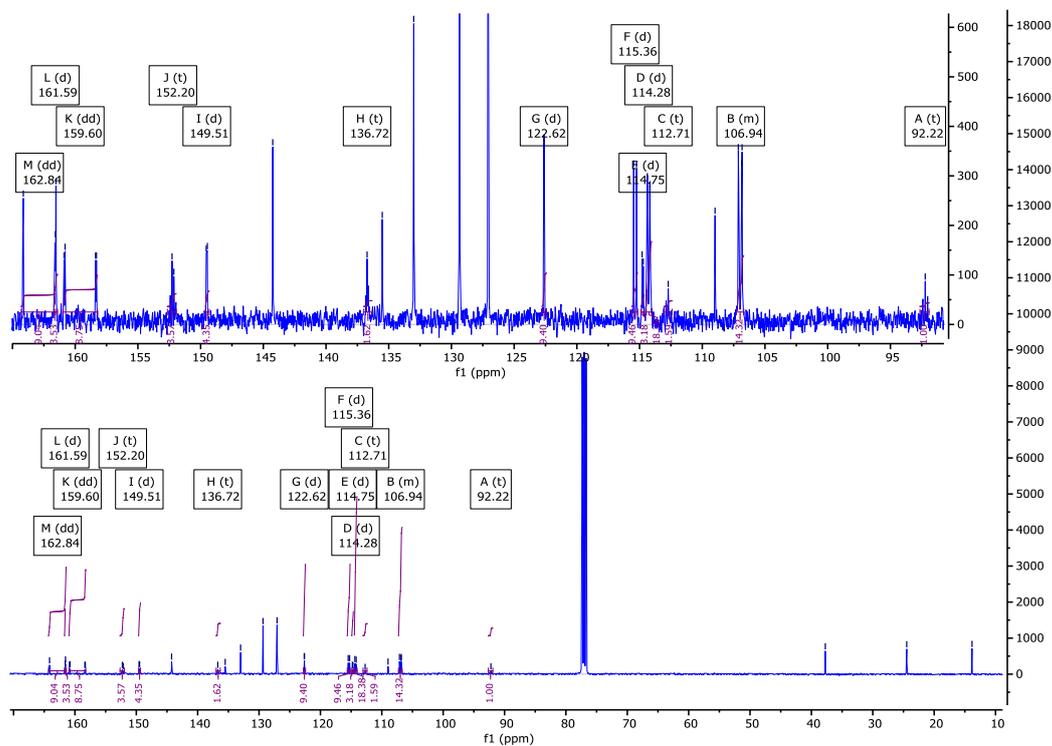

**Fig. S61**	$^{13}$C{$^1$H} NMR spectra of **NC2210** in CDCl$_3$.

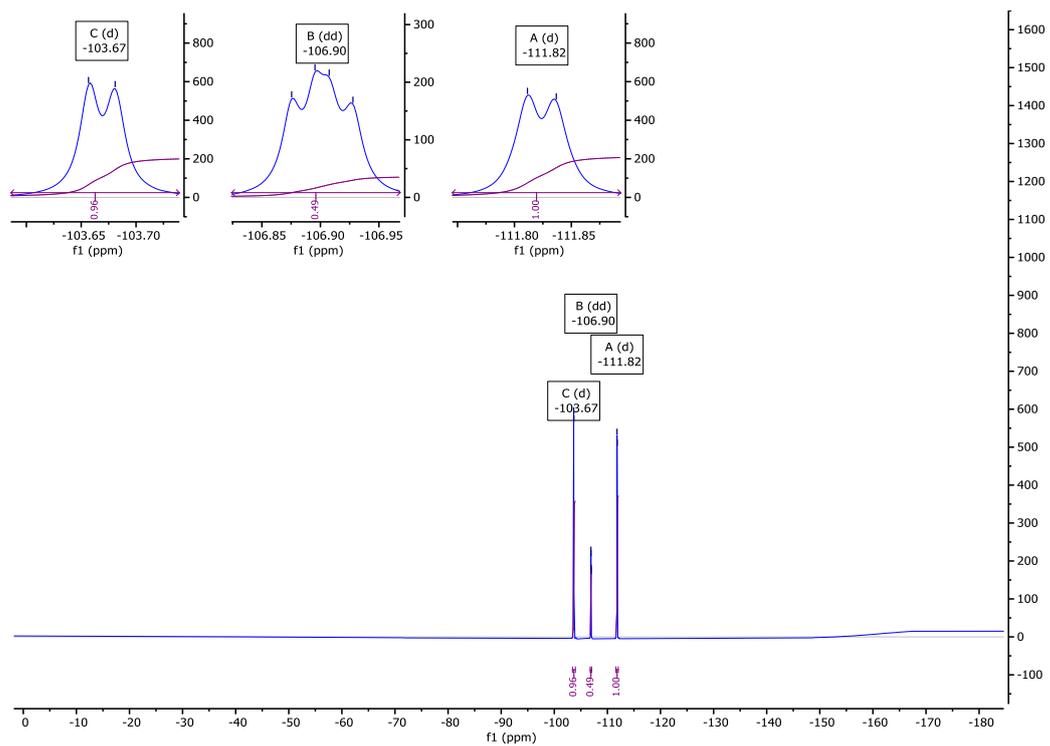

**Fig. S62** 19F NMR spectra of **NC2210** in CDCl$_3$.

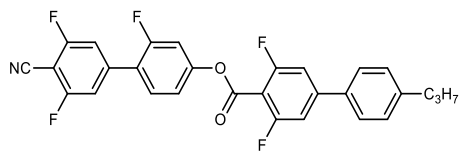

**NC2120**

*4'-cyano-2,3',5'-trifluoro-[1,1'-biphenyl]-4-yl 3,5-difluoro-4'-propyl-[1,1'-biphenyl]-4-carboxylate*

| | |
|---|---|
| Yield: | (white solid) 203 mg, 80 % |
| Re-crystallisation solvent: | MeCN |
| $^1$H NMR (400 MHz): | 7.56 – 7.47 (m, 3H, Ar-**H**)*, 7.34 – 7.22 (m, 8H, Ar-**H**)*†, 2.66 (t, *J* = 7.4 Hz, 2H, Ar-C**H$_2$**-CH$_2$), 1.69 (h, *J* = 7.3 Hz, 2H, CH$_2$-C**H$_2$**-CH$_3$), 0.98 (t, *J* = 7.3 Hz, 3H, CH$_2$-C**H$_3$**). *Overlapping Signals, † Overlapping CDCl$_3$. |
| $^{13}$C{$^1$H} NMR (101 MHz): | 164.39 (dd, *J* = 260.5, 5.3 Hz), 162.35 (dd, *J* = 258.3, 5.8 Hz), 159.26 (d, *J* = 252.3 Hz), 152.20 (d, *J* = 11.4 Hz), 148.15 (t, *J* = 10.4 Hz), 144.66, 143.07 (t, *J* = 10.2 Hz), 134.77, 130.61 (d, *J* = 3.5 Hz), 129.43, 126.89, 123.26 (d, *J* = 12.7 Hz), 118.61 (d, *J* = 3.7 Hz), 112.68 (dt, *J* = 20.6, 3.4 Hz), 111.04 (d, *J* = 25.8 Hz), 110.50 (dd, *J* = 23.4, 3.0 Hz), 109.11, 107.12 (t, *J* = 16.9 Hz), 91.53 (t, *J* = 19.3 Hz), 37.72, 24.43, 13.81. |
| $^{19}$F NMR (376 MHz): | -103.45 (d, *J$_{F-H}$* = 9.1 Hz, 2F, Ar-**F**), -108.08 (d, *J$_{F-H}$* = 10.7 Hz, 2F, Ar-**F**), -113.21 (t, *J$_{F-H}$* = 9.7 Hz, 1F, Ar-**F**). |

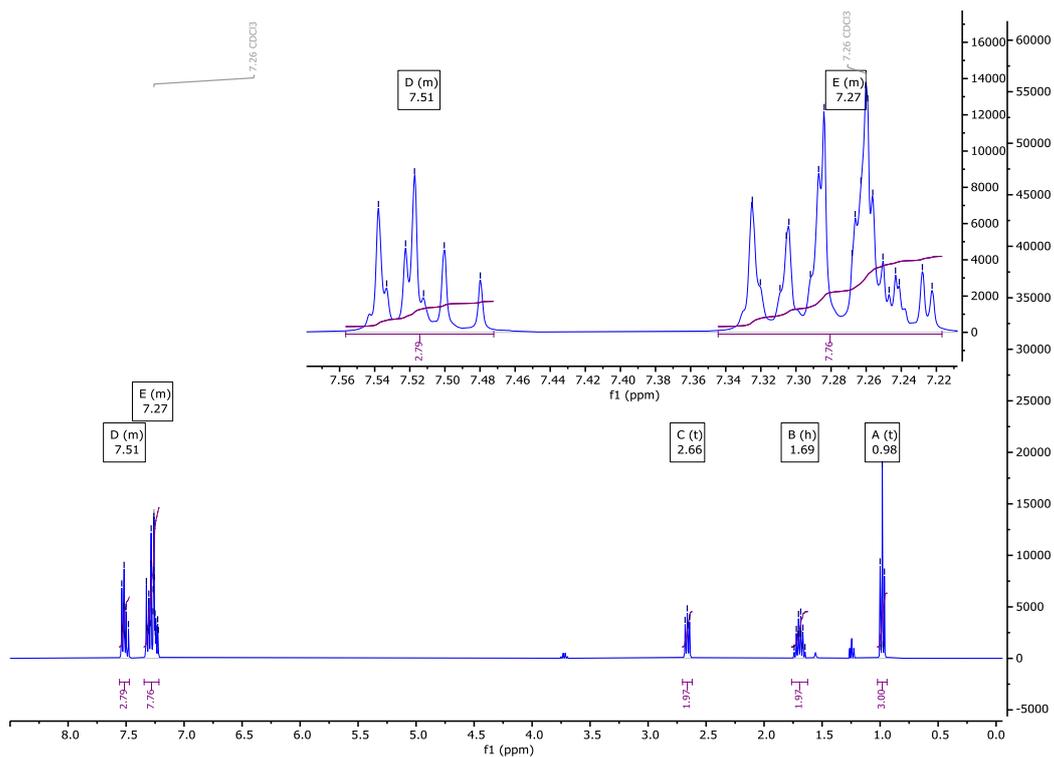

**Fig. S63**  $^1$H NMR spectra of **NC2120** in CDCl$_3$.

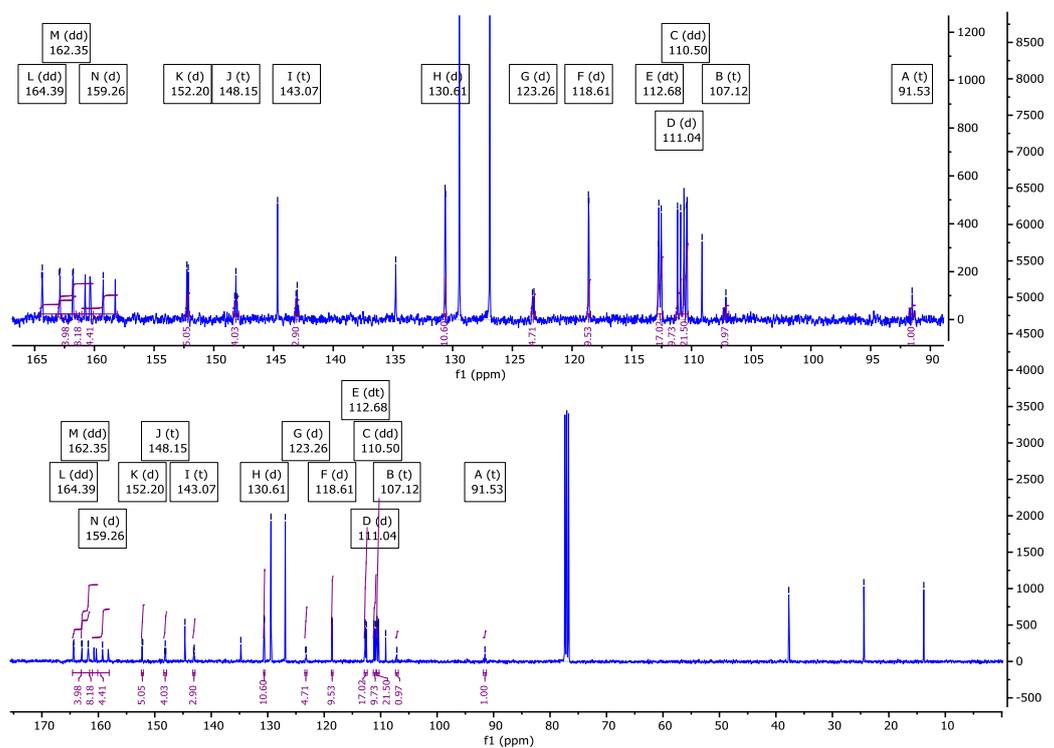

**Fig. S64**  $^{13}$C{$^1$H} NMR spectra of **NC2120** in CDCl$_3$.

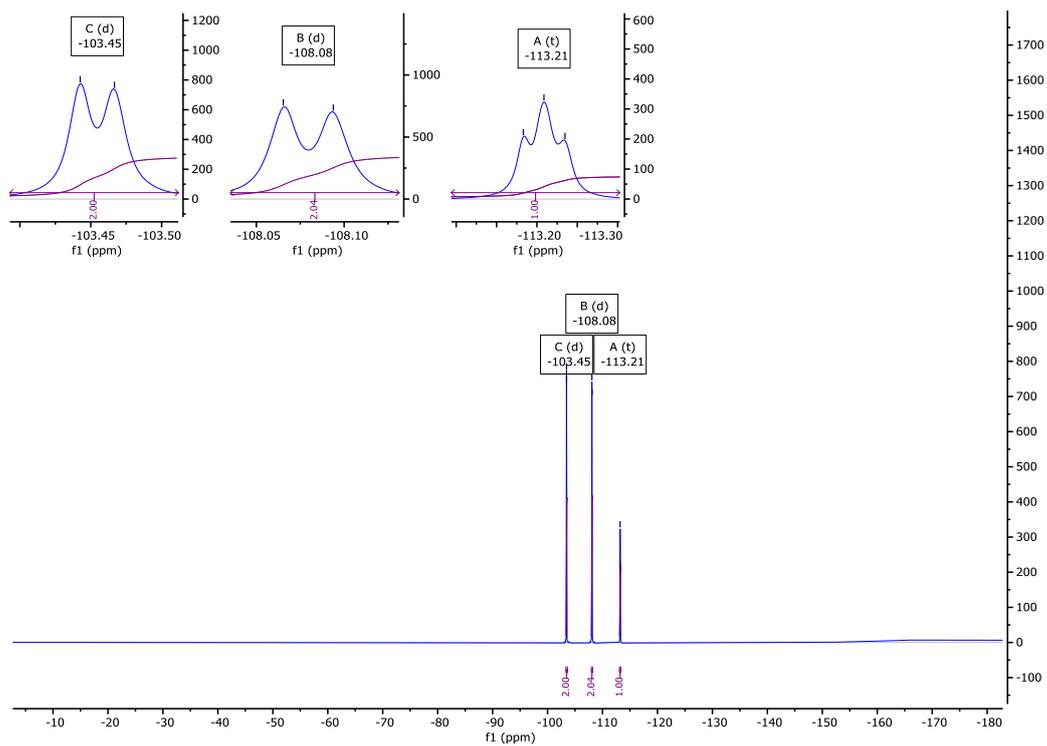

**Fig. S65** ¹⁹F NMR spectra of **NC2120** in CDCl$_3$.

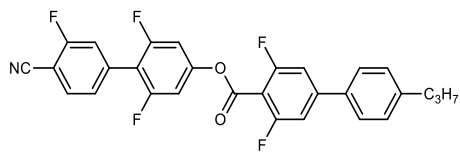

**NC1220**

*4'-cyano-2,3',6-trifluoro-[1,1'-biphenyl]-4-yl 3,5-difluoro-4'-propyl-[1,1'-biphenyl]-4-carboxylate*

| | |
|---|---|
| Yield: | (white needles) 175 mg, % |
| Re-crystallisation solvent: | MeCN |
| $^1$H NMR (400 MHz): | 7.73 (t, *J* = 7.3 Hz, 1H, Ar-**H**), 7.52 (d, *J* = 8.0 Hz, 2H, Ar-**H**), 7.39 (m, 2H, Ar-**H**), 7.34 – 7.27 (m, 4H, Ar-**H**)*, 7.07 (ddd, *J* = 8.5, 5.9, 3.5 Hz, 2H, Ar-**H**), 2.66 (t, *J* = 7.6 Hz, 2H, Ar-C**H$_2$**-CH$_2$), 1.69 (h, *J* = 7.4 Hz, 2H, CH$_2$-C**H$_2$**-CH$_3$), 0.98 (t, *J* = 7.4 Hz, 3H, CH$_2$-C**H$_3$**). *Overlapping Signals. |
| $^{13}$C{$^1$H} NMR (101 MHz): | 162.83 (d, *J* = 259.1 Hz), 161.67 (dd, *J* = 258.7, 6.3 Hz), 159.55 (dd, *J* = 251.3, 9.7 Hz), 158.95, 151.42 (t, *J* = 14.4 Hz), 148.36 (t, *J* = 10.4 Hz), 144.72, 135.90 (d, *J* = 8.6 Hz), 134.72, 133.27, 129.44, 126.89, 118.51 (d, *J* = 20.9 Hz), 118.40, 113.70 (d, *J* = 8.2 Hz), 110.53 (dd, *J* = 23.1, 3.0 Hz), 107.18 – 106.16 (m), 101.34 (d, *J* = 15.1 Hz), 37.72, 24.43, 13.81. |
| $^{19}$F NMR (376 MHz): | -106.08 (t, *J$_{F-H}$* = 8.1 Hz, 1F, Ar-**F**), -107.88 (d, *J$_{F-H}$* = 10.7 Hz, 2F, Ar-**F**), -111.88 (d, *J$_{F-H}$* = 9.1 Hz, 2F, Ar-**F**). |

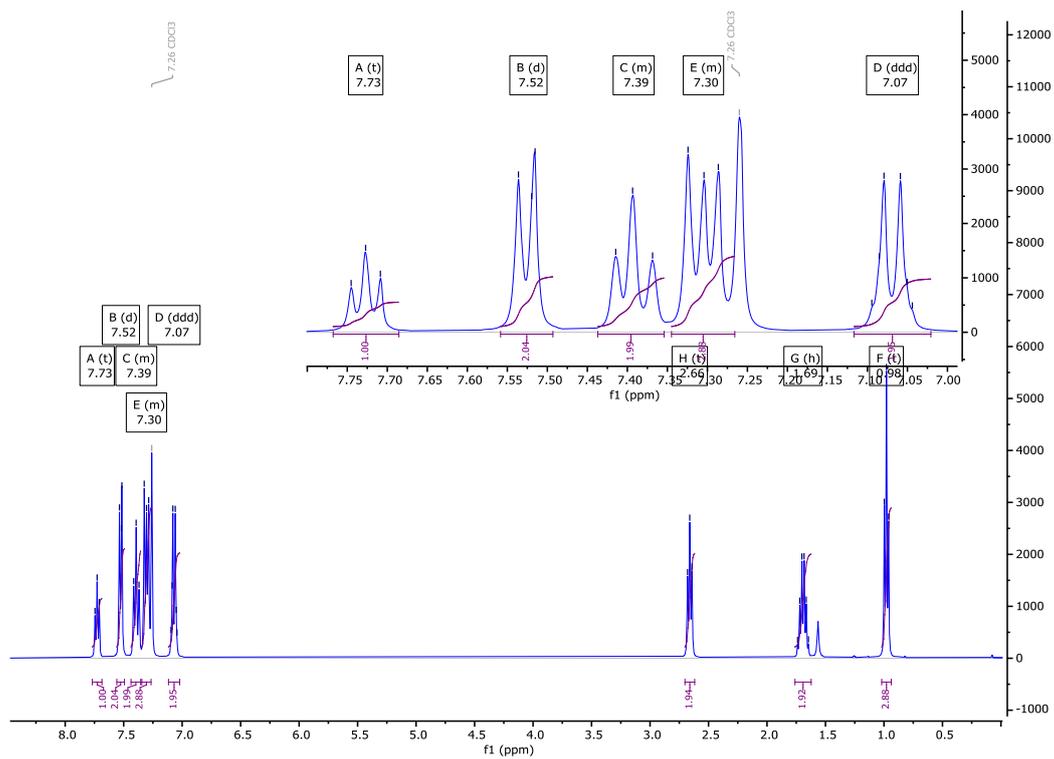

**Fig. S66** $^{1}$H NMR spectra of **NC1220** in CDCl$_3$.

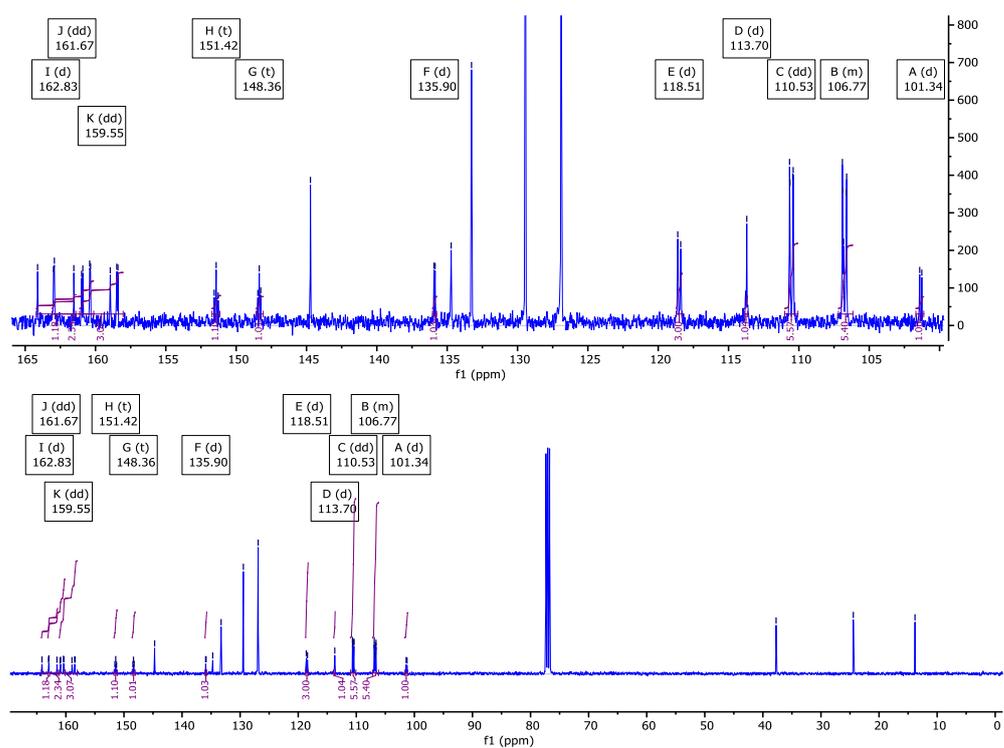

**Fig. S67** $^{13}$C{$^{1}$H} NMR spectra of **NC1220** in CDCl$_3$.

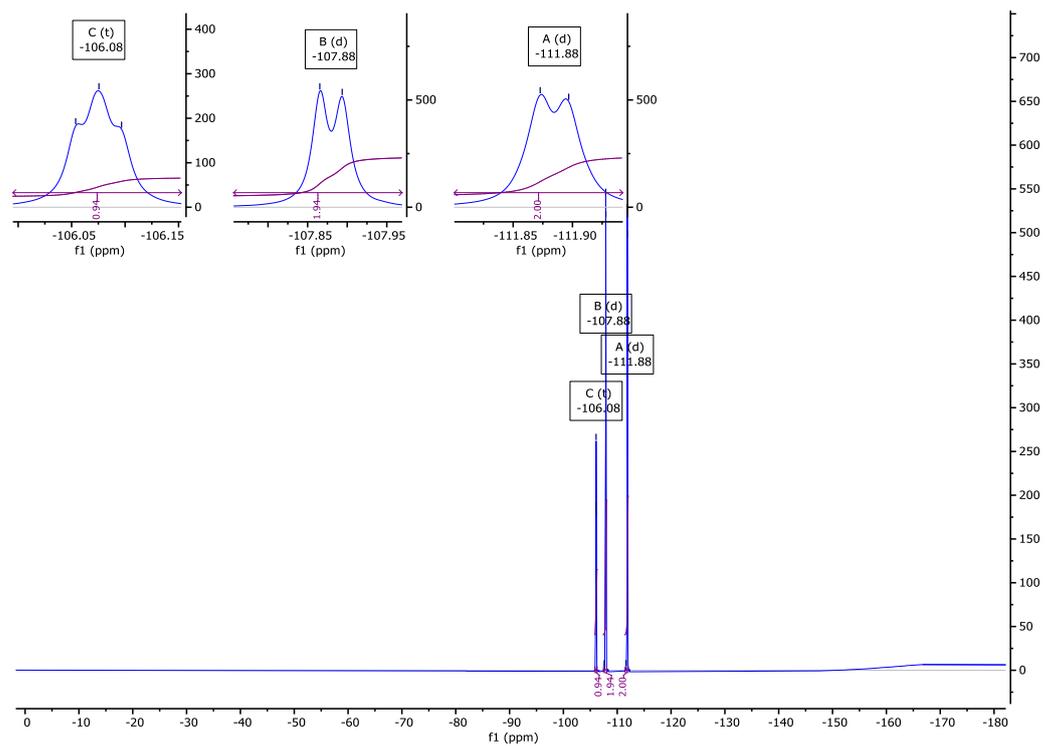

**Fig. S68** ¹⁹F NMR spectra of **NC1220** in CDCl$_3$.

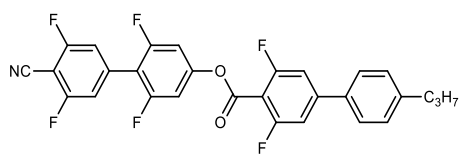

**NC2220**

*4'-cyano-2,3',5',6-tetrafluoro-[1,1'-biphenyl]-4-yl 3,5-difluoro-4'-propyl-[1,1'-biphenyl]-4-carboxylate*

| | |
|---|---|
| Yield: | (white crystalline solid) 190 mg, 72 % |
| Re-crystallisation solvent: | MeCN |
| $^1$H NMR (400 MHz): | 7.53 (ddd, *J* = 8.2, 1.9, 1.6 Hz, 2H, Ar-**H**), 7.32 (d, *J* = 8.3 Hz, 2H, Ar-**H**), 7.30 – 7.25 (m, 2H, Ar-**H**)†, 7.22 (d, *J* = 8.8 Hz, 2H, Ar-**H**), 7.09 (ddd, *J* = 8.2, 5.7, 3.2 Hz, 2H, Ar-**H**), 2.66 (t, *J* = 7.5 Hz, 2H, Ar-C**H$_2$**-CH$_2$), 1.69 (h, *J* = 7.5 Hz, 2H, CH$_2$-C**H$_2$**-CH$_3$), 0.98 (t, *J* = 7.3 Hz, 3H, CH$_2$-C**H$_3$**). † Overlapping CDCl$_3$. |
| $^{13}$C{$^1$H} NMR (101 MHz): | 162.90 (dd, *J* = 261.4, 4.4 Hz), 160.85 (dd, *J* = 259.1, 8.1 Hz), 159.27 (dd, *J* = 252.6, 8.6 Hz), 158.85, 151.86 (t, *J* = 14.3 Hz), 148.45 (t, *J* = 10.5 Hz), 144.75, 136.68 (t, *J* = 10.7 Hz), 134.68, 129.45, 126.89, 114.28 (dd, *J* = 20.6, 2.6 Hz), 113.06 (t, *J* = 17.6 Hz), 110.55 (dd, *J* = 23.1, 3.1 Hz), 108.96, 107.19 – 106.61 (m), 92.28 (t, *J* = 19.2 Hz), 37.72, 24.43, 13.81. |
| $^{19}$F NMR (376 MHz): | -103.62 (d, *J$_{F-H}$* = 9.0 Hz, 2F, Ar-**F**), -107.81 (d, *J$_{F-H}$* = 10.9 Hz, 2F, Ar-**F**), -111.57 (d, *J$_{F-H}$* = 9.4 Hz, 2F, Ar-**F**). |

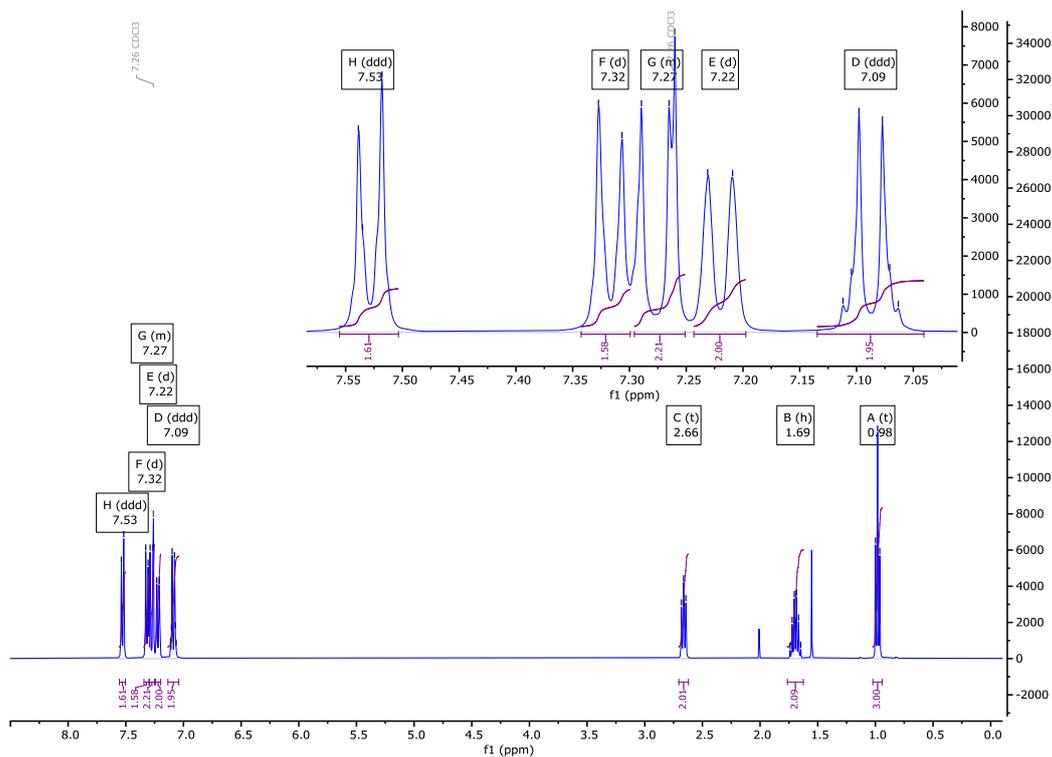

**Fig. S69** $^1$H NMR spectra of **NC2220** in CDCl$_3$.

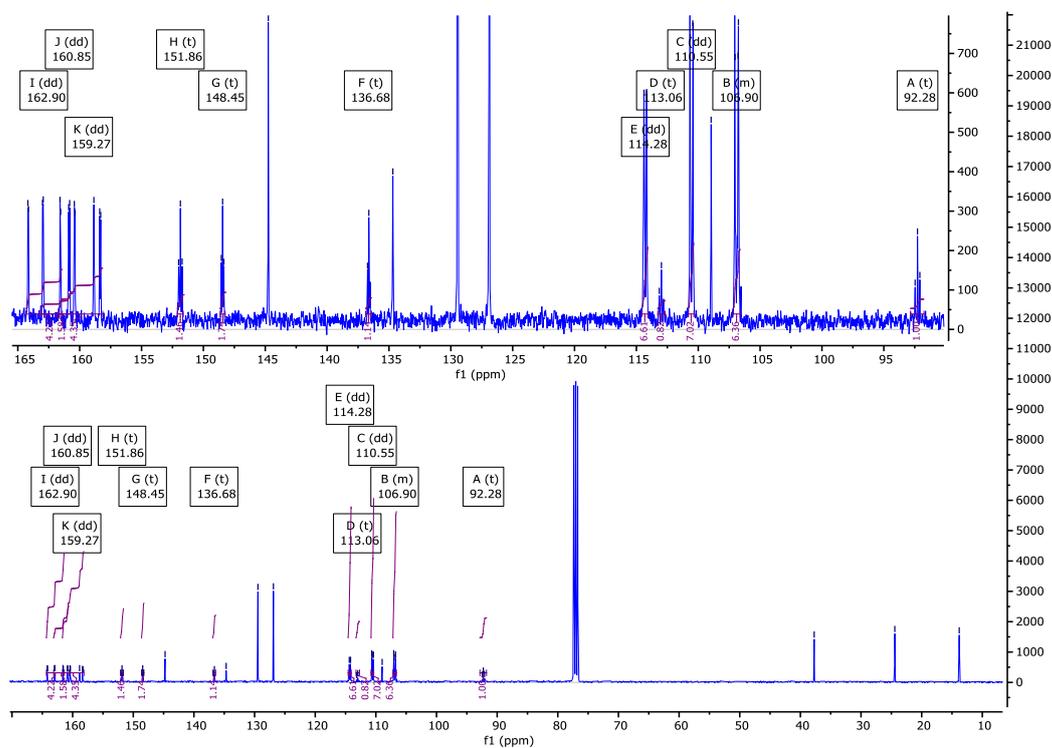

**Fig. S70** $^{13}$C{$^1$H} NMR spectra of **NC2220** in CDCl$_3$.

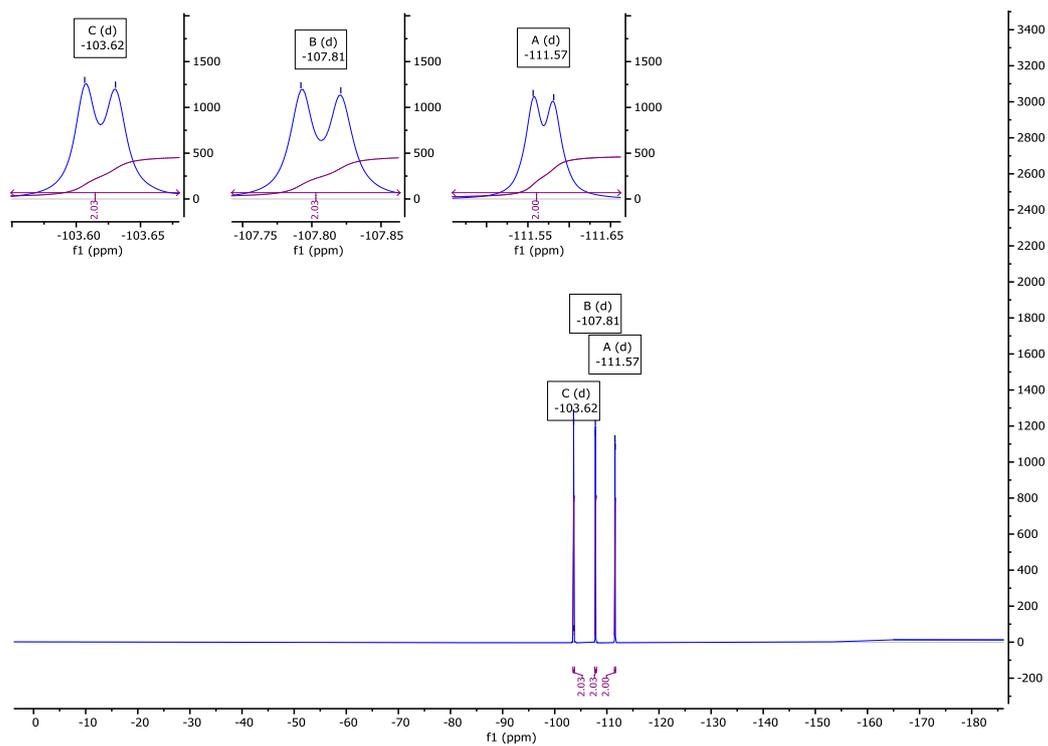

**Fig. S71**  $^{19}$F NMR spectra of **NC2220** in CDCl$_3$.

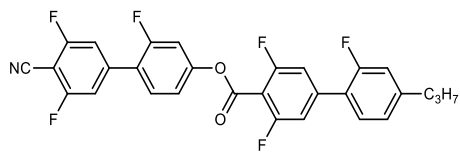

**NC2121**

*4'-cyano-2,3',5'-trifluoro-[1,1'-biphenyl]-4-yl 2',3,5-trifluoro-4'-propyl-[1,1'-biphenyl]-4-carboxylate*

| | |
|---|---|
| Yield: | (white solid) 158 mg, 60 % |
| Re-crystallisation solvent: | MeCN |
| $^1$H NMR (400 MHz): | 7.50 (t, *J* = 8.6 Hz, 1H, Ar-**H**), 7.37 (t, *J* = 8.0 Hz, 1H, Ar-**H**), 7.31 – 7.20 (m, 6H, Ar-**H**)†, 7.09 (dd, *J* = 7.9, 1.7 Hz, 1H, Ar-**H**), 7.03 (dd, *J* = 12.0, 1.6 Hz, 1H, Ar-**H**), 2.65 (t, *J* = 7.4 Hz, 2H, Ar-C**H$_2$**-CH$_2$), 1.69 (h, *J* = 7.4 Hz, 2H, CH$_2$-C**H$_2$**-CH$_3$), 0.98 (t, *J* = 7.3 Hz, 3H, CH$_2$-C**H$_3$**). †Overlapping CDCl$_3$. |
| $^{13}$C{$^1$H} NMR (101 MHz): | 162.83 (dd, *J* = 261.0, 4.7 Hz), 160.47 (dd, *J* = 258.8, 5.6 Hz), 159.15 (d, *J* = 2.9 Hz), 158.30 (dt, *J* = 252.7, 9.2 Hz), 152.15 (d, *J* = 10.9 Hz), 147.10 (d, *J* = 8.1 Hz), 143.24 – 142.58 (m), 130.63 (d, *J* = 3.6 Hz), 129.74 (d, *J* = 2.9 Hz), 125.06 (d, *J* = 3.0 Hz), 123.24 (d, *J* = 12.4 Hz), 122.64 (d, *J* = 12.4 Hz), 118.58 (d, *J* = 3.7 Hz), 116.48 (d, *J* = 22.0 Hz), 112.66 (m), 111.02 (d, *J* = 25.9 Hz), 109.11, 107.76 (t, *J* = 17.4 Hz), 91.54 (t, *J* = 18.9 Hz), 37.55, 24.13, 13.72. |
| $^{19}$F NMR (376 MHz): | -103.44 (d, *J$_{F-H}$* = 9.2 Hz, 2F, Ar-**F**), -108.64 (d, *J$_{F-H}$* = 10.7 Hz, 2F, Ar-**F**), -113.17 (t, *J$_{F-H}$* = 9.7 Hz, 1F, Ar-**F**), -117.55 (t, *J$_{F-H}$* = 9.4 Hz, 1F, Ar-**F**). |

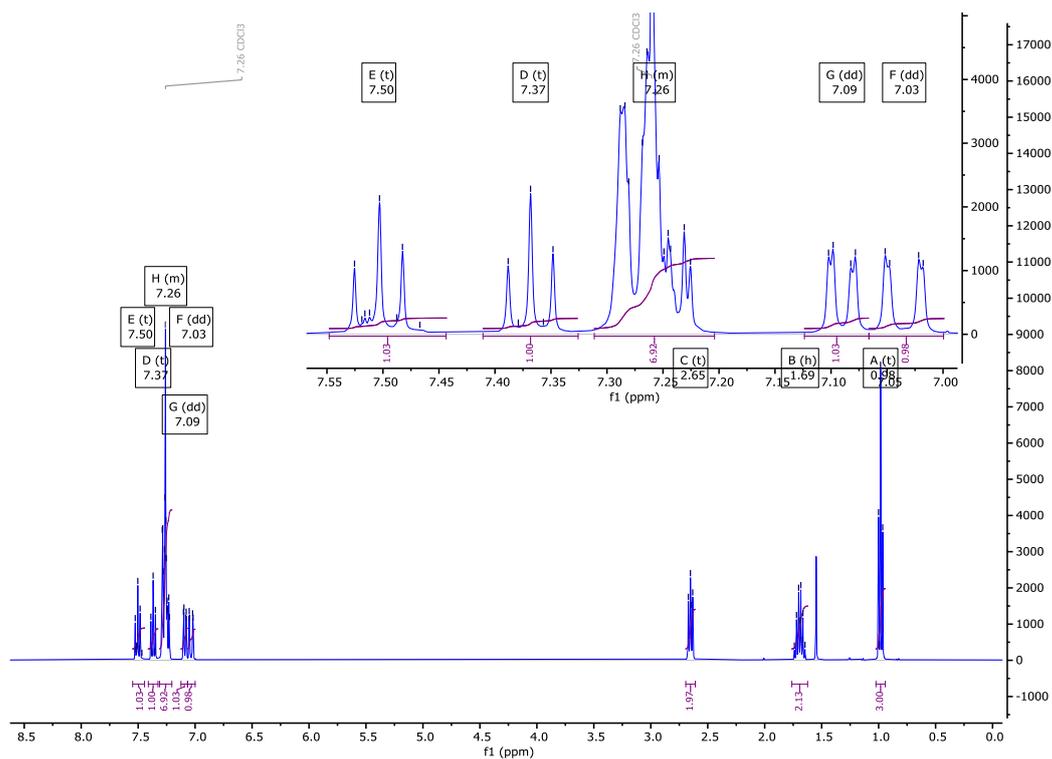

**Fig. S72**  $^1$H NMR spectra of **NC2121** in CDCl$_3$.

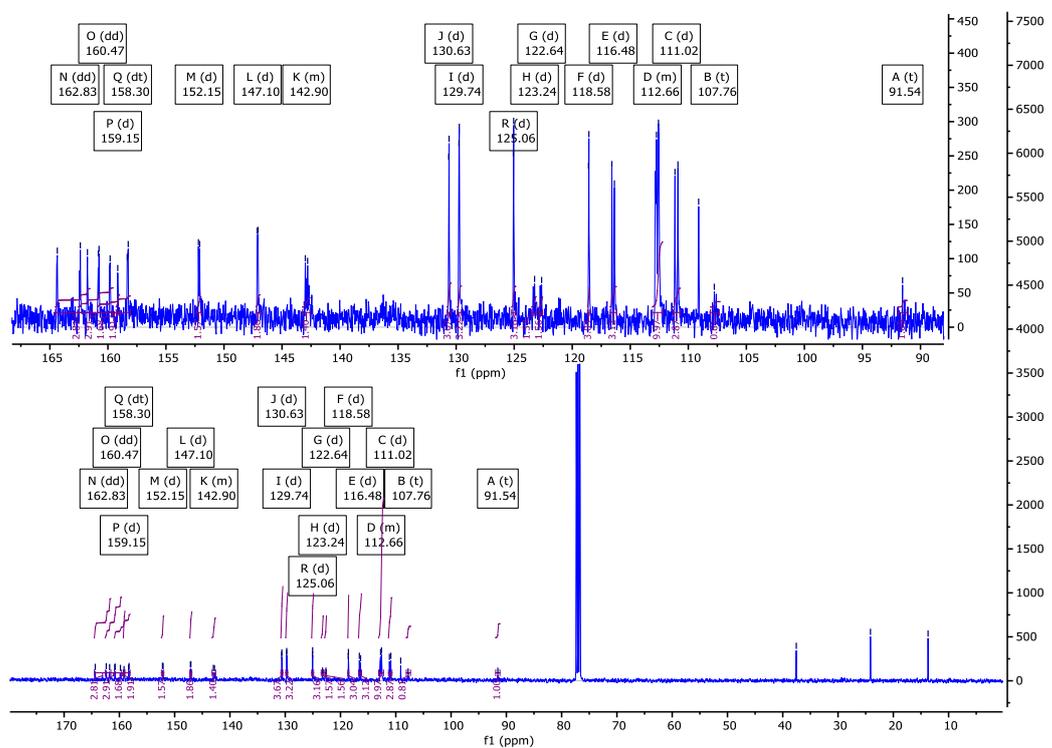

**Fig. S73**  $^{13}$C{$^1$H} NMR spectra of **NC2121** in CDCl$_3$.

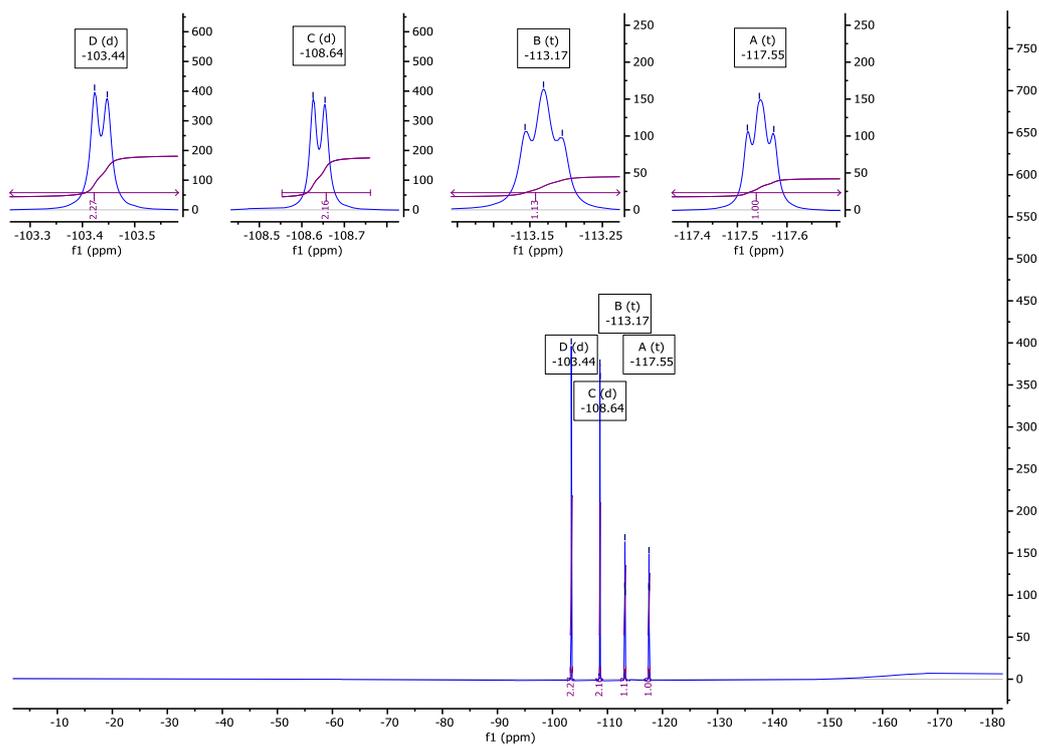

**Fig. S74** ¹⁹F NMR spectra of **NC2121** in CDCl$_3$.

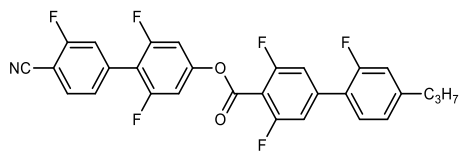

**NC1221**

*4'-cyano-2,3',6-trifluoro-[1,1'-biphenyl]-4-yl 2',3,5-trifluoro-4'-propyl-[1,1'-biphenyl]-4-carboxylate*

| | |
|---|---|
| Yield: | (white crystalline solid) 171 mg, 65 % |
| Re-crystallisation solvent: | MeCN |
| $^1$H NMR (400 MHz): | 7.73 (t, *J* = 7.3 Hz, 1H, Ar-**H**), 7.38 (m, 3H, Ar-**H**)*, 7.27 (d, *J* = 10.9 Hz, 2H, Ar-**H**)†, 7.12 – 6.99 (m, 4H, Ar-**H**)*, 2.65 (t, *J* = 7.6 Hz, 2H, Ar-C**H₂**-CH₂), 1.69 (h, *J* = 7.4 Hz, 2H, CH₂-C**H₂**-CH₃), 0.98 (t, *J* = 7.3 Hz, 3H, CH₂-C**H₃**). *Overlapping signals, † Overlapping CDCl₃. |
| $^{13}$C{$^1$H} NMR (101 MHz): | 162.64 (d, *J* = 259.3 Hz), 160.50 (dd, *J* = 258.4, 5.8 Hz), 159.85, 158.84 (dd, *J* = 251.4, 8.7 Hz), 158.34 (d, *J* = 250.2 Hz), 151.30 (t, *J* = 14.4 Hz), 147.17 (d, *J* = 7.9 Hz), 143.02 (d, *J* = 11.0 Hz), 135.83 (d, *J* = 8.4 Hz), 129.74 (d, *J* = 3.0 Hz), 125.08 (d, *J* = 3.1 Hz), 122.66 (d, *J* = 12.0 Hz), 118.51 (d, *J* = 20.7 Hz), 116.49 (d, *J* = 22.0 Hz), 113.70 (d, *J* = 13.4 Hz), 112.76 (dt, *J* = 23.1, 3.6 Hz), 107.46 (t, *J* = 16.6 Hz), 106.74 (m), 101.35 (d, *J* = 15.5 Hz), 37.55, 24.13, 13.72. |
| $^{19}$F NMR (376 MHz): | -106.07 (t, *J*$_{F-H}$ = 8.2 Hz, 1F, Ar-**F**), -108.44 (d, *J*$_{F-H}$ = 10.6 Hz, 2F, Ar-**F**), -111.83 (d, *J*$_{F-H}$ = 9.0 Hz, 2F, Ar-**F**), -117.52 (t, *J*$_{F-H}$ = 9.4 Hz, 1F, Ar-**F**). |

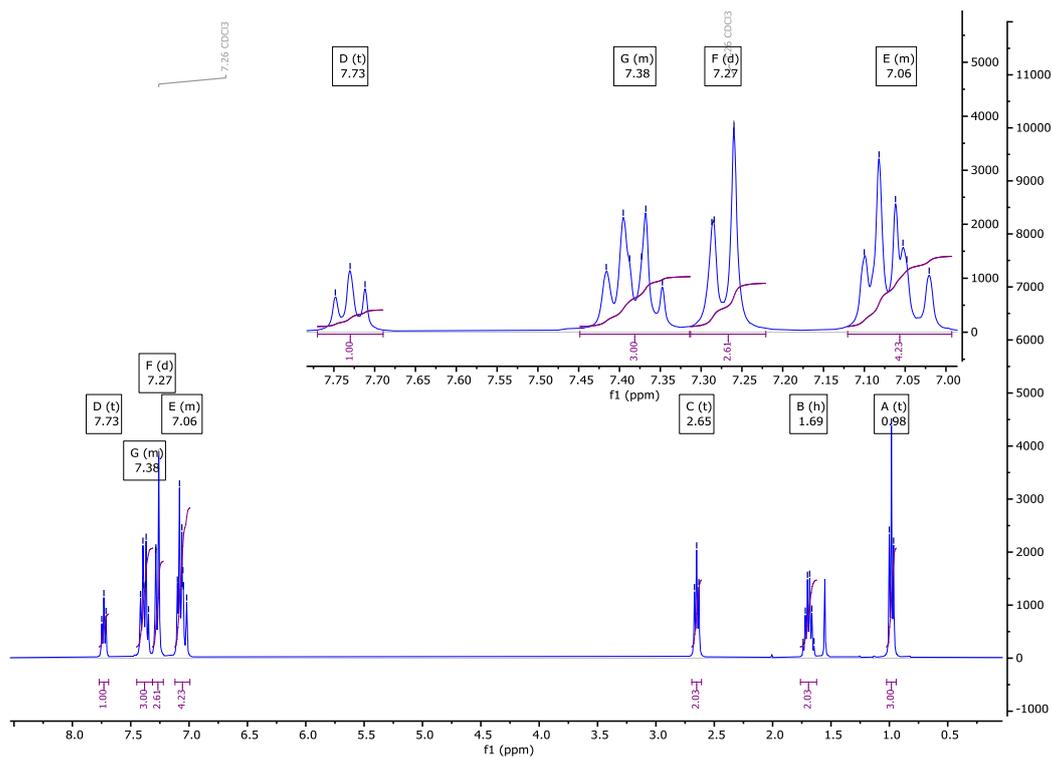

**Fig. S75** $^1$H NMR spectra of **NC1221** in CDCl$_3$.

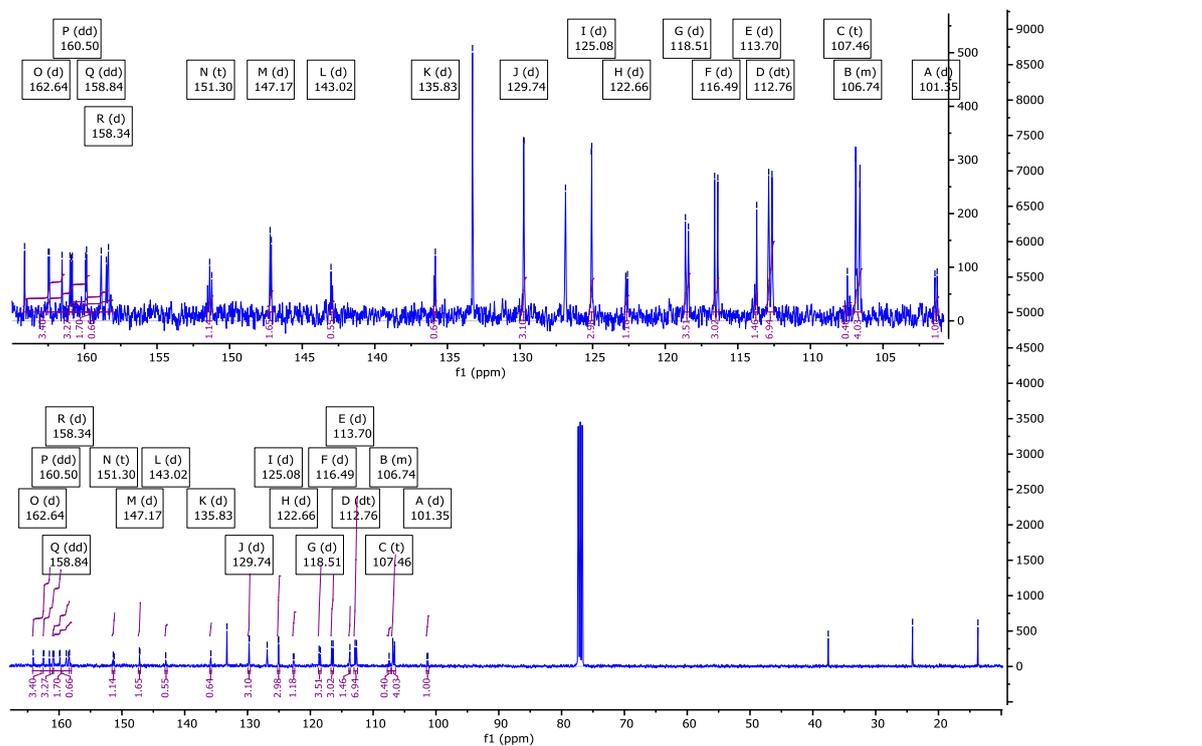

**Fig. S76** $^{13}$C{$^1$H} NMR spectra of **NC1221** in CDCl$_3$.

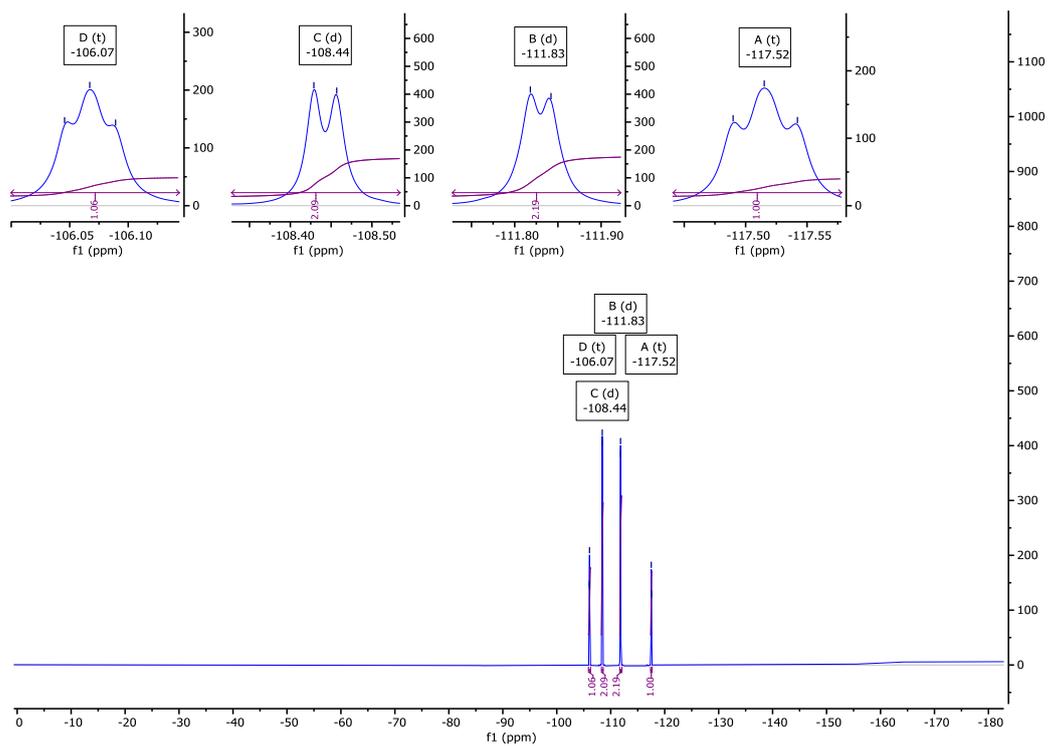

**Fig. S77** $^{19}$F NMR spectra of **NC1221** in CDCl$_3$.

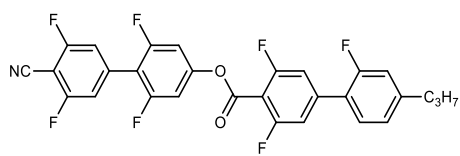

**NC2221**

*4'-cyano-2,3',5',6-tetrafluoro-[1,1'-biphenyl]-4-yl 2',3,5-trifluoro-4'-propyl-[1,1'-biphenyl]-4-carboxylate*

| | |
|---|---|
| Yield: | (white needles) 136 mg, 50 % |
| Re-crystallisation solvent: | MeCN |
| $^1$H NMR (400 MHz): | 7.37 (t, *J* = 8.0 Hz, 1H, Ar-**H**), 7.31 – 7.19 (m, 4H, Ar-**H**)*†, 7.13 – 7.06 (m, 3H, Ar-**H**)*, 7.04 (dd, *J* = 11.9, 1.6 Hz, 1H, Ar-**H**), 2.65 (t, *J* = 7.3 Hz, 2H, Ar-C**H₂**-CH₂), 1.69 (h, *J* = 7.4 Hz, 2H, CH₂-C**H₂**-CH₃), 0.98 (t, *J* = 7.3 Hz, 3H, CH₂-C**H₃**). * Overlapping Signals, † Overlapping CDCl₃. |
| $^{13}$C{$^1$H} NMR (101 MHz): | 162.73 (dd, *J* = 262.0, 5.1 Hz), 160.38 (dd, *J* = 258.9, 6.4 Hz), 160.96 – 158.14 (m)* 158.75, 151.81 (t, *J* = 14.4 Hz), 147.20 (d, *J* = 7.9 Hz), 143.11 (t, *J* = 10.9 Hz), 136.60 (t, *J* = 10.7 Hz), 129.73 (d, *J* = 3.1 Hz), 125.08 (d, *J* = 3.1 Hz), 122.63 (d, *J* = 12.5 Hz), 116.50 (d, *J* = 22.0 Hz), 114.28 (dd, *J* = 21.0, 3.0 Hz), 113.37 – 112.54 (m), 108.95, 107.34 (t, *J* = 16.6 Hz), 107.15 – 106.58 (m), 92.19 (t, *J* = 19.2 Hz). * Overlapping Signals. |
| $^{19}$F NMR (376 MHz): | -103.61 (d, *J*$_{F-H}$ = 9.1 Hz, 2F, Ar-**H**), -108.37 (d, *J*$_{F-H}$ = 10.6 Hz, 2F, Ar-**F**), -111.51 (d, *J*$_{F-H}$ = 9.4 Hz, 2F, Ar-**F**), -117.51 (t, *J*$_{F-H}$ = 10.0 Hz, 1F, Ar-**F**). |

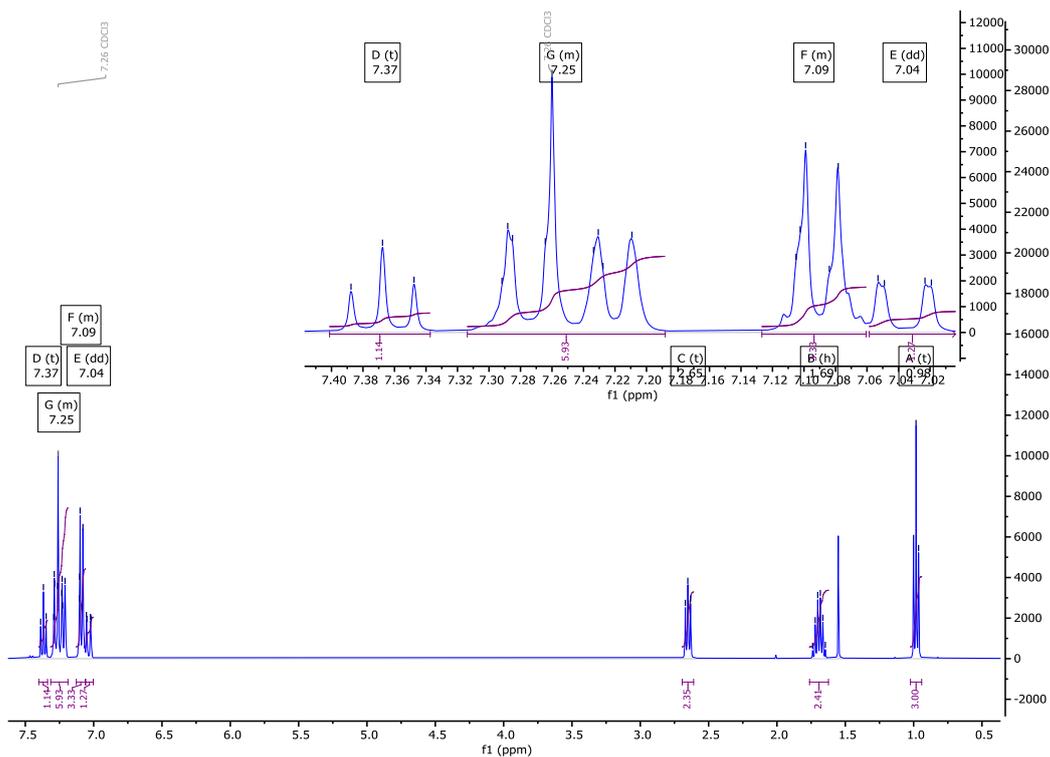

**Fig. S78** $^1$H NMR spectra of **NC2221** in CDCl$_3$.

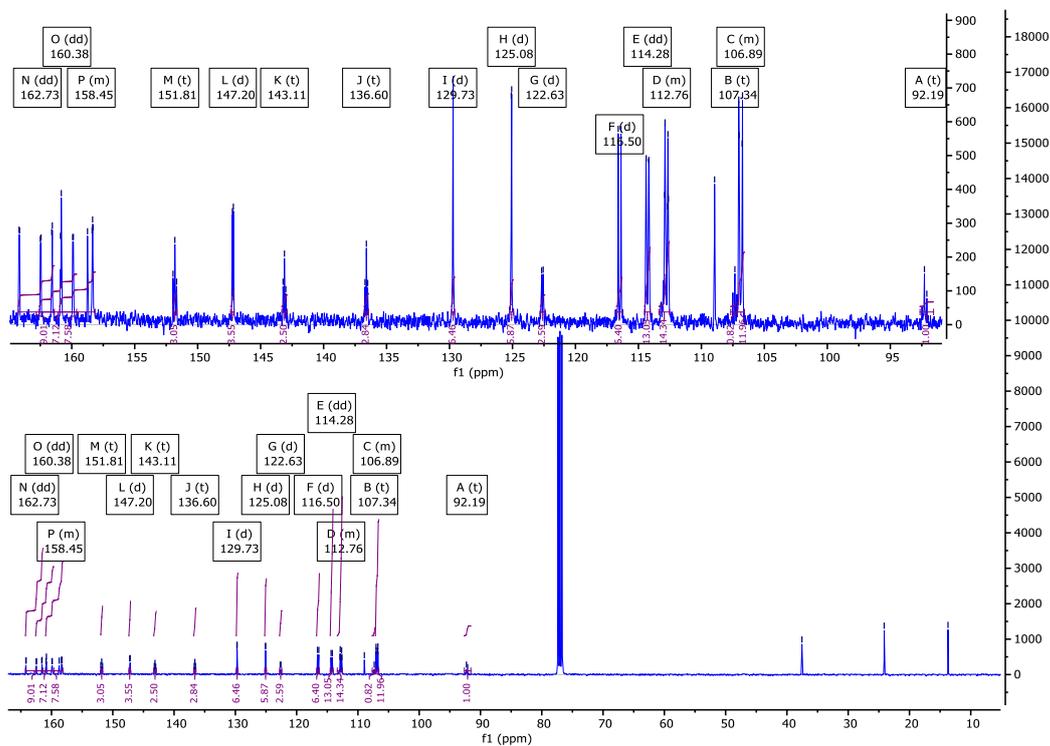

**Fig. S79** $^{13}$C{$^1$H} NMR spectra of **NC2221** in CDCl$_3$.

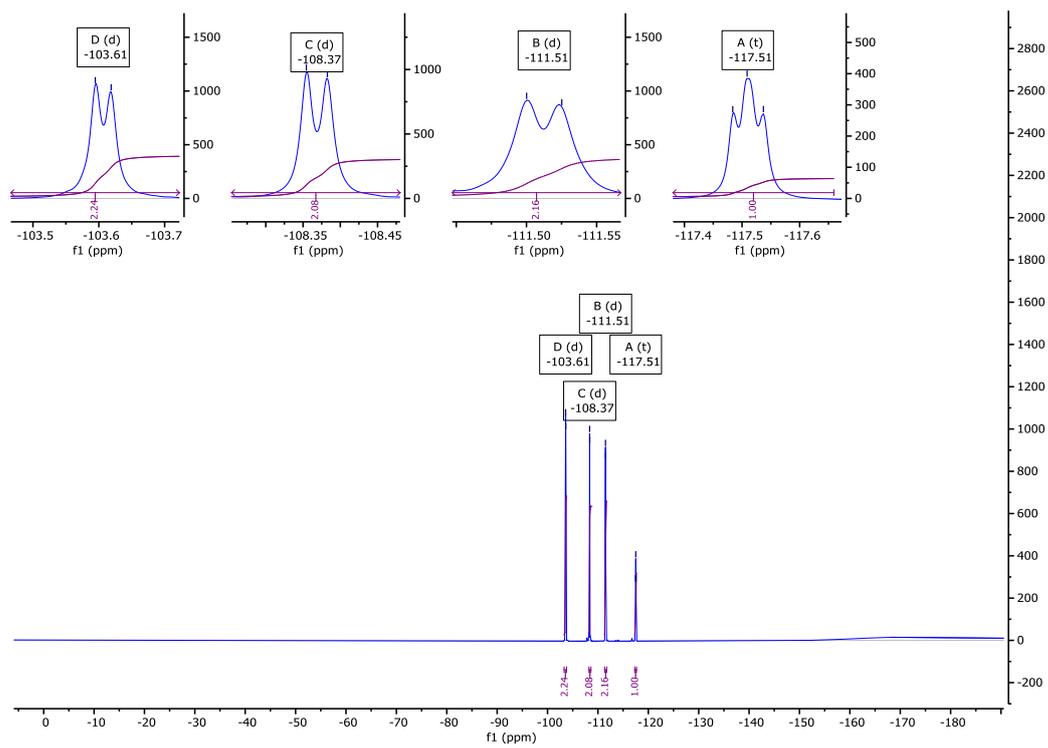

**Fig. S80**     19F NMR spectra of **NC2221** in CDCl$_3$.

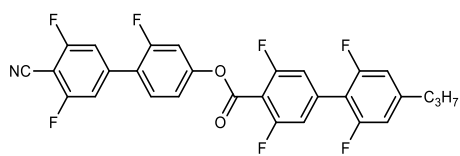

**NC2122**

*4'-cyano-2,3',5'-trifluoro-[1,1'-biphenyl]-4-yl 2',3,5,6'-tetrafluoro-4'-propyl-[1,1'-biphenyl]-4-carboxylate*

| | |
|---|---|
| Yield: | (white crystalline solid) 163 mg, 60 % |
| Re-crystallisation solvent: | MeCN |
| $^1$H NMR (400 MHz): | 7.51 (t, $J$ = 8.6 Hz, 1H, Ar-**H**), 7.32 – 7.16 (m, 7H, Ar-**H**)*†, 6.87 (d, $J$ = 9.1 Hz, 2H, Ar-**H**), 2.64 (t, $J$ = 7.6 Hz, 2H, Ar-C**H$_2$**-CH$_2$), 1.69 (h, $J$ = 7.4 Hz, 2H, CH$_2$-C**H$_2$**-CH$_3$), 0.99 (t, $J$ = 7.3 Hz, 3H, CH$_2$-C**H$_3$**). *Overlapping Signals, † Overlapping CDCl$_3$. |
| $^{13}$C{$^1$H} NMR (101 MHz): | 162.75 (dd, $J$ = 260.9, 5.0 Hz), 160.23 (dd, $J$ = 257.8, 6.8 Hz), 160.69 – 158.05 (m)*, 152.10 (d, $J$ = 11.2 Hz), 147.33 (t, $J$ = 9.5 Hz), 142.98 (t, $J$ = 9.8 Hz), 136.45 (tz, $J$ = 11.3 Hz), 130.65 (d, $J$ = 3.7 Hz), 123.36 (d, $J$ = 12.5 Hz), 118.56 (d, $J$ = 3.8 Hz), 114.45 (dd, $J$ = 23.7, 2.1 Hz), 112.60 (dt, $J$ = 21.0, 3.9 Hz), 112.24 – 111.74 (m), 111.00 (d, $J$ = 25.9 Hz), 108.57, 108.49 (t, $J$ = 16.5 Hz), 91.56 (t, $J$ = 19.1 Hz), 37.69, 23.85, 13.63. *Overlapping Signals. |
| $^{19}$F NMR (376 MHz): | -103.43 (d, $J_{F-H}$ = 9.2 Hz, 1F, Ar-**F**), -109.11 (d, $J_{F-H}$ = 9.9 Hz, 1F, Ar-**F**), -113.12 (t, $J_{F-H}$ = 9.8 Hz, 1F, Ar-**F**), -114.95 (d, $J_{F-H}$ = 9.7 Hz, 1F, Ar-**F**). |

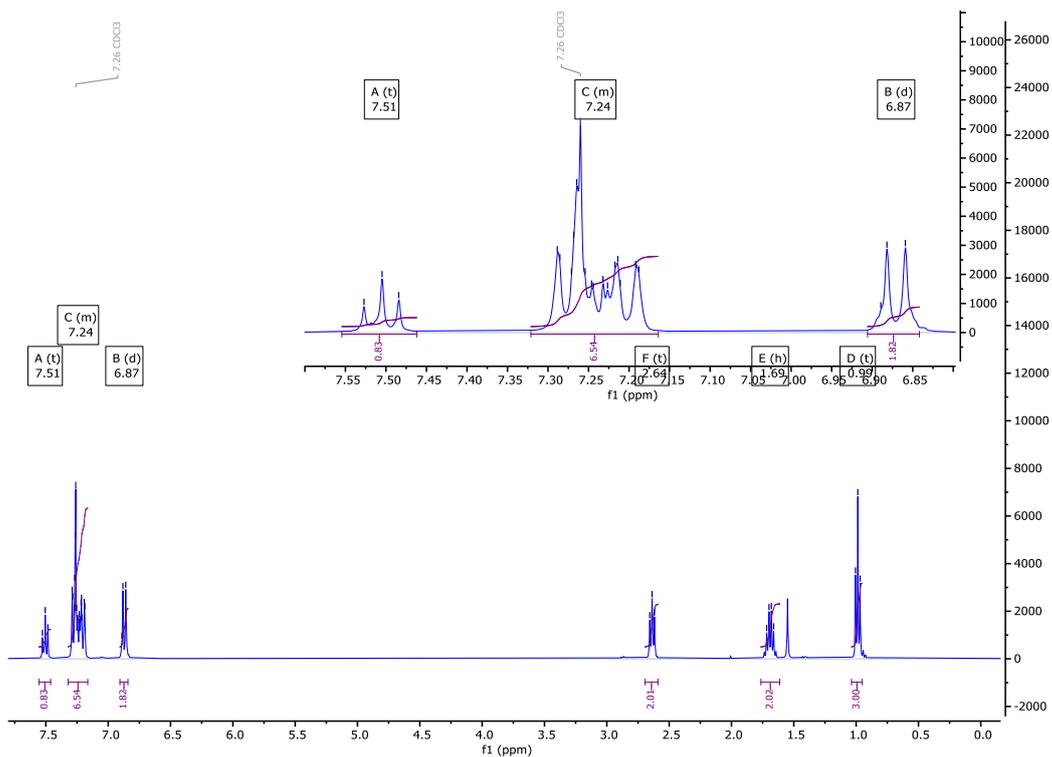

**Fig. S81**      $^1$H NMR spectra of **NC2122** in CDCl$_3$.

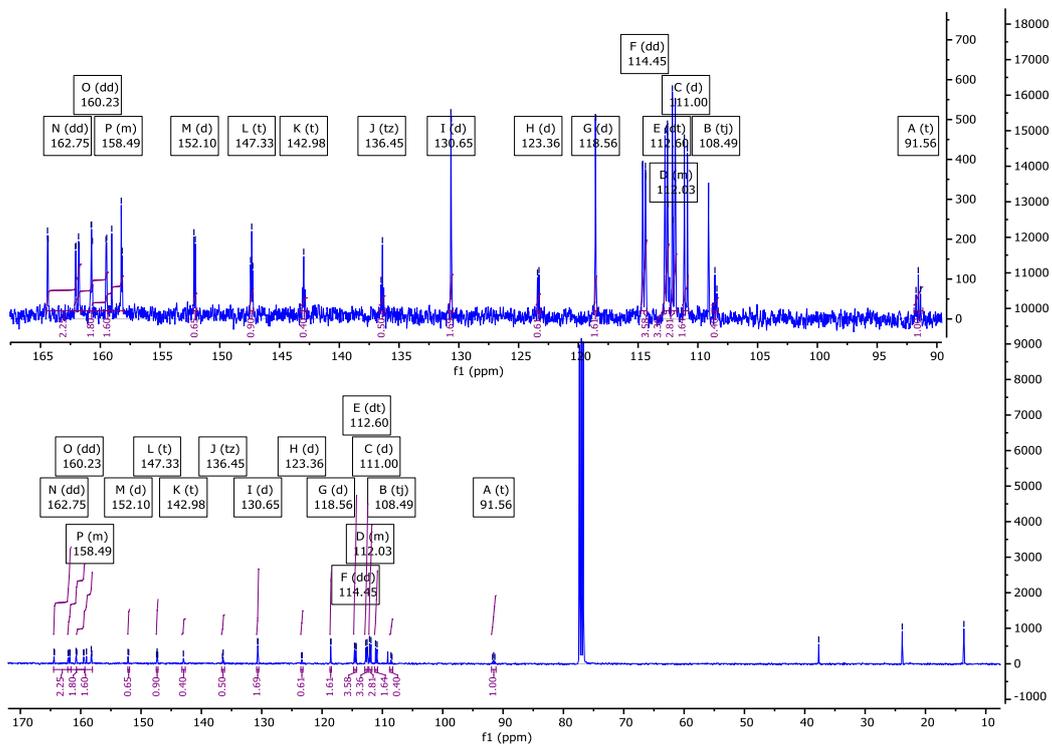

**Fig. S82**      $^{13}$C{$^1$H} NMR spectra of **NC2122** in CDCl$_3$.

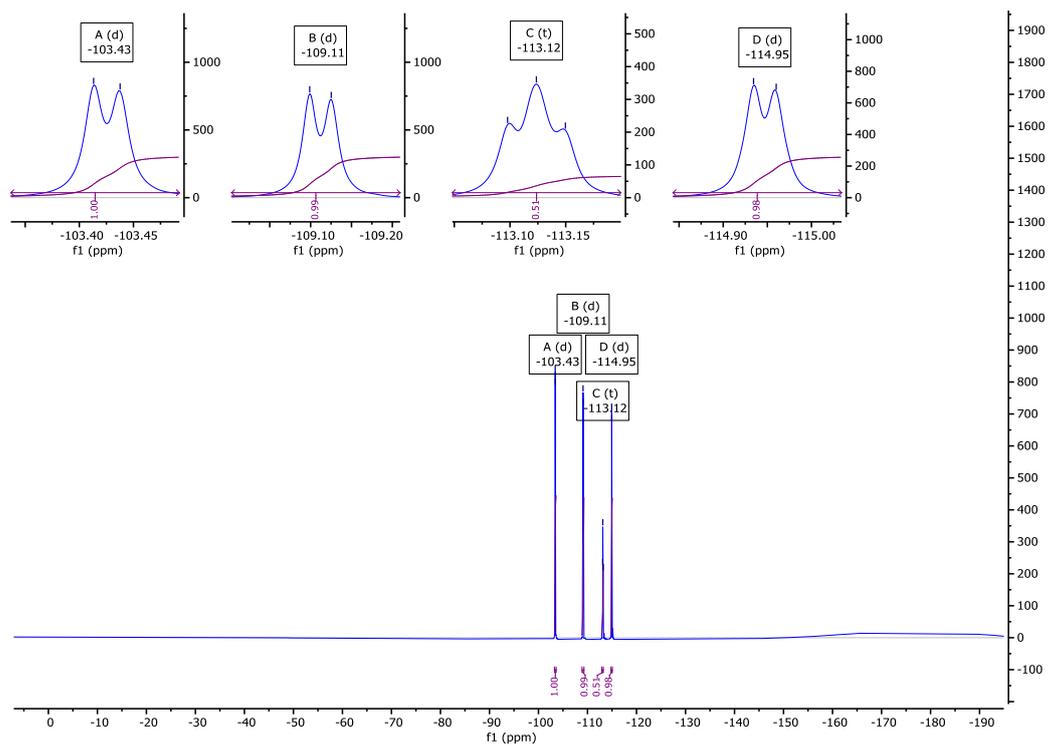

**Fig. S83** $^{19}$F NMR spectra of **NC2122** in CDCl$_3$.

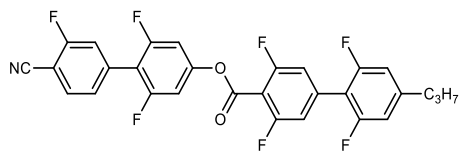

**NC1222**

*4'-cyano-2,3',6-trifluoro-[1,1'-biphenyl]-4-yl 2',3,5,6'-tetrafluoro-4'-propyl-[1,1'-biphenyl]-4-carboxylate*

| | |
|---|---|
| Yield: | (white crystalline solid) 180 mg, 55 % |
| Re-crystallisation solvent: | MeCN |
| $^1$H NMR (400 MHz): | 7.73 (t, *J* = 8.2 Hz, 1H, Ar-**H**), 7.40 (t, *J* = 8.1 Hz, 2H, Ar-**H**), 7.20 (d, *J* = 10.7 Hz, 2H, Ar-**H**), 7.07 (ddd, *J* = 8.1, 6.2, 3.1 Hz, 2H, Ar-**H**), 6.87 (ddd, *J* = 9.1, 5.1, 2.9 Hz, 2H, Ar-**H**), 2.63 (t, *J* = 7.6 Hz, 2H, Ar-C**H$_2$**-CH$_2$), 1.68 (h, *J* = 7.0 Hz, 2H, CH$_2$-C**H$_2$**-CH$_3$), 0.98 (t, *J* = 7.3 Hz, 3H, CH$_2$-C**H$_3$**). |
| $^{13}$C{$^1$H} NMR (101 MHz): | 162.60 (d, *J* = 259.8 Hz), 160.40 (dd, *J* = 258.9, 6.1 Hz), 159.56, 158.55 (dd, *J* = 251.1, 8.0 Hz), 158.16 (dd, *J* = 250.4, 6.6 Hz), 151.32 (t, *J* = 14.8 Hz), 147.39 (t, *J* = 10.1 Hz), 136.62 (t, *J* = 11.6 Hz), 135.86 (d, *J* = 8.9 Hz), 133.28, 126.88 (d, *J* = 2.2 Hz), 118.51 (d, *J* = 21.3 Hz), 114.53 (dd, *J* = 23.1, 2.9 Hz), 112.02 (dd, *J* = 24.6, 5.9 Hz), 108.26 (d, *J* = 16.8 Hz), 107.02 – 106.31 (m), 101.35 (d, *J* = 15.3 Hz), 37.69, 23.84, 13.62. |
| $^{19}$F NMR (376 MHz): | -106.07 (t, *J$_{F-H}$* = 8.2 Hz, 1F, Ar-**F**), -108.92 (d, *J$_{F-H}$* = 10.3 Hz, 2F, Ar-**F**), -111.77 (d, *J$_{F-H}$* = 9.0 Hz, 2F, Ar-**F**), -114.94 (d, *J$_{F-H}$* = 9.8 Hz, 2F, Ar-**F**). |

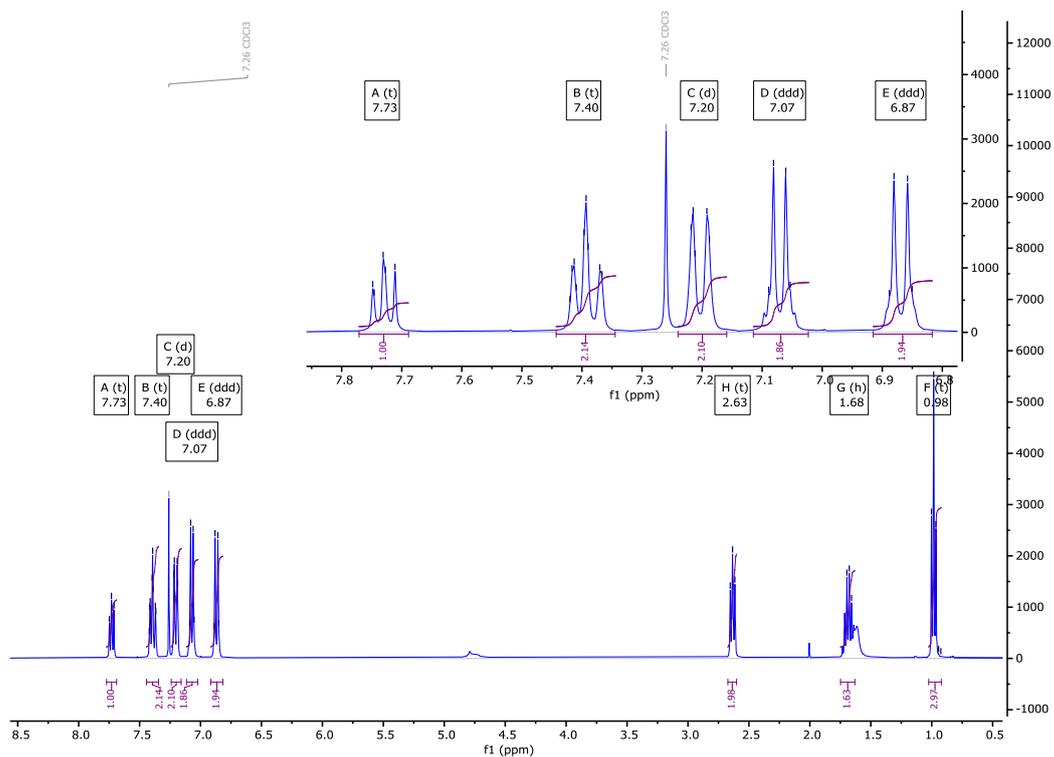

**Fig. S84**    $^1$H NMR spectra of **NC1222** in CDCl$_3$.

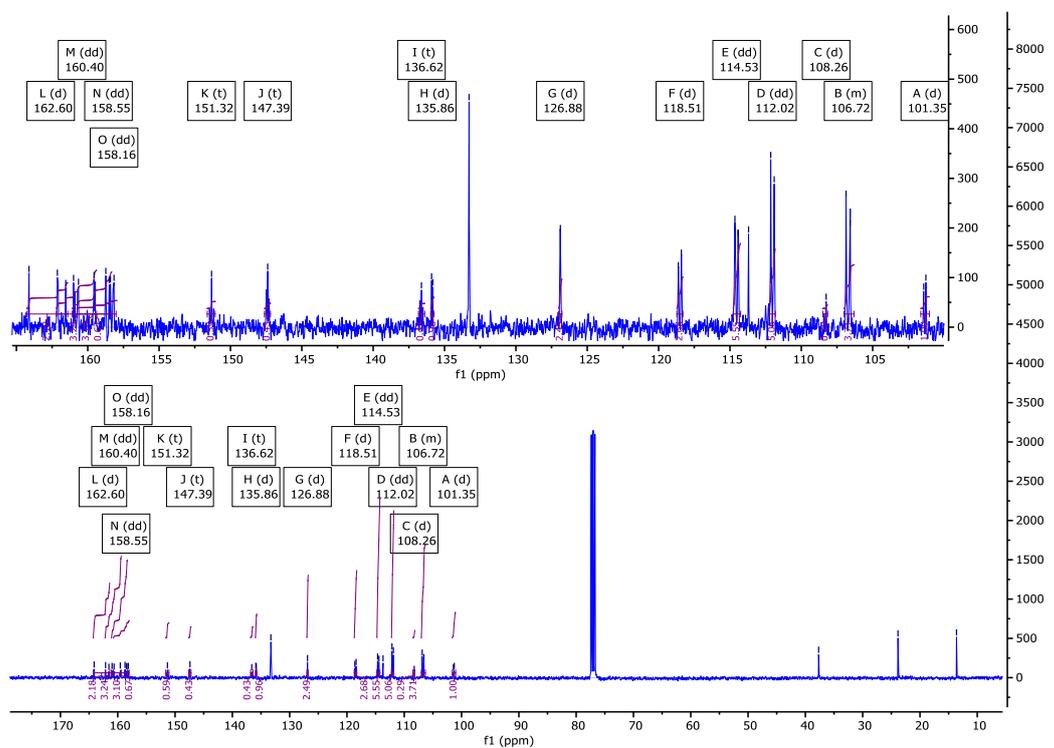

**Fig. S85**    $^{13}$C{$^1$H} NMR spectra of **NC1222** in CDCl$_3$.

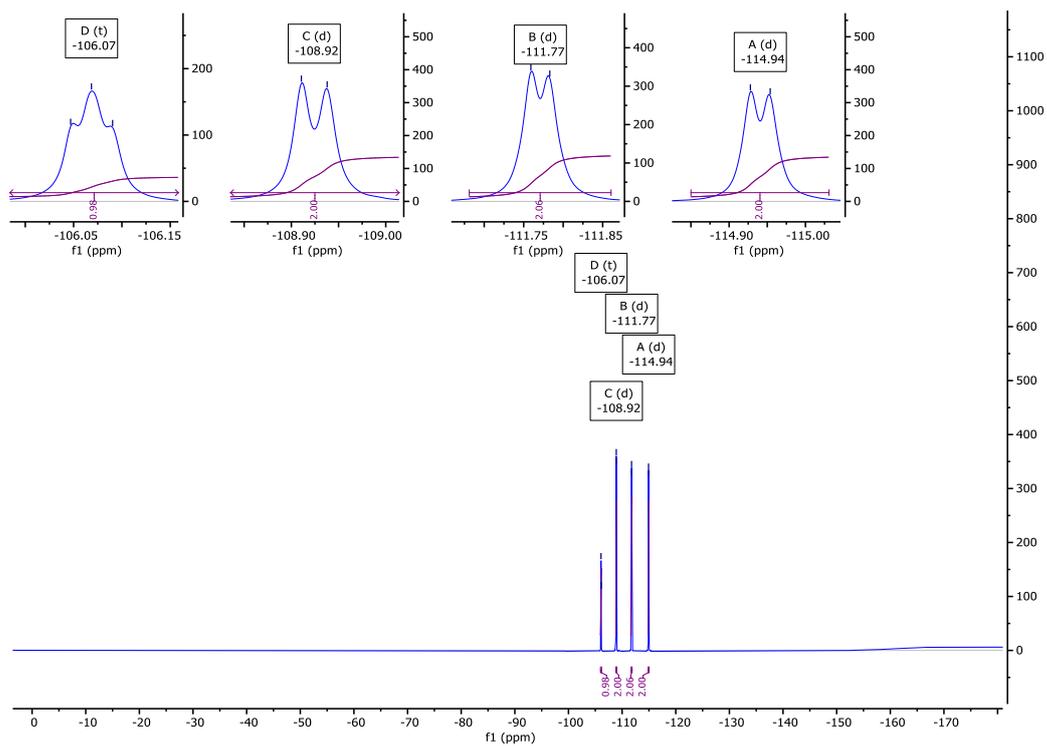

**Fig. S86** $^{19}$F NMR spectra of **NC1222** in CDCl$_3$.

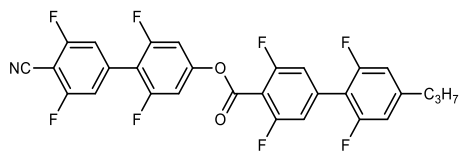

**NC2222**

*4'-cyano-2,3',5',6-tetrafluoro-[1,1'-biphenyl]-4-yl 2',3,5,6'-tetrafluoro-4'-propyl-[1,1'-biphenyl]-4-carboxylate*

| | |
|---|---|
| Yield: | (white needles) 135 mg, 48 % |
| Re-crystallisation solvent: | MeCN |
| $^1$H NMR (400 MHz): | 7.25 – 7.17 (m, 4H, Ar-**H**)*, 7.09 (ddd, *J* = 8.2, 5.7, 2.9 Hz, 2H, Ar-**H**), 6.87 (ddd, *J* = 9.1, 5.0, 2.8 Hz, 2H, Ar-**H**), 2.64 (t, *J* = 7.4 Hz, 2H, Ar-C**H$_2$**-CH$_2$), 1.69 (h, *J* = 7.4 Hz, 2H, CH$_2$-C**H$_2$**-CH$_3$), 0.99 (t, *J* = 7.3 Hz, 3H, CH$_2$-C**H$_3$**). *Overlapping Signals. |
| $^{13}$C{$^1$H} NMR (126 MHz): | 163.02 (dd, *J* = 256.7, 4.6 Hz), 161.97 (dd, *J* = 255.5, 2.9 Hz), 160.74 (dd, *J* = 252.2, 8.3 Hz), 159.08 (dd, *J* = 250.0, 6.8 Hz), 151.90 (t, *J* = 14.4 Hz), 147.56 (t, *J* = 9.6 Hz), 137.21 – 136.48 (m), 114.89 – 114.24 (m), 113.21 (t, *J* = 17.6 Hz), 112.43 (tj, *J* = 17.8 Hz), 112.16 (dd, *J* = 20.5, 4.9 Hz), 108.28 (t, *J* = 16.4 Hz), 107.25 – 106.82 (m), 92.44 (t, *J* = 19.2 Hz), 37.83, 23.98, 13.77. |
| $^{19}$F NMR (376 MHz): | -103.61 (d, *J$_{F-H}$* = 8.8 Hz, 2F, Ar-**F**), -108.85 (d, *J$_{F-H}$* = 10.2 Hz, 2F, Ar-**F**), -111.46 (d, *J$_{F-H}$* = 9.4 Hz, 2F, Ar-**F**), -114.94 (d, *J$_{F-H}$* = 9.7 Hz, 2F, Ar-**F**). |

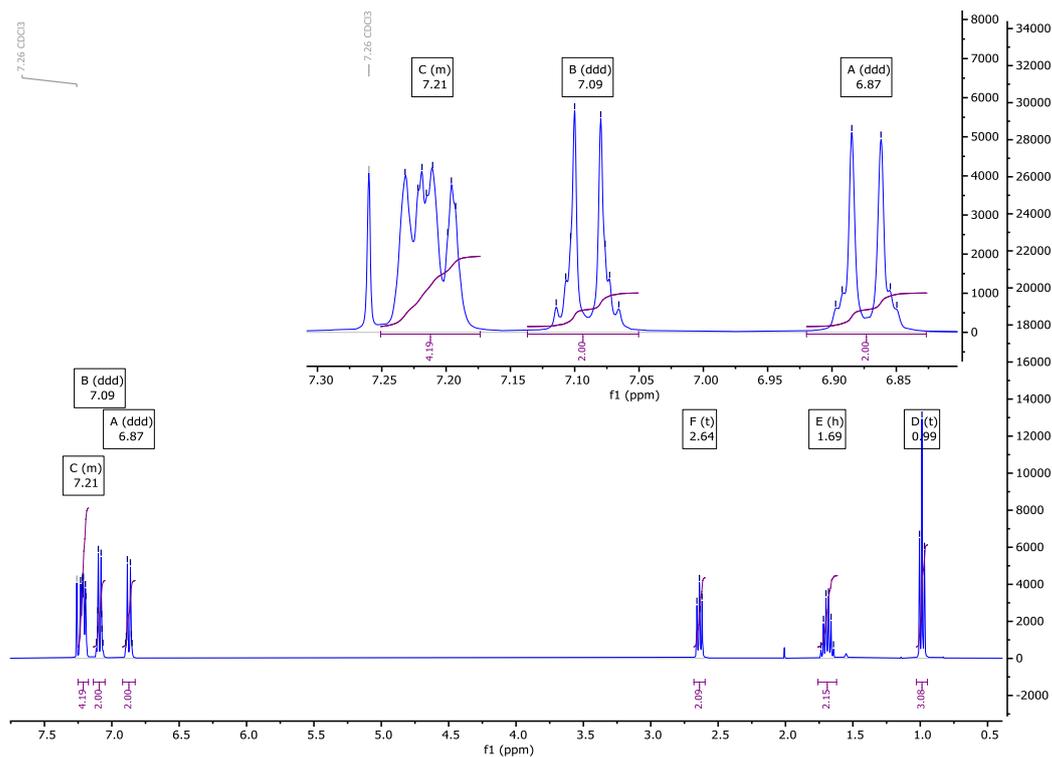

**Fig. S87** $^1$H NMR spectra of **NC2222** in CDCl$_3$.

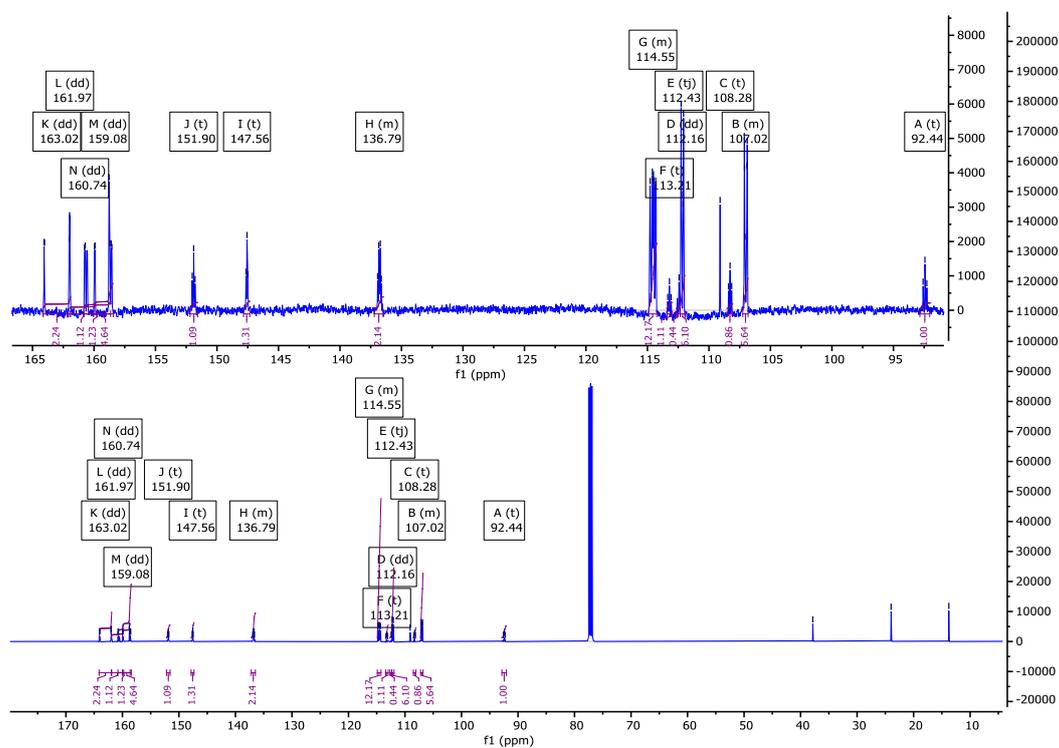

**Fig. S88** $^{13}$C{$^1$H} NMR spectra of **NC2222** in CDCl$_3$.

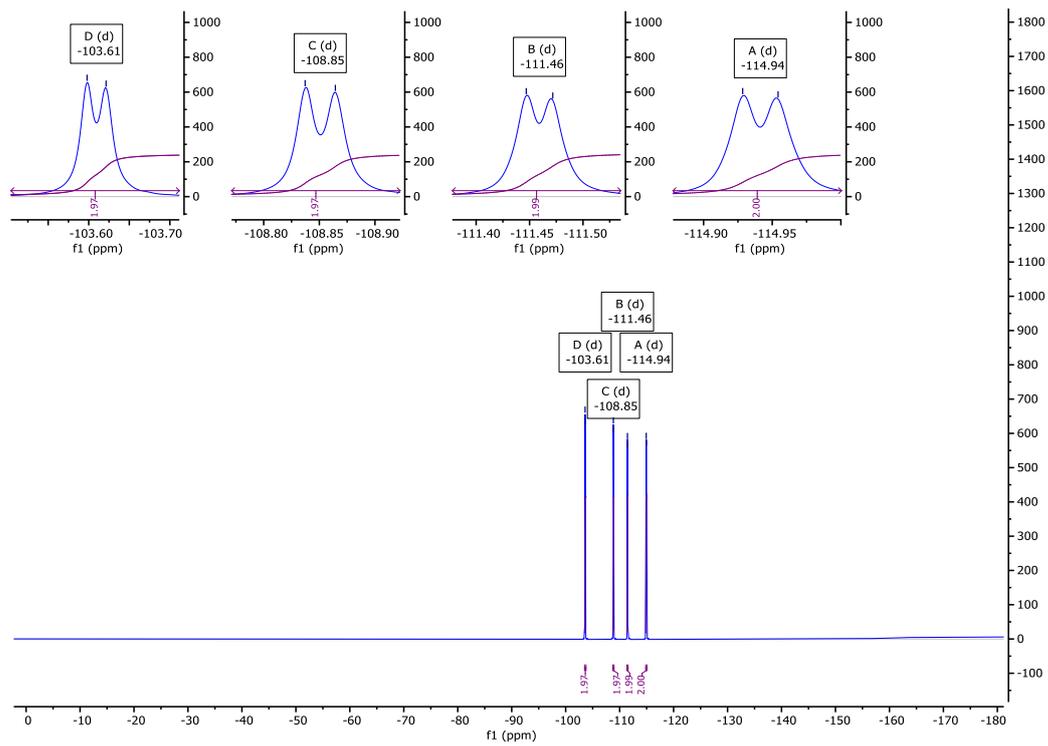

**Fig. S89** $^{19}$F NMR spectra of **NC2222** in CDCl$_3$.

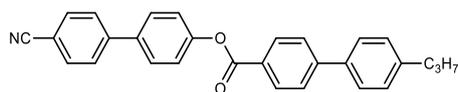

**NC0000**

*4'-cyano-[1,1'-biphenyl]-4-yl 4'-propyl-[1,1'-biphenyl]-4-carboxylate*

| | |
|---|---|
| Yield: | (white solid) 175 mg, 84 % |
| Re-crystallisation solvent: | EtOH |

$^1$H NMR (400 MHz): 8.28 (d, *J* = 8.2 Hz, 2H, Ar-**H**), 7.79 – 7.62 (m, 8H, Ar-**H**)[*†], 7.60 (d, *J* = 7.8 Hz, 2H, Ar-**H**), 7.37 (d, *J* = 8.7 Hz, 2H, Ar-**H**), 7.31 (d, *J* = 7.9 Hz, 2H, Ar-**H**), 2.67 (t, *J* = 7.6 Hz, 2H, Ar-C**H$_2$**-CH$_2$), 1.71 (h, *J* = 7.5 Hz, 2H, CH$_2$-C**H$_2$**-CH$_3$), 1.00 (t, *J* = 7.3 Hz, 3H, CH$_2$-C**H$_3$**).

$^{13}$C{$^1$H} NMR (101 MHz): 165.05, 151.52, 146.56, 144.83, 143.29, 137.08, 136.90, 132.69, 130.78, 129.19, 128.43, 127.73, 127.62, 127.18, 127.08, 122.54, 118.90, 111.08, 37.75, 24.54, 13.89.

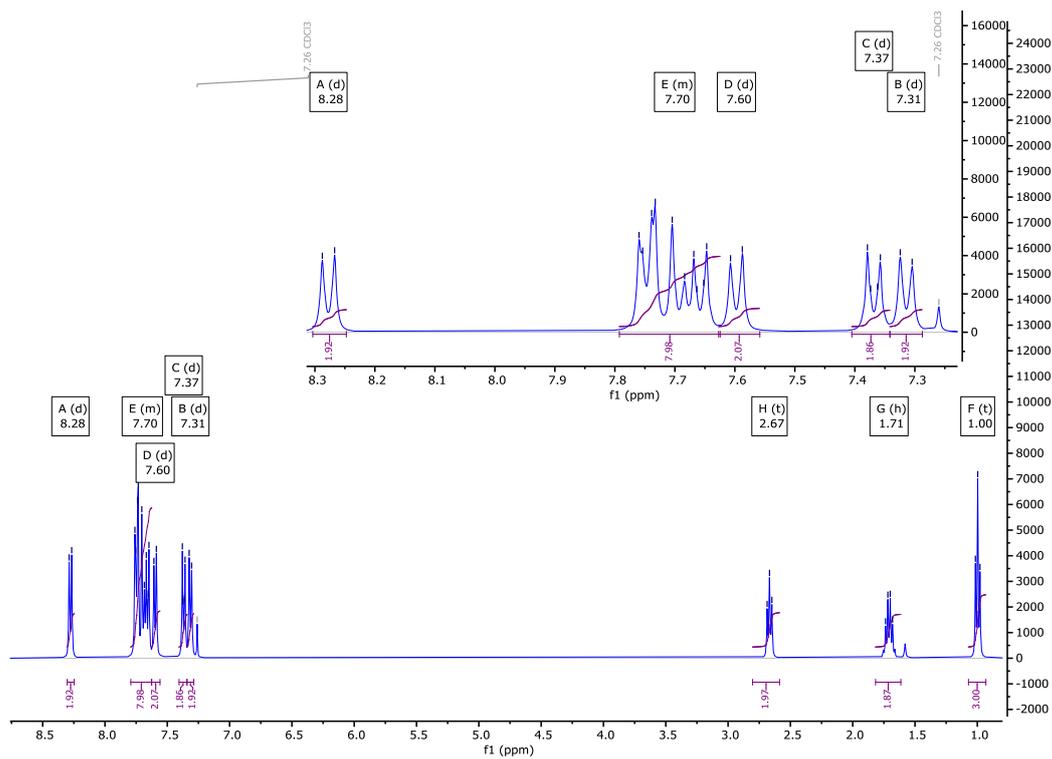

**Fig. S90** $^1$H NMR spectra of **NC0000** in CDCl$_3$.

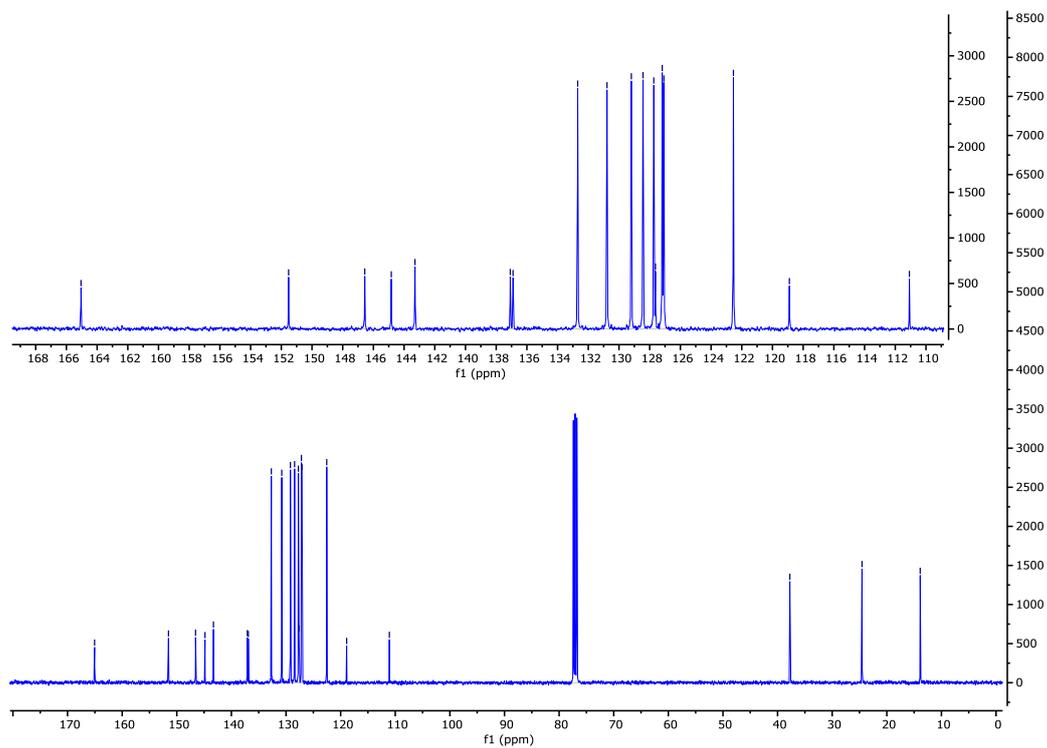

**Fig. S91** $^{13}$C{$^1$H} NMR spectra of **NC0000** in CDCl$_3$.

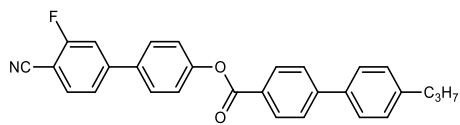

**NC1000**

*4'-cyano-3'-fluoro-[1,1'-biphenyl]-4-yl 4'-propyl-[1,1'-biphenyl]-4-carboxylate*

| | |
|---|---|
| Yield: | (white solid) 174 mg, 80 % |
| Re-crystallisation solvent: | EtOH |
| $^1$H NMR (400 MHz): | 8.27 (ddd, *J* = 8.5, 1.9, 1.8 Hz, 2H, Ar-**H**), 7.75 (ddd, *J* = 8.6, 2.0, 1.9 Hz, 2H, Ar-**H**), 7.70 (dd, *J* = 6.7, 1.4 Hz, 1H, Ar-**H**), 7.65 (ddd, *J* = 8.7, 2.1, 2.1 Hz, 2H, Ar-**H**), 7.59 (ddd, *J* = 8.2, 2.2, 1.8 Hz, 2H, Ar-**H**), 7.50 (dd, *J* = 8.0, 1.5 Hz, 1H, Ar-**H**), 7.44 (dd, *J* = 10.1, 1.6 Hz, 1H, Ar-**H**), 7.38 (ddd, *J* = 8.7, 2.8, 2.1 Hz, 2H, Ar-**H**), 7.31 (d, *J* = 8.2 Hz, 2H, Ar-**H**), 2.66 (t, *J* = 7.3 Hz, 2H, Ar-C**H$_2$**-CH$_2$), 1.70 (h, *J* = 7.5 Hz, 2H, CH$_2$-C**H$_2$**-CH$_3$), 0.99 (t, *J* = 7.3 Hz, 3H, CH$_2$-C**H$_3$**). |
| $^{13}$C{$^1$H} NMR (101 MHz): | 164.97, 163.49 (d, *J* = 258.8 Hz), 151.98, 147.69 (d, *J* = 7.9 Hz), 146.64, 143.31, 137.05, 135.72 (d, *J* = 1.4 Hz), 130.79, 129.19, 128.43, 127.50, 127.17, 127.09, 123.38 (d, *J* = 3.2 Hz), 122.72, 114.82 (d, *J* = 20.2 Hz), 114.04, 99.98 (d, *J* = 15.9 Hz), 37.74, 24.54, 13.88. |
| $^{19}$F NMR (376 MHz): | -105.93 (t, *J$_{F-H}$* = 8.4 Hz, 1F, Ar-**F**). |

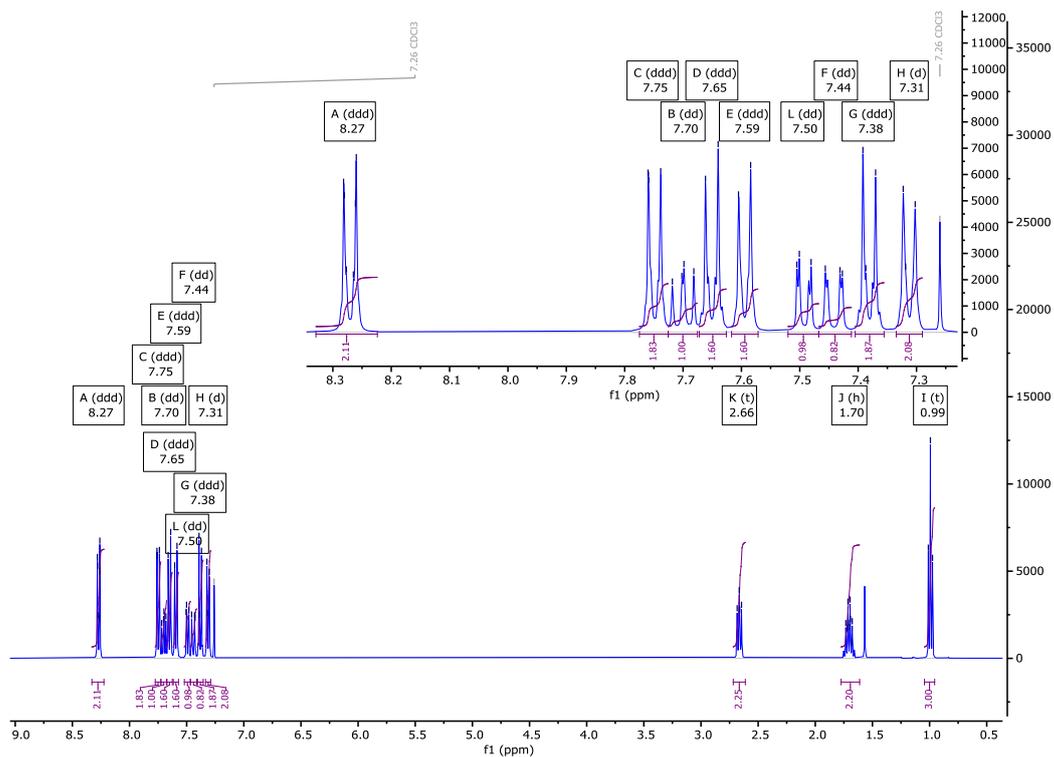

**Fig. S92**  $^1$H NMR spectra of **NC1000** in CDCl$_3$.

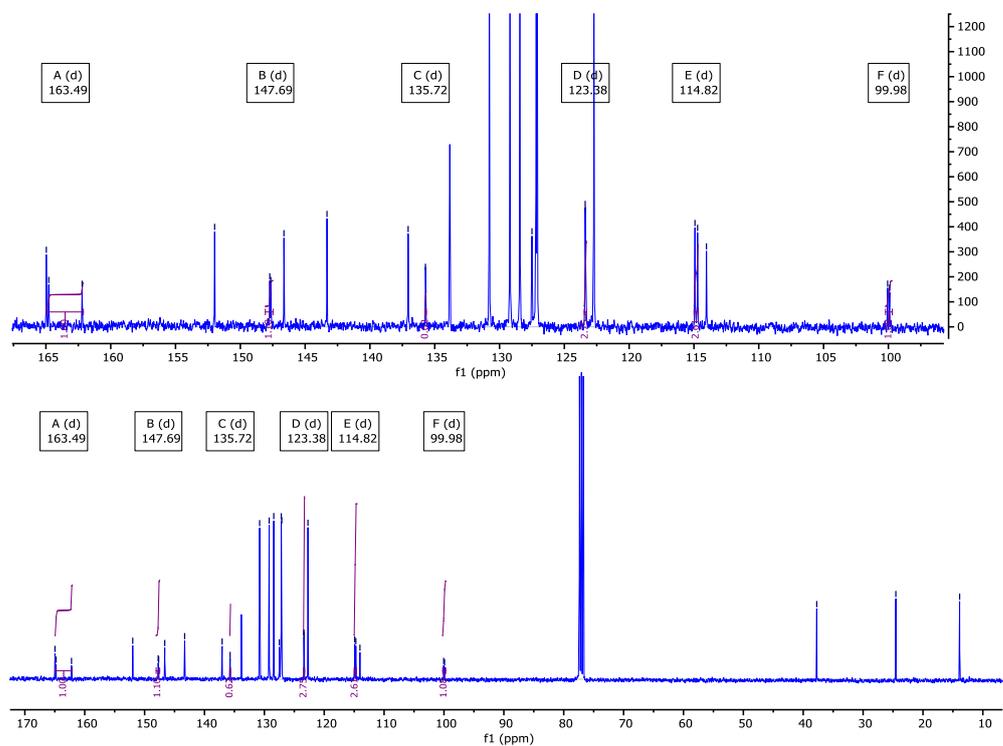

**Fig. S93**  $^{13}$C{$^1$H} NMR spectra of **NC1000** in CDCl$_3$.

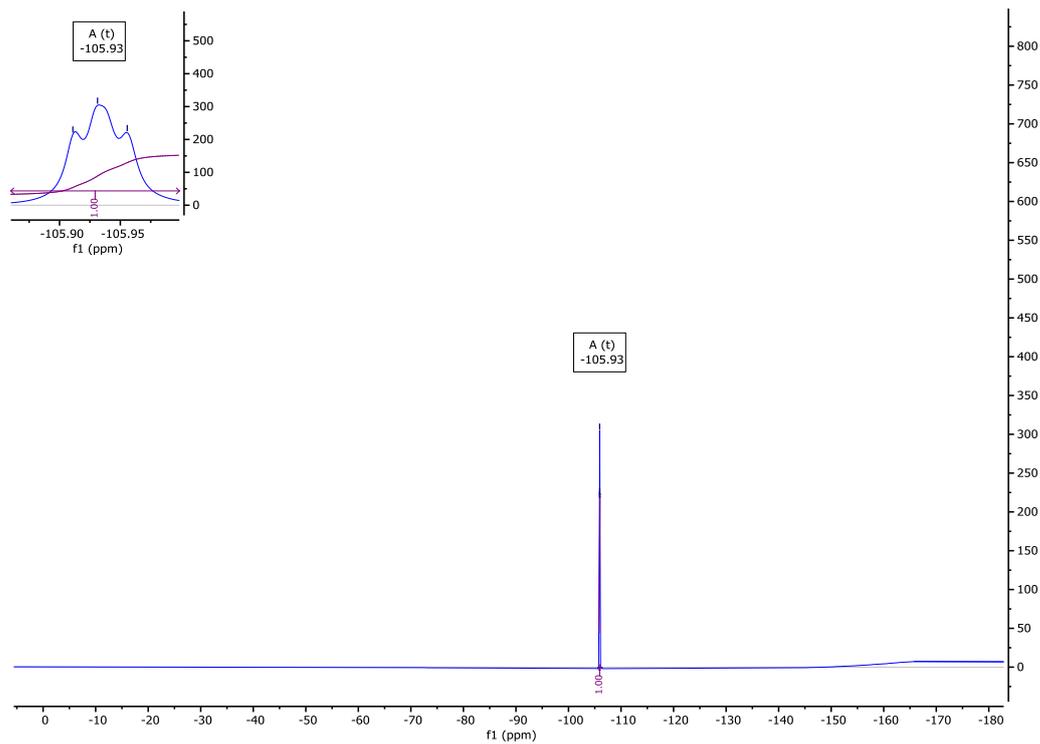

**Fig. S94** 19F NMR spectra of **NC1000** in CDCl₃.

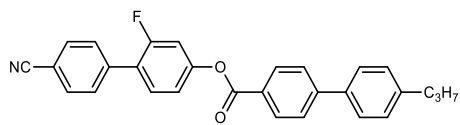

**NC0100**

*4'-cyano-2-fluoro-[1,1'-biphenyl]-4-yl 4'-propyl-[1,1'-biphenyl]-4-carboxylate*

| | |
|---|---|
| Yield: | (white solid) 170 mg, 78 % |
| Re-crystallisation solvent: | EtOH |
| $^1$H NMR (400 MHz): | 8.26 (ddd, *J* = 8.6, 1.9, 1.9 Hz, 2H, Ar-**H**), 7.79 – 7.72 (m, 4H, Ar-**H**)*, 7.70 – 7.65 (m, 2H, Ar-**H**)*, 7.59 (ddd, *J* = 8.2, 2.0, 1.8 Hz, 2H, Ar-**H**), 7.50 (t, *J* = 8.6 Hz, 1H, Ar-**H**), 7.31 (ddd, *J* = 8.3, 1.9, 1.8 Hz, 2H, Ar-**H**), 7.21 – 7.15 (m, 2H, Ar-**H**)*, 2.67 (t, *J* = 7.3 Hz, 2H, Ar-C**H$_2$**-CH$_2$), 1.70 (h, *J* = 7.4 Hz, 2H, CH$_2$-C**H$_2$**-CH$_3$), 0.99 (t, *J* = 7.3 Hz, 3H, CH$_0$-C**H$_3$**). |
| $^{13}$C{$^1$H} NMR (101 MHz): | 164.65, 160.86, 158.36, 152.04 (d, *J* = 11.0 Hz), 146.79, 143.37, 139.81, 136.99, 132.34, 130.93 – 130.77 (m), 129.67 (d, *J* = 3.3 Hz), 129.21, 127.18, 127.13, 124.88 (d, *J* = 13.1 Hz), 118.39 (d, *J* = 3.6 Hz), 118.37, 111.57, 110.81 (d, *J* = 25.8 Hz), 37.75, 24.53, 13.88. |
| $^{19}$F NMR (376 MHz): | -114.39 (t, *J$_{F-H}$* = 9.7 Hz, 1F, Ar-**F**). |

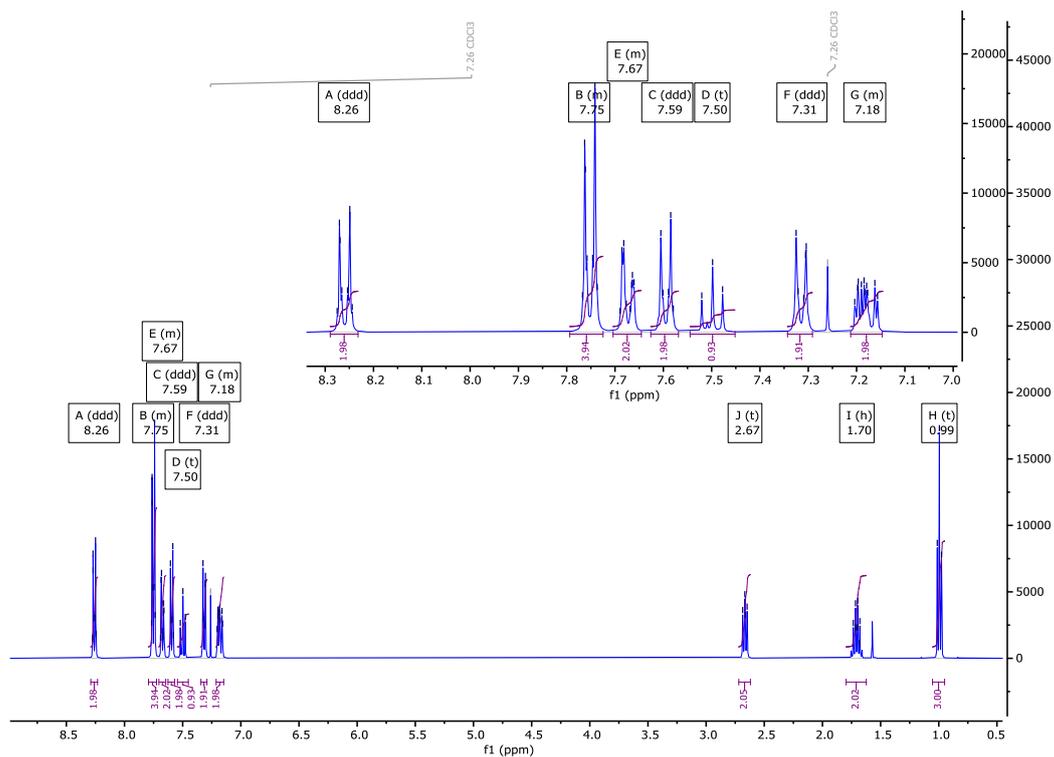

**Fig. S95**          $^1$H NMR spectra of **NC0100** in CDCl$_3$.

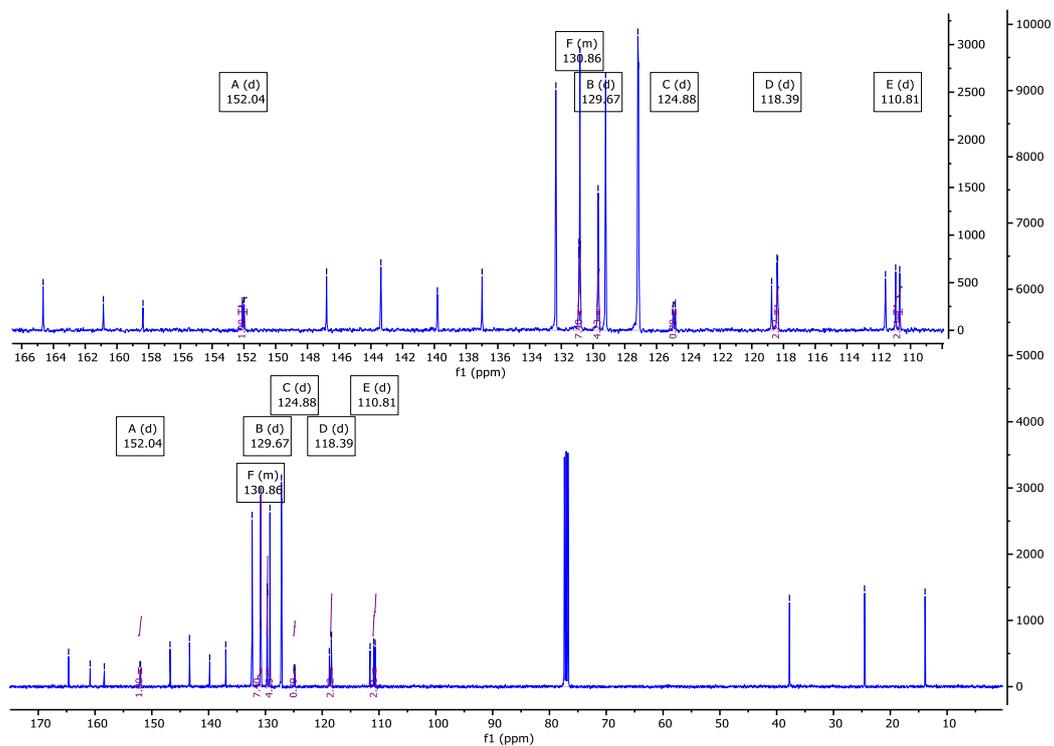

**Fig. S96**          $^{13}$C{$^1$H} NMR spectra of **NC0100** in CDCl$_3$.

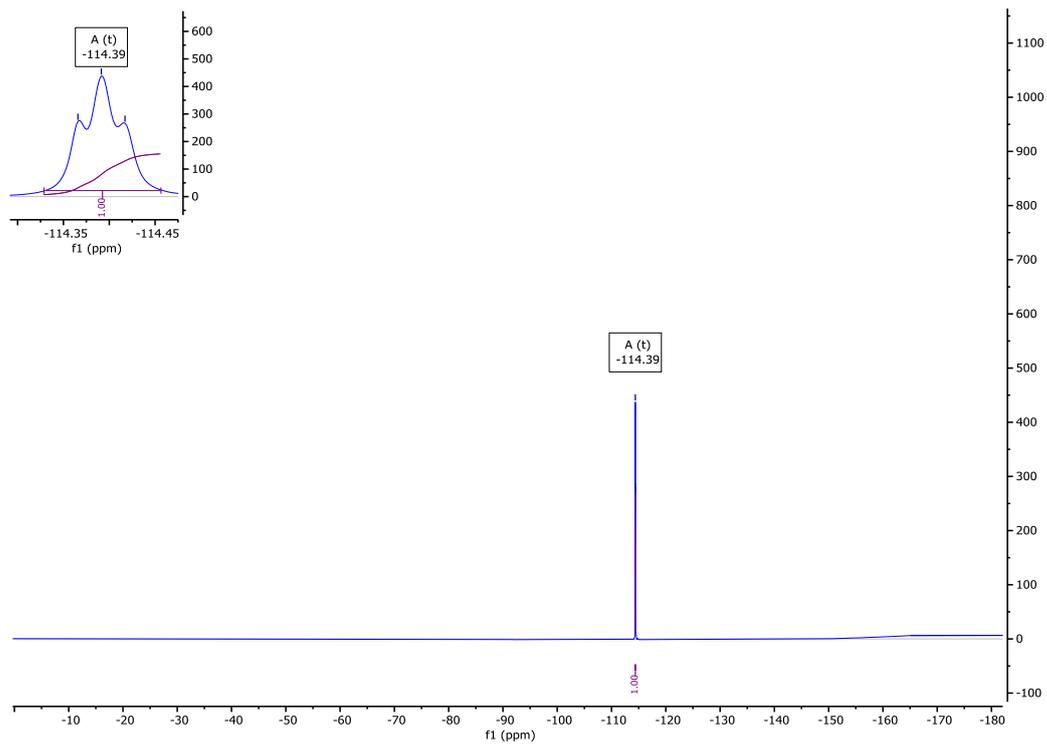

**Fig. S97** $^{19}$F NMR spectra of **NC0100** in CDCl$_3$.

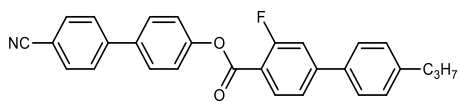

**NC0010**

*4'-cyano-[1,1'-biphenyl]-4-yl 3-fluoro-4'-propyl-[1,1'-biphenyl]-4-carboxylate*

| | |
|---|---|
| Yield: | (white solid) 170 mg, 78 % |
| Re-crystallisation solvent: | EtOH |
| $^1$H NMR (400 MHz): | 8.17 (t, *J* = 7.9 Hz, 1H, Ar-**H**), 7.77 – 7.68 (m, 4H, Ar-**H**)*, 7.66 (ddd, *J* = 8.7, 2.8, 2.0 Hz, 2H, Ar-**H**), 7.57 (ddd, *J* = 8.2, 2.0, 1.8 Hz, 2H, Ar-**H**), 7.52 (dd, *J* = 8.2, 1.7 Hz, 1H, Ar-**H**), 7.45 (dd, *J* = 12.1, 1.7 Hz, 1H, Ar-**H**), 7.38 (ddd, *J* = 8.5, 2.8, 2.0 Hz, 2H, Ar-**H**), 7.31 (ddd, *J* = 8.2, 2.3, 1.5 Hz, 2H, Ar-**H**), 2.66 (t, *J* = 7.5 Hz, 2H, Ar-C**H₂**-CH₂), 1.70 (h, *J* = 7.4 Hz, 2H, CH₂-C**H₂**-CH₃), 0.99 (t, *J* = 7.3 Hz, 3H, CH₂-C**H₃**). * Overlapping Signals. |
| $^{13}$C{$^1$H} NMR (126 MHz): | 163.80 (d, *J* = 261.7 Hz), 162.64 (d, *J* = 4.3 Hz), 151.17, 148.83 (d, *J* = 9.0 Hz), 144.82, 144.01, 137.05, 135.74, 132.99, 132.69, 129.31, 128.43, 127.75, 127.06, 122.50, 122.50 (d, *J* = 1.5 Hz), 118.89, 115.76 (d, *J* = 9.5 Hz), 115.27 (d, *J* = 23.1 Hz), 111.09, 37.74, 24.49, 13.85. |
| $^{19}$F NMR (376 MHz): | -107.56 (dd, *J*<sub>F-H</sub> = 12.2 Hz, *J*<sub>F-H</sub> = 7.6 Hz, 1F, Ar-**F**). |

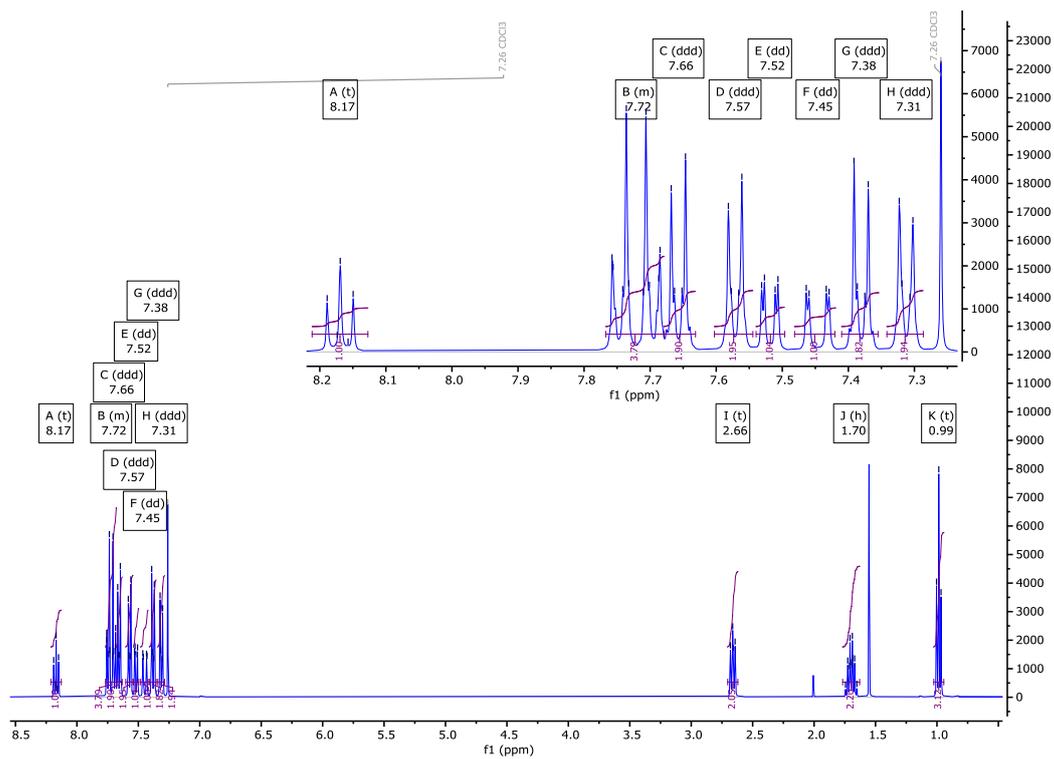

**Fig. S98**  $^1$H NMR spectra of **NC0010** in CDCl$_3$.

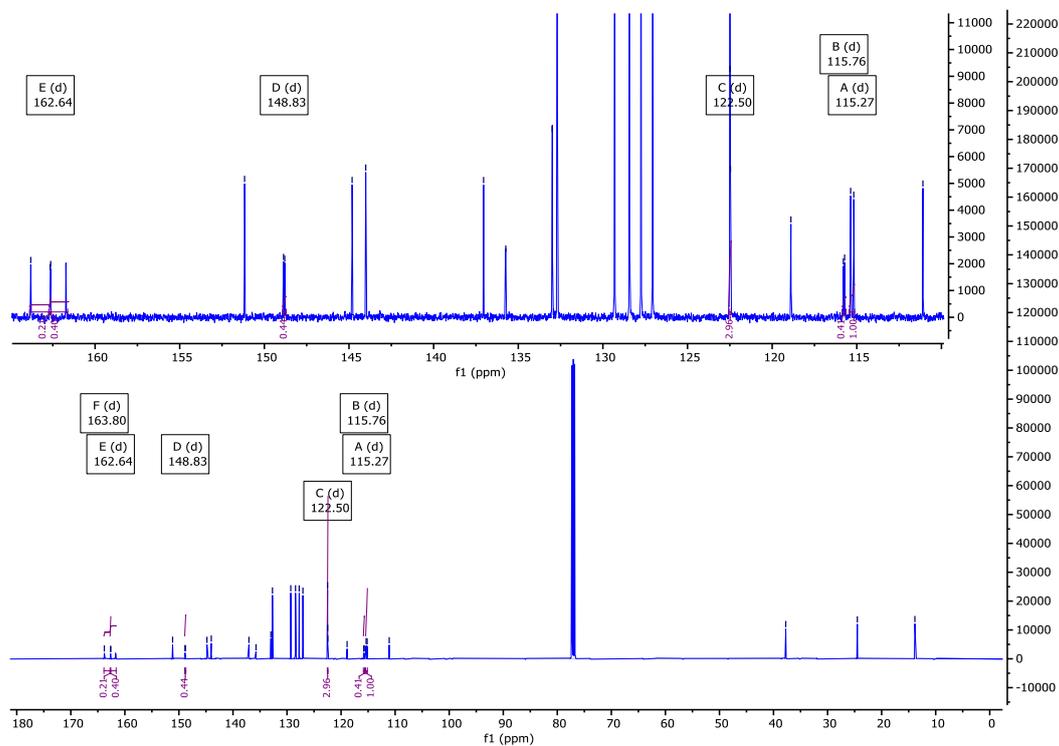

**Fig. S99**  $^{13}$C{$^1$H} NMR spectra of **NC0010** in CDCl$_3$.

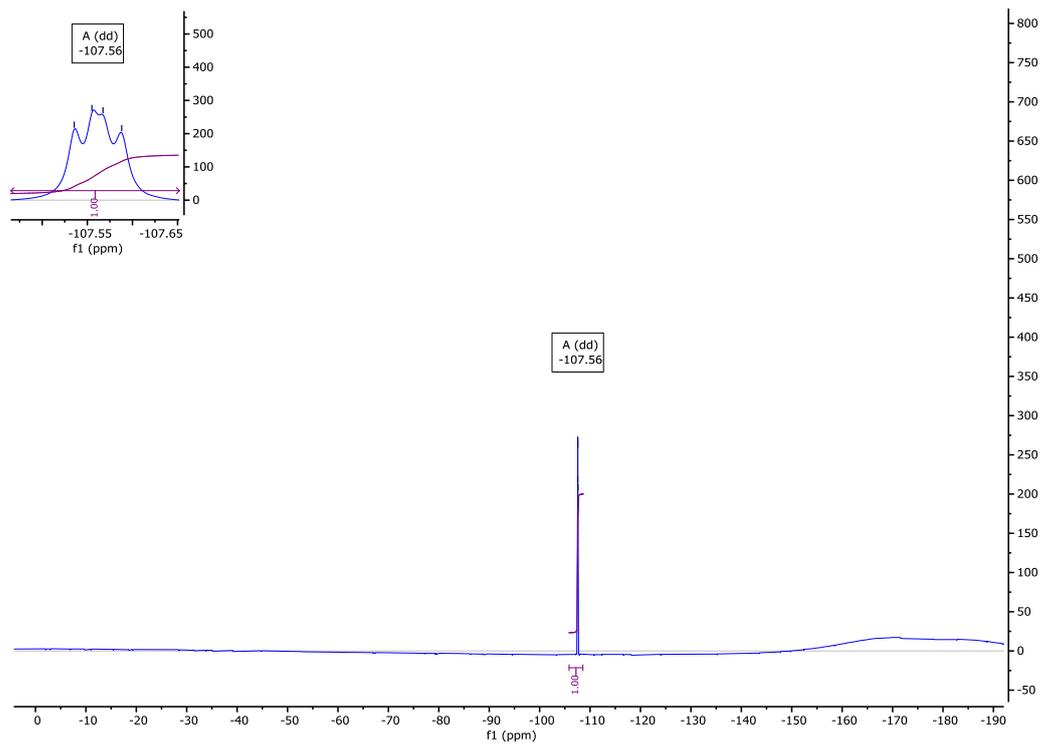

**Fig. S100**  $^{19}$F NMR spectra of **NC0010** in CDCl$_3$.

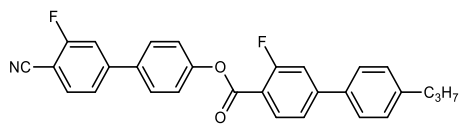

**NC1010**

*4'-cyano-3'-fluoro-[1,1'-biphenyl]-4-yl 3-fluoro-4'-propyl-[1,1'-biphenyl]-4-carboxylate*

| | |
|---|---|
| Yield: | (white solid) 156 mg, 69 % |
| Re-crystallisation solvent: | EtOH |
| $^1$H NMR (400 MHz): | 8.19 (t, *J* = 7.9 Hz, 1H, Ar-**H**), 7.73 (dd, *J* = 8.1, 6.6 Hz, 1H, Ar-**H**), 7.67 (ddd, *J* = 8.7, 2.8, 2.1 Hz, 2H, Ar-**H**), 7.60 (ddd, *J* = 8.2, 2.2, 1.6 Hz, 2H, Ar-**H**), 7.57 – 7.44 (m, 4H, Ar-**H**)[*], 7.42 (ddd, *J* = 8.7, 2.9, 2.0 Hz, 2H, Ar-**H**), 7.34 (ddd, *J* = 8.1, 2.2, 1.8 Hz, 2H, Ar-**H**), 2.69 (t, *J* = 7.6 Hz, 2H, Ar-C**H$_2$**-CH$_2$), 1.72 (h, *J* = 7.4 Hz, 2H, CH$_2$-C**H$_2$**-CH$_3$), 1.01 (t, *J* = 7.3 Hz, 3H, CH$_2$-C**H$_3$**). [*] Overlapping Signals. |
| $^{13}$C{$^1$H} NMR (126 MHz): | 163.60 (d, *J* = 258.4 Hz), 162.55 (d, *J* = 4.2 Hz), 161.73 (d, *J* = 262.0 Hz), 151.63, 148.91 (d, *J* = 8.9 Hz), 147.65 (d, *J* = 8.2 Hz), 144.04, 135.78 (dd, *J* = 16.5, 1.9 Hz)z, 133.84, 132.99, 129.32, 128.43, 127.06, 123.39 (d, *J* = 3.1 Hz), 122.69, 122.49 (d, *J* = 3.2 Hz), 115.64 (d, *J* = 9.6 Hz), 115.28 (d, *J* = 23.1 Hz), 114.84 (d, *J* = 20.3 Hz), 114.03, 100.00 (d, *J* = 15.7 Hz), 37.74, 24.49, 13.85. |
| $^{19}$F NMR (376 MHz): | -105.93 (t, $J_{F-H}$ = 7.4 Hz, 1F, Ar-**F**), -107.50 (dd, $J_{F-H}$ = 12.3 Hz, $J_{F-H}$ = 7.6 Hz, 1F, Ar-**F**). |

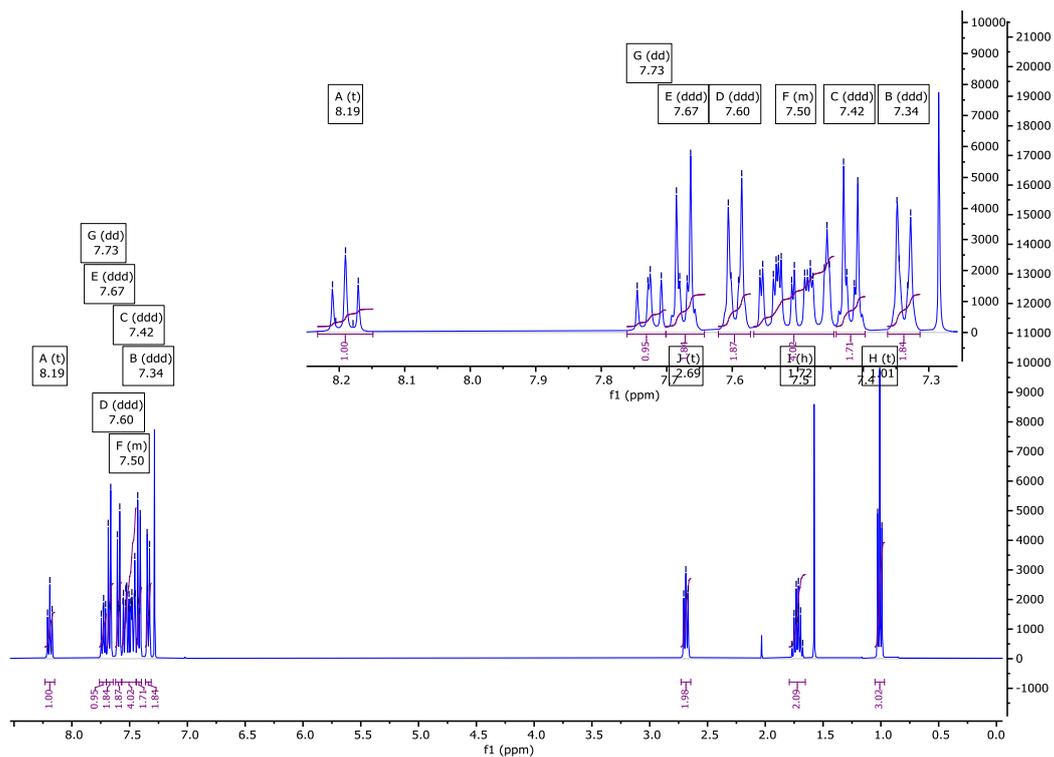

**Fig. S101**  $^1$H NMR spectra of **NC1010** in CDCl$_3$.

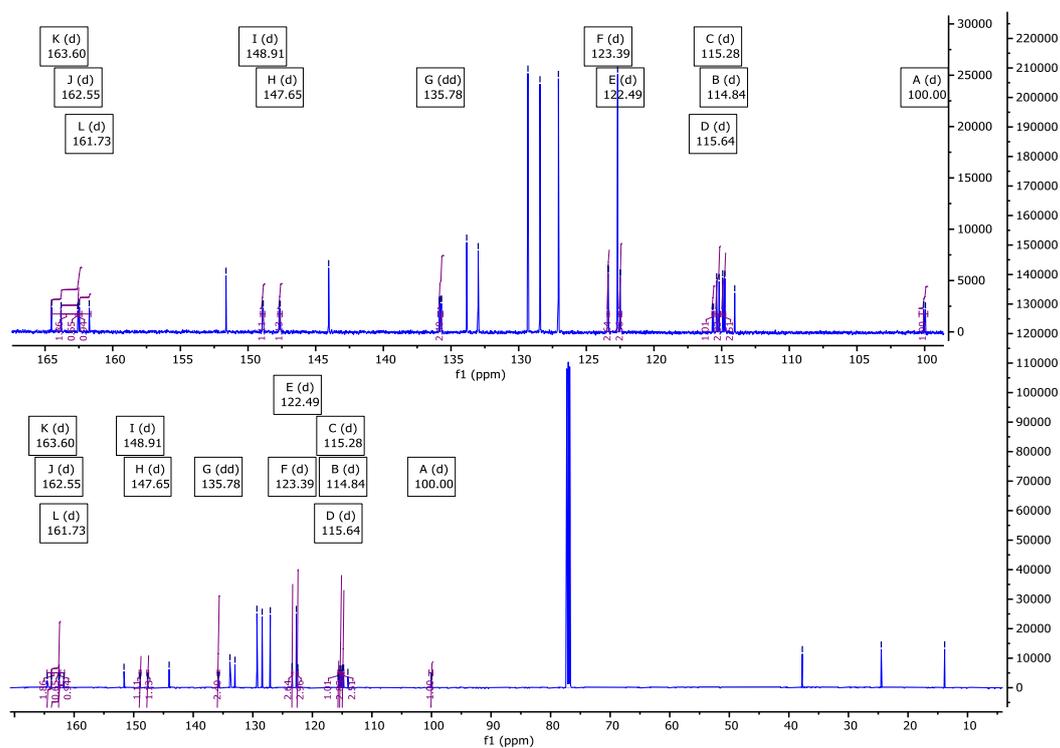

**Fig. S102**  $^{13}$C{$^1$H} NMR spectra of **NC1010** in CDCl$_3$.

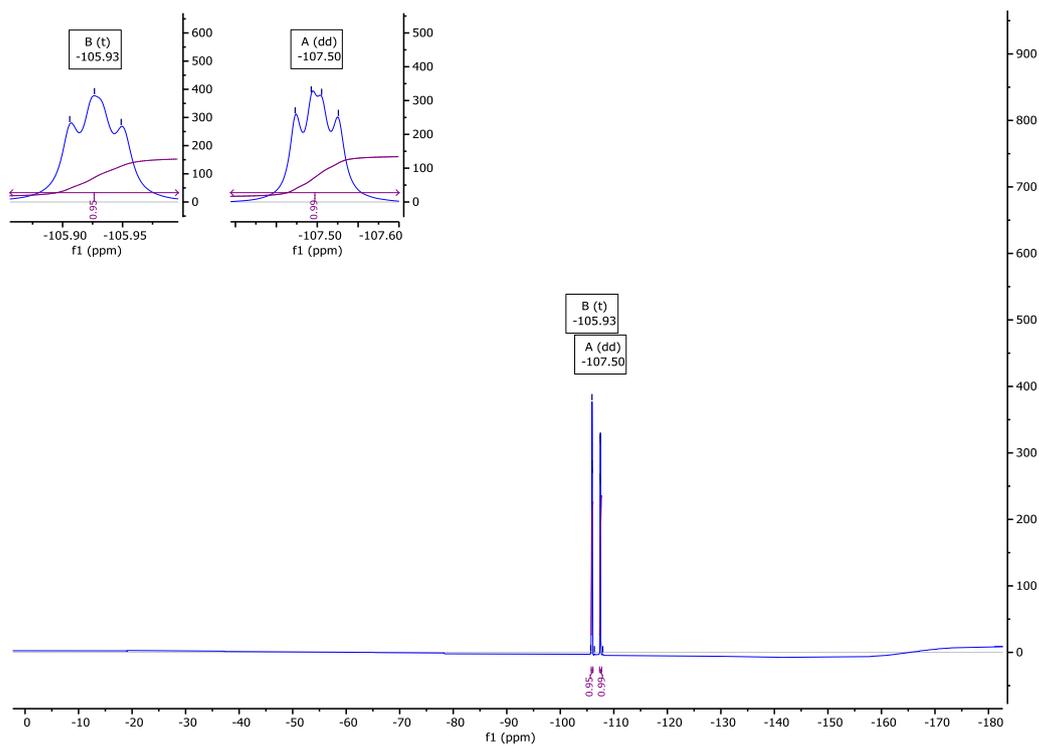

**Fig. S103** $^{19}$F NMR spectra of **NC1010** in CDCl$_3$.

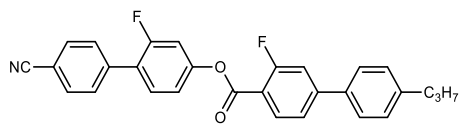

**NC0110**

*4'-cyano-2-fluoro-[1,1'-biphenyl]-4-yl 3-fluoro-4'-propyl-[1,1'-biphenyl]-4-carboxylate*

| | |
|---|---|
| Yield: | (white crystalline solid) 175 mg, 77 % |
| Re-crystallisation solvent: | EtOH |
| $^1$H NMR (400 MHz): | 8.15 (t, *J* = 7.9 Hz, 1H, Ar-**H**), 7.76 (ddd, *J* = 8.0, 2.2, 1.3 Hz, 2H, Ar-**H**), 7.70 – 7.64 (m, 2H, Ar-**H**), 7.57 (ddd, *J* = 8.1, 2.0, 1.7 Hz, 2H, Ar-**H**), 7.54 – 7.47 (m, 2H, Ar-**H**)*, 7.45 (dd, *J* = 12.1, 1.7 Hz, 1H, Ar-**H**), 7.31 (ddd, *J* = 8.1, 2.0, 1.8 Hz, 2H, Ar-**H**), 7.23 – 7.16 (m, 2H, Ar-**H**)*, 2.66 (t, *J* = 7.4 Hz, 2H, Ar-C**H$_2$**-CH$_2$), 1.70 (h, *J* = 7.4 Hz, 2H, CH$_2$-C**H$_2$**-CH$_3$), 0.98 (t, *J* = 7.3 Hz, 3H, CH$_2$-C**H$_3$**). * Overlapping Signals. |
| $^{13}$C{$^1$H} NMR (126 MHz): | 162.79 (d, *J* = 261.8 Hz), 162.22 (d, *J* = 4.2 Hz), 159.58 (d, *J* = 251.3 Hz), 151.64 (d, *J* = 11.1 Hz), 149.09 (d, *J* = 9.0 Hz), 144.10, 139.77, 135.65, 133.00, 132.35, 130.87 (d, *J* = 4.2 Hz), 129.67 (d, *J* = 3.3 Hz), 129.33, 127.07, 125.03 (d, *J* = 13.2 Hz), 122.53 (d, *J* = 3.2 Hz), 118.75, 118.35 (d, *J* = 3.6 Hz), 115.40, 115.31 (d, *J* = 23.0 Hz), 111.59, 110.78 (d, *J* = 25.9 Hz), 37.74, 24.49, 13.85. |
| $^{19}$F NMR (376 MHz): | -107.30 (dd, $J_{F-H}$ = 12.3 Hz, $J_{F-H}$ = 7.6 Hz, 1F, Ar-**F**), -114.36 (t, $J_{F-H}$ = 9.7 Hz, 1F, Ar-**F**). |

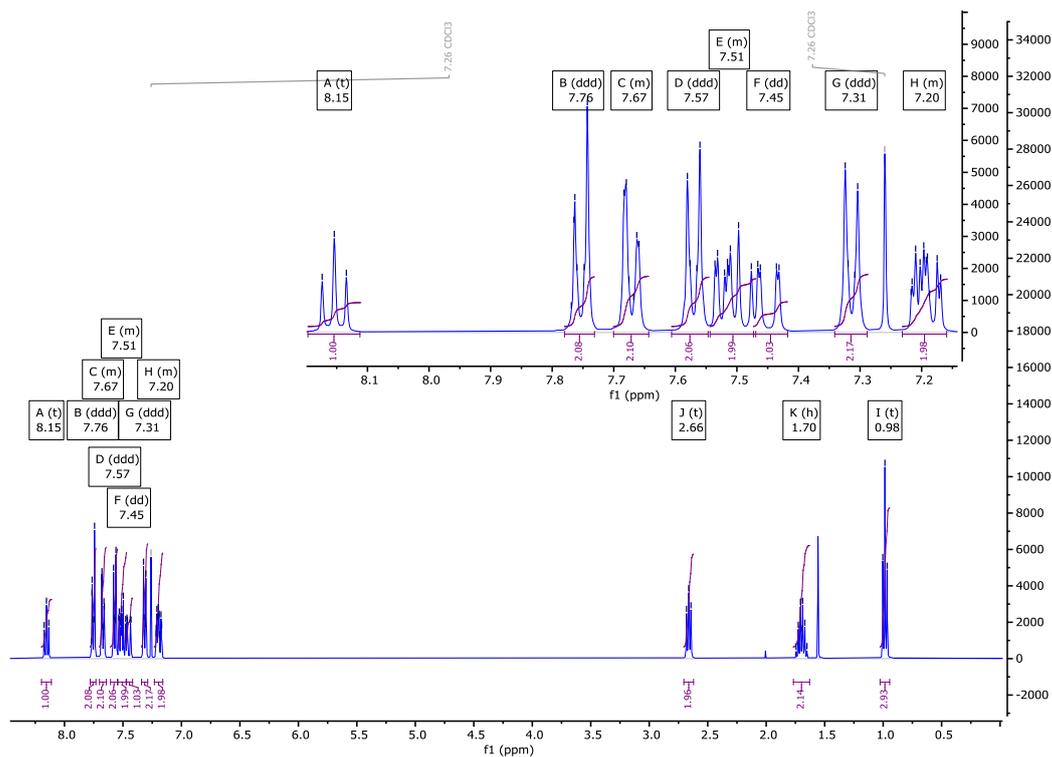

**Fig. S104**          $^1$H NMR spectra of **NC0110** in CDCl$_3$.

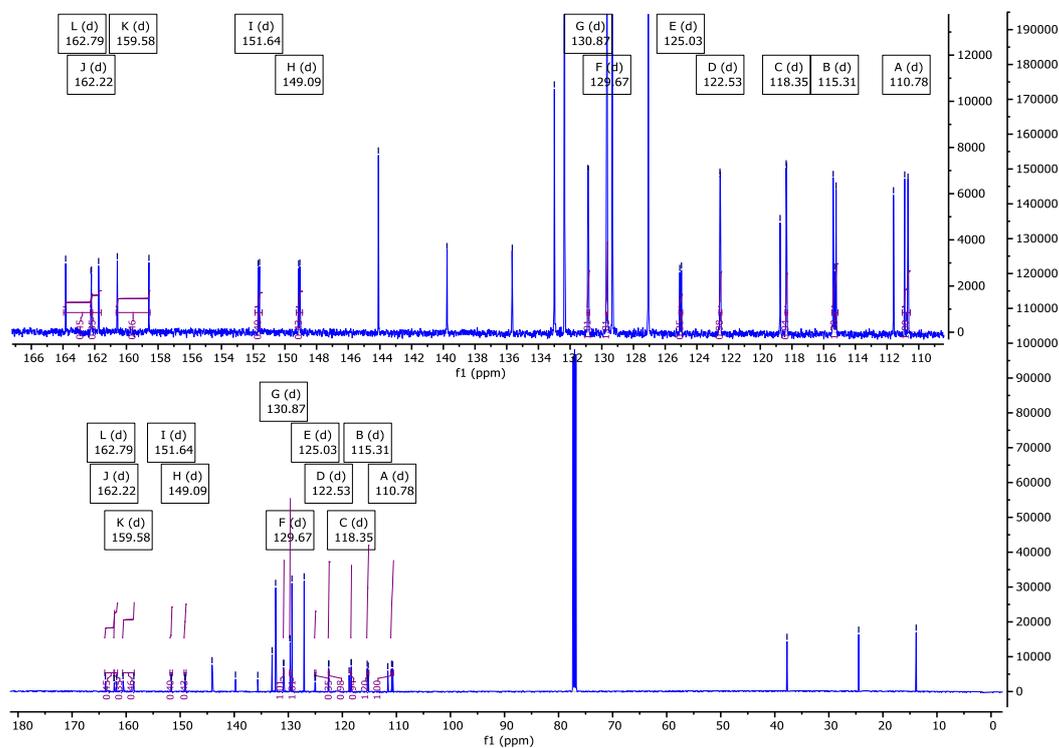

**Fig. S105**          $^{13}$C{$^1$H} NMR spectra of **NC0110** in CDCl$_3$.

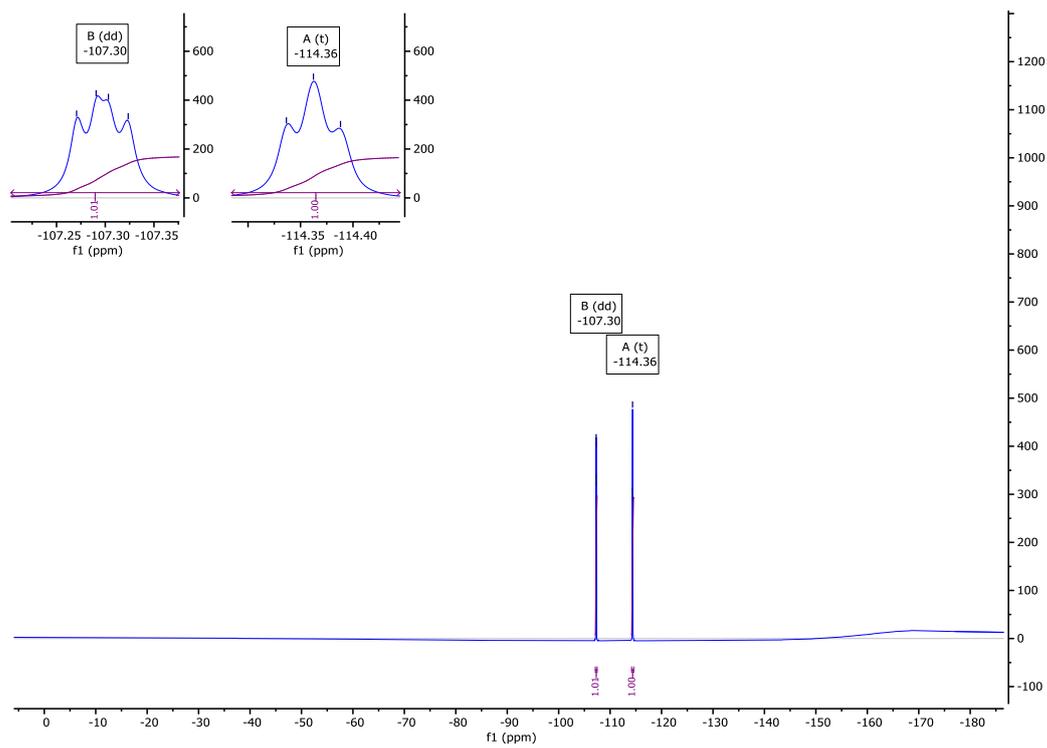

**Fig. S106**         ¹⁹F NMR spectra of **NC0110** in CDCl₃.

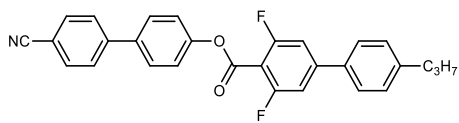

**NC0020**

*4'-cyano-[1,1'-biphenyl]-4-yl 3,5-difluoro-4'-propyl-[1,1'-biphenyl]-4-carboxylate*

| | |
|---|---|
| Yield: | (white solid) 134 mg, 60 % |
| Re-crystallisation solvent: | EtOH |
| $^1$H NMR (400 MHz): | 7.77 – 7.63 (m, 4H, Ar-**H**)*, 7.53 (ddd, *J* = 8.3, 1.9, 1.9 Hz, 2H, Ar-**H**), 7.40 (ddd, *J* = 8.9, 2.8, 2.1 Hz, 2H, Ar-**H**), 7.32 (ddd, *J* = 8.2, 1.6, 1.5 Hz, 2H, Ar-**H**), 7.26 (ddd, *J* = 9.4, 3.4, 2.1 Hz, 2H, Ar-**H**), 2.66 (t, *J* = 7.2 Hz, 2H, Ar-C**H$_2$**-CH$_2$), 1.69 (h, *J* = 7.5 Hz, 2H, CH$_2$-C**H$_2$**-CH$_3$), 0.98 (t, *J* = 7.3 Hz, 3H, CH$_2$-C**H$_3$**). *Overlapping Signals. |
| $^{13}$C{$^1$H} NMR (101 MHz): | 161.52 (dd, *J* = 257.6, 6.6 Hz), 159.88, 150.87, 147.66 (t, *J* = 9.8 Hz), 144.73, 144.49, 137.32, 134.93, 132.69, 129.39, 128.48, 127.77, 126.89, 122.39, 118.86, 111.17, 110.43 (dd, *J* = 23.4, 2.8 Hz), 107.83 (t, *J* = 15.7 Hz), 37.73, 24.45, 13.82. |
| $^{19}$F NMR (376 MHz): | -108.53 (d, *J$_{F-H}$* = 10.4 Hz, 2F, Ar-**F**). |

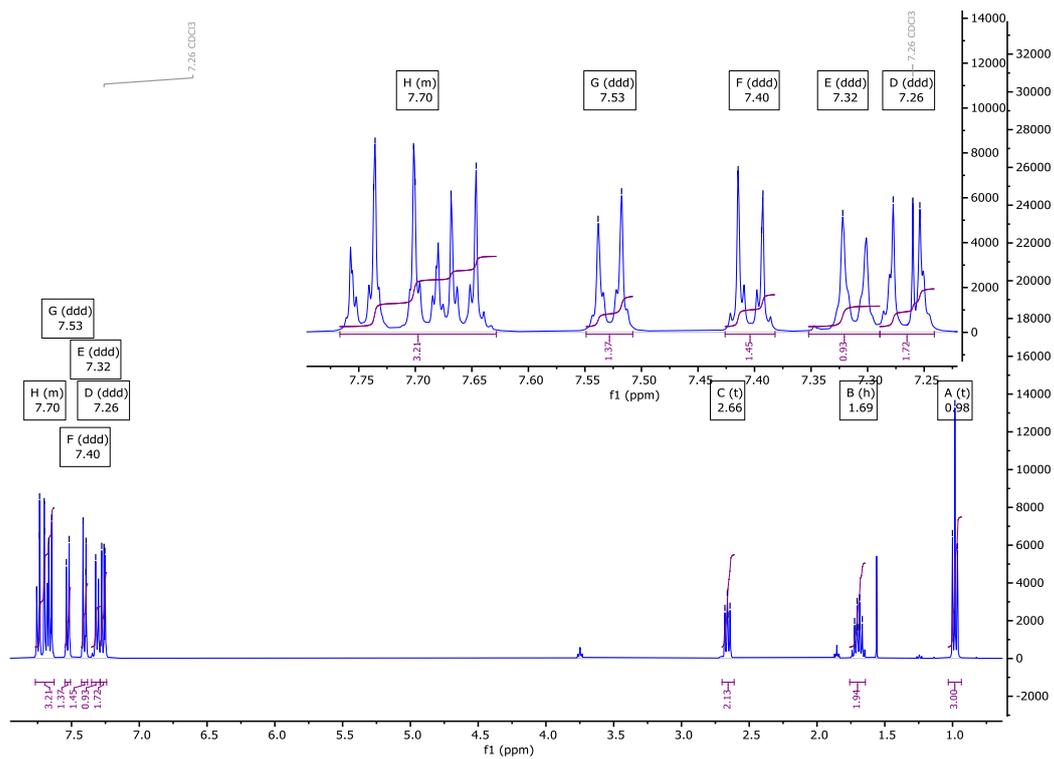

**Fig. S107** $^1$H NMR spectra of **NC0020** in CDCl$_3$.

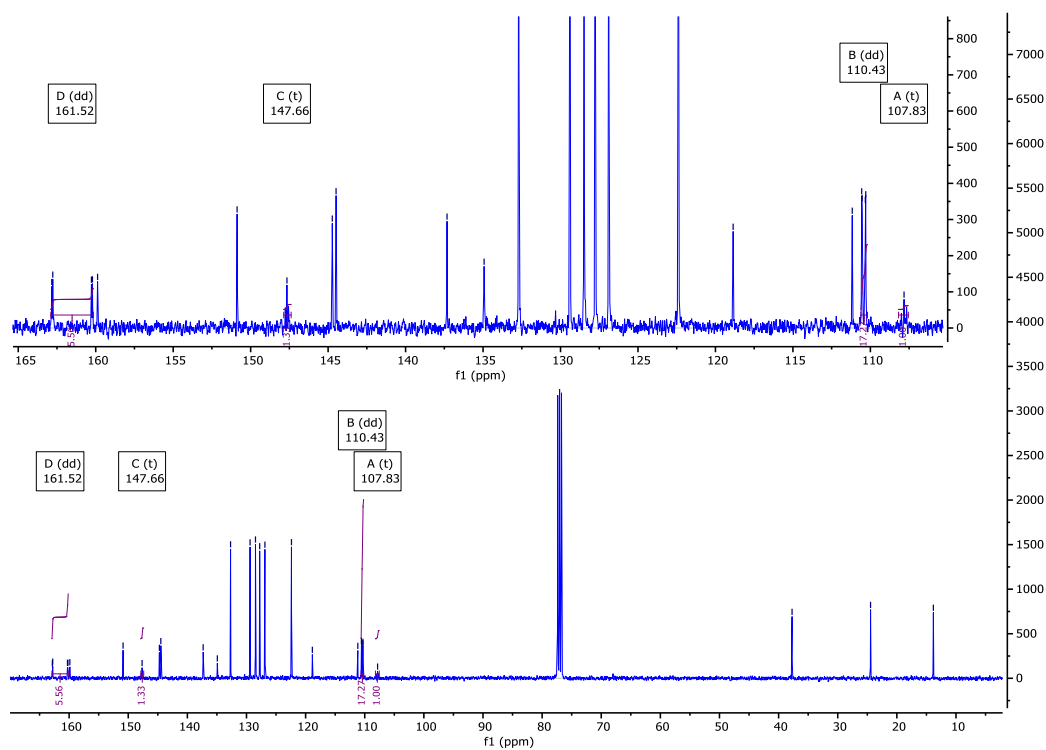

**Fig. S108** $^{13}$C{$^1$H} NMR spectra of **NC0020** in CDCl$_3$.

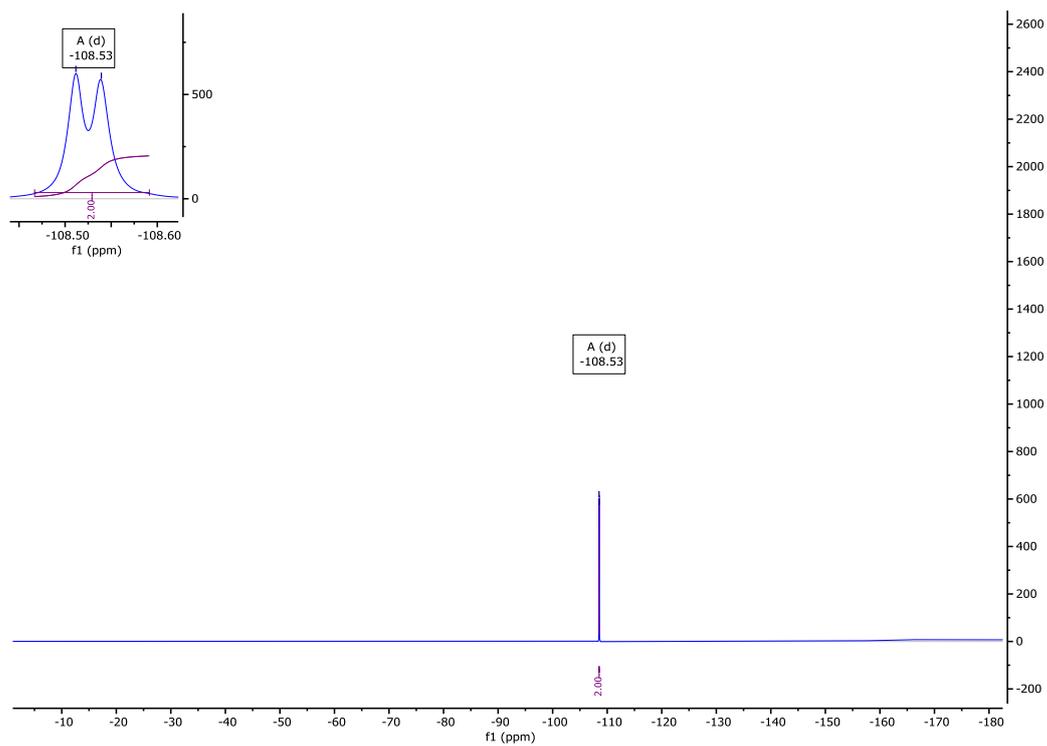

**Fig. S109** ¹⁹F NMR spectra of **NC0020** in CDCl₃.

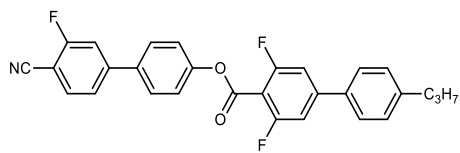

**NC1020**

*4'-cyano-3'-fluoro-[1,1'-biphenyl]-4-yl 3,5-difluoro-4'-propyl-[1,1'-biphenyl]-4-carboxylate*

| | |
|---|---|
| Yield: | (white solid) 172 mg, 73 % |
| Re-crystallisation solvent: | EtOH |
| $^1$H NMR (400 MHz): | 7.70 (dd, *J* = 8.1, 6.6 Hz, 1H, Ar-**H**), 7.65 (ddd, *J* = 8.6, 2.8, 1.9 Hz, 2H, Ar-**H**), 7.53 (ddd, *J* = 8.3, 1.8, 1.8 Hz, 2H, Ar-**H**), 7.49 (dd, *J* = 8.1, 1.6 Hz, 1H, Ar-**H**), 7.46 – 7.39 (m, 3H, Ar-**H**)*, 7.31 (ddd, *J* = 8.2, 2.0 Hz, 2H, Ar-**H**), 7.26 (d, *J* = 9.4 Hz, 2H, Ar-**H**)†, 2.66 (t, *J* = 7.4 Hz, 2H, Ar-C**H₂**-CH₂), 1.69 (h, *J* = 7.5 Hz, 2H, CH₂-C**H₂**-CH₃), 0.98 (t, *J* = 7.3 Hz, 3H, CH₂-C**H₃**). *Overlapping Signals, †Overlapping CDCl₃. |
| $^{13}$C{$^1$H} NMR (101 MHz): | 163.28 (d, *J* = 259.4 Hz), 160.23 (dd, *J* = 258.0, 6.4 Hz), 159.79, 151.32, 147.97 – 147.46 (m), 144.52, 136.13, 134.91, 133.86, 129.40, 128.48, 126.89, 123.41 (d, *J* = 3.1 Hz), 122.58, 114.87 (d, *J* = 20.4 Hz), 114.00, 110.45 (dd, *J* = 23.7, 2.1 Hz), 107.70 (t, *J* = 17.6 Hz), 100.08 (d, *J* = 15.7 Hz), 88.00 (t, *J* = 17.6 Hz), 37.72, 24.44, 13.82. |
| $^{19}$F NMR (376 MHz): | -105.88 (dd, *J*$_{F-H}$ = 9.1 Hz, *J*$_{F-H}$ = 7.0 Hz, 1F, Ar-**F**), -108.46 (d, *J* = 10.5 Hz, 2F, Ar-**F**). |

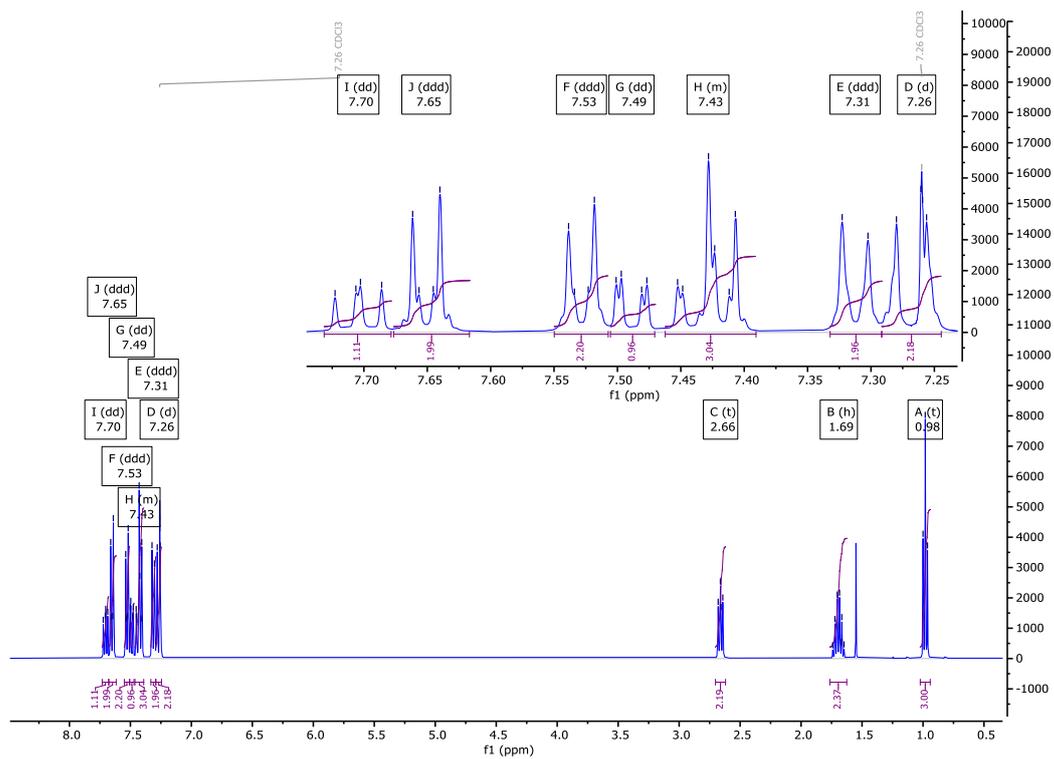

**Fig. S110**  $^1$H NMR spectra of **NC1020** in CDCl$_3$.

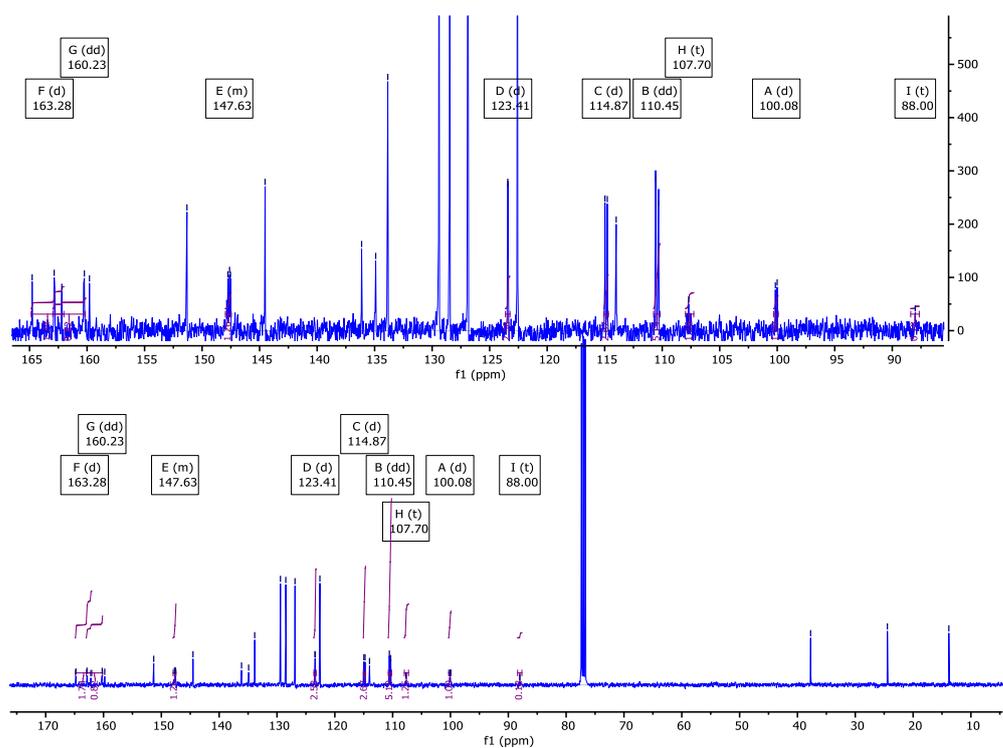

**Fig. S111**  $^{13}$C{$^1$H} NMR spectra of **NC1020** in CDCl$_3$.

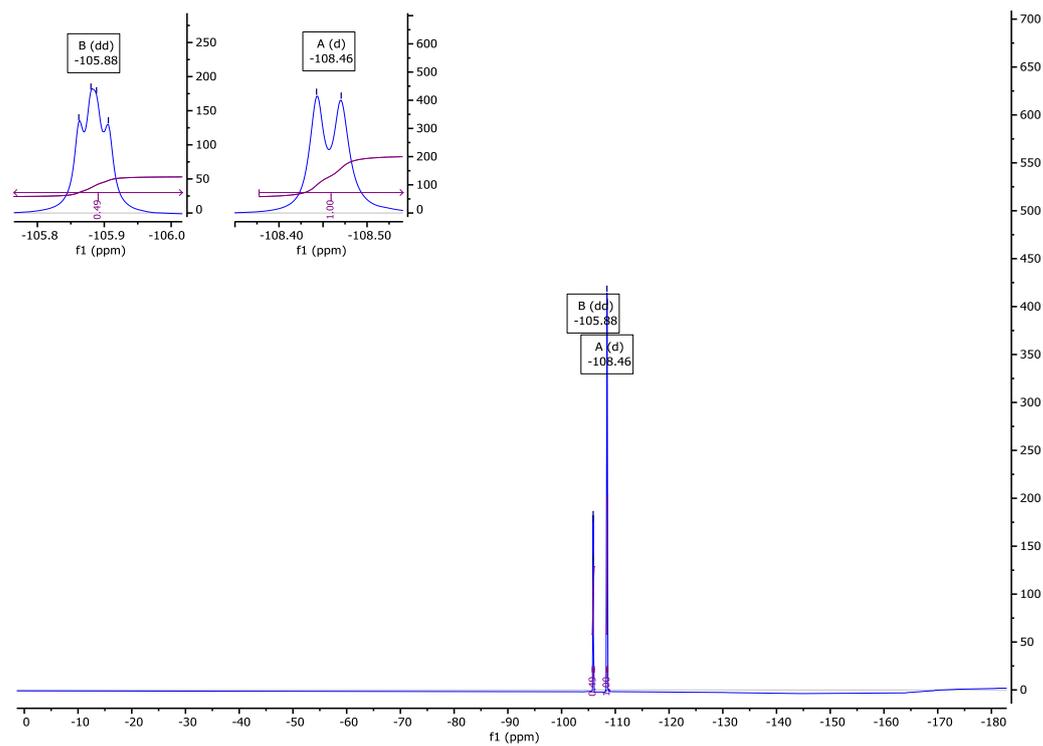

**Fig. S112**  $^{19}$F NMR spectra of **NC1020** in CDCl$_3$.

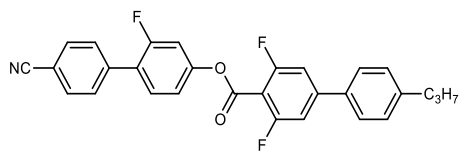

**NC0120**

*4'-cyano-2-fluoro-[1,1'-biphenyl]-4-yl 3,5-difluoro-4'-propyl-[1,1'-biphenyl]-4-carboxylate*

| | |
|---|---|
| Yield: | (white solid) 146 mg, 62 % |
| Re-crystallisation solvent: | EtOH |
| $^1$H NMR (400 MHz): | 7.80 – 7.63 (m, 4H, Ar-**H**)*, 7.57 – 7.46 (m, 3H, Ar-**H**)*, 7.35 – 7.18 (m, 6H, Ar-**H**)*†, 2.66 (t, *J* = 7.6 Hz, 2H, Ar-C**H$_2$**-CH$_2$), 1.70 (h, *J* = 7.4 Hz, 2H, CH$_2$-C**H$_2$**-CH$_3$), 0.98 (t, *J* = 7.3 Hz, 3H, CH$_2$-C**H$_3$**). *Overlapping Signals, † Overlapping CDCl$_3$. |
| $^{13}$C{$^1$H} NMR (101 MHz): | 161.44 (dd, *J* = 258.3, 6.6 Hz), 158.89 (d, *J* = 251.7 Hz), 158.31, 151.28 (d, *J* = 11.0 Hz), 147.91 (t, *J* = 10.6 Hz), 144.59, 139.68, 134.84, 132.36, 130.93 (d, *J* = 4.2 Hz), 129.68 (d, *J* = 3.2 Hz), 129.42, 126.89, 125.31 (d, *J* = 13.3 Hz), 118.72, 118.24 (d, *J* = 3.8 Hz), 111.66, 110.95 – 110.18 (m), 107.37 (t, *J* = 16.3 Hz), 37.73, 24.44, 13.82. |
| $^{19}$F NMR (376 MHz): | -108.23 (d, *J$_{F-H}$* = 10.5 Hz, 2F, Ar-**F**), -114.15 (t, *J$_{F-H}$* = 9.8 Hz, 1F, Ar-**F**). |

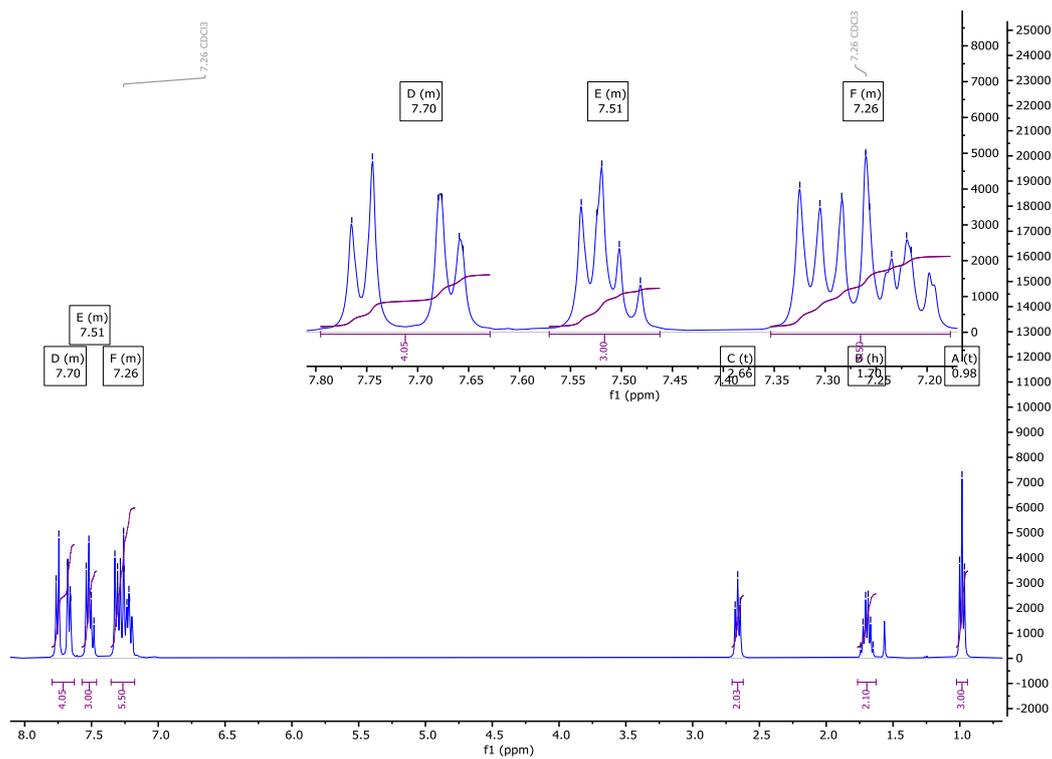

**Fig. S113** $^1$H NMR spectra of **NC0120** in CDCl$_3$.

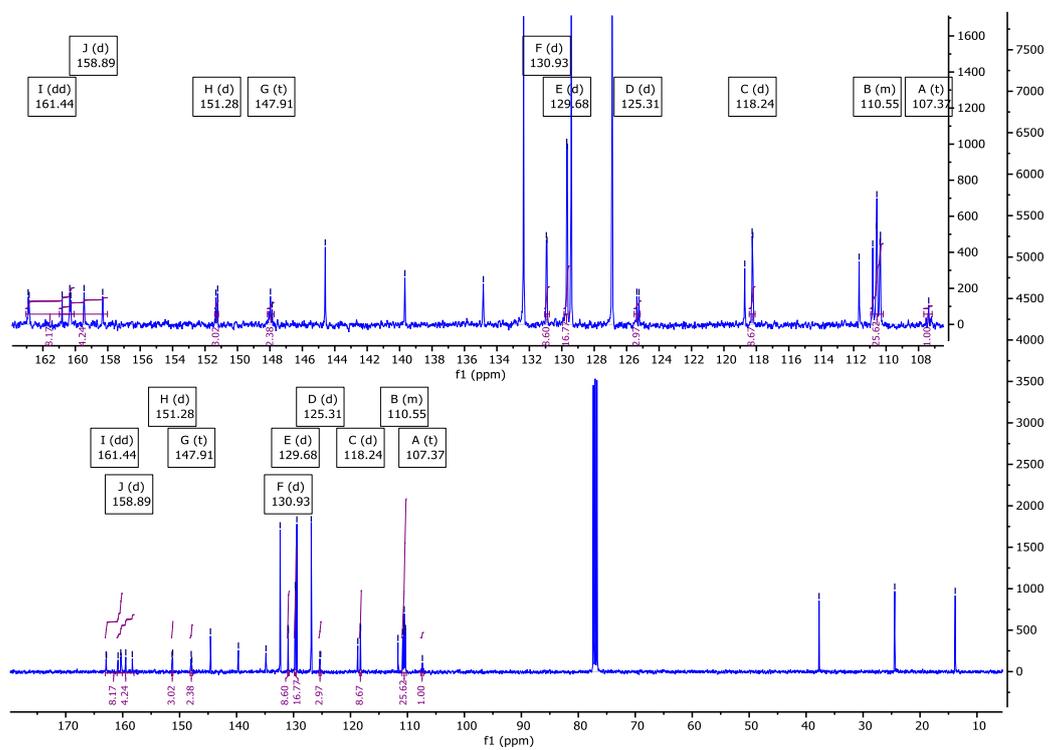

**Fig. S114** $^{13}$C{$^1$H} NMR spectra of **NC0120** in CDCl$_3$.

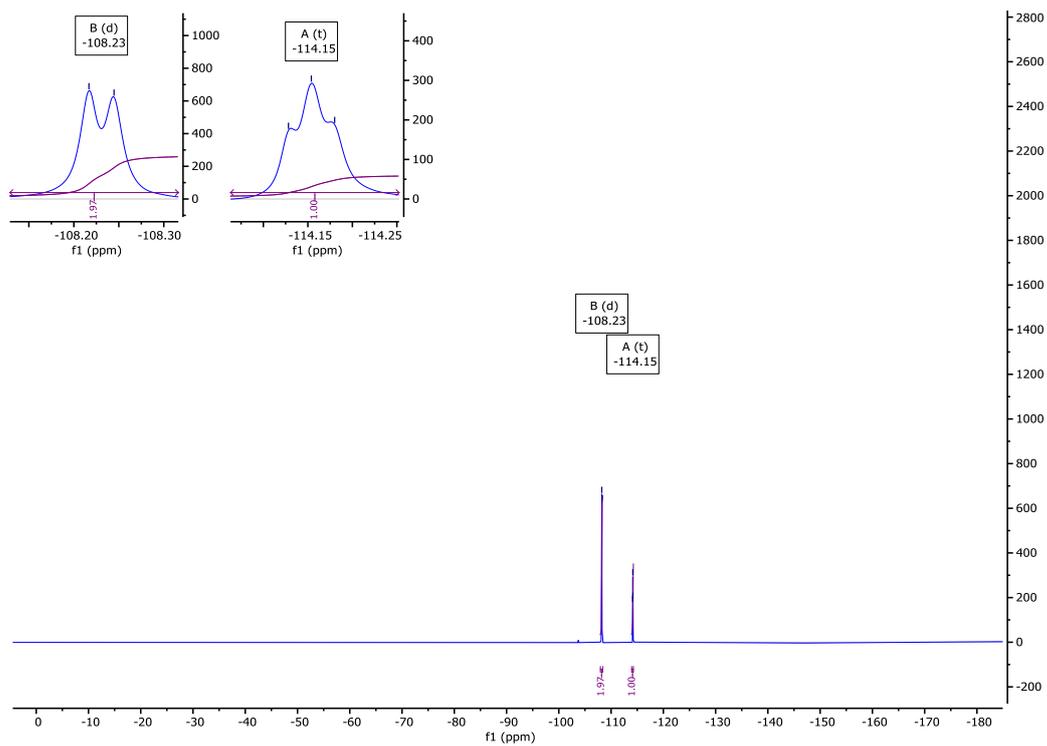

**Fig. S115** 19F NMR spectra of **NC0120** in CDCl$_3$.

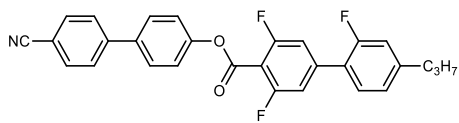

**NC0021**

*4'-cyano-[1,1'-biphenyl]-4-yl 2',3,5-trifluoro-4'-propyl-[1,1'-biphenyl]-4-carboxylate*

| | |
|---|---|
| Yield: | (white crystalline solid) 118 mg, 50 % |
| Re-crystallisation solvent: | EtOH |
| $^1$H NMR (400 MHz): | 7.77 – 7.67 (m, 2H, Ar-**H**), 7.66 (ddd, *J* = 8.6, 2.6, 1.9 Hz, 2H, Ar-**H**), 7.41 (ddd, *J* = 8.6, 2.8, 1.9 Hz, 2H, Ar-**H**), 7.37 (t, *J* = 8.2 Hz, 1H, Ar-**H**), 7.29 – 7.23 (m, 2H, Ar-**H**)†, 7.09 (dd, *J* = 7.9, 1.6 Hz, 1H, Ar-**H**), 7.03 (dd, *J* = 12.0, 1.6 Hz, 1H, Ar-**H**), 2.65 (t, *J* = 7.6 Hz, 2H, Ar-C**H$_2$**-CH$_2$), 1.69 (h, *J* = 7.3 Hz, 2H, CH$_2$-C**H$_2$**-CH$_3$), 0.98 (t, *J* = 7.3 Hz, 3H, CH$_2$-C**H$_3$**). † Overlapping CDCl$_3$. |
| $^{13}$C{$^1$H} NMR (101 MHz): | 160.96 (dd, *J* = 257.1, 6.8 Hz), 159.78, 158.34 (d, *J* = 250.4 Hz), 158.22, 146.94 (d, *J* = 7.8 Hz), 144.73, 142.30 (d, *J* = 12.1 Hz), 137.37, 132.70, 129.77 (d, *J* = 3.0 Hz), 128.49, 127.77, 125.02 (d, *J* = 3.1 Hz), 122.37, 118.86, 116.45 (d, *J* = 22.0 Hz), 112.65 (dt, *J* = 23.3, 3.3 Hz), 111.18, 37.54, 24.14, 13.72. |
| $^{19}$F NMR (376 MHz): | -109.08 (d, *J$_{F-H}$* = 10.3 Hz, 2F, Ar-**F**), -117.59 (t, *J$_{F-H}$* = 9.9 Hz, 1F, Ar-**F**). |

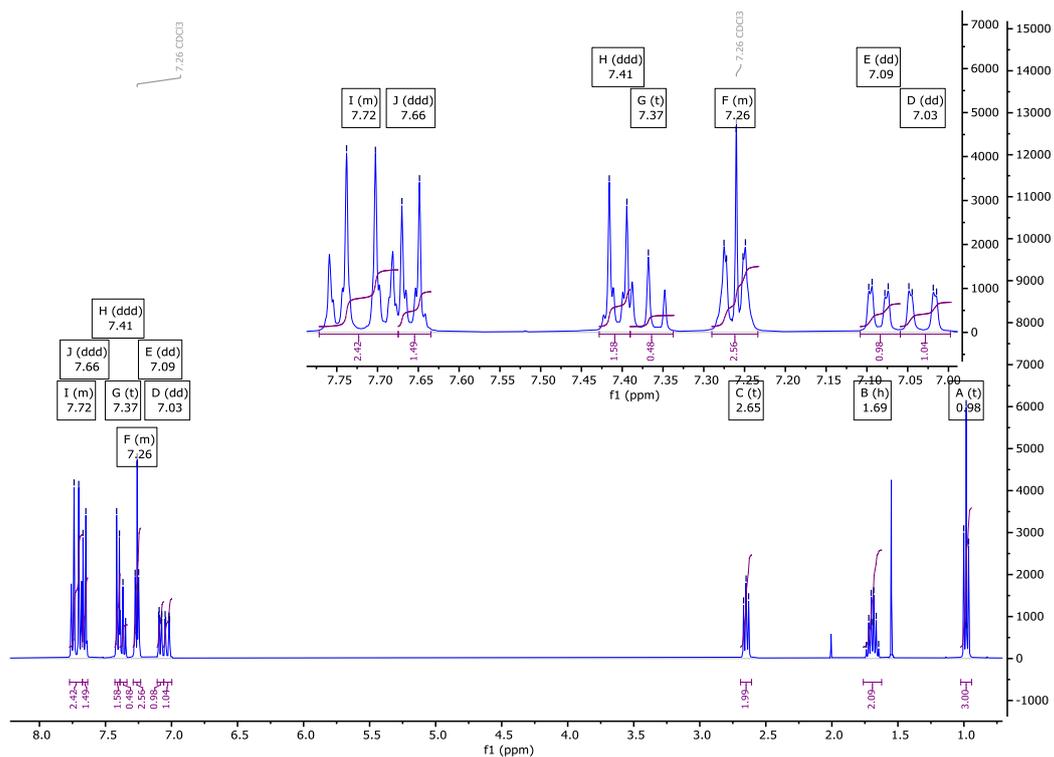

**Fig. S116** ¹H NMR spectra of **NC0021** in CDCl₃.

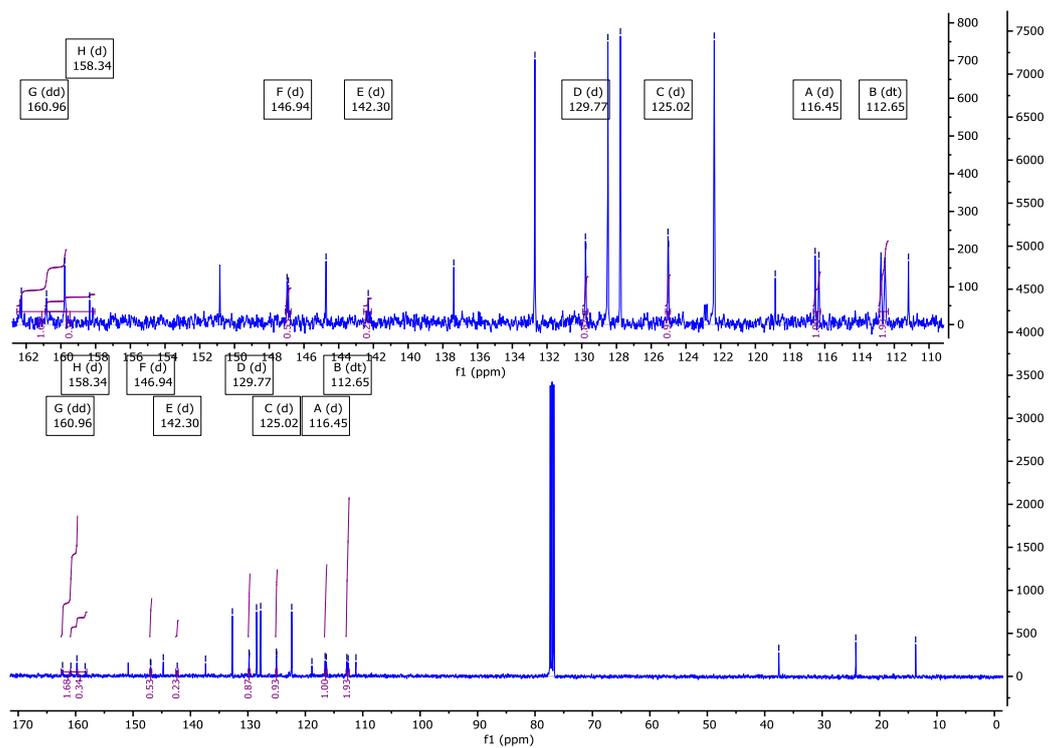

**Fig. S117** ¹³C{¹H} NMR spectra of **NC0021** in CDCl₃.

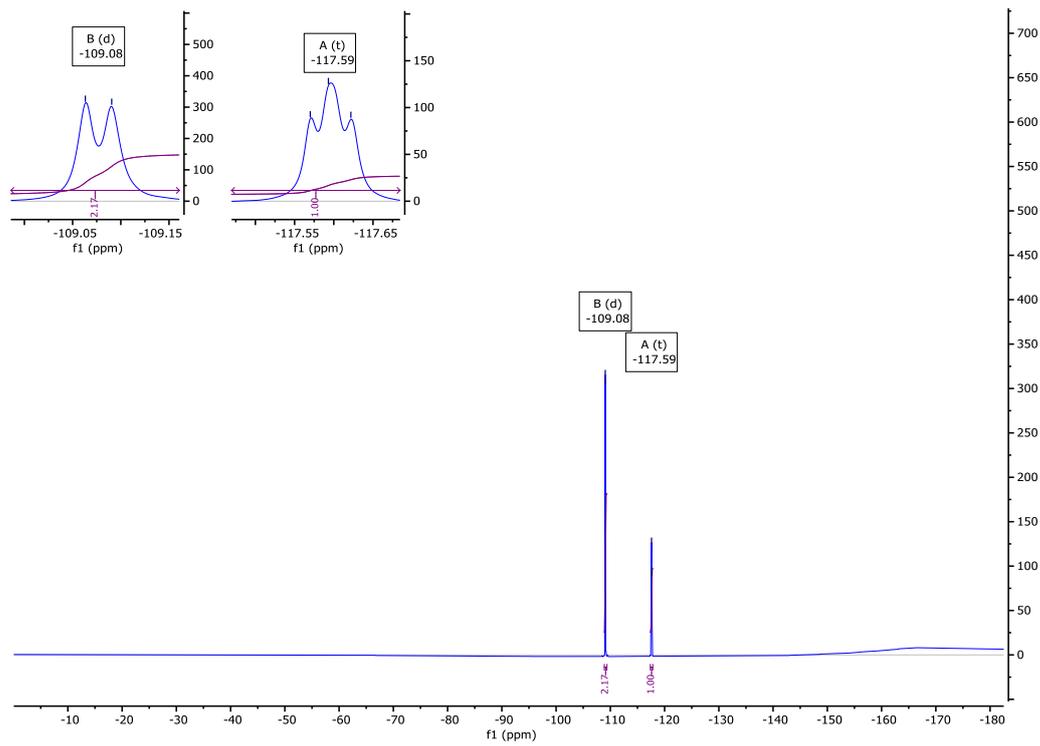

**Fig. S118**   $^{19}$F NMR spectra of **NC0021** in CDCl$_3$.

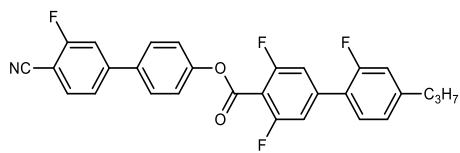

**NC1021**

*4'-cyano-3'-fluoro-[1,1'-biphenyl]-4-yl 2',3,5-trifluoro-4'-propyl-[1,1'-biphenyl]-4-carboxylate*

| | |
|---|---|
| Yield: | (white solid) 110 mg, 45 % |
| Re-crystallisation solvent: | EtOH |
| $^1$H NMR (400 MHz): | 7.70 (dd, *J* = 8.1, 6.7 Hz, 1H, Ar-**H**), 7.65 (ddd, *J* = 8.7, 2.7, 2.0 Hz, 2H, Ar-**H**), 7.49 (dd, *J* = 8.1, 1.6 Hz, 1H, Ar-**H**), 7.46 – 7.40 (m, 3H, Ar-**H**)*, 7.37 (t, *J* = 8.0 Hz, 1H, Ar-**H**), 7.30 – 7.23 (m, 2H, Ar-**H**)†, 7.09 (dd, *J* = 7.9, 1.7 Hz, 1H, Ar-**H**), 7.03 (dd, *J* = 12.0, 1.7 Hz, 1H, Ar-**H**), 2.65 (t, *J* = 7.5 Hz, 2H, Ar-C**H$_2$**-CH$_2$), 1.69 (h, *J* = 7.4 Hz, 2H. CH$_2$-C**H$_2$**-CH$_3$), 0.98 (t, *J* = 7.3 Hz, 3H, CH$_2$-C**H$_3$**). |
| $^{13}$C{$^1$H} NMR (101 MHz): | 164.77 (d, *J* = 258.9 Hz), 161.20 (dd, *J* = 258.0, 6.7 Hz), 159.01 (d, *J* = 249.9 Hz), 159.68, 151.29, 147.55 (d, *J* = 8.0 Hz), 146.97 (d, *J* = 8.0 Hz), 142.34 (t, *J* = 11.1 Hz), 136.17 (d, *J* = 1.5 Hz), 133.86, 129.77 (d, *J* = 3.0 Hz), 128.49, 125.03 (d, *J* = 3.1 Hz), 123.42 (d, *J* = 3.2 Hz), 122.56, 116.45 (d, *J* = 22.0 Hz), 114.87 (d, *J* = 20.3 Hz), 112.77, 112.66 (dt, *J* = 23.1, 3.4 Hz), 108.33 (t, *J* = 16.8 Hz), 100.09 (d, *J* = 15.6 Hz), 37.54, 24.14, 13.72. |
| $^{19}$F NMR (376 MHz): | -105.88 (t, *J$_{F-H}$* = 7.4 Hz, 1F, Ar-**F**), -109.00 (d, *J$_{F-H}$* = 10.3 Hz, 2F, Ar-**F**), -117.59 (d, *J$_{F-H}$* = 9.9 Hz, 1F, Ar-**F**). |

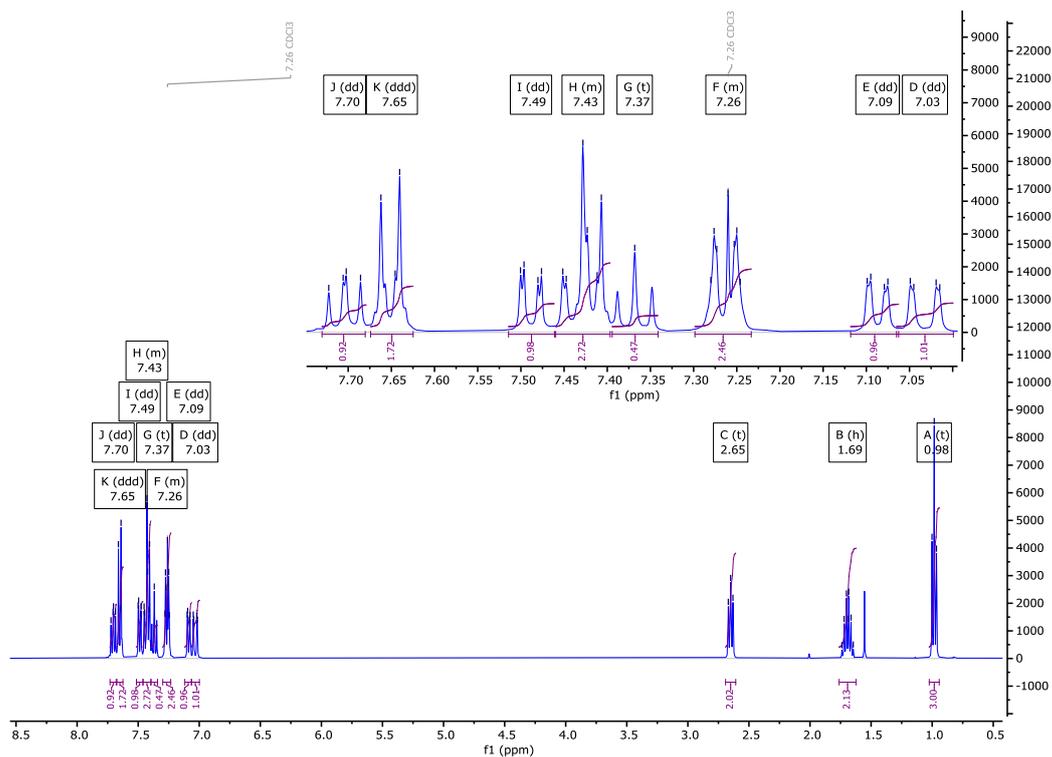

**Fig. S119** $^1$H NMR spectra of **NC1021** in CDCl$_3$.

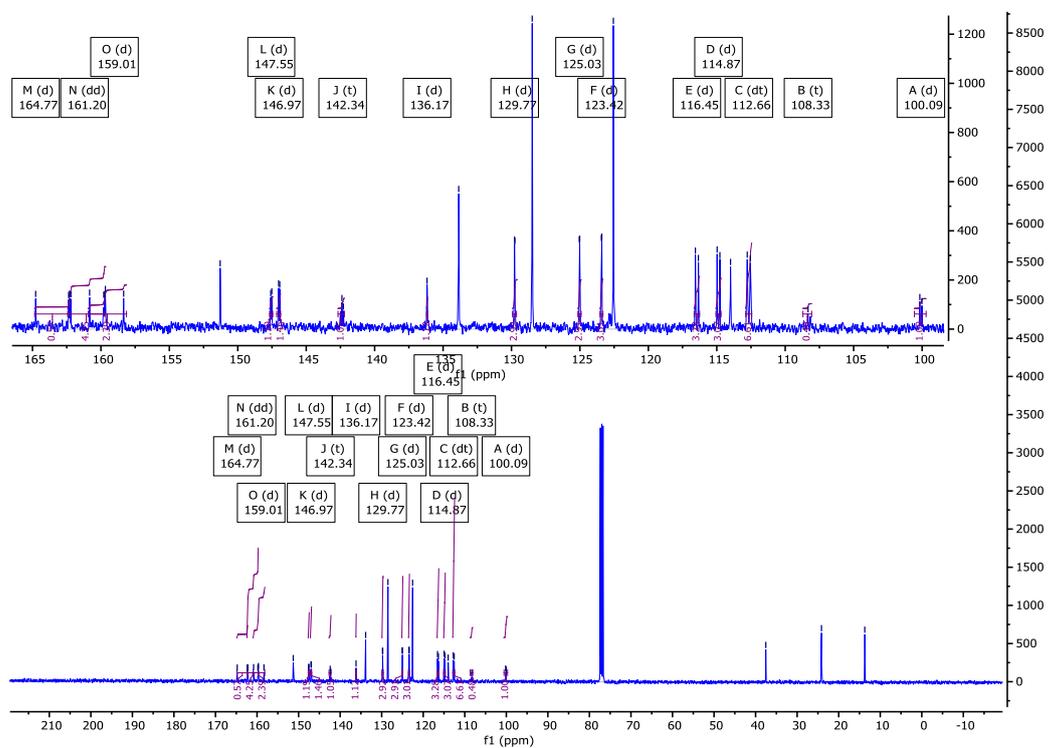

**Fig. S120** $^{13}$C{$^1$H} NMR spectra of **NC1021** in CDCl$_3$.

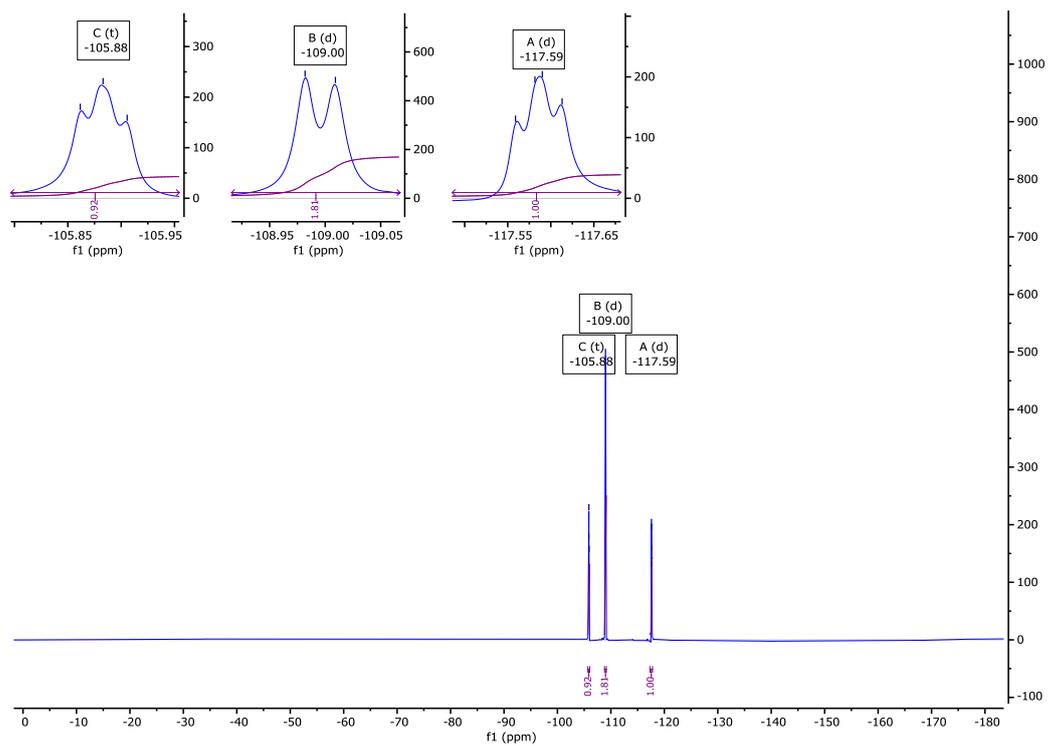

**Fig. S121**  $^{19}$F NMR spectra of **NC1021** in CDCl$_3$.

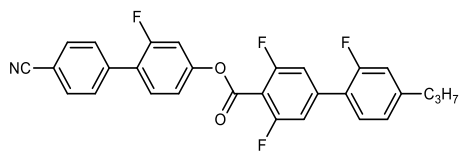

**NC0121**

*4'-cyano-2-fluoro-[1,1'-biphenyl]-4-yl 2',3,5-trifluoro-4'-propyl-[1,1'-biphenyl]-4-carboxylate*

| | |
|---|---|
| Yield: | (white solid) 100 mg, 41 % |
| Re-crystallisation solvent: | EtOH |
| $^1$H NMR (400 MHz): | 7.76 (ddd, *J* = 8.7, 1.7, 1.7 Hz, 2H, Ar-**H**), 7.70 – 7.63 (m, 2H, Ar-**H**), 7.50 (t, *J* = 8.6 Hz, 1H, Ar-**H**), 7.37 (t, *J* = 8.0 Hz, 1H, Ar-**H**), 7.30 – 7.18 (m, 4H, Ar-**H**)*†, 7.09 (dd, *J* = 8.0, 1.6 Hz, 1H, Ar-**H**), 7.03 (dd, *J* = 11.9, 1.6 Hz, 1H, Ar-**H**), 2.65 (t, *J* = 7.6 Hz, 2H, Ar-C**H$_2$**-CH$_2$), 1.69 (h, *J* = 7.4 Hz, 2H, CH$_2$-C**H$_2$**-CH$_3$), 0.98 (t, *J* = 7.3 Hz, 3H, CH$_2$-C**H$_3$**). *Overlapping Signals, †Overlapping CDCl$_3$. |
| $^{13}$C{$^1$H} NMR (126 MHz): | 162.23 (dd, *J* = 257.8, 6.3 Hz), 160.67, 160.28 (dd, *J* = 250.2, 4.4 Hz), 151.35 (d, *J* = 11.0 Hz), 147.16 (d, *J* = 7.9 Hz), 142.73 (t, *J* = 10.7 Hz), 139.78, 132.47, 131.07 (d, *J* = 4.1 Hz), 129.88 (d, *J* = 3.0 Hz), 129.79 (d, *J* = 3.3 Hz), 125.47 (d, *J* = 13.1 Hz), 125.17 (d, *J* = 3.1 Hz), 122.87 (d, *J* = 12.5 Hz), 118.34 (d, *J* = 3.7 Hz), 116.58 (d, *J* = 21.9 Hz), 112.81 (dt, *J* = 23.0, 3.7 Hz), 111.78, 110.79 (d, *J* = 26.1 Hz), 108.12 (t, *J* = 16.7 Hz), 37.66, 24.26, 13.84. |
| $^{19}$F NMR (376 MHz): | -108.80 (d, *J$_{F-H}$* = 10.4 Hz, 2F, Ar-**F**), -114.12 (t, *J$_{F-H}$* = 9.8 Hz, 1F, Ar-**F**), -117.56 (t, *J$_{F-H}$* = 9.4 Hz, 1F, Ar-**F**). |

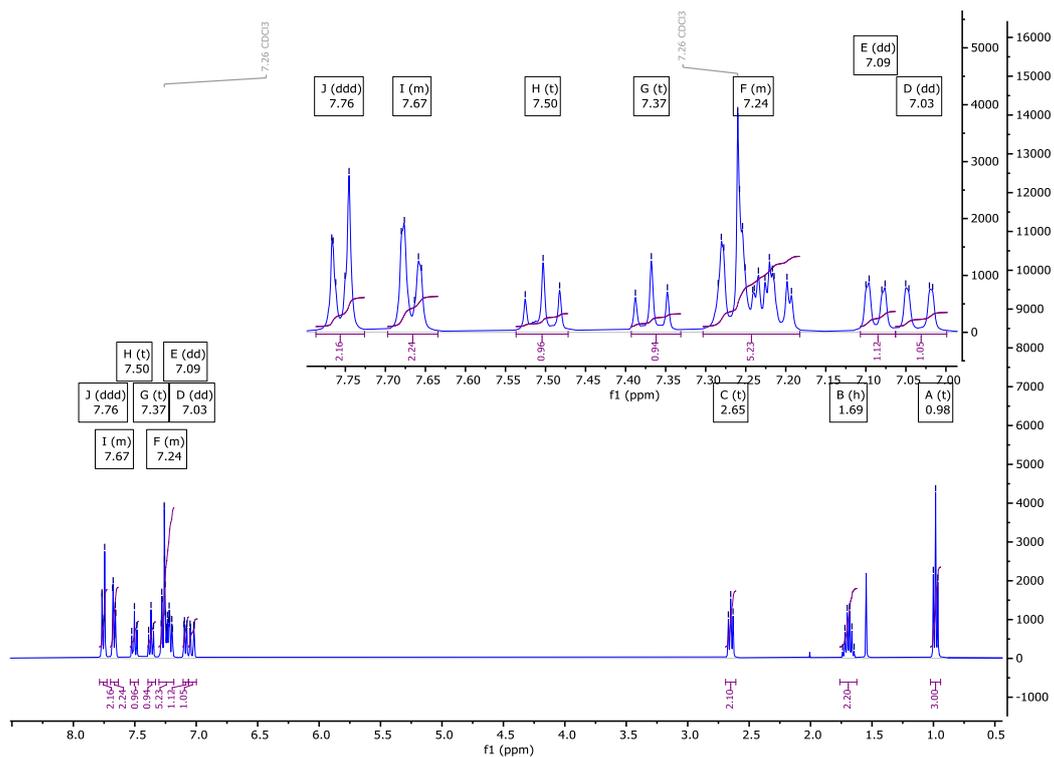

**Fig. S122** $^1$H NMR spectra of **NC0121** in CDCl$_3$.

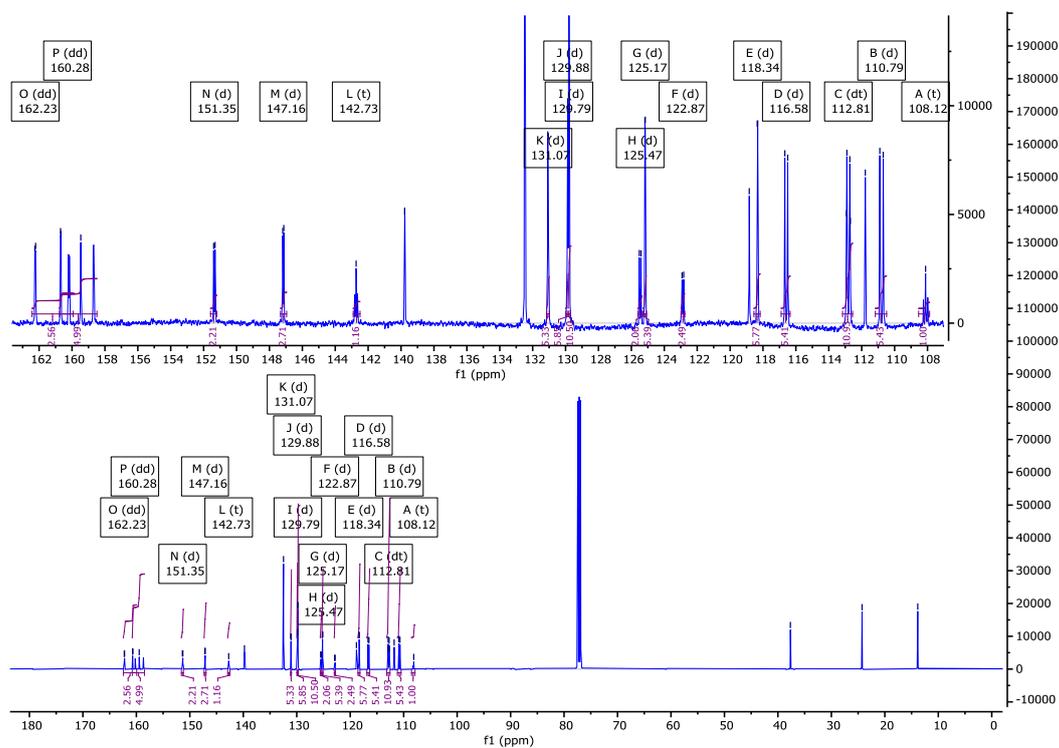

**Fig. S123** $^{13}$C{$^1$H} NMR spectra of **NC0121** in CDCl$_3$.

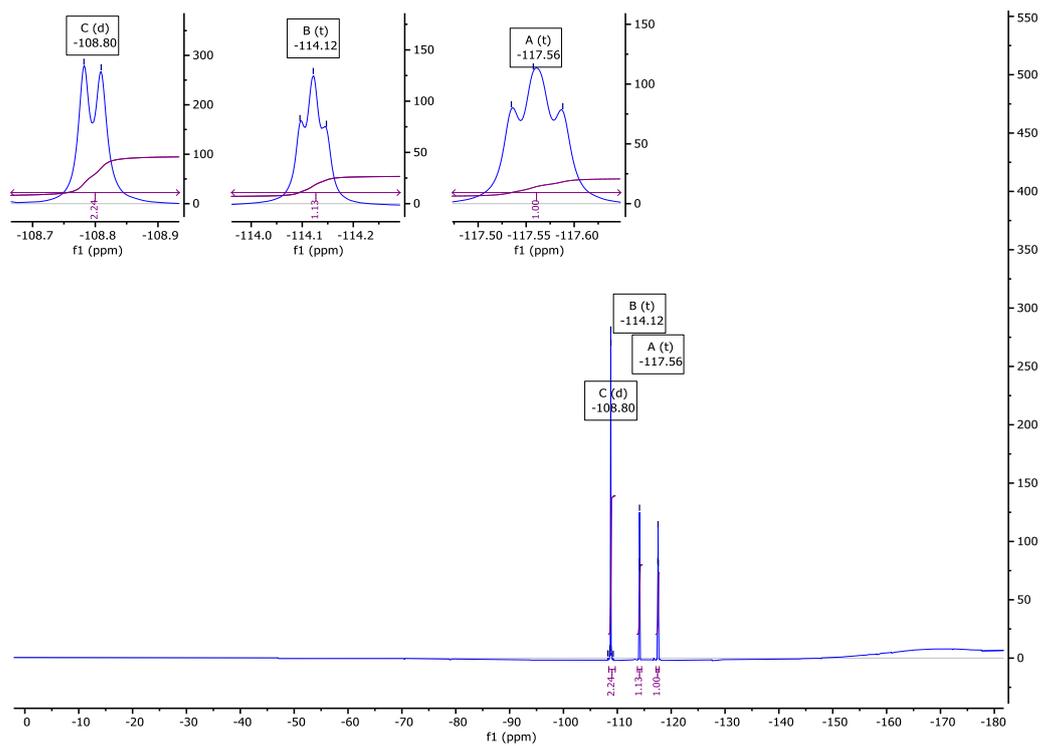

**Fig. S124** $^{19}$F NMR spectra of **NC0121** in CDCl$_3$.

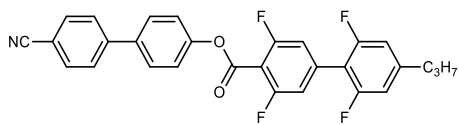

**NC0022**

*4'-cyano-[1,1'-biphenyl]-4-yl 2',3,5,6'-tetrafluoro-4'-propyl-[1,1'-biphenyl]-4-carboxylate*

| | |
|---|---|
| Yield: | (white crystalline solid) 125 mg, 51 % |
| Re-crystallisation solvent: | EtOH |
| $^1$H NMR (501 MHz): | 7.77 – 7.68 (m, 4H, Ar-**H**)*, 7.66 (ddd, *J* = 8.8, 2.7, 2.1 Hz, 2H, Ar-**H**), 7.40 (ddd, *J* = 8.6, 2.7, 2.0 Hz, 2H, Ar-**H**), 7.19 (d, *J* = 8.9 Hz, 2H, Ar-**H**), 6.87 (ddd, *J* = 9.0, 3.0, 2.3 Hz, 2H, Ar-**H**), 2.63 (t, *J* = 6.7 Hz, 2H, Ar-C**H$_2$**-CH$_2$), 1.69 (h, *J* = 7.4 Hz, 2H, CH$_2$-C**H$_2$**-CH$_3$), 0.99 (t, *J* = 7.4 Hz, 3H, CH$_2$-C**H$_3$**). *Overlapping Signals. |
| $^{13}$C{$^1$H} NMR (126 MHz): | 161.90 – 159.64 (m)*, 159.59 (dd, *J* = 250.5, 7.1 Hz), 150.93, 147.30 (t, *J* = 9.5 Hz), 144.83, 137.53, 135.99 (t, *J* = 11.3 Hz), 132.82, 128.63, 127.90, 122.47, 118.98, 114.49 (d, *J* = 23.1 Hz), 112.62 (t, *J* = 17.9 Hz), 112.11 (dd, *J* = 25.3, 5.1 Hz), 111.31, 109.38 (t, *J* = 17.1 Hz), 37.80, 23.98, 13.76. *Overlapping Signals. |
| $^{19}$F NMR (376 MHz): | -109.53 (d, *J$_{F-H}$* = 9.9 Hz, 2F, Ar-**F**), -114.95 (d, *J$_{F-H}$* = 9.6 Hz, 2F, Ar-**F**). |

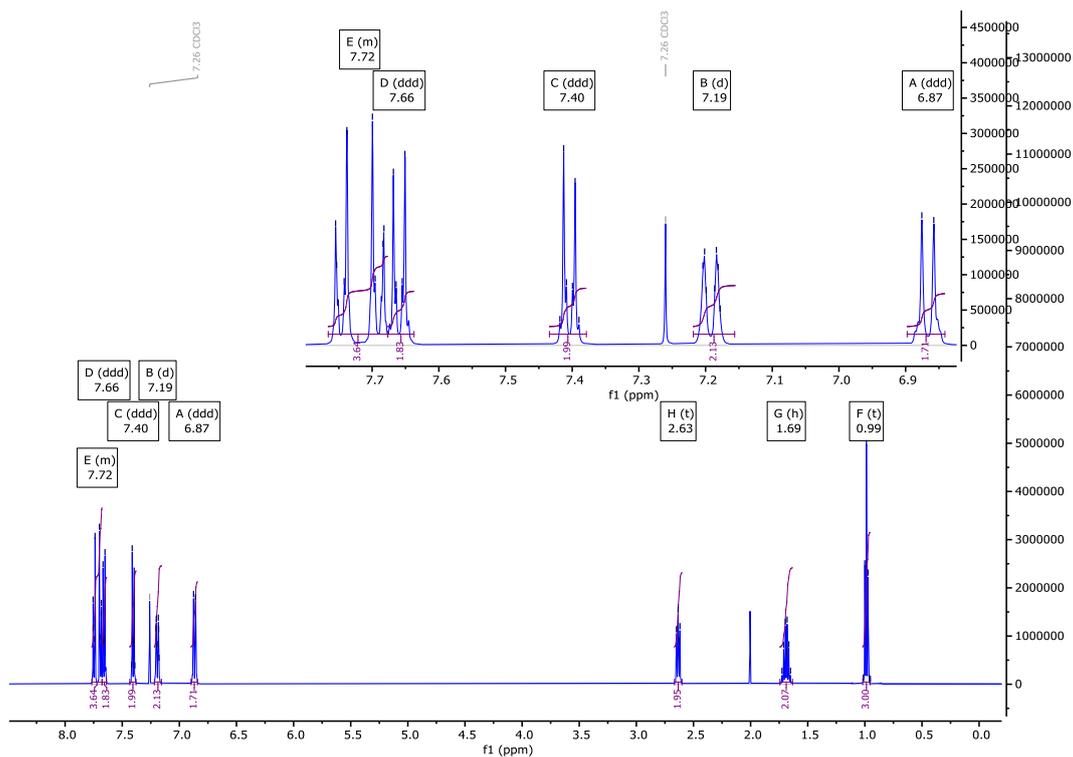

**Fig. S125** $^1$H NMR spectra of **NC0022** in CDCl$_3$.

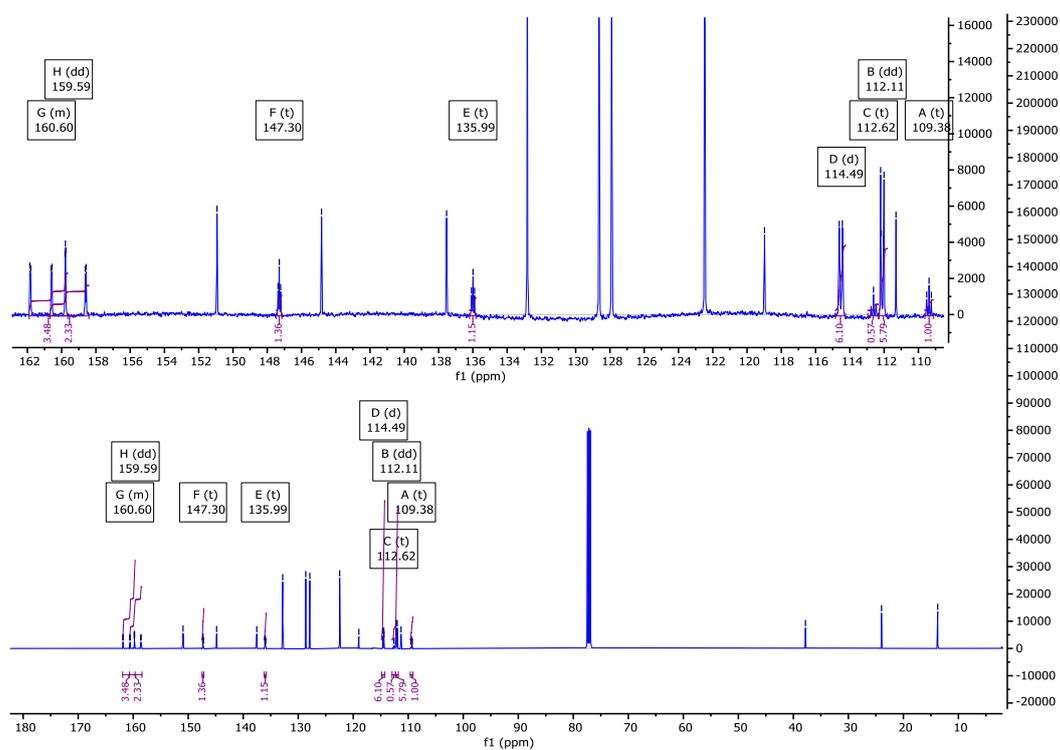

**Fig. S126** $^{13}$C{$^1$H} NMR spectra of **NC0022** in CDCl$_3$.

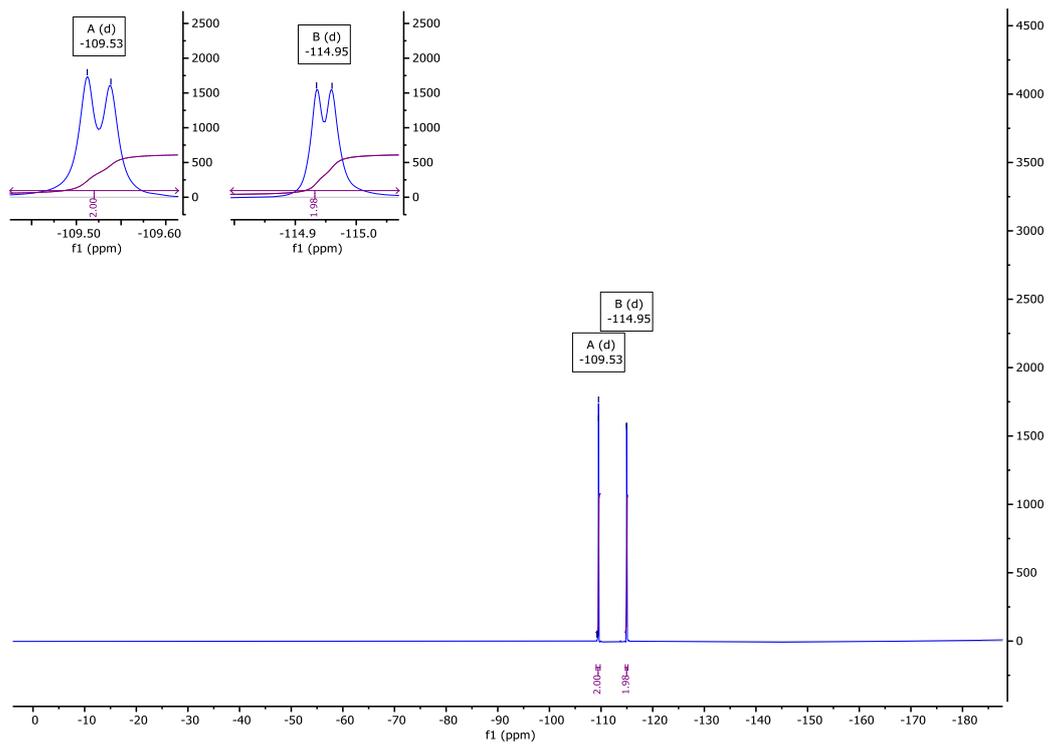

**Fig. S127** $^{19}$F NMR spectra of **NC0022** in CDCl$_3$.

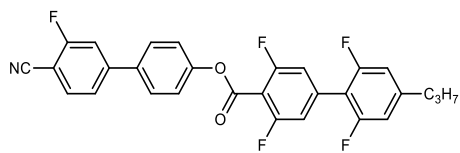

**NC1022**

*4'-cyano-3'-fluoro-[1,1'-biphenyl]-4-yl 2',3,5,6'-tetrafluoro-4'-propyl-[1,1'-biphenyl]-4-carboxylate*

| | |
|---|---|
| Yield: | (white crystalline solid) 96 mg, 38 % |
| Re-crystallisation solvent: | EtOH |
| $^1$H NMR (501 MHz): | 7.71 (dd, *J* = 8.1, 6.6 Hz, 1H, Ar-**H**), 7.65 (ddd, *J* = 8.8, 2.6, 2.1 Hz, 2H, Ar-**H**), 7.49 (dd, *J* = 8.1, 1.7 Hz, 1H, Ar-**H**), 7.46 – 7.40 (m, 3H)*, 7.19 (d, *J* = 9.1 Hz, 2H, Ar-**H**), 6.87 (ddd, *J* = 9.0, 2.9, 2.1 Hz, 2H Ar-**H**), 2.63 (t, *J* = 6.8 Hz, 2H, Ar-C**H$_2$**-CH$_2$), 1.69 (h, *J* = 7.5 Hz, 2H, CH$_2$-C**H$_2$**-CH$_3$), 0.99 (t, *J* = 7.3 Hz, 3H, CH$_2$-C**H$_3$**). *Overlapping Signals. |
| $^{13}$C{$^1$H} NMR (126 MHz): | 163.62 (d, *J* = 259.1 Hz), 160.75 (dd, *J* = 257.9, 6.2 Hz), 160.62, 159.70 (dd, *J* = 250.4, 7.3 Hz), 151.38, 147.68 (d, *J* = 7.9 Hz), 147.33 (t, *J* = 9.5 Hz), 136.34, 136.10 (t, *J* = 11.2 Hz), 134.00, 128.64, 123.57, 122.67, 115.02 (d, *J* = 20.3 Hz), 114.56 (dd, *J* = 23.6, 2.6 Hz), 114.13, 112.61 (t, *J* = 17.7 Hz), 112.12 (dd, *J* = 24.8, 4.5 Hz), 109.20 (t, *J* = 17.2 Hz), 100.24 (d, *J* = 15.5 Hz), 37.82, 23.99, 13.77. |
| $^{19}$F NMR (376 MHz): | -105.87 (dd, *J$_{F-H}$* = 10.2 Hz, *J$_{F-H}$* = 6.7 Hz, 1F, Ar-**F**), -109.46 (d, *J$_{F-H}$* = 10.3 Hz, 2F, Ar-**F**), -114.95 (d, *J$_{F-H}$* = 9.6 Hz, 2F, Ar-**F**). |

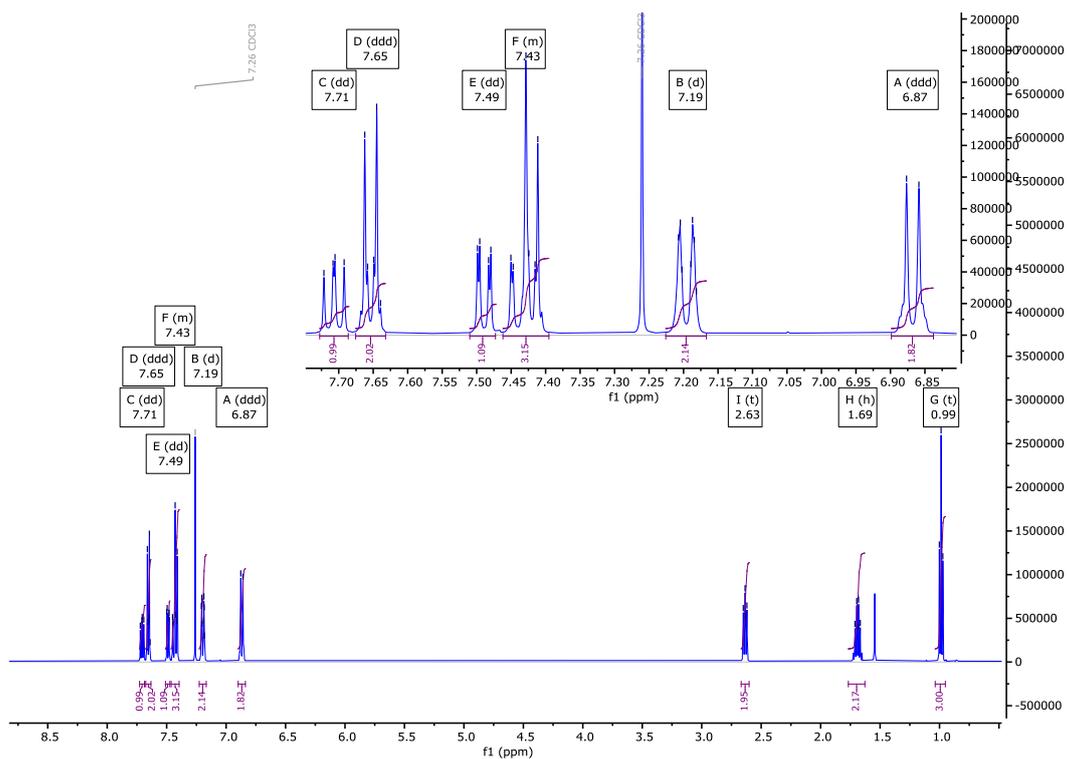

**Fig. S128** $^1$H NMR spectra of **NC1022** in CDCl$_3$.

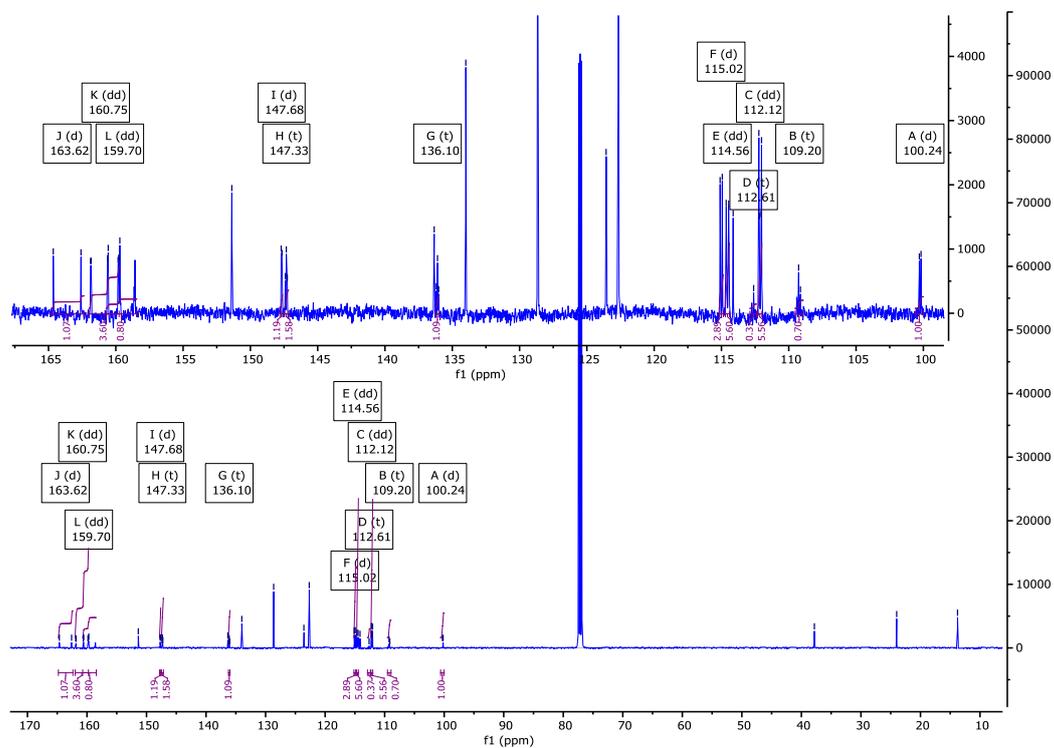

**Fig. S129** $^{13}$C{$^1$H} NMR spectra of **NC1022** in CDCl$_3$.

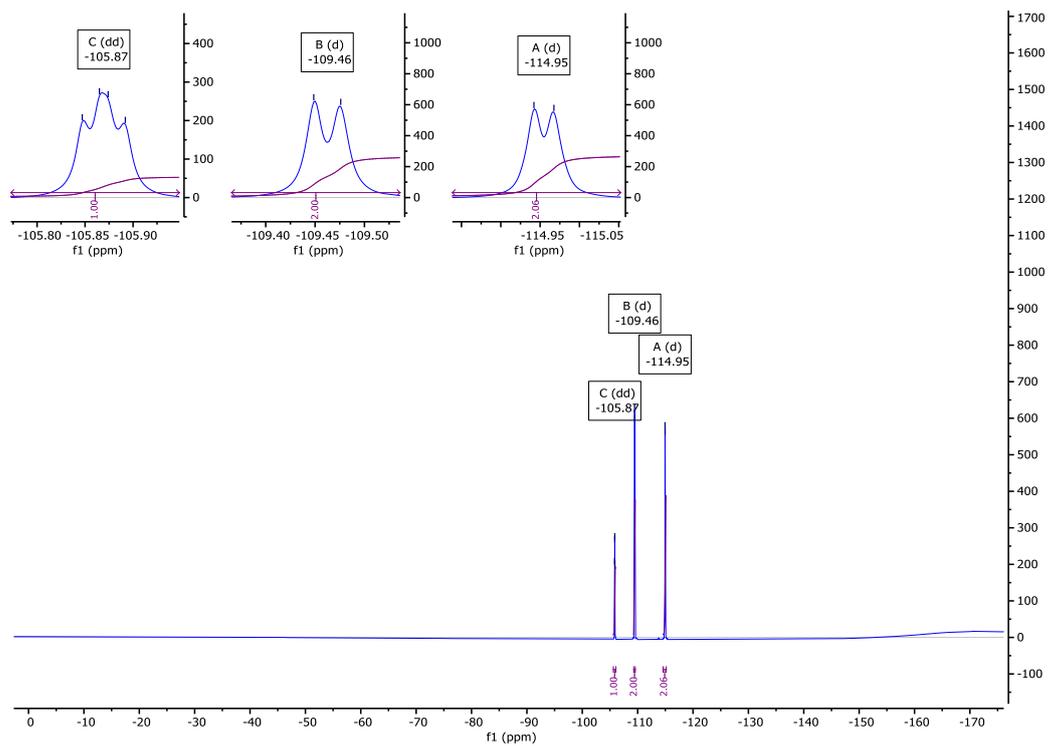

**Fig. S130** ¹⁹F NMR spectra of **NC1022** in CDCl$_3$.

## 4 Supplemental References

1   Frisch, M. J. *et al*. Gaussian 16 Rev. C.01 (Wallingford, CT, 2016).

2   Becke, A. D. Density-functional thermochemistry. III. The role of exact exchange. J. Chem. Phys. 98, 5648 (1993).

3   Lee, C., Weitao, Y. & Parr, R. G. Development of the Colic-Salvetti correlation-energy formula into a functional of the electron density. Phys. Rev. B 37, 785 (1988).

4   Dunning, T. H., Jr. Gaussian basis sets for use in correlated molecular calculations. I. The atoms boron through neon and hydrogen. J. Chem. Phys. 90, 1007-1023 (1989).

5   Wang, J., Wolf, R. M., Caldwell, J. W., Kollman, P. A. & Case, D. A. Development and testing of a general amber force field. J. Comp. Chem. 25, 1157-1174 (2004).

6   Boyd, N. J. & Wilson, M. R. Validating an optimized GAFF force field for liquid crystals: TNI predictions for bent-core mesogens and the first atomistic predictions of a dark conglomerate phase. Phys. Chem. Chem. Phys. 20, 1485-1496 (2018).

7   Boyd, N. J. & Wilson, M. R. Optimization of the GAFF force field to describe liquid crystal molecules: the path to a dramatic improvement in transition temperature predictions. Phys. Chem. Chem. Phys. 17, 24851-24865 (2015).

8   Wang, J., Cieplak, P. & Kollman, P. A. How well does a restrained electrostatic potential (RESP) model perform in calculating conformational energies of organic and biological molecules? J. Comp. Chem. 21, 1049-1074 (2000).

9   Case, D. A. et al. AmberTools. J Chem. Info. Mod. 63, 6183-6191 (2023).

10   Sousa da Silva, A. W. & Vranken, W. F. ACPYPE - AnteChamber PYthon Parser interfacE. BMC Res. Notes 5, 367 (2012).

11   Abraham, M. J. et al. GROMACS: High performance molecular simulations through multi-level parallelism from laptops to supercomputers. SoftwareX 1-2, 19-25 (2015).

12   Hess, B., Bekker, H., Berendsen, H. J. C. & Fraaije, J. G. E. M. LINCS: A linear constraint solver for molecular simulations. J. Comp. Chem. 18, 1463-1472 (1997).

13   McGibbon, Robert T. et al. MDTraj: A Modern Open Library for the Analysis of Molecular Dynamics Trajectories. Biophys. J. 109, 1528-1532 (2015).

14   Mandle, R. J., Sebastián, N., Martinez-Perdiguero, J. & Mertelj, A. On the molecular origins of the ferroelectric splay nematic phase. Nat. Commun. 12, 4962 (2021).

15   Mandle, R. J. Implementation of a cylindrical distribution function for the analysis of anisotropic molecular dynamics simulations. PLOS ONE 17, e0279679 (2022).

16      Hobbs, J., Gibb, C. J. & Mandle, R. J. Emergent Antiferroelectric Ordering and the Coupling of Liquid Crystalline and Polar Order. Small Sci. 4, 2400189 (2024).

17      Hobbs, J. et al. Polar Order in a Fluid Like Ferroelectric with a Tilted Lamellar Structure – Observation of a Polar Smectic C (SmC$_P$) Phase. Angew. Chem. In. Ed. 64, e202416545 (2025).

18      Gibb, C. J., Hobbs, J. & Mandle, R. J. Systematic Fluorination Is a Powerful Design Strategy toward Fluid Molecular Ferroelectrics. J. Am Chem. Soc. 147, 4571-4577 (2025).

19      Gibb, C. J. et al. Spontaneous symmetry breaking in polar fluids. Nat. Commun. 15, 5845 (2024).